\documentclass[aps,twocolumn,showpacs,amsmath,amssymb,superscriptaddress]{revtex4-1}

\usepackage[margin=0.7in]{geometry}
\usepackage{float}
\usepackage{amsmath}
\usepackage{subfig}
\usepackage{amsfonts}
\usepackage{amssymb}
\usepackage{chngpage}
\usepackage{graphicx}
\usepackage{dcolumn}
\usepackage{bm}
\usepackage{lipsum}
\usepackage{mathptmx}
\usepackage[usenames, dvipsnames]{color}
\usepackage[pdftex]{hyperref}

\begin{document}
\widetext

\title{Comparison of $\mathbf{\nu_\mu-}$Ar multiplicity distributions observed by MicroBooNE to GENIE model predictions}





\affiliation{Universit{\"a}t Bern, Bern CH-3012, Switzerland}
\affiliation{Brookhaven National Laboratory (BNL), Upton, NY, 11973, USA}
\affiliation{University of Cambridge, Cambridge CB3 0HE, United Kingdom}
\affiliation{University of Chicago, Chicago, IL, 60637, USA}
\affiliation{University of Cincinnati, Cincinnati, OH, 45221, USA}
\affiliation{Colorado State University, Fort Collins, CO, 80523, USA}
\affiliation{Columbia University, New York, NY, 10027, USA}
\affiliation{Fermi National Accelerator Laboratory (FNAL), Batavia, IL 60510, USA}
\affiliation{Harvard University, Cambridge, MA 02138, USA}
\affiliation{Illinois Institute of Technology (IIT), Chicago, IL 60616, USA}
\affiliation{Kansas State University (KSU), Manhattan, KS, 66506, USA}
\affiliation{Lancaster University, Lancaster LA1 4YW, United Kingdom}
\affiliation{Los Alamos National Laboratory (LANL), Los Alamos, NM, 87545, USA}
\affiliation{The University of Manchester, Manchester M13 9PL, United Kingdom}
\affiliation{Massachusetts Institute of Technology (MIT), Cambridge, MA, 02139, USA}
\affiliation{University of Michigan, Ann Arbor, MI, 48109, USA}
\affiliation{New Mexico State University (NMSU), Las Cruces, NM, 88003, USA}
\affiliation{Otterbein University, Westerville, OH, 43081, USA}
\affiliation{University of Oxford, Oxford OX1 3RH, United Kingdom}
\affiliation{Pacific Northwest National Laboratory (PNNL), Richland, WA, 99352, USA}
\affiliation{University of Pittsburgh, Pittsburgh, PA, 15260, USA}
\affiliation{Saint Mary's University of Minnesota, Winona, MN, 55987, USA}
\affiliation{SLAC National Accelerator Laboratory, Menlo Park, CA, 94025, USA}
\affiliation{Syracuse University, Syracuse, NY, 13244, USA}
\affiliation{Tel Aviv University, Tel Aviv, Israel, 69978}
\affiliation{University of Tennessee, Knoxville, TN, 37996, USA}
\affiliation{University of Texas, Arlington, TX, 76019, USA}
\affiliation{TUBITAK Space Technologies Research Institute, METU Campus, TR-06800, Ankara, Turkey}
\affiliation{Tufts University, Medford, MA, 02155, USA}
\affiliation{Center for Neutrino Physics, Virginia Tech, Blacksburg, VA, 24061, USA}
\affiliation{Yale University, New Haven, CT, 06520, USA}


\author{C.~Adams}
\affiliation{Harvard University, Cambridge, MA 02138, USA}


\author{R.~An}
\affiliation{Illinois Institute of Technology (IIT), Chicago, IL 60616, USA}

\author{J.~Anthony}
\affiliation{University of Cambridge, Cambridge CB3 0HE, United Kingdom}

\author{J.~Asaadi}
\affiliation{University of Texas, Arlington, TX, 76019, USA}

\author{M.~Auger}
\affiliation{Universit{\"a}t Bern, Bern CH-3012, Switzerland}

\author{S.~Balasubramanian}
\affiliation{Yale University, New Haven, CT, 06520, USA}

\author{B.~Baller}
\affiliation{Fermi National Accelerator Laboratory (FNAL), Batavia, IL 60510, USA}

\author{C.~Barnes}
\affiliation{University of Michigan, Ann Arbor, MI, 48109, USA}

\author{G.~Barr}
\affiliation{University of Oxford, Oxford OX1 3RH, United Kingdom}

\author{M.~Bass}
\affiliation{University of Oxford, Oxford OX1 3RH, United Kingdom}
\affiliation{Brookhaven National Laboratory (BNL), Upton, NY, 11973, USA}

% ...


 \collaboration{\textbf{The MicroBooNE Collaboration}}

\date{\today}

\begin{abstract}
We measure a large set of observables in inclusive charged current muon neutrino scattering on argon with the MicroBooNE liquid argon time projection chamber operating at Fermilab. We evaluate three neutrino interaction models based on the widely used GENIE event generator using these observables. The measurement uses a data set consisting of neutrino interactions with a final state muon candidate fully contained within the MicroBooNE detector. \ These data were collected in 2016 with the Fermilab Booster Neutrino Beam, which has an average
neutrino energy of $800$ MeV, using an exposure corresponding to $%
5.0\times10^{19}$ protons-on-target. \ The analysis employs fully automatic
event selection and charged particle track reconstruction and uses a
data-driven technique to separate neutrino interactions from cosmic ray
background events. We find that GENIE models consistently describe the shapes of a large number of kinematic distributions for fixed observed multiplicity.
\end{abstract}

\maketitle  

\section{\label{Introduction}\textbf{Introduction}}

A growing number of neutrino physics experiments use liquid argon as a neutrino interaction target nucleus in a time projection chamber~\textcolor{Blue}{\cite{Rubbia LArTPC}}. Experiments that use or will use liquid argon time projection chamber (LArTPC) technology include those in the Short-Baseline Neutrino (SBN) program~\textcolor{Blue}{\cite{SBN}} at Fermilab, centered on searches for non-standard neutrino oscillations, and the long-baseline DUNE experiment~\textcolor{Blue}{\cite{DUNE}}. The SBN program consists of the MicroBooNE experiment~\textcolor{Blue}{\cite{MicroBooNE detector}}, an upgraded ICARUS experiment~\textcolor{Blue}{\cite{ICARUS detector}}, and the new SBND experiment~\textcolor{Blue}{\cite{SBND}}. The DUNE experiment seeks to establish the mass ordering of the three standard model neutrinos and the charge parity violation parameter phase $\delta_{CP}$
in the PMNS neutrino mixing matrix~\textcolor{Blue}{\cite{PDG}}.

All LArTPC neutrino oscillation-related measurements require a precise
understanding of neutrino scattering physics and the measured response of
the LArTPC detector to final state particles. These depend on: (a)
the neutrino flux seen by the experiment, (b) the neutrino scattering cross
sections, (c) the interaction physics of scattering final state particles with
argon, (d) transport and instrumentation effects of charge and light in the LArTPCs, and (e) software reconstruction algorithms. In practice item (a) is
determined by a combination of hadron production cross section measurements
and precise descriptions of neutrino beamline components. Item (b) is most commonly provided by the GENIE~\textcolor{Blue}{\cite{GENIE reference}} neutrino event generation model for neutrino-argon scattering. Items (c)-(e) are incorporated into a detailed suite of GEANT4-based~\textcolor{Blue}{\cite{Geant4
reference}} simulation and event reconstruction products called LArSoft~\textcolor{Blue}{\cite%
{LArSoft reference}}.

GENIE has been built up from models of the most important physical
scattering neutrino-nucleon mechanisms for the SBN and DUNE energy regimes ($0.5-5$ GeV) : quasi-elastic (QE) scattering $\nu_{\ell}N\rightarrow\ell^{-}N^{%
\prime},\nu_{\ell}N^{\prime}$, resonance production (RES) $\nu_{\ell}N\rightarrow%
\ell^{-}R,\nu_{\ell}R^{\prime}$, and non-resonant multi-hadron production
referred to as deep inelastic scattering (DIS): $\nu_{\ell }N\rightarrow\ell^{-}X,\nu_{\ell}X^{\prime}$~\textcolor{Blue}{\cite{Sam XC Review}}.
\ The underlying neutrino-nucleon scattering processes receive significant
modification from the nuclear environment, including the effects of Fermi motion
of target nucleons, many-nucleon effects, and final state interactions
(FSI)~\textcolor{Blue}{\cite{Nuclear_effects}}. While GENIE has received a fair amount of \textquotedblleft
tuning\textquotedblright\ (the process of finding a set of GENIE parameters chosen to optimize agreement with a particular data set) from previous electron and neutrino scattering
measurements, considerable uncertainties remain in the modeling of both the
underlying neutrino-nucleon scattering and the nuclear environment~%
\textcolor{Blue}{\cite{NuSTEC}}. 

Relatively few neutrino scattering measurements on argon exist~\textcolor{Blue}{\cite%
{Arneodo:2006ug,Anderson:2011ce,Acciarri:2014isz,Acciarri:2014gev,Acciarri:2014eit,Acciarri:2016sli}}, especially for the recoil hadronic system. Most of these report low-statistics exclusive final states. Nearly all existing neutrino scattering constraints on GENIE models derive measurements on scattering from carbon, which has $30\%$ of argon's atomic mass
number and a $22\%$ lower neutron-to-proton ratio. We take a step in
improving the empirical understanding of neutrino scattering from argon here
by performing a large set of comparisons of observed inclusive properties of
charged current scattering, measured at the MicroBooNE experiment in the
Fermilab Booster Neutrino Beam (BNB)~\textcolor{Blue}{\cite{BNB reference}} (MicroBooNE and MiniBooNE share the same beam), to predictions
from several variants of GENIE. These comparisons are generated by applying fully automated event reconstruction and signal selection tools to a subset of MicroBooNE's first collected data. While this analysis must focus in large part on reducing cosmic ray backgrounds, sensitivity to GENIE model parameters remains. \\

In Sec.~\textcolor{Blue}{\ref{Formalism}}, we introduce the formalism that relates
interaction models to our measurements. \ In Sec.~\textcolor{Blue}{\ref{Apparatus}}, we
describe the MicroBooNE\ detector and Booster Neutrino Beam. \ Section~\textcolor{Blue}{\ref%
{CRbkg}} describes the cosmic ray backgrounds in MicroBooNE. \ Section~\textcolor{Blue}{\ref%
{Selection}} summarizes the event selection procedure and the main software
tools used in the analysis. \ Section~\textcolor{Blue}{\ref{Signal}} presents the procedure
for discriminating BNB\ neutrino interactions from cosmic ray background
events. \ Section~\textcolor{Blue}{\ref{Error Analysis}} shows the results of a systematic
uncertainty analysis. \ Section~\textcolor{Blue}{\ref{Results}} presents the comparison of
observed charged particle multiplicity distributions and charged track
kinematic distributions for each multiplicity, to predictions from GENIE. \
Section~\textcolor{Blue}{\ref{Discussion}} provides a discussion of the result, and Sec.~\textcolor{Blue}{\ref{Conclusion}} gives an overall conclusion.

\section{\label{Formalism}Introduction to Observed Charged Particle Multiplicity and
Kinematic Distributions}

Neutrino interactions in the MicroBooNE detector produce charged particles
that can be reconstructed as tracks in the liquid argon medium of the
MicroBooNE LArTPC. These interactions can be characterized by a number of
inclusive properties. The charged particle multiplicity, or number of
primary charged particles, $n$, is a simple observable characterizing
final states in high-energy-collision processes, including neutrino
interactions. We note that in MicroBooNE the observable charged particle
multiplicity corresponds to that of charged particles exiting the target
nucleus participating in the neutrino interaction. 

The charged particle multiplicity distributions (CPMD) comprise the set of probabilities, $P_{n}$%
, associated with producing $n$ charged particles in an event, either in
full phase space or in restricted phase space domains. In addition to the
observed CPMD, kinematic properties of all charged particle tracks for each
multiplicity can be examined. Determination of inclusive event properties
such as the CPMD and of individual track kinematic properties at Fermilab
BNB neutrino energies naturally fits into the modern strategy~\textcolor{Blue}{\cite{Sam XC
Review}} of presenting neutrino interaction measurements in the form of
directly observable quantities.

Inclusive measurements expand the empirical knowledge of neutrino-argon
scattering that will be required by the DUNE experiment and the Fermilab SBN
program. As physical observables, the CPMD and other distributions can also
be used to test models, or particular tunes of models such as GENIE. These
models are typically constructed from a set of exclusive cross section
channels, and tests of inclusive distributions can provide independent
checks.

We describe here an evaluation of several variants of GENIE against \emph{%
observed} charged particle distributions, including the observed CPMD in
MicroBooNE data collected in 2016 in the Fermilab
BNB. For the observed CPMD, we mean the conditional probability, after application of
certain detector selection requirements, of observing a neutrino interaction
with $n$ charged tracks relative to the probability of observing a neutrino
interaction with at least one charged track:%
\begin{equation}
O_{n}=\frac{N_{\text{obs,}n}}{\sum\limits_{m=1}^{\infty}N_{%
\text{obs,}m}},
\end{equation}
where $N_{\text{obs,}n}$ is the number of neutrino interaction events with $n
$ observed tracks.

Our analysis requires at least one of the charged tracks to be consistent
with a muon; hence the $O_{n}$ are effectively observed CPMD for $\nu _{\mu }
$ charged current ($\nu _{\mu }$ CC) interactions. The $\nu _{\mu }$ NC, $\nu _{e}$, $\bar{\nu}_{e}$, and $\bar{\nu}_{\mu }$ backgrounds, in total,
are expected to be less than $10\%$ of the final sample. The muon candidate
is included in the charged particle multiplicity, and all events thus have $%
n\geq 1$. For each multiplicity, we have available the kinematic properties
of charged tracks. These can in principle be related to the 4-vector
components of each track; however, we choose distributions of directly observable quantities in the detector: visible track length and track angles.

Values for $O_{n}$ depend on cross sections for producing a multiplicity $n$%
, $\sigma_{CC,n}$, as well as the BNB neutrino flux and detector acceptance
and efficiency:%
\begin{equation}
\begin{split}
N_{\text{obs,}n} & = \sum_{\nu}\sum_{n^{\prime}}
\int dE_{\nu}\Phi_{\nu}\left( E_{\nu}\right) \\
& .\int d\Pi_{n^{\prime}}\frac{d\sigma_{CC,n^{\prime}}\left(
E_{\nu},\Pi_{n^{\prime}}\right) }{d\Pi_{n^{\prime}}}\epsilon_{n,n^{\prime}}%
\left( E_{\nu},\Pi_{n^{\prime}}\right) ,  \label{N-observed}
\end{split}%
\end{equation}
where $E_{\nu}$ is the neutrino energy, $\Phi_{\nu}\left( E_{\nu}\right) $
is the neutrino flux summed over $\nu_{\mu}$, $\bar{\nu}_{\mu}$, $\nu_{e}$, and $\bar{\nu}_{e}$ species, $%
d\Pi_{n^{\prime}}$ represents the $n^{\prime}$-particle final state phase space, $\epsilon
_{n,n^{\prime}}\left( E_{\nu},\Pi_{n^{\prime}}\right) $ is an acceptance and
efficiency matrix that gives the probability that an $n^{\prime}$ charged
particle final state produced in phase space element $d\Pi_{n^{\prime}}$ is
observed as an $n$-particle final state in the detector, and $%
d\sigma_{CC,n^{\prime}}\left( E_{\nu},\Pi_{n^{\prime}}\right) /d\Pi_{n^{\prime}}$ are the
differential cross sections for producing a multiplicity $n^{\prime}$. One
can likewise express the distribution of any observed kinematic distribution 
$X_{n}$ corresponding to an observed multiplicity $n$ as%
\begin{equation}
\begin{split}
dN_{\text{obs,}n} & =\sum_{\nu}\sum_{n^{\prime}}\int dE_{\nu}\Phi_{\nu}\left( E_{\nu}\right) \\
& .\int d\Pi_{n^{\prime}}\frac{d\sigma_{CC,n^{\prime}}\left(
E_{\nu},\Pi_{n^{\prime}}\right) }{d\Pi_{n^{\prime}}}\hat{\epsilon}%
_{n,n^{\prime}}\left( E_{\nu},\Pi_{n^{\prime}}\rightarrow X_{n}\right) ,
\label{Xn-observed}
\end{split}%
\end{equation}
where $\hat{\epsilon}_{n,n^{\prime}}\left(
E_{\nu},\Pi_{n^{\prime}}\rightarrow X_{n}\right) $ is the probability that
an $n^{\prime}$ charged particle final state produced in phase space element 
$d\Pi_{n^{\prime}}$ produces the observed value $X_{n}$ of the kinematic
variable in the detector. In practice we obtain the $O_{n}$ and
distributions of $X_{n}$ directly from data and compare these to values
derived from evaluating Eqs.~\textcolor{Blue}{\ref{N-observed}} and \textcolor{Blue}{\ref{Xn-observed}} using a
Monte Carlo simulation that includes GENIE neutrino interaction event
generators coupled to detailed GEANT-based models of the Fermilab BNB and the detector.

The observed CPMD and inclusive observed kinematic distributions have
several desirable attributes. The $\sigma _{CC,n}$ are all large up to $n\lesssim 4$ at these neutrino energies (see Sec.~\ref{Apparatus}); therefore only modest event statistics are required.
Only minimal kinematic properties of the final state are imposed (the track
definition implies an effective minimum kinetic energy), and complexities
associated with particle identification and photon reconstruction are avoided. At the same time, the observed quantities reveal much of the
power of the LArTPC in identifying and characterizing complex
neutrino interactions. The observed CPMD and associated kinematic distribution ratios will have reduced sensitivity to systematic normalization uncertainties
associated with flux and efficiency compared to absolute cross section
measurements. \ 

A disadvantage of the use of observed CPMD and other kinematic quantities is
their lack of portability. One must have access to the full MicroBooNE
simulation suite to use the $O_{n}$ to test other models. 

\section{\label{Apparatus}The MicroBooNE\ Detector and the Booster Neutrino Beam}

\begin{figure}[!hptb]
\centering
\includegraphics[width=0.5\textwidth]{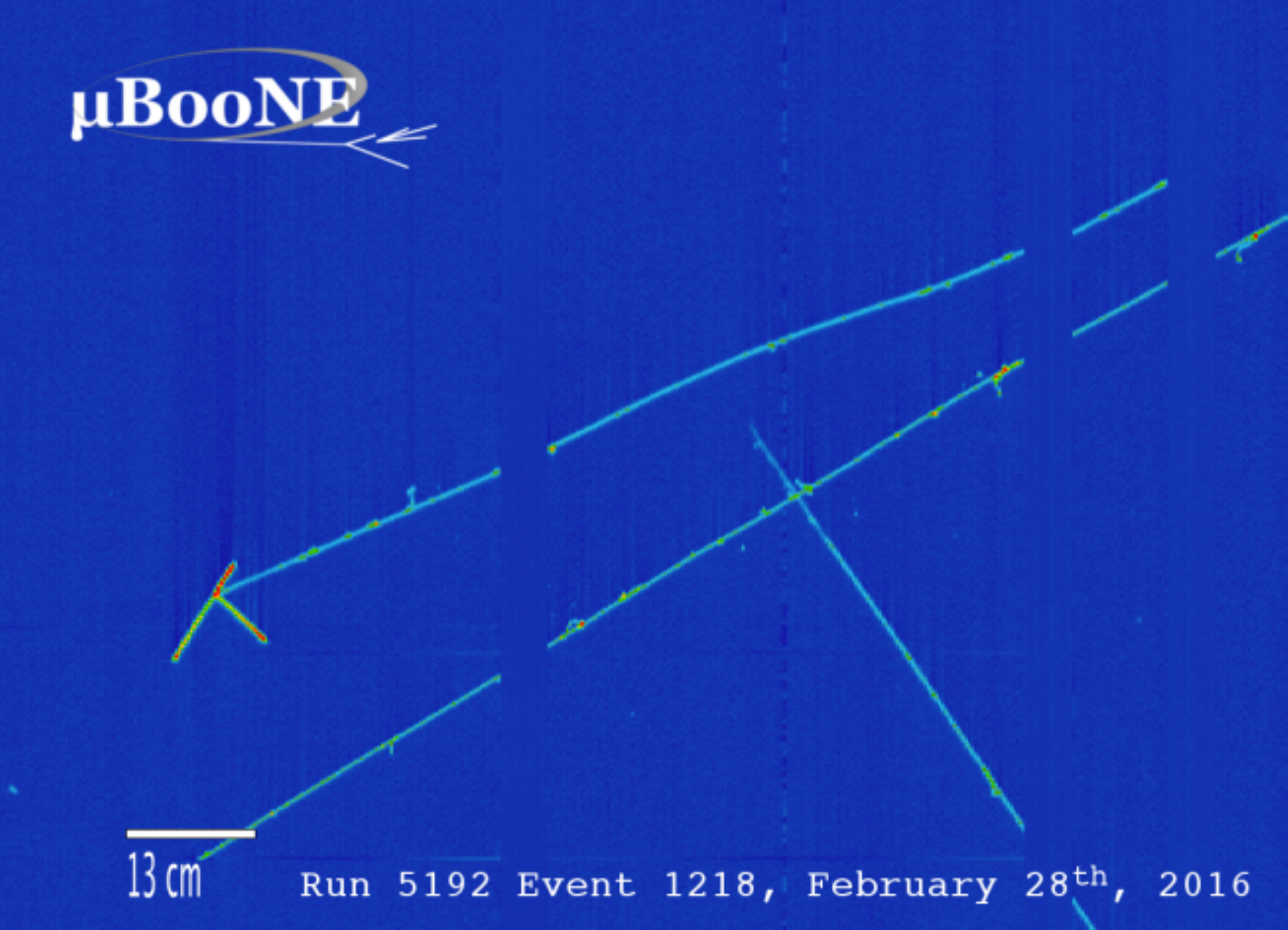} 
\caption{Event display showing raw data for a region of the collection plane associated with a candidate high-multiplicity neutrino event. Wire-number is
represented on the horizontal axis, and time on the vertical. Color is
associated with the charge deposition on each wire. Two perpendicularly
crossing tracks are cosmic tracks which is the dominant background. The gaps in tracks are due to non-responsive wires in the detector~\textcolor{Blue}{\cite{Noise Filtering}}.}
\label{img:Event display}
\end{figure}

\begin{figure}[!hptb]
\centering
\includegraphics[width=0.5\textwidth]{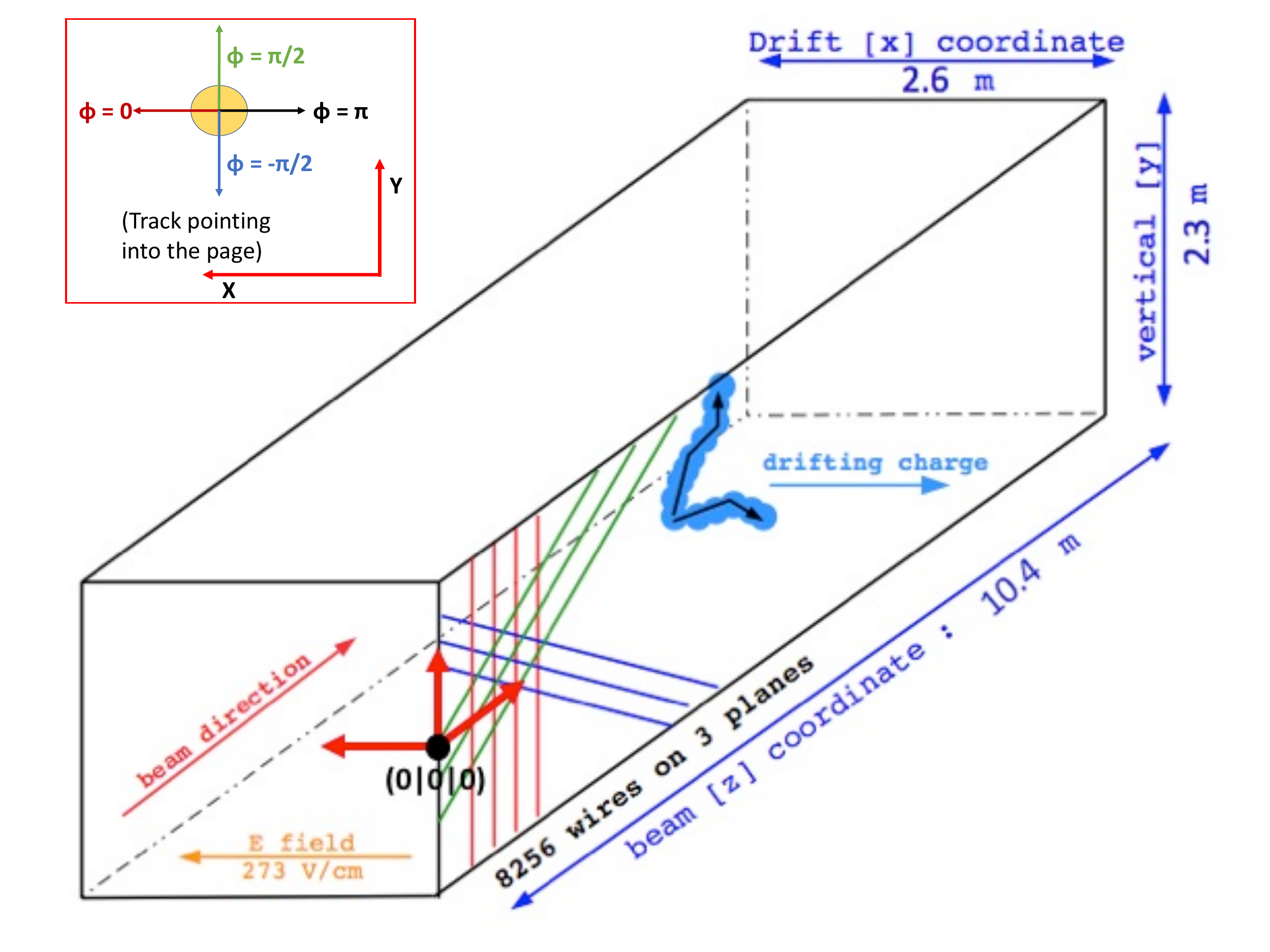} 
\caption{A schematic of the MicroBooNE TPC showing the coordinate system and
wire planes.}
\label{img:MicroBooNE picture}
\end{figure}

The MicroBooNE detector is a LArTPC installed on the Fermilab BNB. It is a high-resolution surface detector designed to accurately identify neutrino interactions~\textcolor{Blue}{\cite{MicroBooNE detector}}. It began collecting neutrino beam
data in October of 2015. \ Figure~\textcolor{Blue}{\ref{img:Event display}} shows an image of
a high multiplicity event from MicroBooNE\ data.

The MicroBooNE TPC (Fig.~\textcolor{Blue}{\ref{img:MicroBooNE picture}}) has an active mass of about 85 tons (85 Mg) of liquid argon. It
is 10.4 meters long in the beam direction, 2.3 meters tall, and 2.6 meters
in the electron drift direction. Electrons require 2.3 ms to drift across
the full width of the TPC from cathode (at $-70$ kV) to anode (at $\sim0$ kV). Events are read
out on three anode wire planes with $3$ mm spacing between wires. Drifting
electrons pass through the first two wire planes, which are oriented at $\pm
60$ degrees relative to vertical, producing bipolar induction signals. The
third wire plane, the collection plane, has its
wires oriented vertically and collects the charge of the drifting electrons
in the form of a unipolar signal. The MicroBooNE readout electronics allow
for measurement of both the time and charge created by drifting electrons on
each wire. The amplified, shaped waveforms from 8256 wires from induction and collection planes are digitized at $2$ MHz using 12-bit ADCs. A data acquisition system readout window consisting of $9600$ recorded samples ($4.8$ msec) for all wires is then noise-filtered and deconvolved utilizing offline software algorithms. Reconstruction algorithms are then used on these output waveforms to reconstruct the times and amplitudes of charge depositions (hits) on the wires from particle-induced ionization in the TPC bulk.

While all three anode planes are used for track reconstruction, the collection plane
provides the best signal-to-noise performance and charge resolution. The
analysis presented here excludes regions of the detector that have
non-functional collection plane channels ($\sim10\%$). It also imposes
requirements on the minimum number of collection plane hits$-$current pulses
processed through noise filtering~\textcolor{Blue}{\cite{Noise Filtering}}, deconvolution, and
calibration$-$associated with the reconstructed tracks. All
charged particle track candidates are required to have at least 15 collection plane hits,
and the longest muon track candidate is required to have at least 80 collection plane
hits. Furthermore, as described in Sec.~\textcolor{Blue}{\ref{event_selection}}, we use two
discriminants to extract the neutrino interaction and cosmic ray background
contributions to our data sample that are based on collection plane hits.

A light collection system consisting of 32 8-inch photomultiplier tubes (PMTs) with
nanosecond timing resolution enables precise determination of the time of the neutrino interaction, which crucially aids in the reduction of cosmic ray backgrounds.  

The BNB employs protons from the Fermilab Booster synchrotron impinging on a
beryllium target. The proton beam comes in bunches of protons called ``beam spills", has a kinetic energy of 8 GeV, a repetition rate of up to 5 Hz, and is capable of producing $5\times10^{12}$
protons-per-spill. Secondary pions and kaons decay producing neutrinos with
an average energy of $800$ MeV. MicroBooNE received $3.6\times10^{20}$
protons-on-target in its first year of running, from fall 2015 through summer
2016. This analysis uses a fraction of that data corresponding
to $5.0\times10^{19}$ protons-on-target.


\section{Cosmic ray Backgrounds}
\label{CRbkg}

The MicroBooNE detector lacks appreciable shielding from cosmic rays (CR) since the detector is at the earth's surface and has little overburden. \ Most events that are recorded and processed through an online software trigger which requires that the total light recorded with the PMT system exceeds 6.5 photoelectrons (PE) during neutrino beam operations
(\textquotedblleft on-beam data\textquotedblright) contain no neutrino
interactions. Triggered events with a neutrino interaction typically
have the products of up to 20 cosmic rays coincident with the beam spill in the event readout window (4.8 msec)
contributing to a recorded event along with the products of the neutrino
collision. \ A large sample of events recorded under identical conditions as
the on-beam data, minus the coincidence requirement with the beam,
(\textquotedblleft off-beam-data\textquotedblright) has been recorded for
use in characterizing cosmic ray backgrounds.\ A straightforward on-beam minus
off-beam background subtraction is difficult, as the off-beam data
does not reproduce all correlated detector effects associated with on-beam
events that contain a neutrino interaction with several overlaid cosmic rays.
\ The situation is particularly complicated with events containing neutrino interactions with $N_{obs,n} = 1$, which share the same topology with the most
common single-muon CR configuration. Monte Carlo simulations of the CR flux
using the CORSIKA package~\textcolor{Blue}{\cite{Corsika}} provide useful guidance; however,
the ability of these simulations to describe the very rare CR topologies
that closely match neutrino interactions is not well known.  \newline

For these reasons this analysis employs a method to separate neutrino
interaction candidates from CR backgrounds that is driven by the data
itself. Even though CR tracks should always appear to at least enter the detector, they can satisfy the experimental condition of being fully contained if a segment of the CR track falls outside the data acquisition readout time window, or if a segment of the track fails to be identified due to instrumentation- or algorithm-related inefficiencies. \ The separation of neutrino
interaction candidates from CR backgrounds rests on the observation that a neutrino $\nu_{\mu}$
CC interaction produces a final state $\mu^{-}$ that slows down as it moves
away from its production point at the neutrino interaction vertex due to
ionization energy loss in the liquid argon. \ As it slows down, its rate of
restricted energy loss~\textcolor{Blue}{\cite{Particle Data Group}}, $dE/dx_{R}$, increases, and deviations from a linear
trajectory due to multiple Coulomb scattering (MCS) become more pronounced.
\ A CR muon track can produce an apparent neutrino interaction vertex if it
comes to rest in the detector or it is not fully reconstructed to the edge of the TPC, but the CR track will exhibit large $%
dE/dx_{R} $ and MCS effects in the vicinity of this vertex. \ Furthermore,
the vast majority of $\nu_{\mu}$ CC muons travel in the neutrino beam direction (\textquotedblleft
upstream\textquotedblright\ to \textquotedblleft
downstream\textquotedblright), whereas CR muons move upstream or downstream
with equal probability. \newline

\section{\label{Selection}Event Selection and Classification}

\subsection{Data}

This analysis uses two data samples:

\begin{itemize}
\item \textquotedblleft On-beam data\textquotedblright, taken only during
periods when a beam spill from the BNB is actually sent. The on-beam data
used in this analysis were recorded from February to April 2016 using data taken in runs in
which the BNB and all detector systems functioned well.
This sample comprises about $15\%$ of the total neutrino data collected by
MicroBooNE in its first running period (October 2015 to summer 2016),

\item \textquotedblleft Off-beam data\textquotedblright\, taken with the same
software trigger settings as the on-beam data, but during periods when no
beam was received. The off-beam data were collected from February to October
2016.
\end{itemize}

\subsection{Simulation}

\label{event_selection}

The LArSoft software framework is used for processing data events and Monte
Carlo simulation (MC) events in the same way. Three simulation samples are used in
this analysis:

\begin{itemize}

\item Neutrino interactions simulated with a default GENIE model overlaid
with CORSIKA CR events (\textquotedblleft MC default\textquotedblright),

\item MC default augmented by the GENIE\ implementation of the Meson
Exchange Current model~\textcolor{Blue}{\cite{MEC}} overlaid
with CORSIKA CR events (\textquotedblleft MC with MEC\textquotedblright),

\item MC default augmented by the GENIE\ implementation of the Transverse
Enhancement Model~\textcolor{Blue}{\cite{TEM}} overlaid
with CORSIKA CR events (\textquotedblleft MC with TEM\textquotedblright).
\end{itemize}

The GENIE models used in this analysis are Relativistic Fermi Gas model~\textcolor{Blue}{\cite{RFG}} for the nuclear momentum, Llewellyn-Smith~\textcolor{Blue}{\cite{LS}} for QE interactions, Rein-Sehgal model~\textcolor{Blue}{\cite{RS}} for resonance interactions, and Empirical model by Dytman~\textcolor{Blue}{\cite{MECDytman}} for MEC interactions. The MC default does not include contributions from the excitation of two particle-two hole (2p2h)~\textcolor{Blue}{\cite{2p2h}} final states in neutrino-nucleus scattering, which may be important for low energy neutrinos. The MC with MEC and TEM include the excitation of 2p2h final states in neutrino-nucleus scattering in alternative ways. In TEM, the empirical superscaling function is modeled with an effective spectral function (ESF)~\textcolor{Blue}{\cite{TEMESF}}.

The generator stage (production of a set of  final state four-vectors for particles originating from the argon nucleus as a result of the $\nu_\mu$-Ar interaction in GENIE) employs GENIE (version $v2.8.6d$ for the MC default and $v2.10.6$ for the MC with MEC and TEM) with overlaid simulated CR backgrounds using CORSIKA version $v7.4003$ with a constant mass composition model (CMC)~\textcolor{Blue}{\cite{CMC}} at $226$ m above sea-level elevation. Simulated secondary particle propagation utilizes GEANT version $v4.9.6.p04d$ using a physics particle list $QGSP\texttt{\underline{{ }}}BIC$ with custom physics list for optical photons, and the detector simulation employs LArSoft version $v4.36.00$. All GENIE samples were processed with the same GEANT and LArSoft versions for detector simulation and reconstruction. These samples thus allow for relative comparison of different GENIE models to the data.

\subsection{Reconstruction}

Event reconstruction makes use of anode plane waveforms from the TPC and
light signals from the PMT\ system. Raw signals from the TPC, recorded on the
wires, first pass through a noise filter~\textcolor{Blue}{\cite{Noise Filtering}} before hits
are extracted from the observed waveforms on each wire. \ A set of
reconstruction algorithms known as Pandora~\textcolor{Blue}{\cite{Pandora-1,Pandora-2}} combine hits connected in space and time into two-dimensional (2D) clusters
for each of the three anode planes. Then a three-dimensional (3D) vertex reconstruction is
performed in which the energy deposition around the vertex and the knowledge of the beam direction is used to preferentially select vertex candidates with more deposited energy and with low $z$ coordinates, respectively. 2D clusters in all three planes are then merged into 3D tracks and
showers. The preferential direction for tracks
and showers is also from the upstream to the downstream end of the detector. \ In
this analysis, for each 3D track candidate, the start position is taken to be the 3D track end 
closest to the candidate neutrino vertex, if the track satisfies the 
requirement that the track start position is sufficiently close to the
vertex. This analysis makes no use of shower objects, which are mainly associated
with electrons and photons.

Light collected on the 32 PMTs in MicroBooNE is used to reconstruct optical
hits. To reduce sensitivity to possible fluctuations in the signal baseline,
a threshold-based hit-reconstruction algorithm requires PMT pulses of a
minimum charge for a hit to be reconstructed. A weighted sum of PMT hits (optical flashes) are reconstructed by requiring a time coincidence of $\sim$1 $\mu s$ between hits on multiple PMTs. 
\ The relative timing and charge collection of optical hits, along with the spatial locations of the PMT,
within an optical flash is then used to associate the flash with
reconstructed tracks in the TPC, a process known as flash-matching.

\subsection{Event Selection}\label{event_sel}

Event selection starts by requiring an optical flash within the $1.6$ $\mu 
\emph{s}$ duration beam spill window and the summed light collected by the PMT to exceed 50 PE. Reconstructed vertices must be contained in
the fiducial volume of the detector, defined as $10$ cm from the border of
the active volume in $x$, $10$ cm from the border of the active volume in $z$%
, and $20$ cm from the border of the active volume in $y$ (see Fig.~\textcolor{Blue}{\ref{img:MicroBooNE picture}} for detector coordinates). At each candidate neutrino interaction vertex, a candidate muon track is identified as the longest of all tracks starting within $5$ cm of the vertex. The candidate muon track is further required to be fully contained within the detector, where containment requires both ends of the track to lie within the same fiducial volume required for an event vertex, to have at least $75$ cm 3D track length, and to
have an event vertex located within 80 cm in $z$ of the PMT-reconstructed
position of an optical flash. Considerable CR backgrounds remain after these
pre-selection procedures, with signal/background $\approx 1/1$.

Pre-selected events then pass through a second stage filter that imposes
further quality conditions on track candidates. Start and end points of the
candidate muon must lie in detector regions with functional collection plane wires.
The candidate muon track must start $46$ cm below the top surface of the TPC in order to suppress CR backgrounds, must start within $3$ cm (reduced from $5$ cm) of the selected vertex position, must have at least $80$ hits in the
collection wire plane, and must not have significant wire gaps in the
start and end $20$ collection plane-hit segments used in the pulse-height (PH) test (Sec.~\textcolor{Blue}{\ref%
{PH}}) and the multiple Coulomb scattering test (Sec.~\textcolor{Blue}{\ref{MCS}}). \newline

Events satisfying all of the above criteria constitute the final data sample.
Table~\textcolor{Blue}{\ref{tab:passingrates}} lists the event passing rates for the on-beam
data, off-beam data, and the MC default samples at different steps of the
event selection. The passing rates in on-beam data are consistent with expectations for a mix of CR-only events, as provided by the off-beam data, and events containing a neutrino interaction in addition to cosmic rays, as provided by the MC. \newline

The observed multiplicity of a selected event is defined to be the number of
particles starting within $3$ cm of the selected vertex that have at least $%
15$ collection plane hits where the Pandora MicroBooNE track reconstruction algorithms
perform optimally. There is no containment requirement for tracks other than the candidate muon track. Table~\textcolor{Blue}{\ref{tab:mult_num}} lists the number of selected
events in each multiplicity bin with relative event rates for on-beam
data, off-beam data, and MC default samples. \newline

The minimum collection plane hit condition corresponds to a minimum range in liquid argon
of $4.5$ cm, and the requirement thus excludes charged particles below a
particle-type-dependent kinetic energy threshold from entering our sample
that ranges from $31$ MeV for a $\pi ^{\pm }$ to $69$ MeV for a proton. \ No
acceptance exists for particles with kinetic energies below these
thresholds, which roughly increase as the secant of the track angle with
respect to the neutrino beam direction.

The average efficiency for the Pandora-based track reconstruction used in this analysis is $%
\left\langle \epsilon\right\rangle \approx45\%$ at the $%
15 $ collection plane hit threshold. This relatively low value, with implicit kinetic
energy thresholds, creates a common occurrence called \textquotedblleft
feed-down\textquotedblright\ wherein events produced with $n$ tracks at the
argon nucleus exit position are reconstructed with an observed multiplicity $%
n^{\prime}<n$. \ For example, $n=1$ is commonly observed because one of the
two tracks in a quasi-elastic-like event fails to be reconstructed due to low
acceptance or tracking efficiency.

The candidate muon containment requirement limits its energy to be $%
\lesssim1.2$ GeV depending on the muon scattering angle. This results in a
sample biased towards relatively higher inelasticity, $E_{H}/E_{\nu}$, with $%
E_{H}$ being the energy transferred from the neutrino to the hadronic system
in the collision.

\begin{figure*}[!ht]
\begin{adjustwidth}{-2cm}{-2cm}
\centering
\subfloat{\includegraphics[width=.35\textwidth]{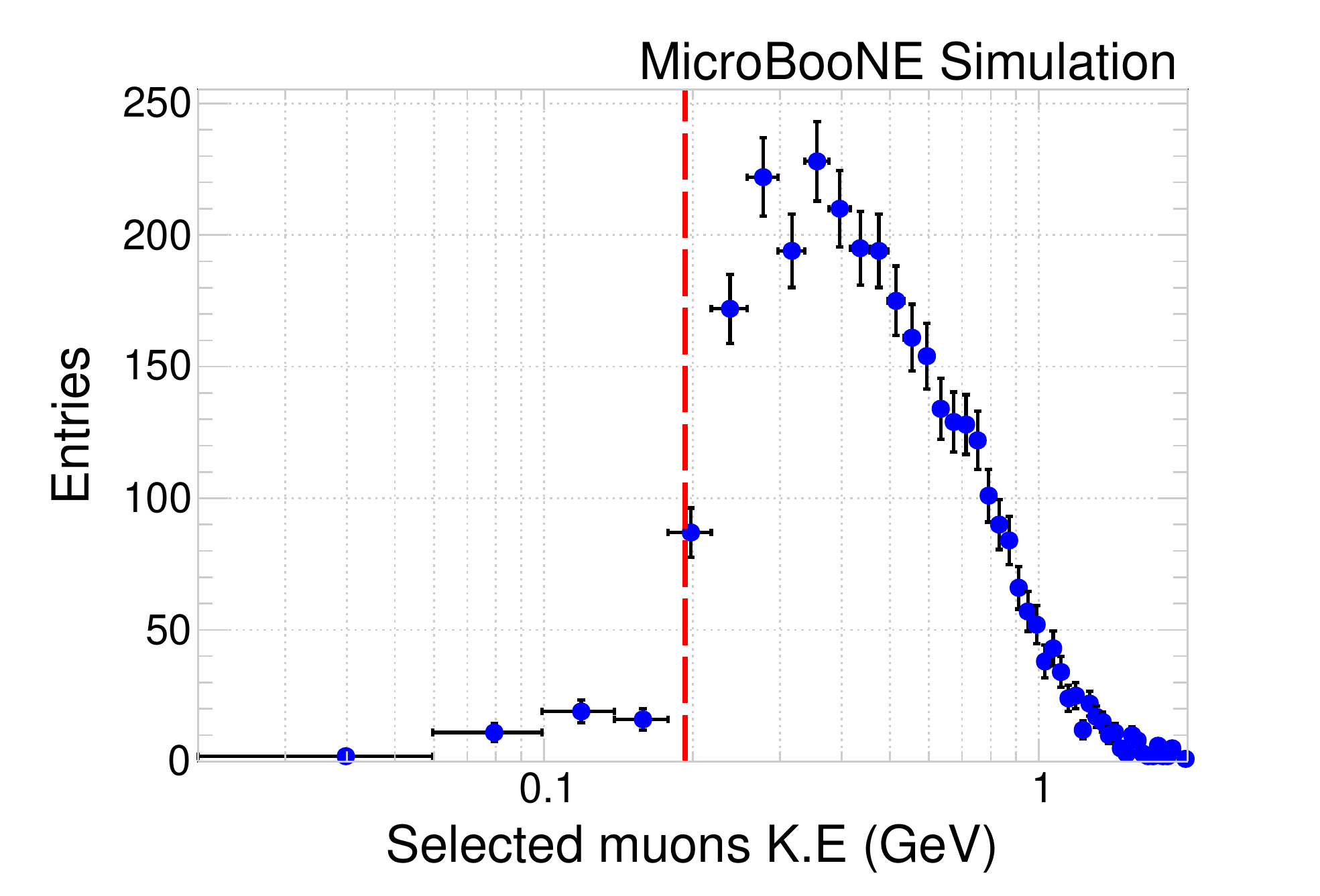}}
\subfloat{\includegraphics[width=.35\textwidth]{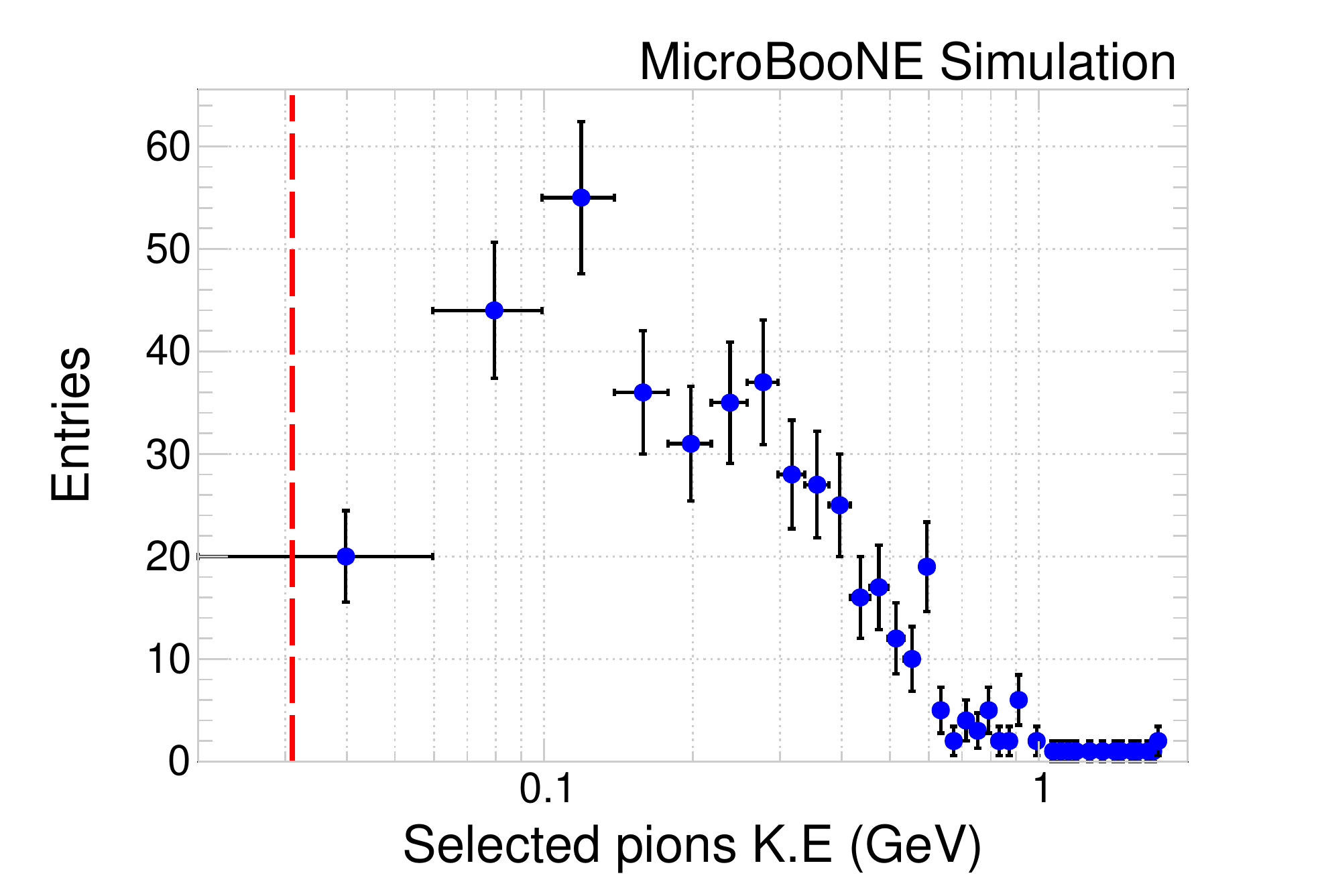}}
\subfloat{\includegraphics[width=.35\textwidth]{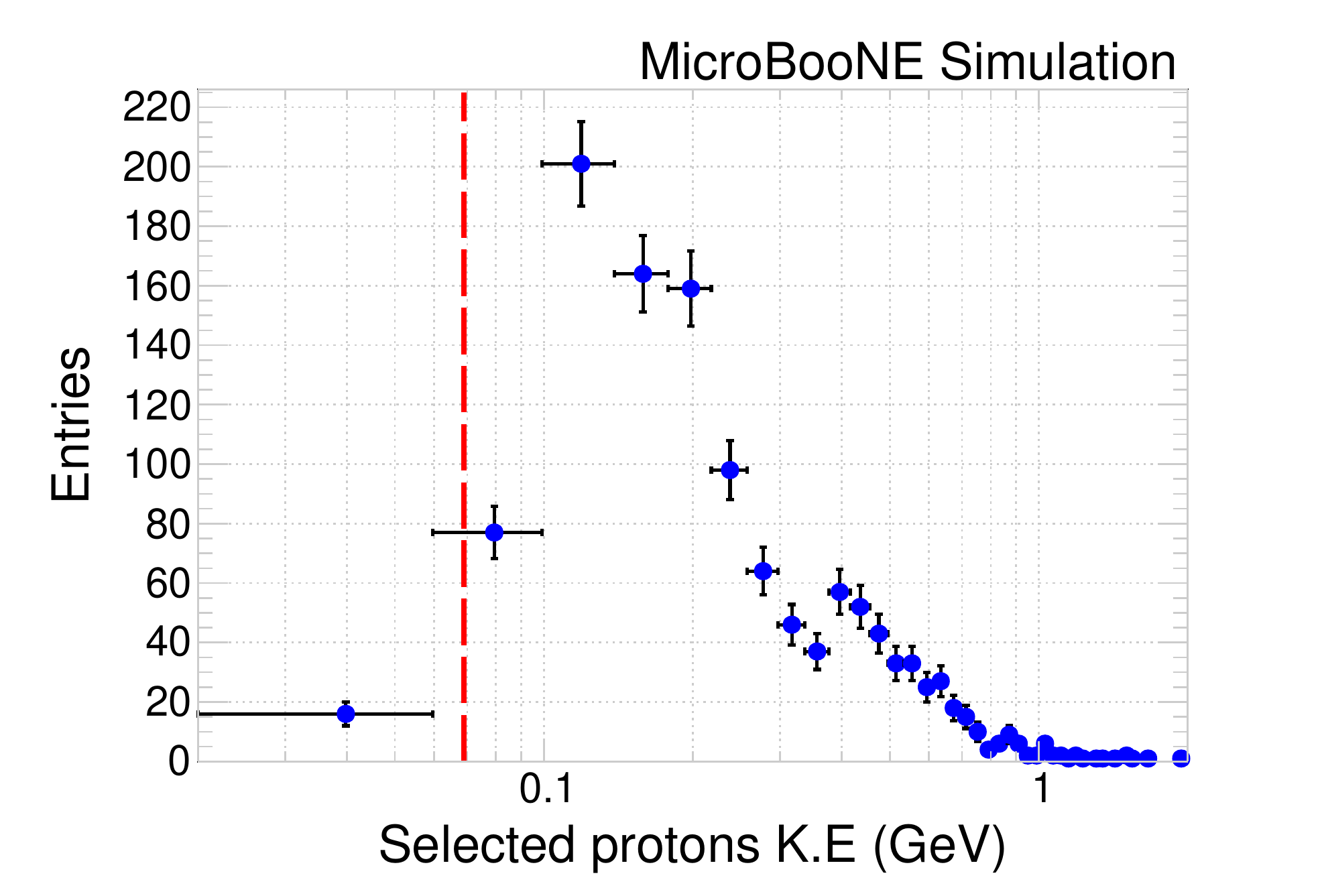}}
\caption{GENIE expectations for true kinetic energy distributions for selected muons, pions, and protons. The kinetic energy thresholds associated with the 15 collection plane hit requirement for short tracks and the 75 cm length requirement for the long track are represented by dashed red lines.}
\label{Acceptance_KE}
\end{adjustwidth}
\end{figure*}

Figure~\textcolor{Blue}{\ref{Acceptance_KE}} shows the GENIE expectations for the true kinetic energy of muons, protons, and pions produced in BNB neutrino interactions in MicroBooNE. The kinetic energy thresholds associated with the 15 collection plane hit requirement for short tracks and the 75 cm 3D track length requirement for the long track are evident.

\begin{table*}[!ht]
\caption{Passing rates for event selection criteria applied to on-beam data,
off-beam data, and MC default samples. Numbers are absolute event counts.
Quantities in parentheses give the relative passing rate with respect to the
step before (first percentage) and the absolute passing rate with respect to
the starting sample (second percentage).}
\label{tab:passingrates}
\centering
\begin{ruledtabular}
\renewcommand{\tabcolsep}{1pt}
\begin{tabular}
[c]{ccccccc}
& \multicolumn{2}{c}{\textbf{On-beam data}} & \multicolumn{2}{c}{\textbf{Off-beam data}}& \multicolumn{2}{c}{\textbf{MC default }}\\\
\textbf{Selection cuts}& \textbf{Events} &\textbf{Passing rates} &\textbf{Events} & \textbf{Passing rates} & \textbf{Events} & \textbf{Passing rates} \\\hline
Total events & 547616 & & 2954586 &  & 188880 & \\
$\nu_{\mu}$ events passing pre-cuts & 4049 & (0.74\%/0.74\%)& 14213 & (0.48\%/0.48\%)& 7106 & (3.8\%/3.8\%)\\
Events passing dead region cut & 3080 & (76\%/0.56\%)& 10507 & (74\%/0.36\%)& 5632 & (79\%/2.9\%) \\
Long track starting $46$ cm below the TPC top surface & 2438 & (79\%/0.44\%)& 7883 & (75\%/0.27\%)& 4795 & (85\%/2.6\%) \\
Long track to vertex distance $<$ 3 cm & 2435 & (99\%/0.44\%)& 7862 & (99\%/0.27\%)& 4781 & (99\%/2.5\%) \\
Events with $\geq 80$ collection plane hits & 1930 & (79\%/0.35\%)& 5279 & (67\%/0.17\%)& 4387 & (92\%/2.3\%) \\
Events passing wire gap cuts & 1795 & (93\%/0.33\%)& 4954 & (94\%/0.16\%)& 4016 & (92\%/2.1\%)\\
\end{tabular}
\end{ruledtabular}
\end{table*}

\begin{table*}[!ht]
\caption{Selected number of events from the on-beam data, off-beam data, and MC default samples and their corresponding acceptance rates on the multiplicity basis.}
\label{tab:mult_num}
\centering
\begin{ruledtabular}
\renewcommand{\tabcolsep}{1pt}
\begin{tabular}
[c]{ccccccc}%
\textbf{Multiplicities} & \multicolumn{2}{c}{\textbf{On-beam data}} &
\multicolumn{2}{c}{\textbf{Off-beam data}} &
\multicolumn{2}{c}{\textbf{MC default}}\\
& \textbf{Events} & \textbf{Event rate} & \textbf{Events} &
\textbf{Event rate} & \textbf{Events} & \textbf{Event
rate}\\\hline
Total events & 1795 &  & 4954 &  & 4016 & \\
mult = 1 & 1379 & 77\% & 4113 & 83\% & 2599 & 65\%\\
mult = 2 & 389 & 22\% & 828 & 17\% & 1186 & 30\%\\
mult = 3 & 26 & 1.4\% & 12 & 0.2\% & 210 & 5\%\\
mult = 4 & 1 & 0.06\% & 1 & 0.2\% & 18 & 0.4\%\\
mult = 5 & 0 & 0\% & 0 & 0\% & 3 & 0.07\%\\
\end{tabular}
\end{ruledtabular}
\end{table*}

\subsection{Event Classification}

Selected events are next classified into four categories based on whether
they pass or fail the PH test and the MCS test described in the following sections. These are the candidate muon track direction-based tests which are used to separate neutrino signal and
CR background contributions in the sample. Table~\textcolor{Blue}{\ref{tab:acceptance_rates}}
lists the event selection rates for the on-beam data, off-beam data, and the MC default samples in each category. The final samples are called 
\emph{neutrino-enriched}, \emph{mixed}, or \emph{background-enriched}
sub-samples depending on whether events pass both tests, pass either one of
the two tests, or fail both tests, respectively.

\begin{table*}[!hpt]
\caption{Final categories from the on-beam data, off-beam data, and
MC default samples. Numbers are absolute event counts. The percentages
correspond to the fraction of events in each category.}
\label{tab:acceptance_rates}
\centering
\begin{ruledtabular}
\renewcommand{\tabcolsep}{1pt}
\begin{tabular}
[c]{ccccccc}%
\textbf{Categories} & \multicolumn{2}{c}{\textbf{On-beam data}} &
\multicolumn{2}{c}{\textbf{Off-beam data}} &
\multicolumn{2}{c}{\textbf{MC default}}\\
\textbf{PH, MCS} & \textbf{Events} & \textbf{Event rate} &
\textbf{Events} & \textbf{Event rate} & \textbf{Events} &
\textbf{Event rate}\\\hline
PASS,  PASS & 802 & 44\% & 1252 & 25\% & 2464 & 61\%\\
PASS, FAIL & 334 & 19\% & 1013 & 20\% & 704 & 18\%\\
FAIL, PASS & 304 & 17\% & 1049 & 21\% & 442 & 11\%\\
FAIL, FAIL & 355 & 20\% & 1640 & 33\% & 406 & 10\%\\
\end{tabular}
\end{ruledtabular}
\end{table*}

\subsubsection{\textbf{Pulse Height Test}}

\label{PH} A neutrino-induced muon from a CC event will exhibit an increasing rate of
energy loss as one moves downstream along its track. A visual diagram for the PH test is shown in Fig.~\textcolor{Blue}{\ref{img:PH test}}.\ We take into account
the expected behavior of the rate of restricted energy loss, $dE/dx_{R}$, with the following procedure:

\begin{figure}[!hptb]
\centering
\includegraphics[width=0.5\textwidth]{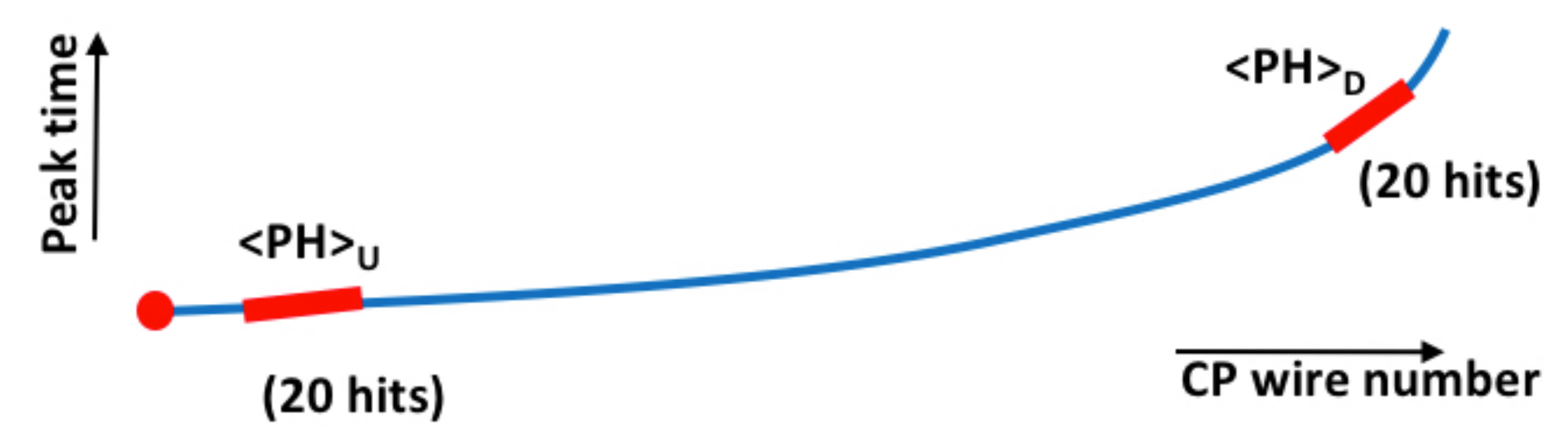} 
\caption{Diagram showing the PH test for a candidate muon track.}
\label{img:PH test}
\end{figure}

\begin{itemize}
\item Compute the truncated mean of the pulse heights deposited in $20$ consecutive
collection plane hits, $\left\langle PH\right\rangle _{U}$, starting $10$ hits away from
the upstream end of the muon track that is taken as a proxy for the upstream
restricted energy loss. \ The truncated mean is formed by taking the average
of the $20$~PH after removing individual PH that do not lie within the range
of $20\%-200\%$ of the average~\textcolor{Blue}{\cite{CCFR 20-200}}:%
\begin{equation}
\left\langle PH\right\rangle _{U}=\frac{\sum\limits_{n=11}^{n=30}PH_{n}%
\left( 0.2\left\langle PH\right\rangle <PH_{n}<2.0\left\langle
PH\right\rangle \right) }{\sum\limits_{n=11}^{n=30}\left( 0.2\left\langle
PH\right\rangle <PH_{n}<2.0\left\langle PH\right\rangle \right) },
\end{equation}
which can be determined iteratively with an initial approximation that $%
\left\langle PH\right\rangle $ is the arithmetic average. \ Use of the
truncated mean PH rather than the average PH minimizes effects of large
energy loss fluctuations, \ 

\item Form a similar quantity from $20$ consecutive collection plane hits that end 10 collection plane
hits away from the downstream end of the track, $\left\langle
PH\right\rangle _{D}$,

\item Form the test $p=\left\langle PH\right\rangle _{U}<\left\langle
PH\right\rangle _{D}$. \ Muons from $\nu_{\mu}$ CC interactions will pass
this test with a probability $P\left( PH\right) $. \ Muons from CR
background can be characterized by the probability that they fail this test, denoted as $Q\left( PH\right) .$
\end{itemize}

\begin{figure}[!hptb]
\centering
\includegraphics[width=0.5\textwidth]{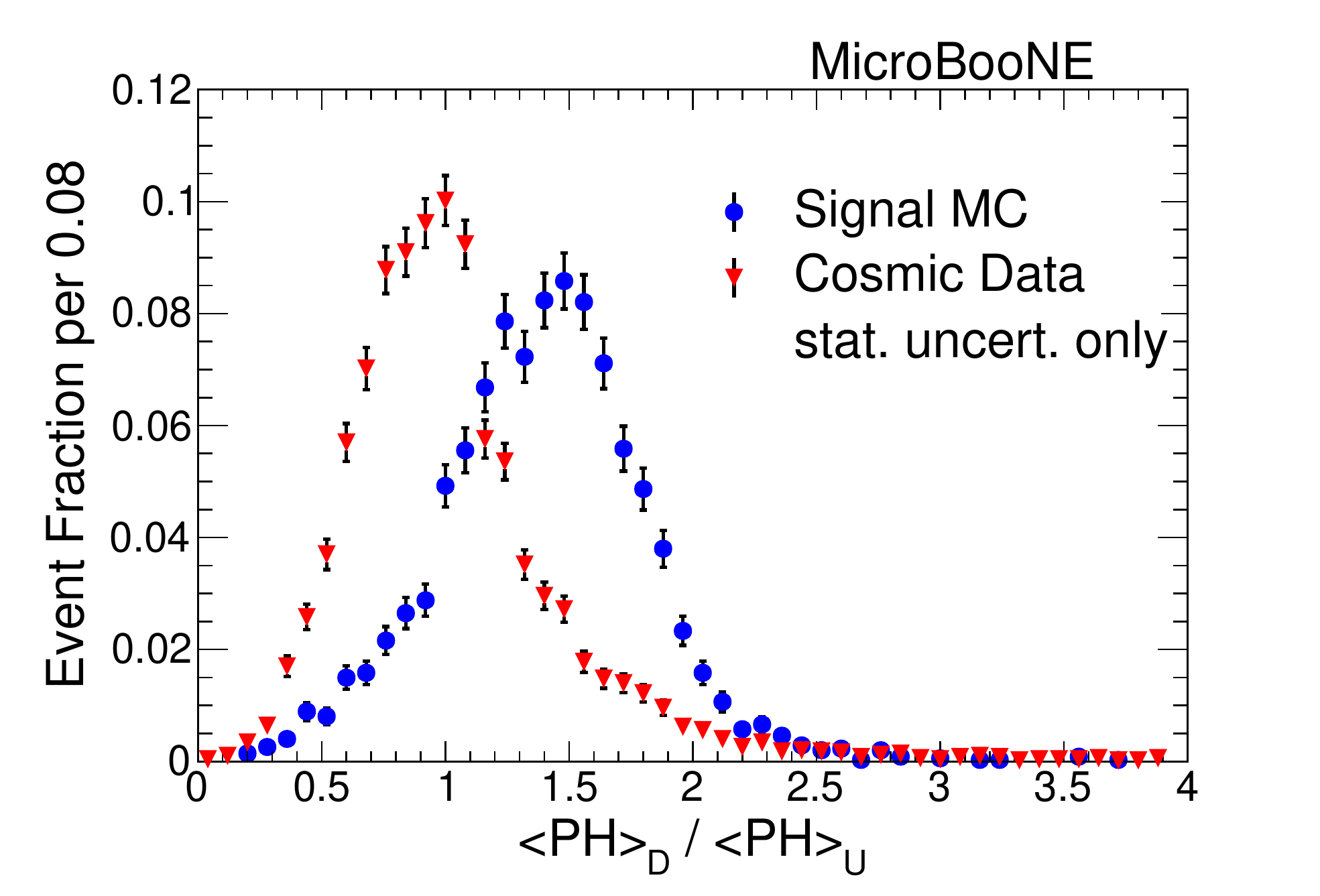} 
\caption{Pulse height (PH) downstream to upstream ratio. Events with $\frac{\left\langle PH\right\rangle_{D}}{\left\langle PH\right\rangle_{U}} > 1$ pass the PH test.}
\label{img:PH_ratio}
\end{figure}

Figure~\textcolor{Blue}{\ref{img:PH_ratio}} presents the PH downstream to upstream
ratio distribution for neutrino events only from MC default (signal MC) and off-beam data
(cosmic data) samples. The expected signal is considerably enriched relative to the background for PH ratios greater than $1$ and we use this value to define the PH test used in the analysis. 

\subsubsection{\textbf{Multiple Coulomb Scattering Test}}\label{MCS} 

A neutrino-induced muon from a CC event will generally exhibit an increasing degree of multiple Coulomb scattering (MCS) about a nominal straight line trajectory as one moves from upstream to downstream along the track. A visual diagram for the MCS test is shown in Fig.~\textcolor{Blue}{\ref{img:MCS test}}.\ We take into account the expected MCS behavior by an independent test with the following
procedure:

\begin{figure}[!hptb]
\centering
\includegraphics[width=0.5\textwidth]{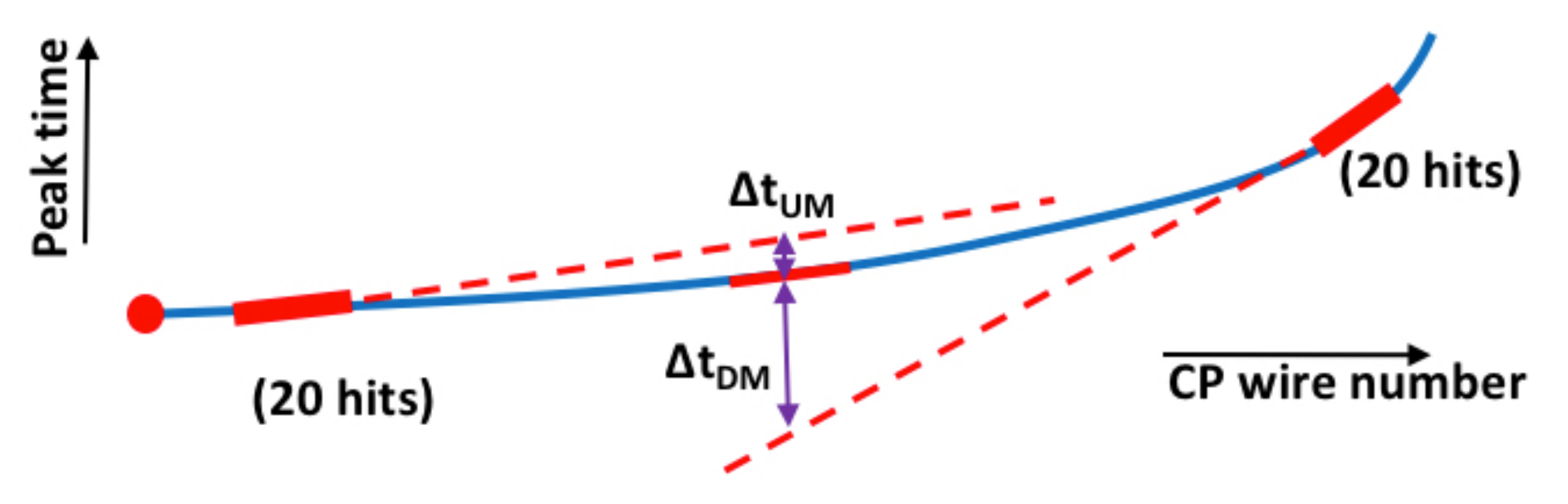} 
\caption{Diagram of the MCS test for a candidate muon track.}
\label{img:MCS test}
\end{figure}

\begin{itemize}
\item Form three 20 collection plane-hit long track segments at the upstream, downstream,
and geometric center of the track. \ The upstream and downstream segments
are displaced by $10$ collection plane hits from the upstream and downstream ends of the
track, respectively,

\item Perform a simple linear least squares fit of hit time vs. (wire)
position using the $20$ contiguous collection plane hits at the upstream end of the track.
\ Denote the determined line as $L_{U}$. Perform a similar fit using the $20$
collection plane hits at the downstream end of the track. \ Denote the determined line as $%
L_{D}.$ Finally perform one more similar fit from the $20$ collection plane hits located
about the geometric center of the track. Denote this line as $L_{M}$,

\item Compare the hit time predicted at the geometric center of the track, $%
t_{C}$, by $L_{M}$, which uses hits about the geometric center, to the time
predicted at the geometric center of the track by the projection of $L_{U}$
from the beginning of the track: 
\begin{equation}
\Delta t_{UM}=\left\vert t_{C}\left( L_{U}\right) -t_{C}\left( L_{M}\right)
\right\vert .
\end{equation}

\item Repeat the process except compare $t_{C}$ from $L_{M}$ to the time
predicted at the geometric center of the track by the projection of $L_{D}$
from the end of the track:%
\begin{equation}
\Delta t_{DM}=\left\vert t_{C}\left( L_{D}\right) -t_{C}\left( L_{M}\right)
\right\vert .
\end{equation}

\item Form the test $q=\Delta t_{UM}<\Delta t_{DM}$. \ Since MCS should
become, on average, more pronounced along the downstream end of the track as
the momentum decreases, this provides a second directional test on the muon
track candidate. \ Muons from $\nu_{\mu}$ CC interactions will pass this
test with a probability $P\left( MCS\right) $. \ Muons from CR
background can be characterized by the probability that they fail this test, denoted as $Q\left( MCS\right) .$
\end{itemize}

\begin{figure}[!hptb]
\centering
\includegraphics[width=0.5\textwidth]{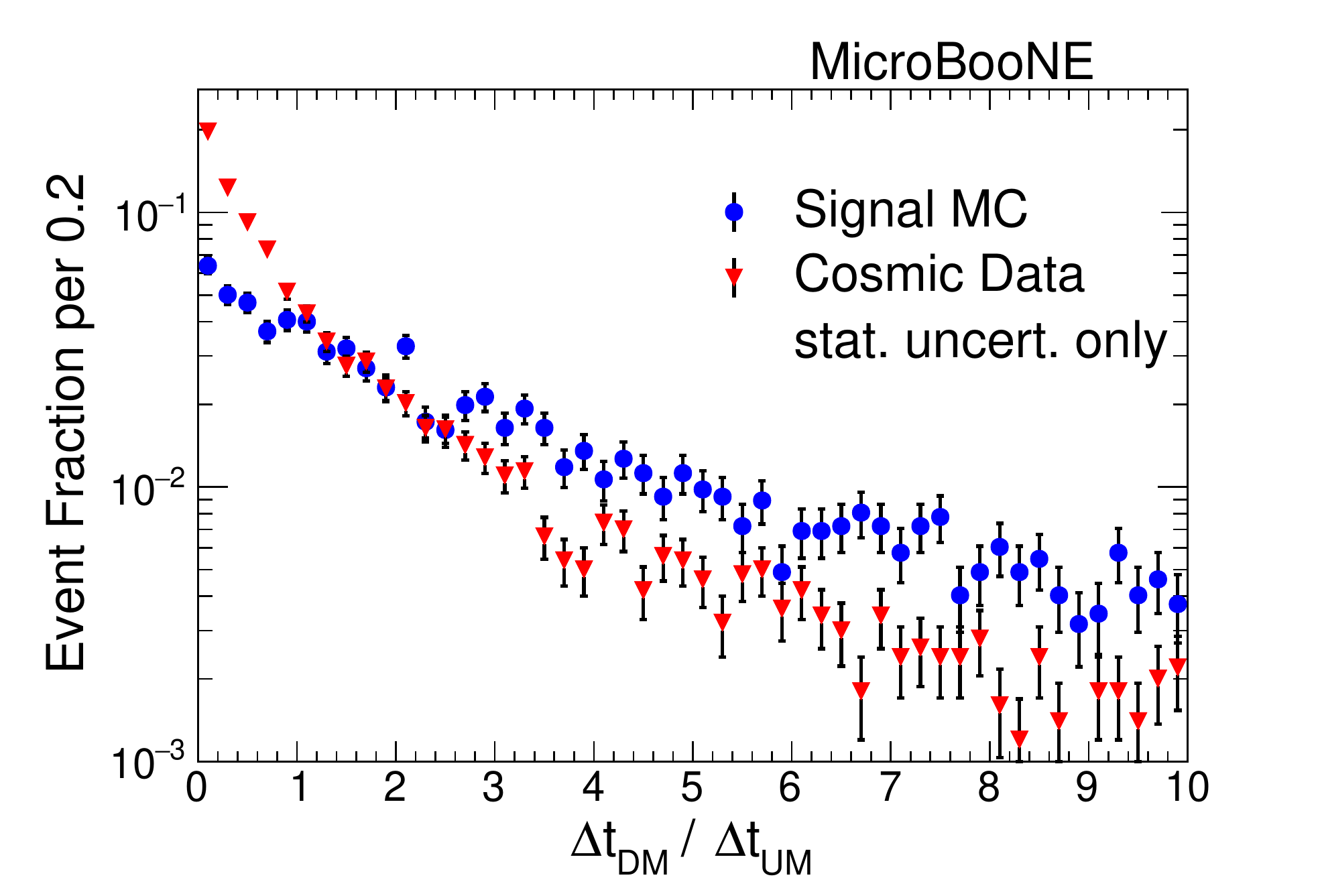} 
\caption{Multiple Coulomb scattering (MCS) downstream to upstream ratio. Events with $\frac{\Delta t_{DM}}{\Delta t_{UM}} > 1$ pass the MCS test.}
\label{img:MCS_ratio}
\end{figure}

Figure~\textcolor{Blue}{\ref{img:MCS_ratio}} presents the MCS downstream to upstream ratio $%
\Delta t_{DM}/\Delta t_{UM}$ distribution for neutrino events only from MC default (signal MC) and off-beam data (cosmic data) samples. We
observe that MCS ratio for the signal dominates over the background for
values greater than $1$ and we use this value to define the MCS test used in this
analysis.

\section{\label{Signal}Signal Extraction}

\subsection{Data-Driven Signal+Background Model}

On-beam data consists of a mixture of neutrino interaction and CR background
events. \ We designate a passing of the PH or MCS test by the symbol $\nu$,
and a failure of either of the two tests by the symbol $CR$, thus creating
the categories \textquotedblleft$\nu\nu$\textquotedblright, \textquotedblleft%
$\nu CR$\textquotedblright, \textquotedblleft$CR\nu$\textquotedblright,
\textquotedblleft$CRCR$\textquotedblright, which contain a corresponding
number of events $N_{\nu\nu}$, $N_{\nu CR}$, $N_{CR\nu}$, and $N_{CRCR}$. \
The \textquotedblleft$\nu\nu$\textquotedblright\ and \textquotedblleft$CRCR$%
\textquotedblright\ categories are expected to have relatively high neutrino or
CR purity, respectively; while the \textquotedblleft$\nu CR$%
\textquotedblright\ and \textquotedblleft$CR\nu$\textquotedblright\ have
mixed purity.

We next build a model for the number of events in each category as follows: 
%
\begin{widetext}
\begin{align}
\hat{N}_{\nu\nu}  & =P\left(  MCS|PH\right)  P\left(  PH\right)  \hat{N}%
_{\nu}\\
&  +\left(  1-Q\left(  PH\right)  -Q\left(  MCS\right)  +Q\left(MCS|PH\right)  Q\left(  PH\right)  \right)  \hat{N}_{CR},\nonumber \\
\hat{N}_{CR\nu}  &  =\left(  1-P\left(  MCS|PH\right)  \right)  P\left(
PH\right)  \hat{N}_{\nu}\\
&  +\left(  Q\left(  MCS\right)  -Q\left(  MCS|PH\right)  Q\left(  PH\right)
\right)  \hat{N}_{CR},\nonumber\\
\hat{N}_{\nu CR}  &  =\left(  P\left(  MCS\right)  -P\left(  MCS|PH\right)
P\left(  PH\right)  \right)  \hat{N}_{\nu}\\
&  +\left(  1-Q\left(  MCS|PH\right)  \right)  Q\left(  PH\right)  \hat
{N}_{CR},\nonumber\\
\hat{N}_{CRCR}  &  =Q\left(  MCS|PH\right)  Q\left(  PH\right)  \hat{N}_{CR} \\
& +\left(  1-P\left(  PH\right)  -P\left(  MCS\right)
+P\left(  MCS|PH\right)  P\left(  PH\right)  \right)  \hat{N}_{\nu}.\nonumber  \\
\end{align}
\end{widetext}
The quantities $\hat{N}_{\nu\nu}$, $\hat{N}_{CR\nu}$, $\hat{N}_{\nu CR}$,
and $\hat{N}_{CRCR}$ are model parameters corresponding to the observed
number of events $N_{\nu\nu}$, $N_{\nu CR}$, $N_{CR\nu}$, and $N_{CRCR}$,
respectively. $\hat{N}_{\nu}$ and $\hat{N}_{CR}$ are the estimated number of
neutrino and CR events, in the sample, to be determined by a
fit described below. \ The quantities $P\left( PH\right) $ and $P\left(
MCS\right) $ represent the average probabilities that a neutrino interaction
muon passes the $PH$ or $MCS$ test condition, while $Q\left(
PH\right) $ and $Q\left( MCS\right) $ denote the mean probabilities that a
cosmic ray muon \emph{fails} one of these tests. The conditional probability 
$P\left( MCS|PH\right) $ denotes the fraction of time that a neutrino
interaction muon event that passes the $MCS$ condition after it has passed
the $PH$ condition, and the conditional probability $Q\left( MCS|PH\right) $
denotes the fraction of time that a cosmic ray event muon fails the $MCS$
test after failing the $PH$ test. 

As the $MCS$ and $PH$ conditions result
from different physical processes (muon-nucleus and muon-electron
scattering, respectively), and the $MCS$ and $PH$ test are formed from
different measurements (time and charge, respectively), the $PH$ and $MCS$
tests are nearly independent with $P\left( MCS|PH\right) \approx$ $P\left(
MCS\right) $ and $Q\left( MCS|PH\right) \approx$ $Q\left( MCS\right) $. \ In
the analysis we find evidence for weak, but non-negligible, correlations
between the tests, and use the conditional probabilities to take these into
account.

We collect data and construct a similar model for off-beam data, which
contains no neutrino content, dividing the events into the same categories
as above, and fitting the observed number of events in each category, $%
N_{\nu\nu }^{\prime}$, $N_{\nu CR}^{\prime}$, $N_{CR\nu}^{\prime}$, and $%
N_{CRCR}^{\prime}$ to the parameterizations:%
\begin{widetext}
\begin{align}
\hat{N}_{\nu\nu}^{\prime}  &  =\left(  1-Q\left(  PH\right)  -Q\left(
MCS\right)  +Q\left(  MCS|PH\right)  Q\left(  PH\right)  \right)  \hat{N}%
_{CR}^{\prime},\\
\hat{N}_{CR\nu}^{\prime}  &  =\left(  Q\left(  MCS\right)  -Q\left(
MCS|PH\right)  Q\left(  PH\right)  \right)  \hat{N}_{CR}^{\prime},\\
\hat{N}_{\nu CR}^{\prime}  &  =\left(  1-Q\left(  MCS|PH\right)  \right)
Q\left(  PH\right)  \hat{N}_{CR}^{\prime},\\
\hat{N}_{CRCR}^{\prime}  &  =Q\left(  MCS|PH\right)  Q\left(  PH\right)
\hat{N}_{CR}^{\prime}.
\end{align}
\end{widetext}
In this case the $\nu\nu$ and $CRCR$ categories are expected to be enriched
samples containing muons characteristic of neutrino interactions and cosmic
rays, respectively, while the $CR$$\nu$ and $\nu$$CR$ samples have a mixed
composition. $\hat{N}_{CR}^{\prime}$ is the estimated CR content of the
sample (in practice the number of events in the sample).

Our algorithm uses the eight categories of events in on-beam and off-beam
data to estimate the neutrino content in each multiplicity bin. To calculate the
MC distributions, we replace the on-beam data with the MC samples and
perform the fit again. The same off-beam data sample was used in both fits.
\ In the absence of correlations, the quantities $\hat{N}_{\nu}$, $\hat{N}%
_{CR}$, $\hat{N}_{CR}^{\prime}$, $P\left( PH\right) $, $P\left( MCS\right) $%
, $Q\left( PH\right) $, and $Q\left( MCS\right) $ can be directly
determined from the data with no model inputs. \ The addition of the two
conditional probabilities $P\left( MCS|PH\right) $ and $Q\left(
MCS|PH\right) $ requires use of a model to determine the correlation between the $PH$
and $MCS$ tests. \ These correlations are implemented through the parameterizations%
\begin{align}
P\left( MCS|PH\right) & =\frac{\alpha_{\nu}P\left( MCS\right) }{1+\left(
\alpha_{\nu}-1\right) P\left( MCS\right) } \textrm{  and} \\
Q\left( MCS|PH\right) & =\frac{\alpha_{\text{CR}}Q\left( MCS\right) }{%
1+\left( \alpha_{\text{CR}}-1\right) Q\left( MCS\right) }.
\end{align}
The two new parameters $\alpha_{\nu}$ and $\alpha_{\text{CR}}$ are obtained
from Monte Carlo simulation of neutrino data and from the off-beam data,
respectively. \ If $\alpha_{\nu}=1$, no correlation would exist between the
tests, whereas a large $\alpha_{\nu}$ would imply near total correlation,
with similar conditions applied to $\alpha_{\text{CR}}$.

\subsection{Fitting Procedure}

We construct a likelihood function based on the probability distribution for
partitioning events into one of four categories of a multinomial distribution,
for both on-beam and off-beam data. \ The multinomial probability of
observing $n_{i}$ events in bin $i$, with $i=1,2,3,4$, with the probability
of a single event landing in bin $i$ equal to $r_{i}$ is

\begin{equation}
\begin{split}
M\left( n_{1},n_{2},n_{3},n_{4};r_{1},r_{2},r_{3},r_{4}\right)  =\frac{%
\left( n_{1}+n_{2}+n_{3}+n_{4}\right) !}{n_{1}!n_{2}!n_{3}!n_{4}!} r_{1}^{n_{1}}r_{2}^{n_{2}}r_{3}^{n3}r_{4}^{n_{4}}. 
\end{split}
\end{equation}
The $n_{i}$ are the observed number of events in each bin, and the $%
r_{i} $ are functions of the model parameters.

The likelihood also incorporates the Poisson statistics of observing $%
n_{1}+n_{2}+n_{3}+n_{4}$ in both the on-beam and off-beam data:%
\begin{align}
P_{on\text{-}beam} & =\frac{\hat{N}^{N}}{N!}e^{-\hat{N}}, \\
P_{off\text{-}beam} & =\frac{\hat{N}^{\prime N^{\prime}}}{N^{\prime}!}e^{-\hat{N}^{\prime}},
\end{align}
with%
\begin{align}
\hat{N} & =\hat{N}_{\nu\nu}+\hat{N}_{CR\nu}+\hat{N}_{\nu CR}+\hat{N}_{CRCR},
\\
\hat{N}^{\prime} & =\hat{N}_{\nu\nu}^{\prime}+\hat{N}_{CR\nu}^{\prime}+\hat{N%
}_{\nu CR}^{\prime}+\hat{N}_{CRCR}^{\prime}, \\
N & =N_{\nu\nu}+N_{CR\nu}+N_{\nu CR}+N_{CRCR}, \\
N^{\prime} & =N_{\nu\nu}^{\prime}+N_{CR\nu}^{\prime}+N_{\nu CR}^{\prime
}+N_{CRCR}^{\prime}.
\end{align}

The final likelihood function is 
\begin{widetext}
\begin{align}
L_{TOT}  &  =M_{on\text{-}beam}\left(  N_{\nu\nu},N_{CR\nu},N_{\nu CR},N_{CRCR};\frac{\hat
{N}_{\nu\nu}}{\hat{N}},\frac{\hat{N}_{CR\nu}}{\hat{N}},\frac{\hat{N}_{\nu CR}%
}{\hat{N}},\frac{\hat{N}_{CRCR}}{\hat{N}}\right) \\
&  \times M_{off\text{-}beam}\left(  N_{\nu\nu}^{\prime},N_{CR\nu}^{\prime},N_{\nu CR}^{\prime
},N_{CRCR}^{\prime};\frac{\hat{N}_{\nu\nu}^{\prime}}{\hat{N}^{\prime}}%
,\frac{\hat{N}_{CR\nu}^{\prime}}{\hat{N}^{\prime}},\frac{\hat{N}_{\nu
CR}^{\prime}}{\hat{N}^{\prime}},\frac{\hat{N}_{CRCR}^{\prime}}{\hat{N}^{\prime}%
}\right) \nonumber\\
&  \times\frac{\hat{N}^{N}}{N!}e^{-\hat{N}}\times\frac{\hat{N}^{\prime
N^{\prime}}}{N^{\prime}!}e^{-\hat{N}^{\prime}}.\nonumber
\end{align}
\end{widetext}

The model parameters $\hat{N}_{\nu}$, $\hat{N}_{CR}$, $\hat{N}_{CR}^{\prime}$%
, $P\left( PH\right) $, $P\left( MCS\right) $, $Q\left( PH\right) $, and $%
Q\left( MCS\right) $ and their statistical uncertainties are estimated via
the maximum likelihood method, implemented by minimizing the
negative-log-likelihood

\begin{equation}
\mathcal{L}_{TOT}=-\ln L_{TOT}\text{,}
\end{equation}
using the MIGRAD minimization in the standard MINUIT~\textcolor{Blue}{\cite{MINIUT}} package
in ROOT~\textcolor{Blue}{\cite{ROOT}}.

The fitting procedure can be used to obtain estimates for $\hat{N}_{\nu}$, $%
\hat{N}_{CR}$, $\hat{N}_{CR}^{\prime}$, $P\left( PH\right) $, $P\left(
MCS\right) $, $Q\left( PH\right) $, and $Q\left( MCS\right) $ for each
multiplicity. \ When the probability parameters $P\left( PH\right) $, $%
P\left( MCS\right) $, $Q\left( PH\right) $, and $Q\left( MCS\right) $ are
consistent between multiplicities, we use all multiplicities together in
their determination for improved statistical precision and vary only the
three parameters $\hat{N}_{\nu}$, $\hat{N}_{CR}$, and $\hat{N}_{CR}^{\prime}$
for each individual multiplicity.

\subsection{Results with Simulated Events}

Maximum likelihood fits were performed on all three GENIE simulation samples
to extract the values of seven parameters $\hat{N}_{\nu}$, $\hat{N}_{CR}$, $%
\hat{N}_{CR}^{\prime}$, $P(PH)$, $Q(PH)$, $P(MCS)$, and $Q(MCS)$). Parameters $\alpha_\nu$ and $\alpha_{CR}$ and their uncertainties were extracted from MC and off-beam data samples and kept fixed for the subsequent fits. As
expected, the $PH$ and $MCS$ probabilities show no statistically significant
difference between the three GENIE\ models considered. \ Table~\textcolor{Blue}{\ref%
{tab:fit_values_BNB+Cosmic_data}} lists the values obtained from the fit for
the above-mentioned parameters in the default MC and the MicroBooNE data.

The number of neutrino events in the simulated data samples were extracted
for each observed multiplicity and compared to the known number from the
event generation. Table~\textcolor{Blue}{\ref{tab:mult_MC+cosmic}} and Fig.~\textcolor{Blue}{\ref%
{img:mult_MC_MCtrue}} summarize this comparison. We find that the fit results
agree within statistics with the known inputs, indicating a lack of bias in
our signal estimation technique. \ We have also verified that our method is
insensitive to the signal-to-background ratio of the sample over a range
corresponding to $0.2-5.0\ $times that estimated in the data.

\begin{table}[!hptb]
\caption{Fit parameter results and corresponding uncertainties for the default MC and data samples. The same off-beam data sample was used in both fits.
\ All uncertainties are from the fit and are purely statistical.}
\label{tab:fit_values_BNB+Cosmic_data}
\begin{center}
\begin{ruledtabular}
\begin{tabular}
[c]{ccc}
& \multicolumn{2}{c}{\textbf{Fit results}}\\
\textbf{Parameters} & \textbf{Default MC} & \textbf{Data}\\\hline
\hline
 multicolumn{3}{c}{\textbf{Floating parameters}}\\
 \hline
 $\hat{N}_{\nu}$ & 3405$\pm$159 & 1023$\pm$170\\
$\hat{N}_{CR}$ & 611$\pm$150 & 782$\pm$169\\
$\hat{N}_{CR}^{\prime}$ & 5002$\pm$71 & 5002$\pm$71\\
$P(PH)$ & 0.848$\pm$0.018 & 0.766$\pm$0.050\\
$P(MCS)$ & 0.770$\pm$0.0123 & 0.730$\pm$ 0.039\\
$Q(PH)$ & 0.542$\pm$0.007 & 0.552$\pm$0.007\\
$Q(MCS)$ & 0.537$\pm$0.007 & 0.534$\pm$ 0.007\\
\hline
\multicolumn{3}{c}{\textbf{Fixed parameters}}\\
\hline
$\alpha_\nu$ & 1.32$\pm$0.05  & 1.32$\pm$0.05\\
$\alpha_{CR}$ & 1.36$\pm$0.04  & 1.36$\pm$0.04\\

\end{tabular}
\end{ruledtabular}
\end{center}
\end{table}

\begin{table}[!hptb]
\caption{Fitted and true number of neutrino events for the MC default sample for different multiplicity bins. The last column shows
good agreement between the fit results and true content for different bins.}
\label{tab:mult_MC+cosmic}
\begin{center}
\begin{ruledtabular}
\begin{tabular}
[c]{cccc}%
\textbf{Multiplicities} & \textbf{Fit $N_{\nu}$} & \textbf{True $N_{\nu}$} &
\textbf{True-Fit $\chi^{2}$/ndf}\\\hline
1 & 2070$\pm$63 & 2152 & 1.7\\
2 & 1112$\pm$44 & 1092 & 0.2\\
3 & 210$\pm$14 & 208 & 0.0\\
4 & 18$\pm$4 & 18 & 0.0\\
5 & 3$\pm$2 & 3 & 0.0\\
\end{tabular}
\end{ruledtabular}
\end{center}
\end{table}

\begin{figure*}[!ht]
\centering
\subfloat
{\includegraphics[width=0.5\linewidth]{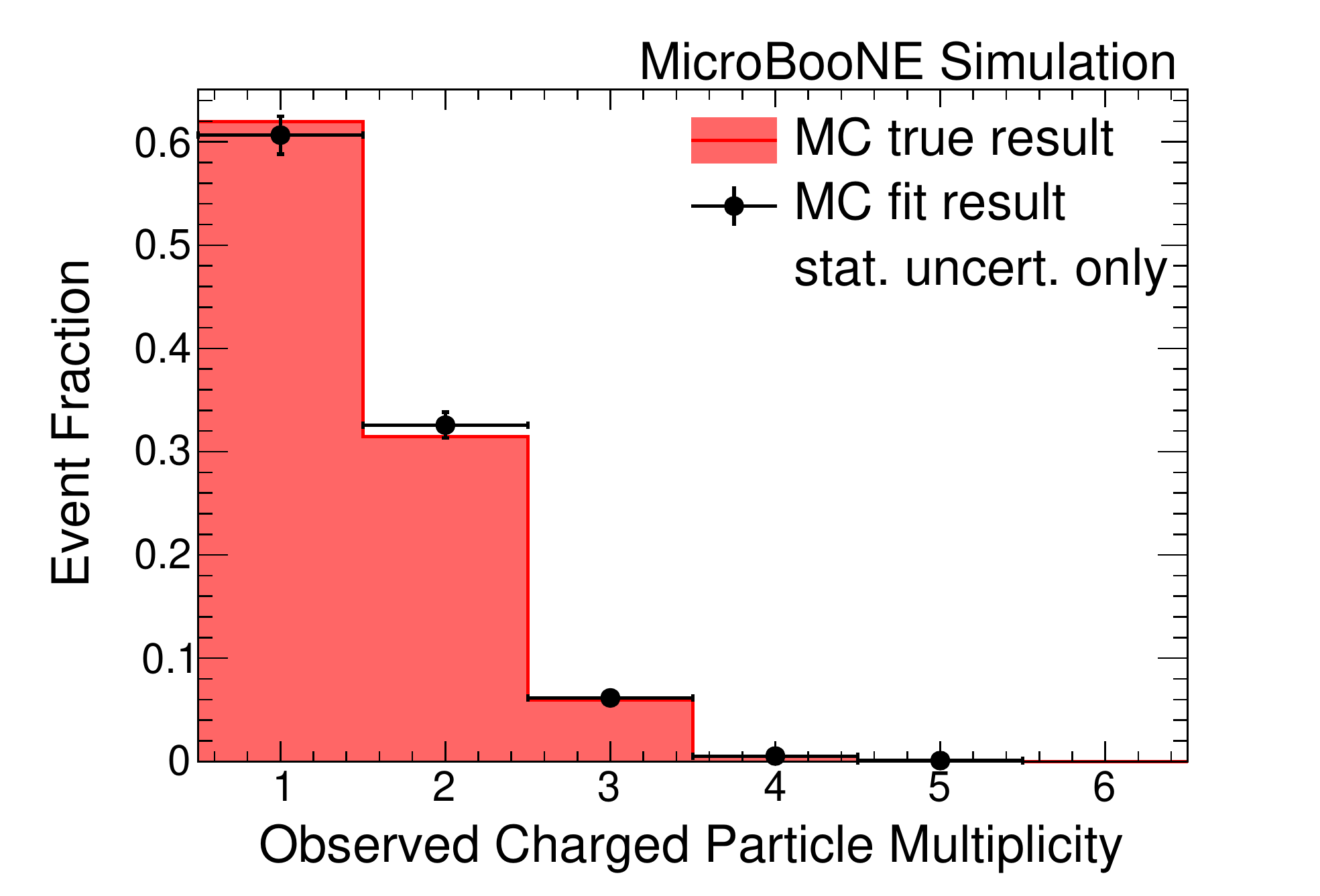}} 
\centering
\subfloat
{\includegraphics[width=0.5\linewidth]{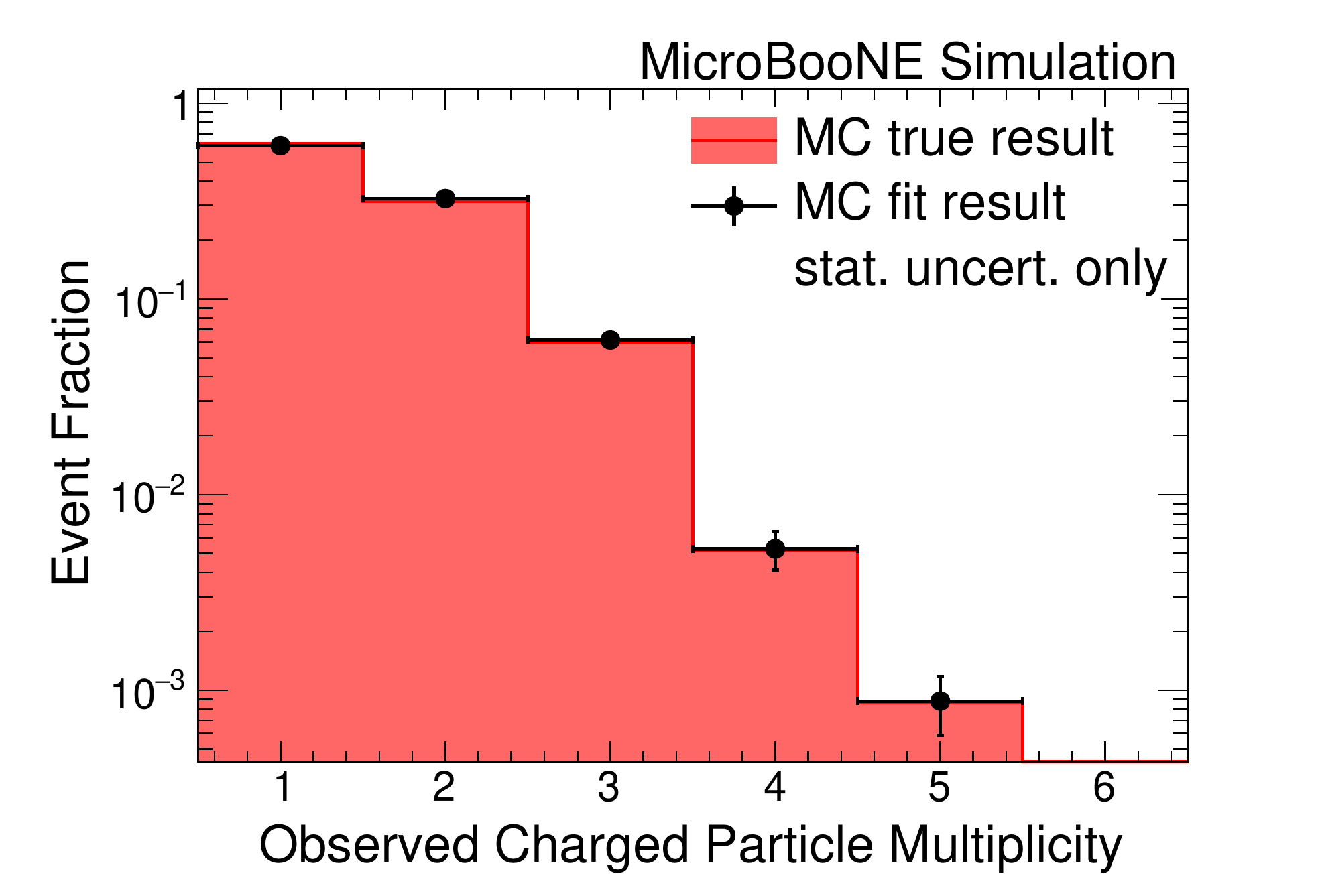}} 
\caption{Overlaid true and fitted observed neutrino multiplicity distributions from the MC default sample in linear scale (left) and in log y scale (right).}
\label{img:mult_MC_MCtrue}
\end{figure*}

\section{\label{Error Analysis}Statistical and Systematic Uncertainty
Estimates}

\begin{figure*}[!ht]
\centering
\includegraphics[width=0.6\textwidth]{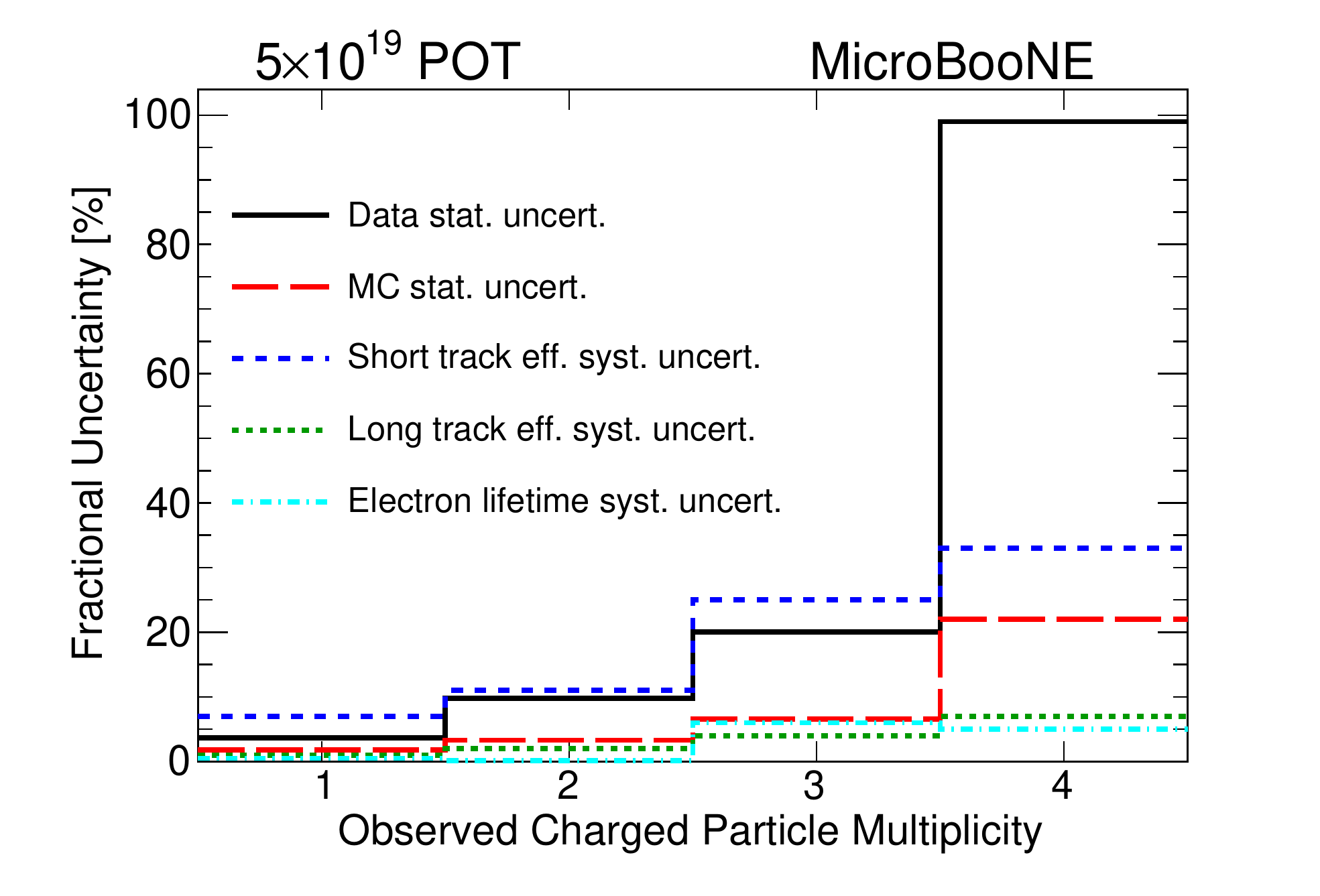} 
\caption{Percentage uncertainty distributions from different systematic and statistical sources as a function of observed charged particle multiplicity.}
\label{img:uncertain}
\end{figure*}

Table~\textcolor{Blue}{\ref{tab:sys_uncer}} presents the percentage estimates for statistical and systematic
uncertainties from different sources. \
Figure~\textcolor{Blue}{\ref{img:uncertain}} presents a plot of each uncertainty source as a
function of observed multiplicity.

\begin{table*}[!ht]
\caption{Statistical and systematic uncertainties estimates from data and MC. }
\label{tab:sys_uncer}
\begin{center}
\begin{ruledtabular}
\begin{tabular}
[c]{cccccc}
& \multicolumn{5}{c}{\textbf{Uncertainty Estimates}}\\
\textbf{Uncertainty Sources} & \textbf{mult=1} & \textbf{mult=2} &
\textbf{mult=3} & \textbf{mult=4} \\\hline
Data statistics & 4\% & 10\% & 20\% & 99\% \\
MC statistics & 2\% & 3\% & 7\% & 22\% \\
Short track efficiency & 7\% & 11\% & 25\% & 33\% \\
Long track efficiency & 1\% & 2\% & 4\% & 7\% \\
Background model systematics & 2\% & 2\% & 0\% & 0\% \\
Flux shape systematics & 0\% & 0.4\% & 0.2\% & 0.5\% \\
Electron lifetime systematics & 0.5\% & 0.1\% & 6\% & 5\% \\
\end{tabular}
\end{ruledtabular}
\end{center}
\end{table*}

\subsection{Statistical Uncertainties}

Statistical uncertainties are returned from the MINUIT package used in our
fitting for both data and MC samples. These uncertainties include
contributions from the CR background in our fitting procedure, and our procedure includes the CR background systematic uncertainty in a contribution to the total statistical uncertainty. \ Both data and MC statistics contribute substantially to the overall uncertainties in our data, as shown
in Fig.~\textcolor{Blue}{\ref{img:uncertain}}. 

\subsection{Short Track Efficiency Uncertainties}

\label{systematic errors} The dominant systematic uncertainty originates from the differences in the efficiency between data and simulation for reconstructing short-length hadron tracks. The overall
efficiencies of the Pandora reconstruction algorithms are
a strong function of the number of hits of the tracks, with a plateau not
being reached until of order of several hundred hits. \ The inclusive
efficiencies for reconstructing protons or pions at the $15$ collection plane hit
threshold is estimated to be $\left\langle \epsilon\right\rangle =0.45\pm0.05
$. \ The absolute efficiency value is not used in this analysis, but we use
this estimate to conservatively assign a mean efficiency uncertainty of $%
\delta=15\%$. 

We then estimate the effect of an efficiency uncertainty on multiplicity by
the following procedure: Consider a track in an event in a multiplicity bin $N$. \ If one lowers the tracking efficiency by the factor $1-\delta $, then there is a $1-\delta $ probability that the
track remained reconstructed and the event stayed in that multiplicity bin, and a
probability $\delta $ that the track would not have been reconstructed and
that the event would thus have a \emph{lower} multiplicity. \ \ If the
overall multiplicity is $N$, with $N-1$ short tracks and one long track,
and each track's reconstruction probability is reduced by a factor $1-\delta 
$, then an overall fraction of events $\left( 1-\delta \right) ^{N-1}$ will
remain in the bin, and a fraction $1-\left( 1-\delta \right) ^{N-1}$ will
migrate to lower multiplicity bins. The fraction of tracks that migrate to
multiplicity $N^{\prime }<N$ from bin $N$, $f\left( N^{\prime };N,\delta
\right) $, is given by binomial statistics:

\begin{equation}
f\left( N^{\prime };N,\delta \right) =\frac{\left( N-1\right) !}{\left(
N^{\prime }-1\right) !\left( N-N^{\prime }\right) !}\left( 1-\delta \right)
^{N^{\prime }-1}\delta ^{N-N^{\prime }}.
\end{equation}

We use this result to generate the expected observed CPMD in simulation that
would emerge from lowering the tracking efficiency by the factor $1-\delta $
compared to the default simulated CPMD. \ The difference between the two
distributions is then taken as the systematic uncertainty assigned to short
track efficiency, with the assumption that the effect of increasing the
default efficiency by a factor $1+\delta $ would produce a symmetric change.
Table~\textcolor{Blue}{\ref{Short track efficiency}} summarizes this study for the three
GENIE\ models used. \ The observed multiplicity $=1$ probability increases because of \textquotedblleft feed
down\textquotedblright\ of events from higher multiplicity, due to the
lowered efficiency, mainly from observed multiplicity $=2$. \ The other
observed multiplicity probabilities decrease accordingly. \ The largest
effects are in high multiplicity bins because the loss of events from
lowering the efficiency by the factor $\left( 1-\delta \right) $ varies as $%
\left( 1-\delta \right) ^{N-1}$ for multiplicity bin $N$. \ Monte Carlo
simulations show that \textquotedblleft fake tracks\textquotedblright\ that
could move events to higher multiplicity are rare. We have observed no statistically significant differences in the shape distributions after adjusting the efficiency by the constant per-track factor implied by the pull factor.

\begin{table}[!hptb]
\caption{Relative change in observed multiplicity probabilities
corresponding to a $-15\%$ uniform reduction in short charged particle
tracking efficiencies for three GENIE models: default, MEC, and TEM. The
missing entry for observed multiplicity 5 in TEM is due to no event being
generated with that observed multiplicity.}
\label{Short track efficiency}\centering
\begin{ruledtabular}
\begin{tabular}[c]{cccc}%
\textbf{Observed multiplicity} & $\frac{\Delta P_{n}}{P_{n}}$\textbf{Default}
& $\frac{\Delta P_{n}}{P_{n}}$\textbf{MEC} & $\frac{\Delta P_{n}}{P_{n}}%
$\textbf{TEM}\\\hline
$1$ & $+7\%$ & $+7\%$ & $+8\%$\\
$2$ & $-11\%$ & $-12\%$ & $-12\%$\\
$3$ & $-25\%$ & $-25\%$ & $-25\%$\\
$4$ & $-33\%$ & $-36\%$ & $-39\%$\\
$5$ & $-44\%$ & $-48\%$ & --\\
\end{tabular}
\end{ruledtabular}
\end{table}

\subsection{Long Track Efficiency Uncertainties}

To first order, the efficiency for reconstructing tracks with length $>$ $75$ cm is not 
expected to affect the observed
multiplicity distribution, as it is common to all multiplicities and cancels
in the ratio when forming observed multiplicity probabilities. \ At second order,
however, a multiplicity dependence that changes the 
distribution of observed multiplicity without affecting the overall number
of events is possible. \ A plausible model for this is that higher multiplicity in an
event helps Pandora better define a vertex, and thus increases the chance that the event passes the $\nu_{\mu}$ CC selection filter.

We estimate the size of this effect by comparing the efficiencies obtained
with the Pandora package for simulated quasi-elastic final states in which
both the proton and muon are reconstructed, to charged pion resonance final
states in which the proton, pion, and muon are all reconstructed. \ From
this study we conclude that the efficiency for finding the muon in final
states where all charged particles are reconstructed could be up to $3\%$
higher for charged pion resonance events (observed multiplicity 3) than
quasi-elastic events (observed multiplicity 2). We then assume, for the
purpose of uncertainty estimation, that this relative enhancement seen for
higher observed multiplicity events in the MC is absent in the
data.

Table~\textcolor{Blue}{\ref{Long track efficiency}} summarizes this study. \ Effects are
generally small compared to those seen in Table~\textcolor{Blue}{\ref{Short track efficiency}}%
. \ No dependence on GENIE variant is found.

\begin{table}[!hptb]
\caption{Relative change in observed multiplicity probabilities
corresponding to increasing the conditional probability for reconstructing
the long track by $3\%$ for each additional track found in the event, as
suggested by Pandora studies of QE and charged pion resonance production for
three GENIE models: default, MEC, and TEM. The missing entry for observed
multiplicity 5 in TEM is due to no event being generated with that observed
multiplicity.}
\label{Long track efficiency}\centering
\begin{ruledtabular}
\begin{tabular}[c]{cccc}%
\textbf{Observed multiplicity} & $\frac{\Delta P_{n}}{P_{n}}$\textbf{Default}
& $\frac{\Delta P_{n}}{P_{n}}$\textbf{MEC} & $\frac{\Delta P_{n}}{P_{n}}%
$\textbf{TEM}\\\hline
$1$ & $-1\%$ & $-1\%$ & $-1\%$\\
$2$ & $+2\%$ & $+2\%$ & $+2\%$\\
$3$ & $+4\%$ & $+4\%$ & $+2\%$\\
$4$ & $+7\%$ & $+7\%$ & $+7\%$\\
$5$ & $+9\%$ & $+9\%$ & --\\
\end{tabular}%
\end{ruledtabular}
\end{table}

\subsection{Background Model Uncertainties}

In the signal extraction fitting procedure, two conditional parameters ($%
\alpha _{\nu }$ and $\alpha _{CR}$) were extracted from the Monte Carlo
simulation and off-beam data. To
calculate the systematic uncertainties on these parameters, their values
were varied by $\pm 1\sigma $ of their statistical uncertainty. Those values were
propagated in the observed charged particle multiplicity distribution. \ \ We
also extracted the $\alpha _{\nu }$ and $\alpha _{CR}$ values separately
from the GENIE default, GENIE+TEM, and GENIE+MEC models. \ The effect from this
systematic variation were found to be very small.

\subsection{Flux Shape Uncertainties}

\begin{figure}[!hptb]
\centering
\includegraphics[width=0.5\textwidth]{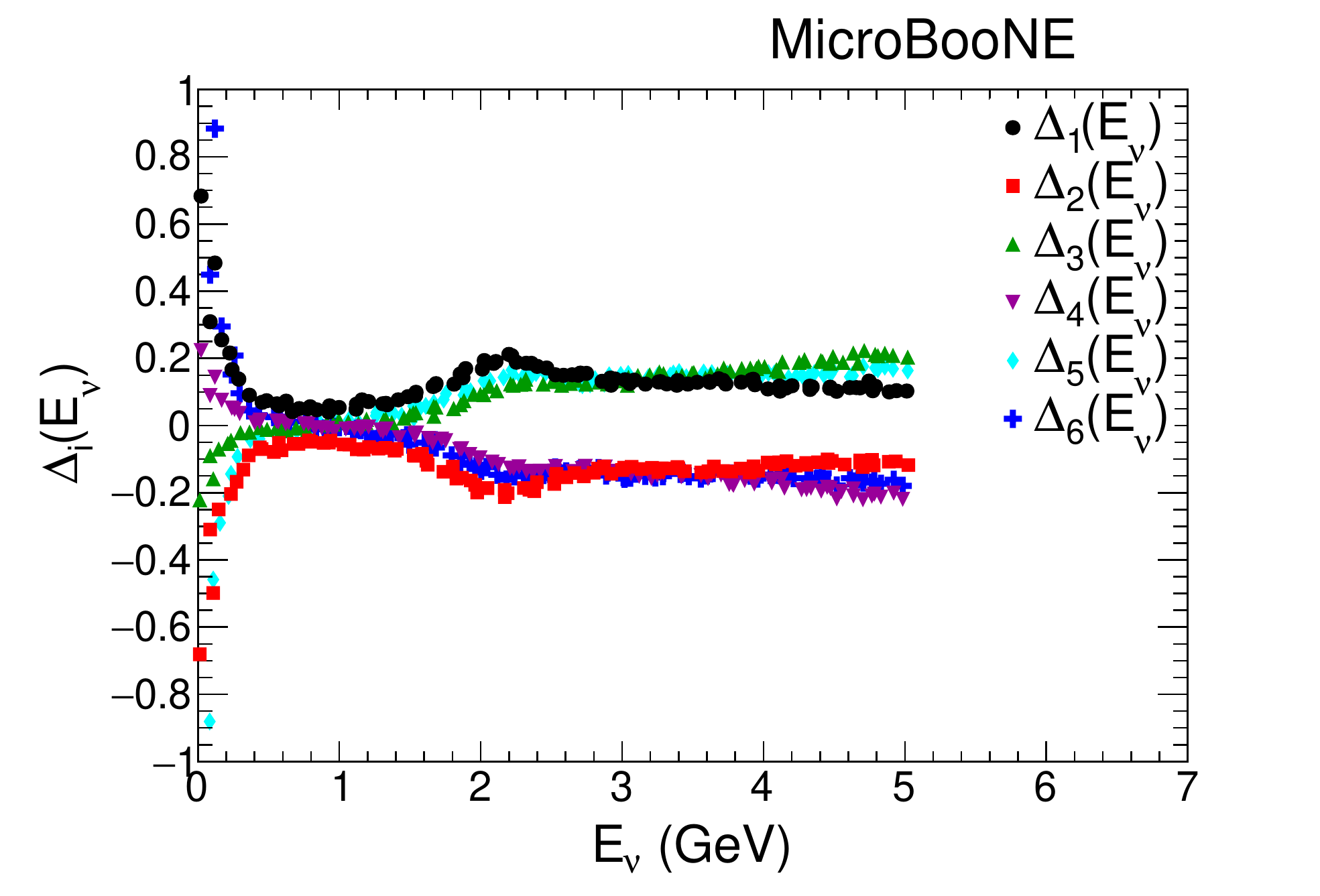} 
\caption{Beam flux shifts for the parameterizations $\Delta_{i}\left( E_{%
\protect\nu }\right) $, $i=1-6$. \ The variations $\Delta_{1}\left( E_{%
\protect\nu}\right) $ and $\Delta_{2}\left( E_{\protect\nu}\right) $ define
the envelope of flux uncertainties for the BNB. }
\label{img:flux shifts}
\end{figure}

Variations in flux can be parameterized by%
\begin{equation}
\Phi\left( E_{\nu}\right) \rightarrow\left( 1+\Delta\left( E_{\nu}\right)
\right) \Phi\left( E_{\nu}\right) ,
\end{equation}
where $\Phi\left( E_{\nu}\right) $ is the neutrino flux at neutrino energy $%
E_{\nu}$ and $\Delta\left( E_{\nu}\right) $ is the fractional uncertainty in
the flux at that energy. An energy-independent $\Delta\left( E_{\nu }\right) 
$ has no effect on the observed multiplicity distributions as this
measurement is independent of absolute normalization. \ On the other hand, raising the high energy flux relative to the low energy flux
could enhance the contributions of higher multiplicity
resonance and DIS\ processes. We confine ourself to considering highly
correlated energy-dependent shifts, denoted as $\Delta_{i}\left(
E_{\nu}\right) $ for $i=1-6$ via an approximate procedure that should be
conservative. \ These shifts, shown in Fig.~\textcolor{Blue}{\ref{img:flux shifts}}, are
allowed to modify the BNB flux within uncertainties determined by the
MiniBooNE\ collaboration~\textcolor{Blue}{\cite{BNB reference}}. \ The first two variations
simply shift all flux values up ($\Delta_{1}\left( E_{\nu}\right) $) or down
($\Delta_{2}\left( E_{\nu }\right) $) according to the flux
uncertainty envelope. \ The next two enhance the high energy flux ($%
\Delta_{3}\left( E_{\nu}\right) $) or low energy flux ($\Delta_{4}\left(
E_{\nu}\right) $) linearly with neutrino energy, with the variation taken to
be zero at the average neutrino energy. \ The final two variations enhance high
energy flux ($\Delta_{5}\left( E_{\nu}\right) $) or low energy flux ($%
\Delta_{6}\left( E_{\nu}\right) $) logarithmically with neutrino energy,
with the variation taken to be zero at the average energy. As expected,
shifts that are positively correlated across all energies produce negligible
differences, but even shifts that produce sizable distortions between high
and low energies contribute systematic uncertainties that are small.

\subsection{Electron Lifetime Uncertainties}

The measured charge from muon-induced ionization can vary within the detector
volume due to the finite probability for drifting electrons to be captured
by electronegative contaminants in the liquid argon. \ This capture
probability can be parameterized by an electron lifetime $\tau $. We perform
our analysis on simulation with two lifetimes that safely bound those measured during detector operating conditions, $\tau =$ $6$ msec
and $\tau =\infty $ msec. \ The resulting distribution of percentage
uncertainty as a function of multiplicity in Fig.~\textcolor{Blue}{\ref{img:uncertain}} shows
that the electron lifetime uncertainties minimally affect the multiplicity.

\subsection{Other Sources of Uncertainty Considered}

A systematic comparison was performed on all kinematic quantities entering
this analysis between off-beam CR data and the CR events simulated with
CORSIKA. No statistically significant discrepancies were observed between
event selection pass rates applied to off-beam data and MC simulation.

A check of possible time-dependent detector response systematics was also
performed by dividing the data into two samples and performing the analysis
separately for each sample. Differences between the two samples are
consistent within statistical fluctuations.

The data are not corrected for $\nu_{\mu}$ NC, $\nu_{e}$, $\bar{\nu}_{e}$,
or $\bar{\nu}_{\mu}$ backgrounds. An assumption is made that the Monte Carlo
simulation adequately describes these non $\nu_{\mu}$ CC backgrounds.
Section~\textcolor{Blue}{\ref{expectations}} shows that these backgrounds, in total, are
expected to be less than $10\%$ of the final sample; their impact on
the final distributions is generally small.


\section{\label{Results}Results}

\subsection{Observed Charged Particle Multiplicity Distribution}

Following the implementation of the signal extraction procedure and
verification through closure test on MC events, we execute the same maximum
likelihood fit on data. Table~\textcolor{Blue}{\ref{tab:fit_values_BNB+Cosmic_data}}
lists the values of the fit parameters obtained for the data; and Table~\textcolor{Blue}{\ref%
{tab:mult_fit_data}} lists the number of neutrino events in different
multiplicity bins for the data. While our method does not require
this to be the case, we note that the fitted PH and MCS test probabilities $%
P(PH)$, $Q(PH)$, $P(MCS)$, and $Q(MCS)$ agree in data and simulation within
statistical uncertainties. This provides evidence that the simulation correctly describes the muon PH and MCS tests used in the analysis.

Area normalized, bin-by-bin fitted multiplicity distributions from three
different GENIE predictions overlaid on data are presented in Fig.~\textcolor{Blue}{\ref%
{img:final_mult_dist}} where data error bars include statistical uncertainties
obtained from the fit and the MC error bands include MC statistical and
systematic uncertainties that are listed in Table~\textcolor{Blue}{\ref{tab:sys_uncer}} added in
quadrature.

In general the three GENIE\ models agree within uncertainties
with one another, and agree qualitatively with the data. 
There are indications that GENIE overestimates the mean charged particle multiplicity relative to the data. We emphasize that no tuning or fitting has been performed to this or any of the other kinematic distributions.  

\begin{table}[!hptb]
\caption{Fitted number of neutrino events for the data sample in
different multiplicity bins. The uncertainties correspond to the statistical
uncertainty estimates obtained from the fit. The percentages correspond to the fraction of events in each category.}
\label{tab:mult_fit_data}
\begin{center}
\begin{ruledtabular}%
\begin{tabular}{ccc}
\textbf{Multiplicities} & \textbf{Fitted $N_{\nu}$} & \textbf{Event fraction} \\ \hline
1 & 732$\pm$53 & 72\% \\ 
2 & 260$\pm$29 & 26\% \\ 
3 & 26$\pm$5 & 2.6\% \\ 
4 & 1$\pm$1 & 0.10\% \\ 
5 & 0$\pm$0 & 0\% \\ 
\end{tabular}
\end{ruledtabular}
\end{center}
\end{table}

\begin{figure*}[!ht]
\centering
\subfloat
{\includegraphics[width=0.5\linewidth]{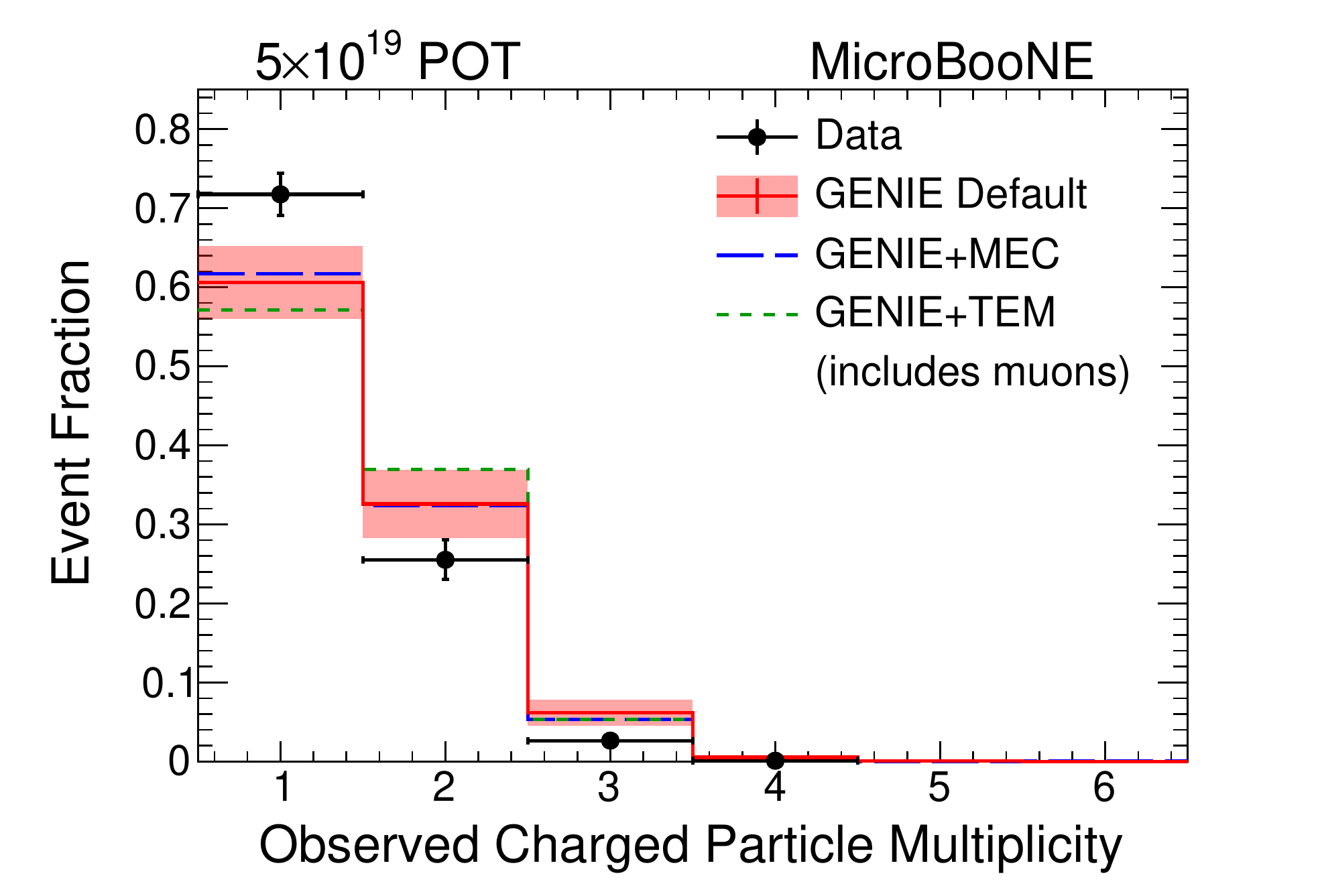}} 
\centering
\subfloat
{\includegraphics[width=0.5\linewidth]{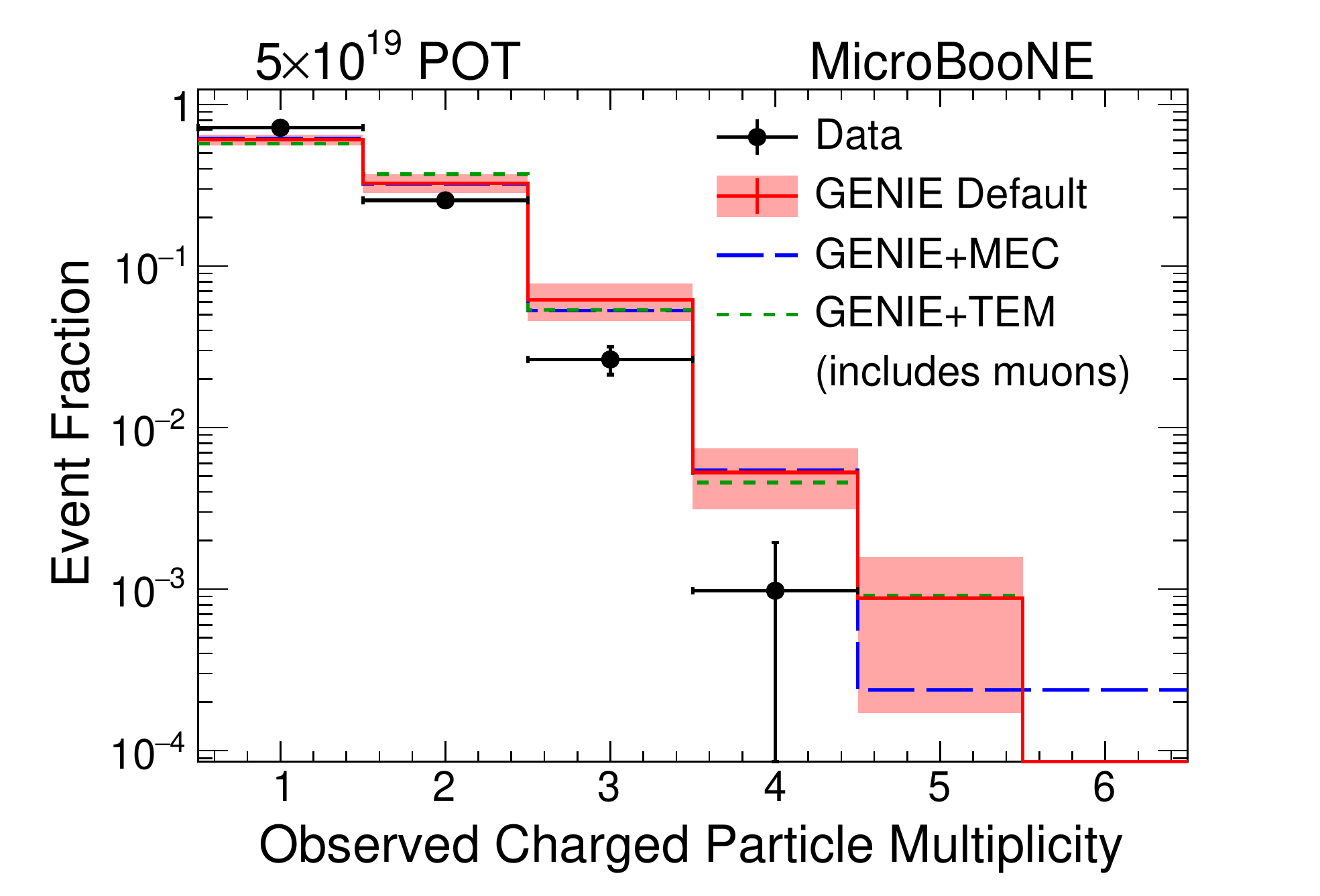}} 
\caption{Bin-by-bin normalized multiplicity distributions using $5\times10^{19}$ POT MicroBooNE data compared with three GENIE predictions (left) in linear scale, (right) in log scale. The data are CR background subtracted. Data error bars include statistical uncertainties obtained from the fit. Monte Carlo error bands include MC statistical uncertainties from the fit and systematic uncertainty contributions added in quadrature.}
\label{img:final_mult_dist}
\end{figure*}

\subsection{Observed Kinematic Distributions}

\label{signal extraction}

A key technical feature of our analysis entails performing tests on the
pulse height and multiple Coulomb scattering behavior of hits on the long
contained track in each event. \ This allows a categorization of events in
each multiplicity into four categories according to whether the long track
passes or fails the PH and MCS tests: $\left( PASS,PASS\right) $, $\left(
PASS,FAIL\right) $, $\left( FAIL,PASS\right) $, and $\left( FAIL,FAIL\right) 
$. We have shown that the $\left( PASS,PASS\right) $ category is
\textquotedblleft neutrino-enriched\textquotedblright\ and the $\left(
FAIL,FAIL\right) $ category is cosmic-ray-dominated. \ The mixed cases $%
\left( PASS,FAIL\right) $ and $\left( FAIL,PASS\right) $ provide samples with intermediate signal-to-background ratios.

Our fit to the distribution of the eight event categories in on-beam and
off-beam data allows us to estimate the number of neutrino events $\hat{N}_{\nu i}$ and the number of corresponding background CR\
events $\hat{N}_{CRi}$ for each observed multiplicity $i$. \
Once $\hat{N}_{\nu n}$ and $\hat{N}_{CRn}$
are established, we can obtain a prediction for the content of any bin $k$
of any kinematic quantity $X_{ij}$ associated with track $j$ in an observed
multiplicity $i$ event in any $\left( PH,MCS\right) $ test combination:%
\begin{widetext}
\begin{equation}
\text{model}\left(  X_{ij},PH,MCS\right)  _{k}=\hat{N}_{\nu i}
\hat{x}_{\nu,ij}\left(  PH,MCS\right)  _{k}+\hat{N}_{CRi}
\hat{x}_{CR,ij}\left(  PH,MCS\right)  _{k}.\label{Model definition}%
\end{equation}
\end{widetext}Here $\hat{x}_{\nu ,ij}\left( PH,MCS\right) _{k}$ is an
area-normalized histogram of $X_{ij}$ for \textquotedblleft true neutrino
events\textquotedblright\ in a given category obtained from a
\textquotedblleft MC\textquotedblright\ sample, and $\hat{x}%
_{CR,ij}\left( PH,MCS\right) _{k}$ is an area-normalized histogram of $X_{ij}
$ for CR events obtained from off-beam data. This distribution can be
compared to the corresponding one for data in each category, data$\left(
X_{ij},PH,MCS\right) _{k}$.

In short, we assume that the observed distribution of events consists
of a mix of neutrino events plus CR\ events. \ The proportions of the mix in
each category are fixed by the output of our fit, which, by construction,
constrains the normalization of the model to equal that of the data. \ We
emphasize that only the PH and MCS\ tests have been used to extract the
neutrino interaction signal sample; no information from any quantity $X_{ij}$
is used.

\subsection{Checks on Distributions lacking Dynamical Significance}

Several kinematic properties of neutrino interactions depend only weakly on the neutrino interaction model; these include the reconstructed vertex positions, the initial and final coordinates of the long track, and the azimuthal angles of individual tracks. \ These distributions provide checks on the overall
signal-to-background separation provided by the test-category fits and flux
and detector modeling. \ They also test for differences between the modeling
of neutrino events, which depend on the GEANT detector simulation, and CR\
events, which use the off-beam data and thus do not depend on detector
simulation.

\begin{figure*}[!hpt]
\centering
\subfloat[][The neutrino-enriched sample (PH pass, MCS pass)]
{\includegraphics[width=.36\textwidth]{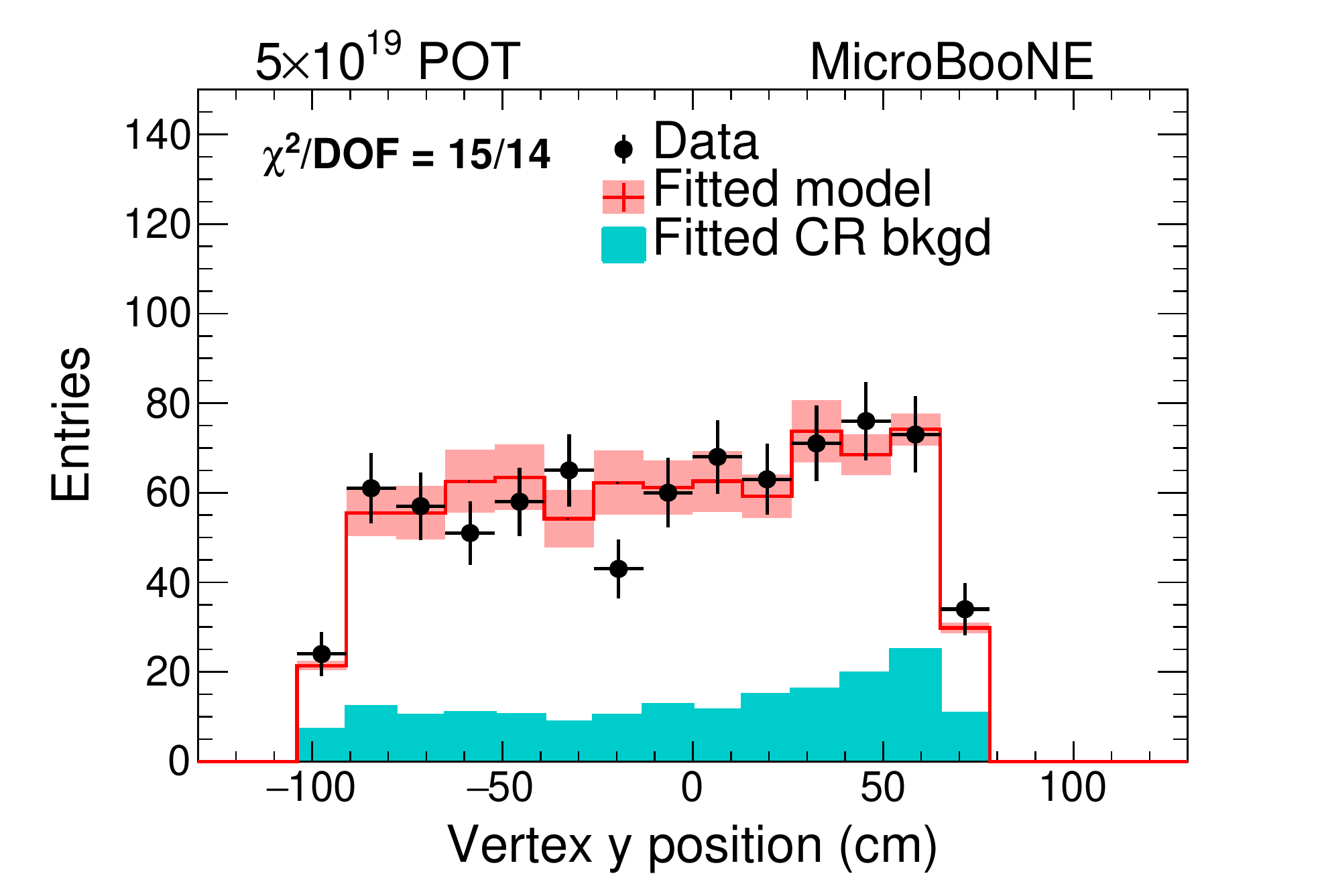}} \quad
\subfloat[][The mixed sample (PH pass, MCS fail)]
{\includegraphics[width=.36\textwidth]{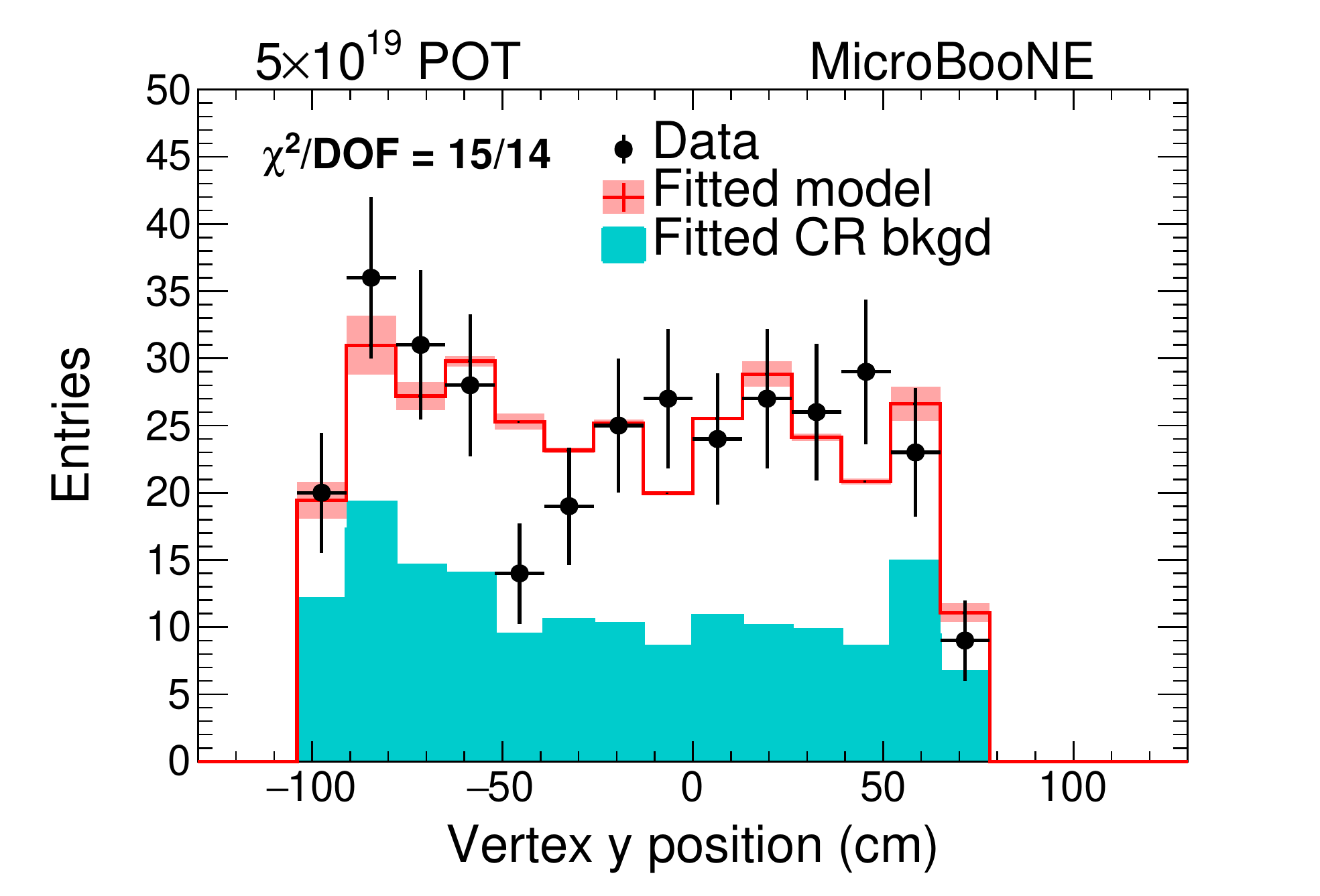}}\quad
\subfloat[][The mixed sample (PH fail, MCS pass)]
{\includegraphics[width=.36\textwidth]{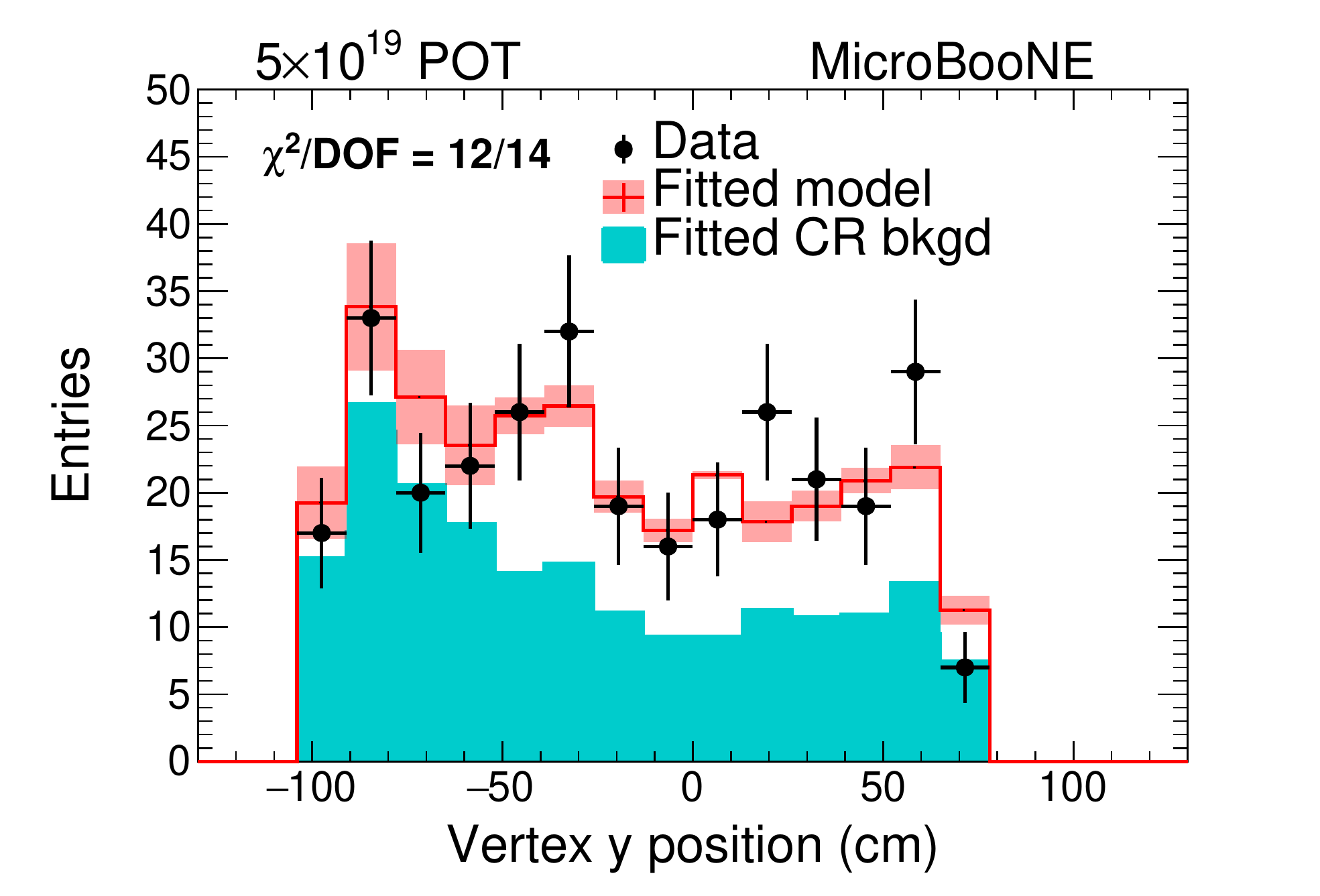}}\quad
\subfloat[][The background-enriched sample (PH fail, MCS fail)]
{\includegraphics[width=.36\textwidth]{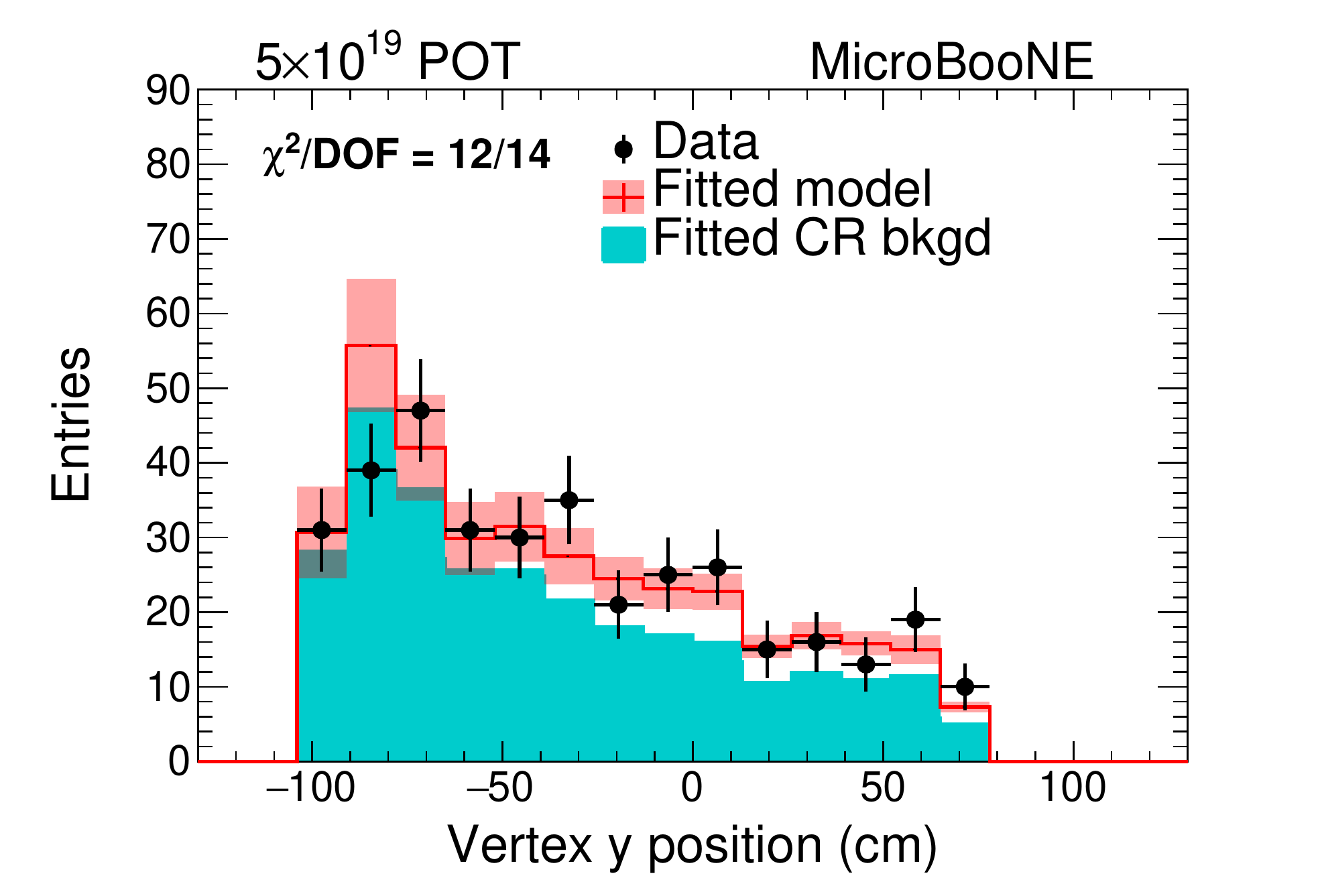}}\quad
\caption{Neutrino interaction reconstructed vertex position along $y$-axis for data and GENIE default MC. Neutrino-enriched sample is nearly flat as expected. The asymmetry in the CR-background-enriched category corresponds to ``upwards-going cosmics" which is a known feature of the selection.}
\label{Dvrtxy}
\end{figure*}

As an example, we show the observed distributions for the
selected vertex $y$ position for the candidate muon track from the full selected sample in Fig.~\textcolor{Blue}{\ref%
{Dvrtxy}}. \ For this and all subsequent distributions, the on-beam data
events are indicated by plotted points with statistical error bars. \ The
model prediction is shown by a colored band (red for GENIE default, green
for GENIE+TEM, and blue for GENIE+MEC) with the width of the band
indicating the correlated statistical plus efficiency systematic uncertainty
from using common $N_{\nu,n},N_{CR,n}$ values for all bins of all
distributions of a given multiplicity bin $n$. The CR\ contribution to a
distribution in a given category is shown by the shaded cyan region. For example, Fig.~\textcolor{Blue}{\ref{Dvrtxy}} compares the on-beam data to GENIE default MC sample and also shows the CR background.

The signal-enriched $\left( PASS,PASS\right) $ category for vertex $y$ has
the nearly flat distribution expected for a neutrino event sample with a
small CR\ background. Note that in our selection, we only allow candidate muon tracks initial $y$ position $<70$ cm. This cut rejects many cosmic rays that produce a downward trajectory in the final selected
sample. The remaining background is dominated by cosmic rays with an apparent upward
trajectory. This can be seen in the background-enriched sample $\left(
FAIL,FAIL\right) $ in the vertex $y$ distribution where a peak at negative $y$
values corresponds to \textquotedblleft upwards-going\textquotedblright CR.
\ 

\begin{figure*}[!hpt]

\centering
\subfloat[][The neutrino-enriched sample (PH pass, MCS pass)]
{\includegraphics[width=.36\textwidth]{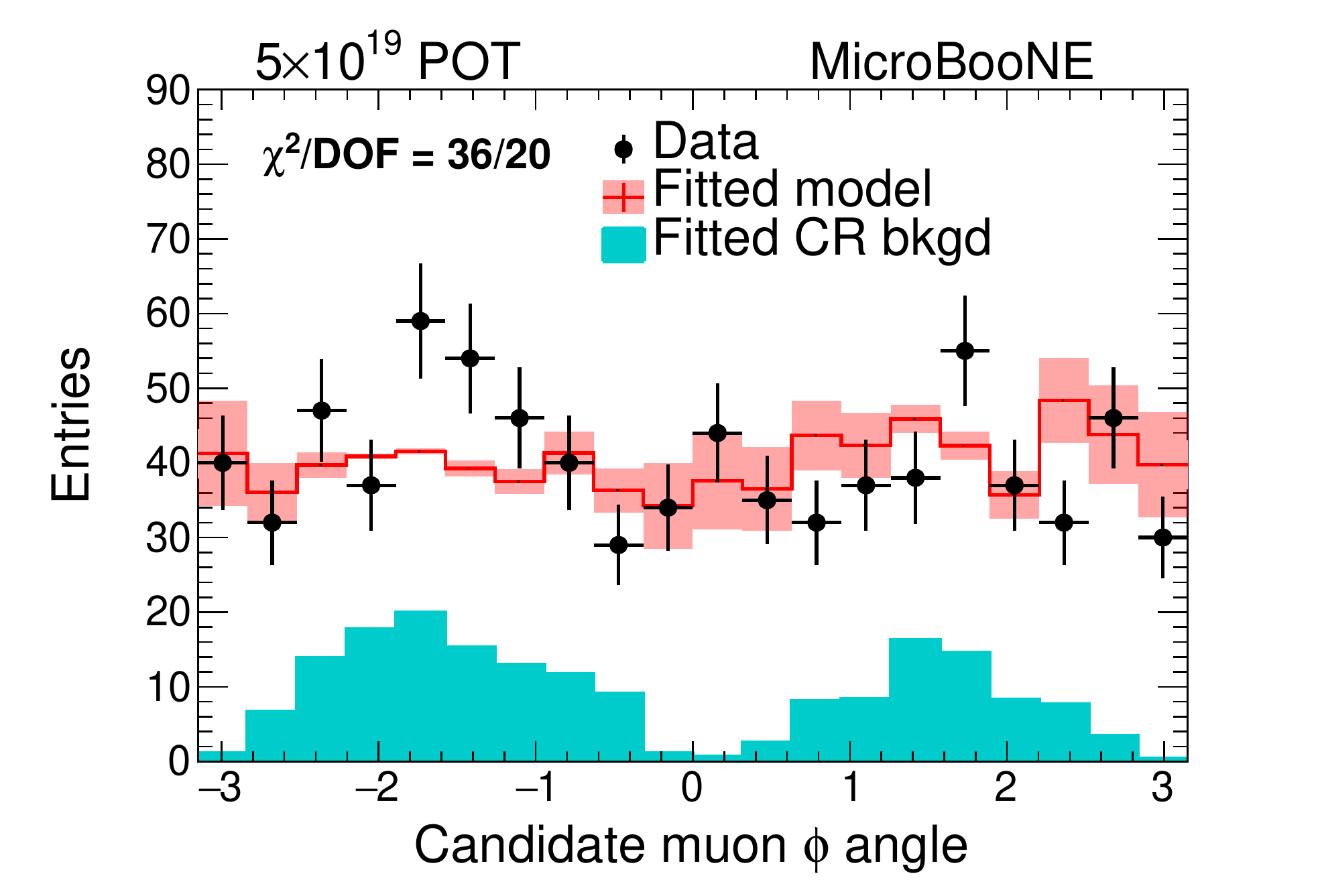}} \quad
\subfloat[][The mixed sample (PH pass, MCS fail)]
{\includegraphics[width=.36\textwidth]{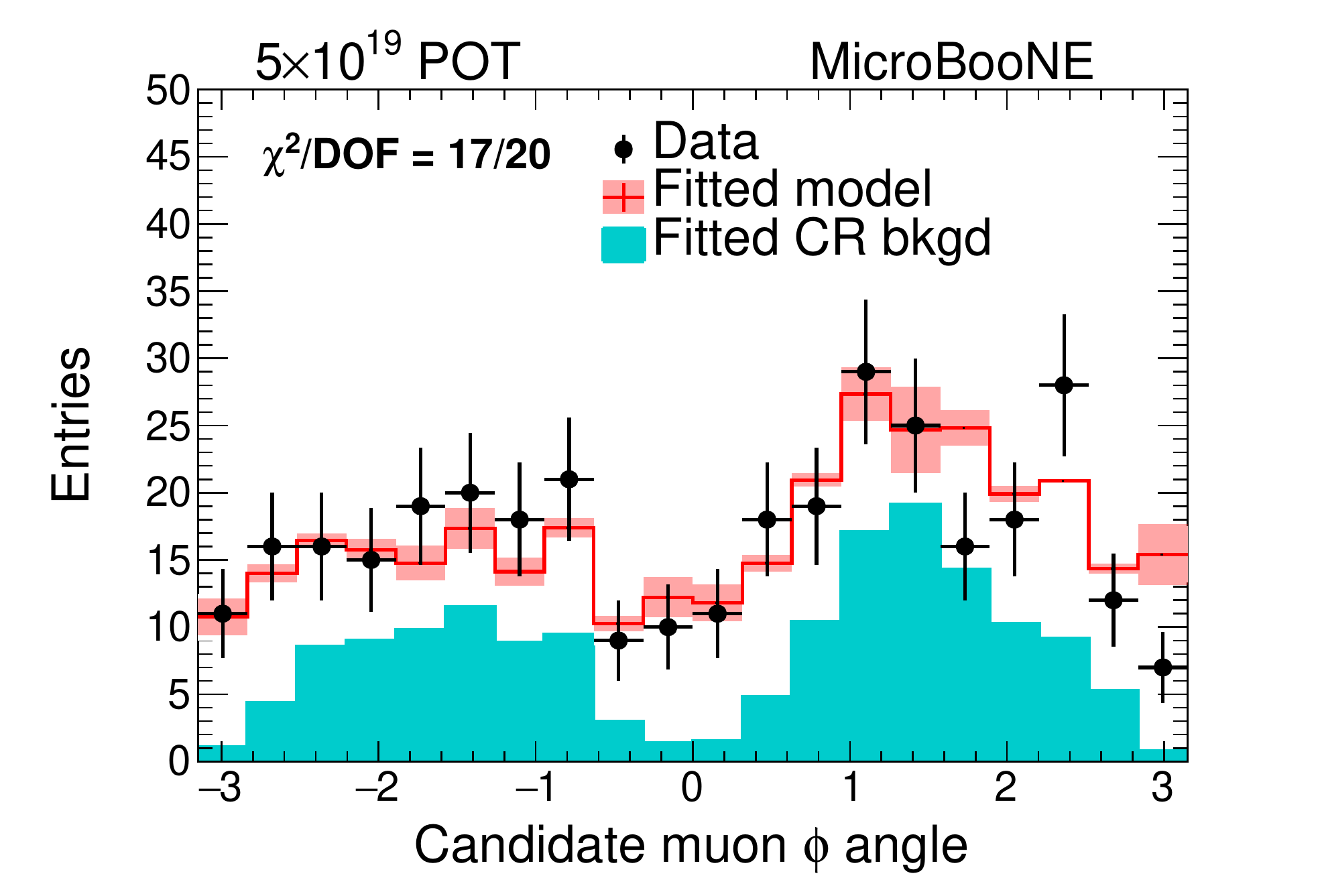}}\quad
\subfloat[][The mixed sample (PH fail, MCS pass)]
{\includegraphics[width=.36\textwidth]{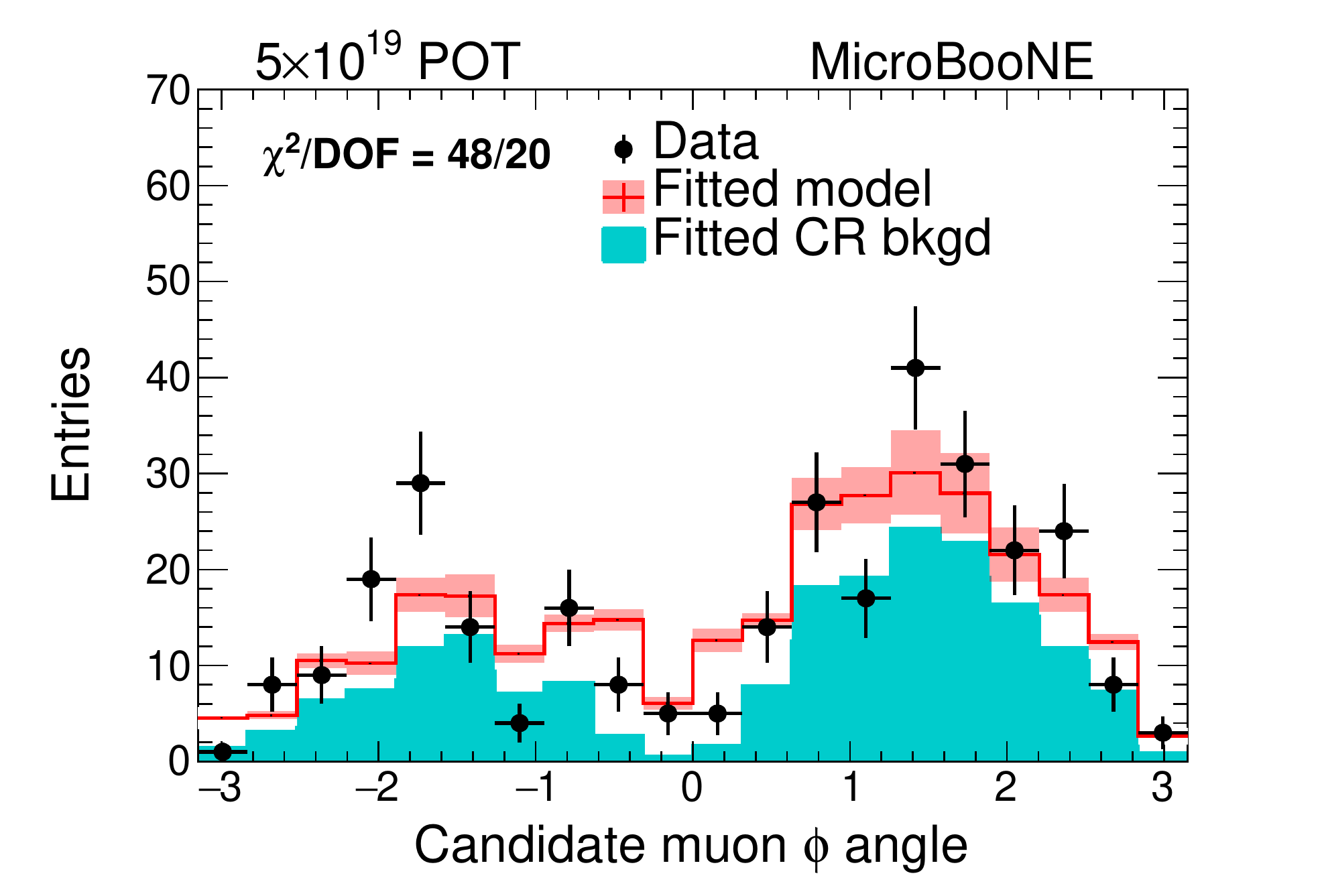}}\quad
\subfloat[][The background-enriched sample (PH fail, MCS fail)]
{\includegraphics[width=.36\textwidth]{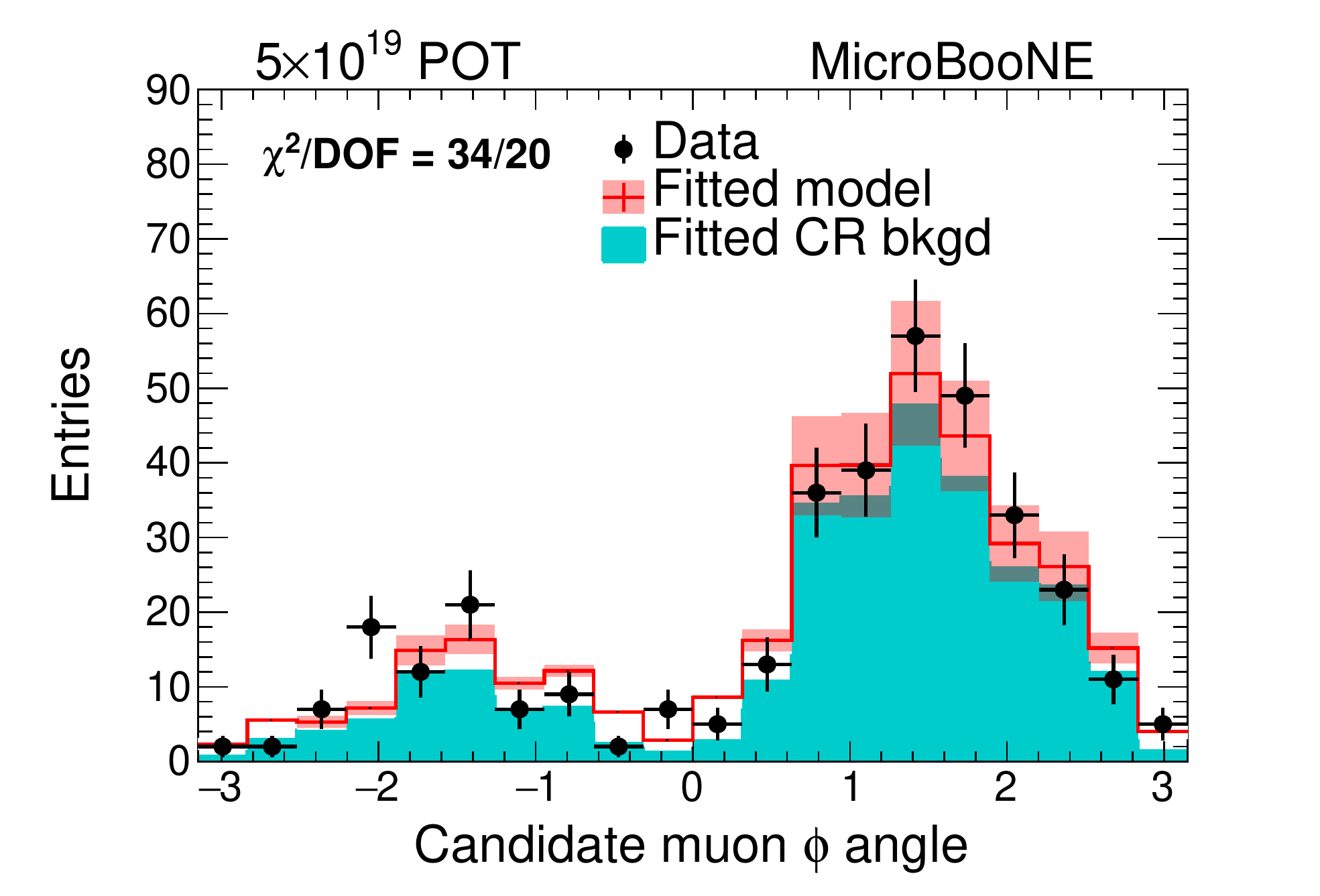}}\quad
\caption{Candidate muon azimuthal angle distribution from the full selected sample for data and GENIE default MC. The neutrino-enriched sample is nearly flat as expected. The CR-background-enriched sample has expected peaks at $\pm\pi/2$.}
\label{Dphi}
\end{figure*}

Figure~\textcolor{Blue}{\ref{Dphi}} shows the distribution of azimuthal angle $\phi$, defined in the plane perpendicular to the beam direction, of the muon candidate
track for the full selected sample. \ The CR-dominated $\left(
FAIL,FAIL\right) $ category shows the expected peaking at $\phi=\pm\pi/2$
from the mainly vertically-oriented CR. The asymmetry in the peak's structure is due to the
requirement on vertex $y$ position described previously in Sec.~\textcolor{Blue}{\ref{event_sel}}.\ By contrast the
signal-enriched $\left( PASS,PASS\right) $ category has the nearly flat
distribution expected for a neutrino event sample with a small CR\
background.

Similar levels of agreement exist between data and simulation for
distributions of the event vertex $x$ and $z$ positions, for the $\left(
x,y,z\right) $ position of the end point of the muon track candidate, and
for the azimuthal angles of individual tracks in multiplicity $2$ and $3$
topologies. \ We thus conclude that the simulation and reconstruction chain
augmented by our method for estimated CR\ backgrounds satisfactorily
describes features of the data that have no dependence on the neutrino
interaction model.

\subsection{Dynamically Significant Distributions}

\label{Dynamic Variables} 

Events with $N$ reconstructed tracks have
potentially $4N$ dynamically significant variables$-$the components of each
particle $4$-vector$-$which will have distributions that depend on the
neutrino interaction model. \ Azimuthal symmetry of the beam eliminates one
of these, leaving $3,7,11,$ and $15$ dynamically significant variables for
multiplicities $1-4$, respectively. In the following, we use the notation $%
X_{ij}$ to label a dynamical variable $x$ associated with track $j$ in an
observed multiplicity $i$ event. For example, $\cos\theta_{11}$ describes
the cosine of polar angle distribution of the only track in multiplicity $1$
events, while $L_{22}$ would describe the length of the second (short) track in
multiplicity 2 events. The notation with three subscripts, $X_{ijk}$, represents a
distribution of the difference in variable $x$ associated with tracks $j$ and $k$ in an observed multiplicity $i$ event. 

For one-track events, three variables exist. \ We use the observed length $%
L_{11}$ as a proxy for kinetic energy, and the cosine of the scattering angle
with respect to the neutrino beam direction $\cos\theta_{11}$. \ The
azimuthal angle $\phi_{11}$ has no dynamical significance and must be uniformly
distributed due to the cylindrical symmetry of the neutrino beam. \ 

Since the particle mass is not determined in our analysis, we are free to
introduce a\ third dynamically significant quantity that is sensitive to
particle mass, which we take to be 
\begin{equation}
\sin\Theta_{11}=\left\vert \hat{s}_{11}\times\hat{t}_{11}\right\vert ,
\end{equation}
where $\hat{s}_{11}$ is a unit vector parallel to the track direction at the
event vertex, and $\hat{t}_{11}$ is a unit vector that points from the start
of the track to the end of the track in the detector. The variable $%
\Theta_{ij}$ measures the angular deflection of a track over its length due
to multiple Coulomb scattering. \ Its dependence on track momentum and
energy differs from that of track length. \ \ For most of the MicroBooNE
kinematic range, we expect light particles ($\pi$ and $\mu$) to scatter more, and thus
produce a broader $\sin\Theta_{ij}$ distribution, than protons over the same
track length.

\begin{figure*}[!hpt]
\centering
\begin{tabular}{cc}
\subfloat{\includegraphics[width=.5\textwidth]{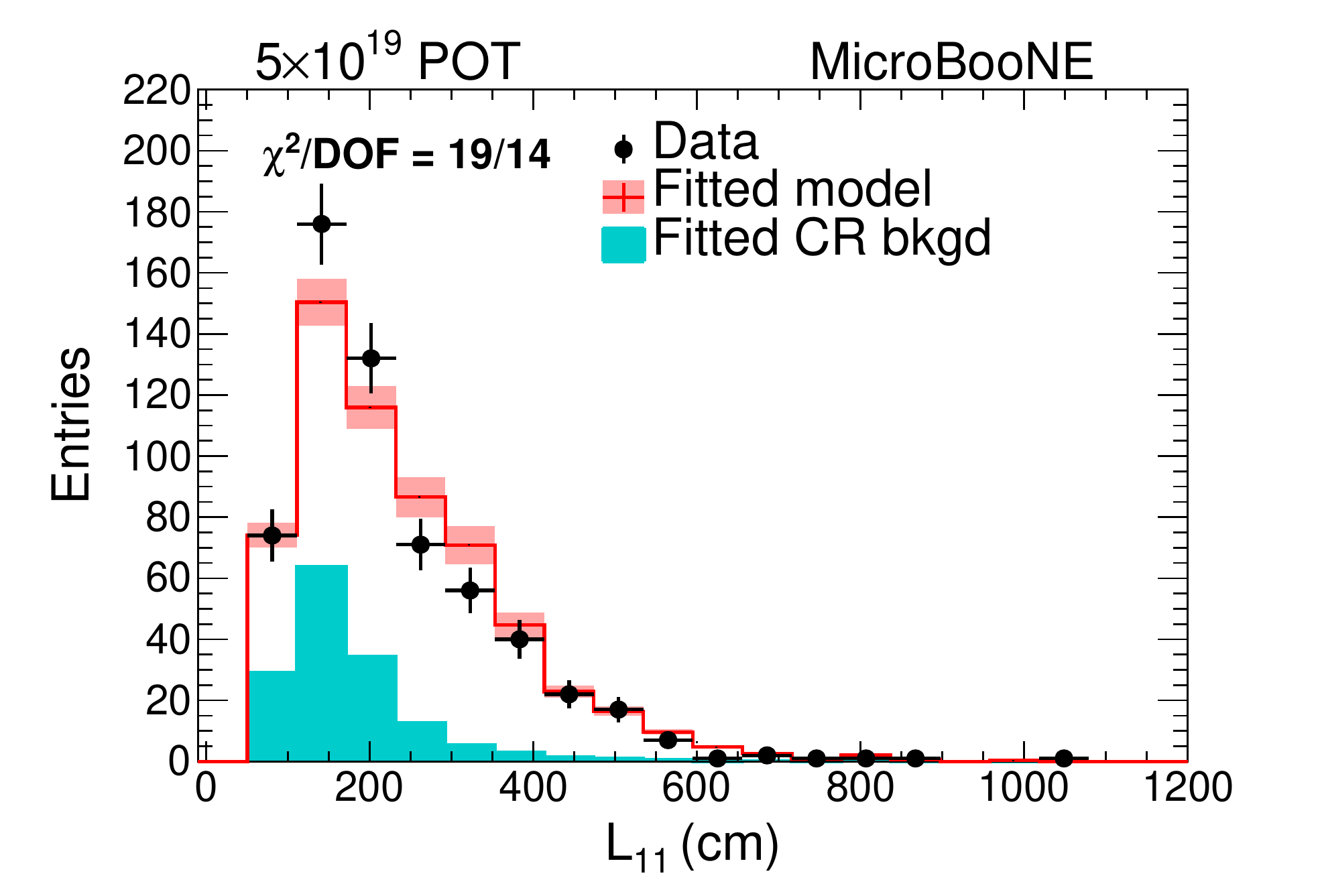}}
& \subfloat{\includegraphics[width=.5%
\textwidth]{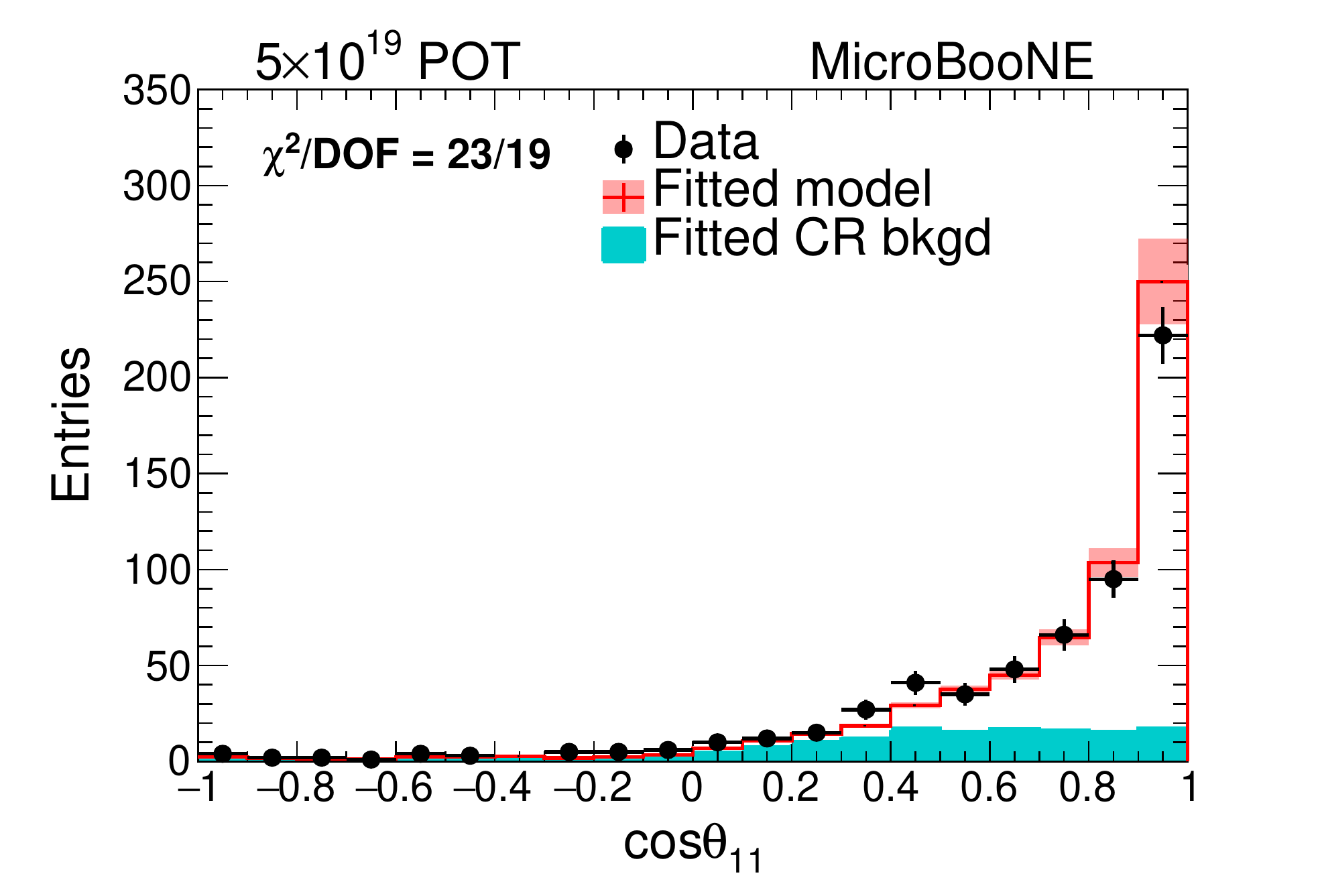}} \\ 
\multicolumn{2}{c}{\subfloat{\includegraphics[width=.5%
\textwidth]{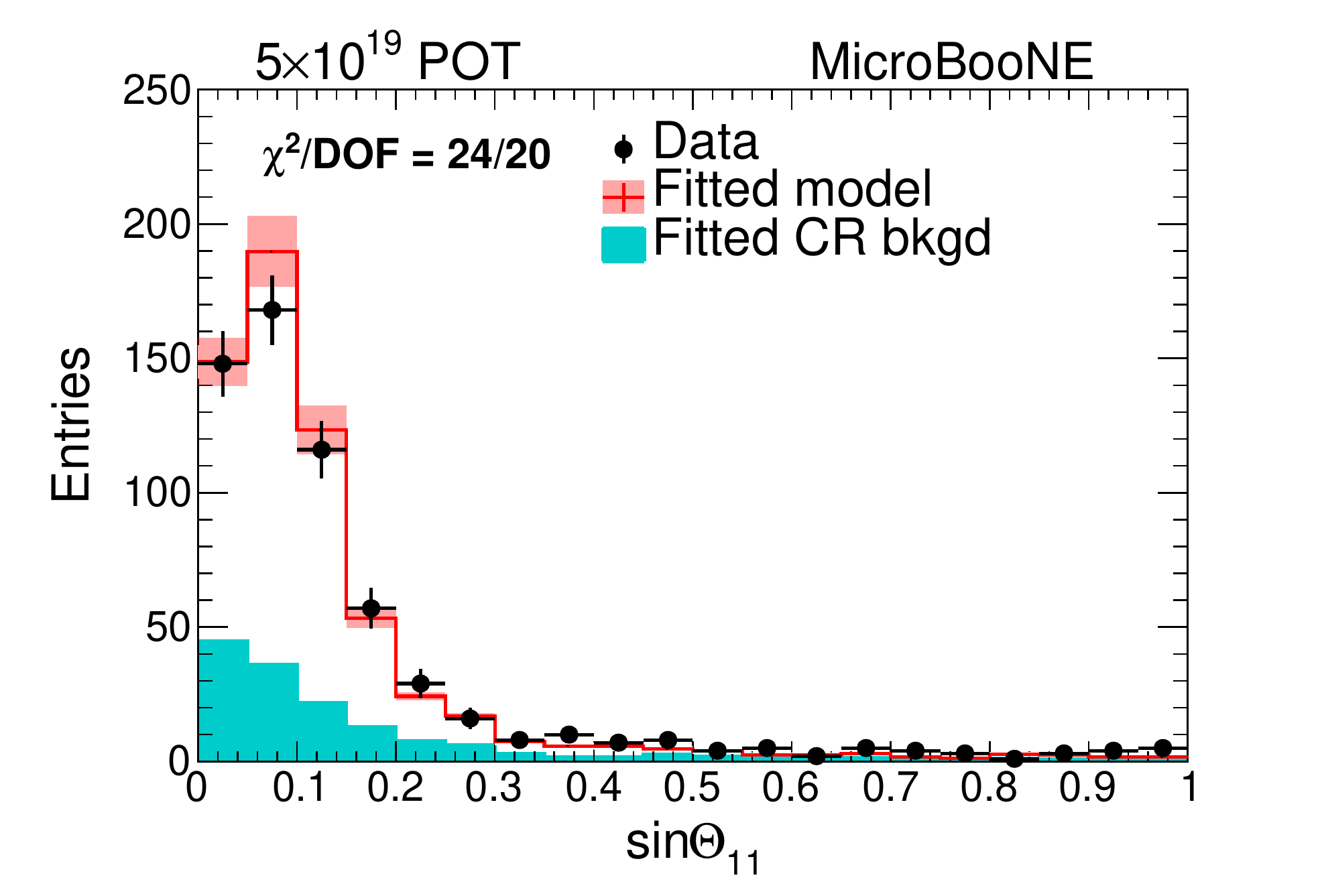}}}%
\end{tabular}
\caption{\textbf{Multiplicity = 1} candidate muon track length, $\cos\protect%
\theta$, and $\sin\Theta$ distributions from neutrino-enriched sample for data and GENIE default MC.}
\label{Dmult1}
\end{figure*}

Figure~\textcolor{Blue}{\ref{Dmult1}} shows the distributions of $L_{11}$, $\cos\theta_{11}$,
and $\sin\Theta_{11}$, from the neutrino-enriched sample compared to the GENIE\
default model. Figure~\textcolor{Blue}{\ref%
{Dlenmult1Others}} presents the $L_{11}$ distribution for the GENIE+MEC and
GENIE+TEM models. This is the distribution where the agreement between data and
GENIE+TEM model, compared to the agreement between data and the other two models, is largest. Figure~\textcolor{Blue}{\ref%
{Dthetamult1Others}} presents the $\cos\theta_{11}$ distribution for
GENIE+MEC and GENIE+TEM models. This is the distribution where the agreement
between data and the GENIE default compares least favorably than to the GENIE+MEC
and GENIE+TEM models.

For brevity in the following, except where noted, we only show comparisons
of data to predictions from the GENIE\ default model. \ Comparisons to
GENIE+TEM and GENIE+MEC show qualitatively similar levels of agreement. \
Differences for specific distributions can be examined in terms of the $%
\chi^{2}$ test statistic values in Table~\textcolor{Blue}{\ref{tab:chi-square}}.

For two track events, seven dynamic variables exist. \ These include
properties of the long track that parallel the choices for one-track events, 
$L_{21}$, $\cos\theta_{21}$, and $\sin\Theta_{21}$, similar quantities for
the second track, $L_{22}$, $\cos\theta_{22}$, and $\sin\Theta_{22}$, plus a
quantity that describes the correlation between the two tracks in the event,
which we take to be the difference in azimuthal variables $%
\phi_{221}=\phi _{22}-\phi_{21}$. \ Since track $2$ can exit the
detector, the meaning of $L_{22}$ and $\sin\Theta_{22}$ differ somewhat from 
$L_{21}$ and $\sin \Theta_{21}$. \ Two other two-track correlated variables
of interest, which are not independent, are the cosine of the opening angle, 
\begin{equation}
\cos\Omega_{221}=\cos\theta_{21}\cos\theta_{22}+\sin\theta_{21}\sin\theta
_{22}\cos\left( \phi_{22}-\phi_{21}\right) ,
\end{equation}
and the cosine of the acoplanarity angle%
\begin{align}
\cos\theta_{A} & =\frac{\hat{s}_{21}\cdot\left( \hat{z}\times\hat{s}%
_{21}\right) }{\left\vert \hat{z}\times\hat{s}_{21}\right\vert } \\
& =\sin\theta_{21}\sin\left( \phi_{22}-\phi_{21}\right) ,
\end{align}
with $\hat{z}$ a unit vector in the neutrino beam direction and $\hat{s}_{21}
$ is a unit vector parallel to the first track direction at the event
vertex. For the scattering of two initial state particles into two final
state particles $\left( 2\rightarrow2\right) $, one expects from momentum
conservation $\phi_{221}=\pm\pi$ and $\cos\theta_{A}=0$. Deviations of $%
\phi_{221}$ from $\pm\pi$ or of $\cos\theta_{A}$ from $0$ could be
caused by undetected tracks in the final state, from
NC\ events in the sample, or from effects of final state interactions in CC
events.

\begin{figure*}[!hpt]
\centering
\subfloat{\includegraphics[width=0.5\linewidth]{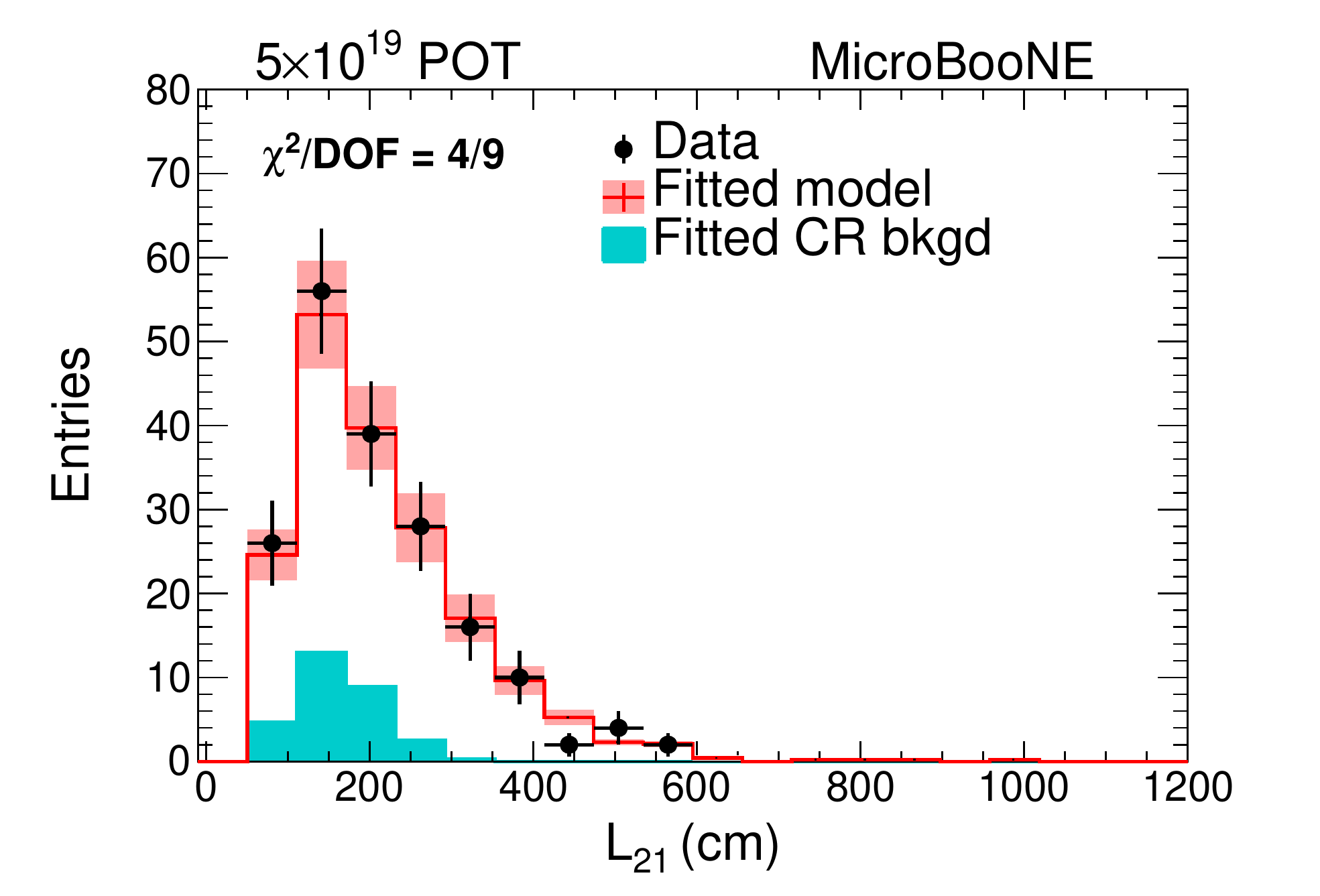}} 
\centering
\subfloat{\includegraphics[width=0.5\linewidth]{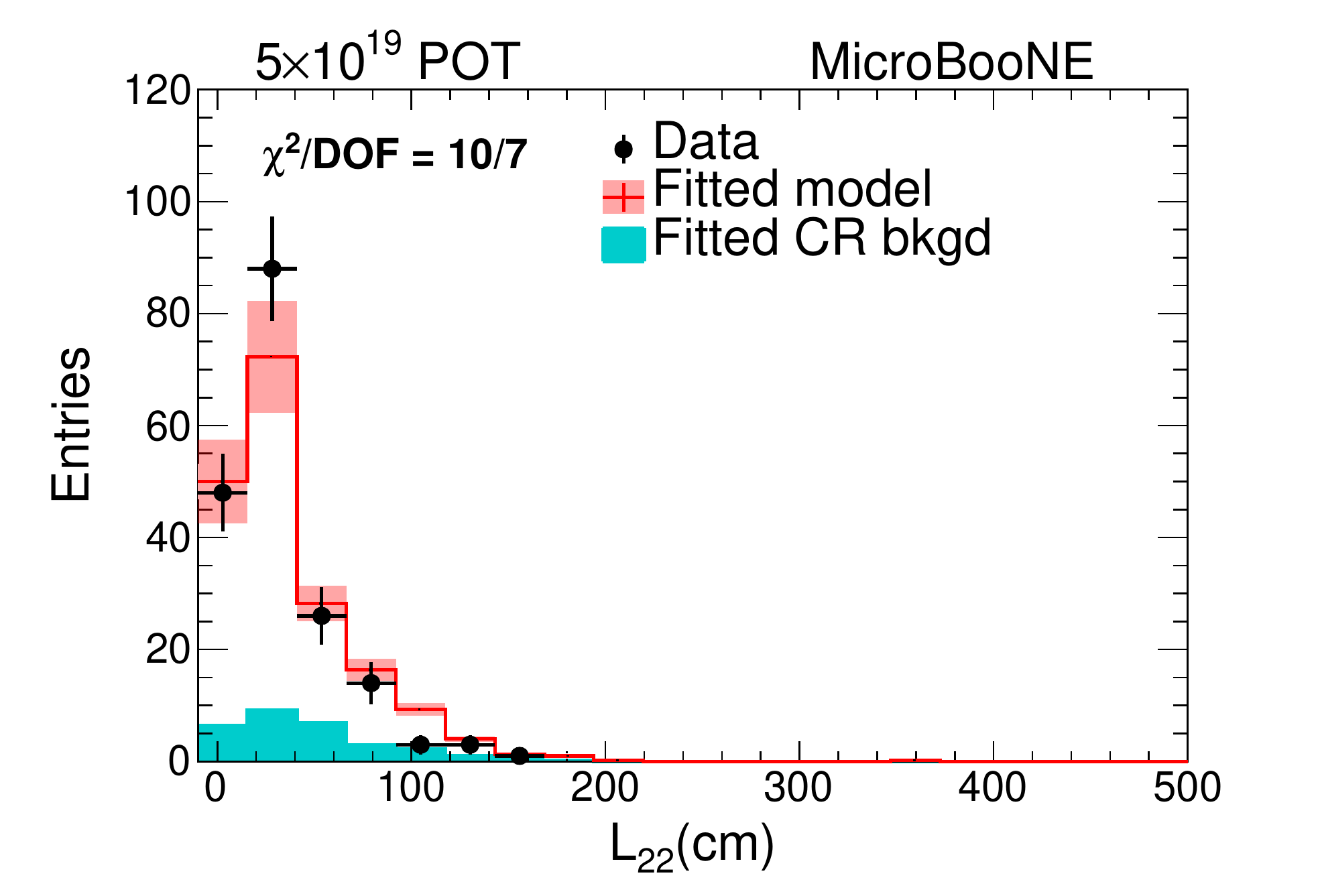}}
\caption{\textbf{Multiplicity = 2} Track length distribution for candidate
muon (left); for second track of the event (right) from neutrino-enriched
sample for data and GENIE default MC.}
\label{DL2}
\end{figure*}

\begin{figure*}[!hpt]
\centering
\subfloat{\includegraphics[width=0.5%
\linewidth]{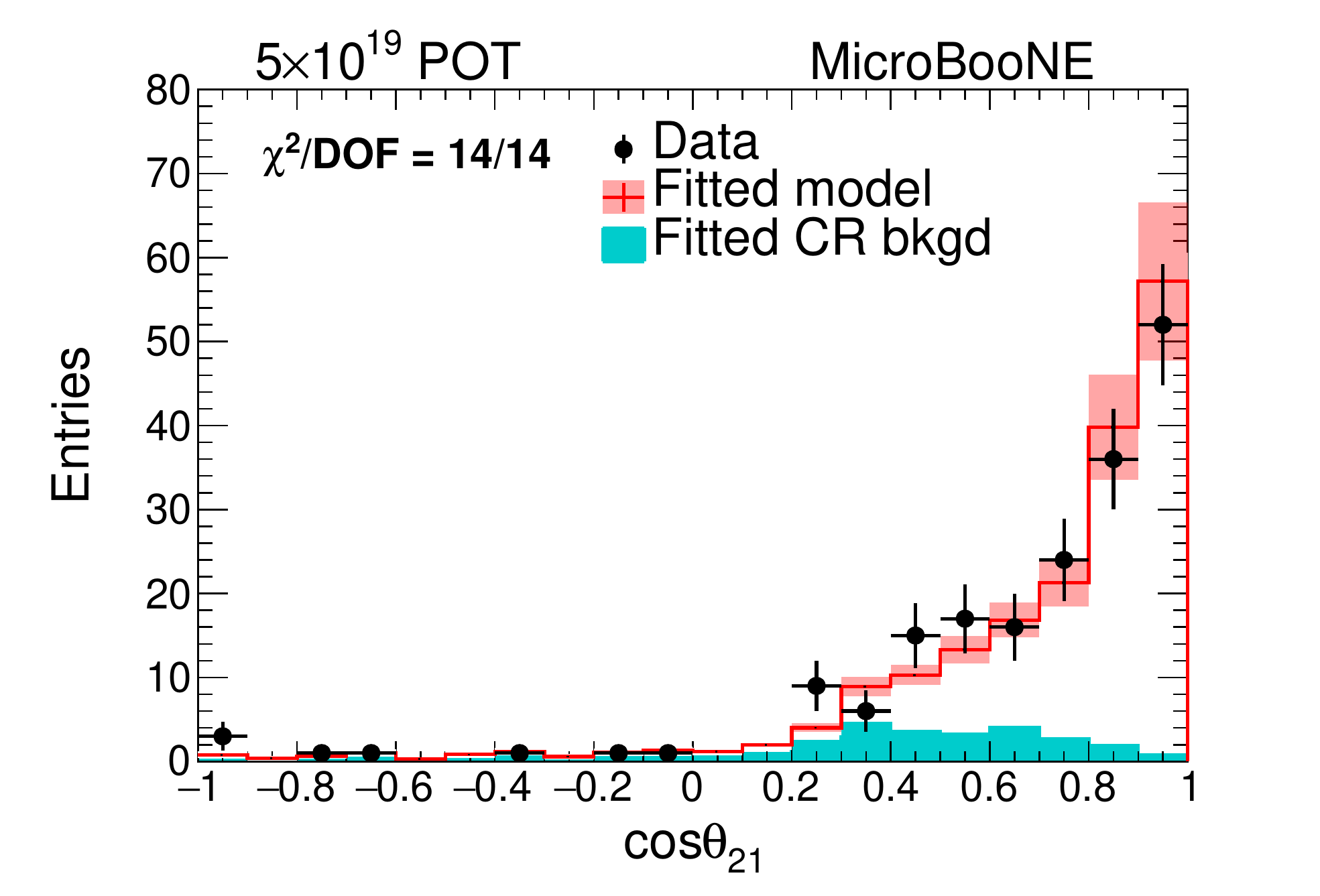}} \centering
\subfloat{\includegraphics[width=0.5%
\linewidth]{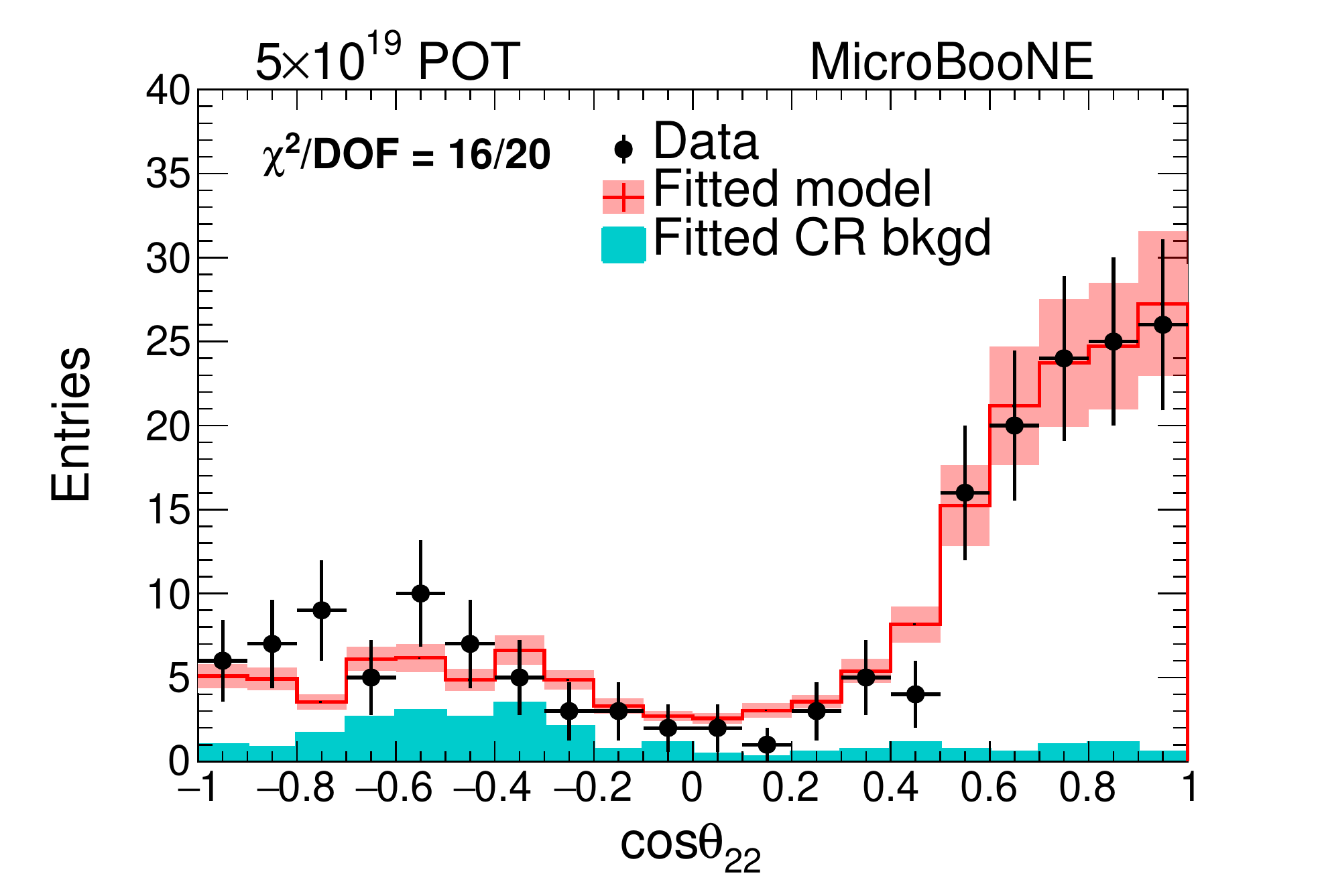}}
\caption{\textbf{Multiplicity = 2} $\cos\protect\theta$ distribution for
candidate muon (left); for second track of the event (right) using GENIE
default model from neutrino-enriched sample for data and GENIE default MC.}
\label{Dcostheta2}
\end{figure*}

\begin{figure*}[!hptb]
\centering
\subfloat{\includegraphics[width=0.5%
\linewidth]{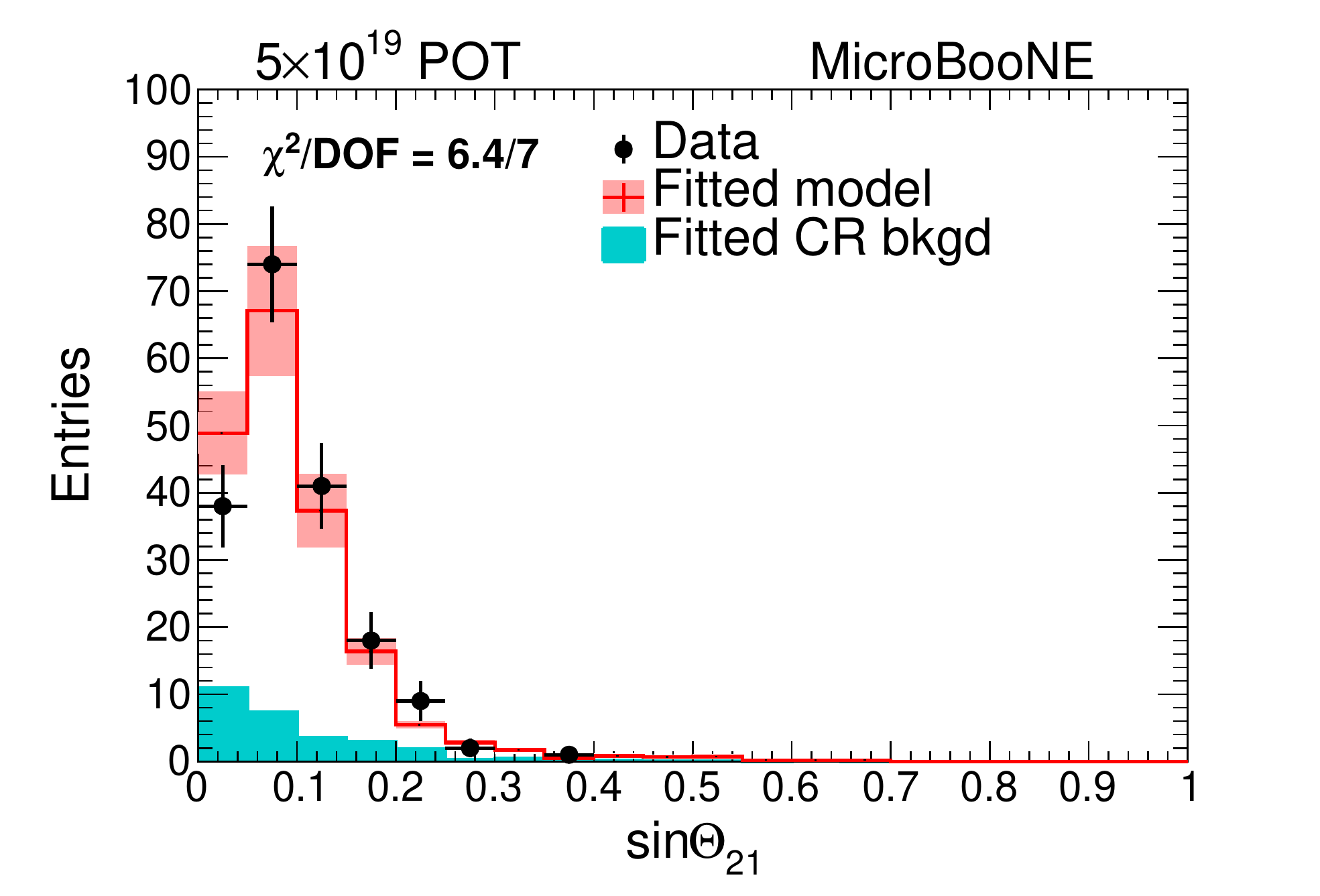}} \centering
\subfloat{\includegraphics[width=0.5%
\linewidth]{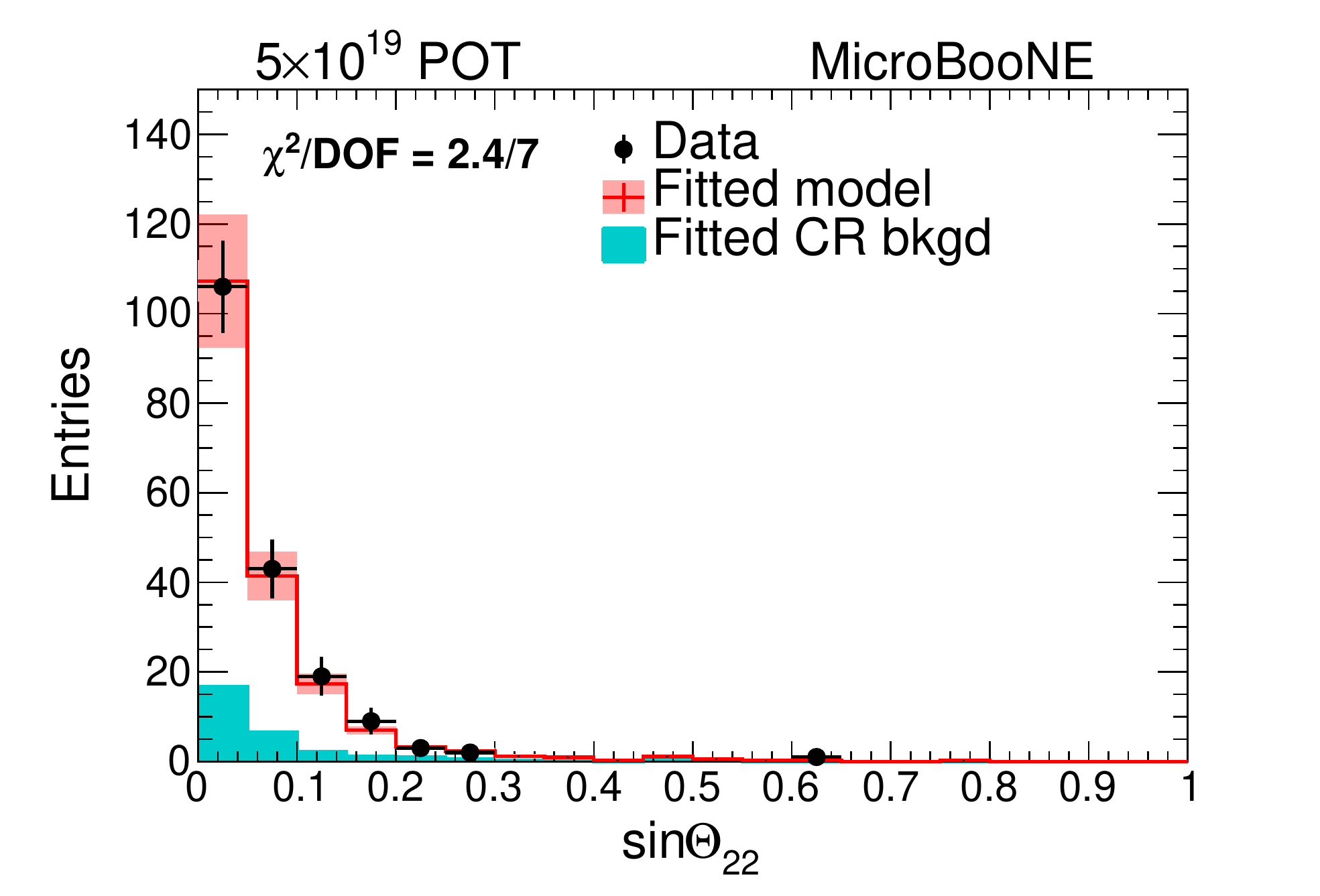}}
\caption{\textbf{Multiplicity = 2} $\sin\Theta$ distribution for candidate
muon (left); for second track of the event (right) from neutrino-enriched
sample for data and GENIE default MC.}
\label{DsinMCS2}
\end{figure*}

\begin{figure*}[!hpt]
\centering
\subfloat{\includegraphics[width=0.5%
\linewidth]{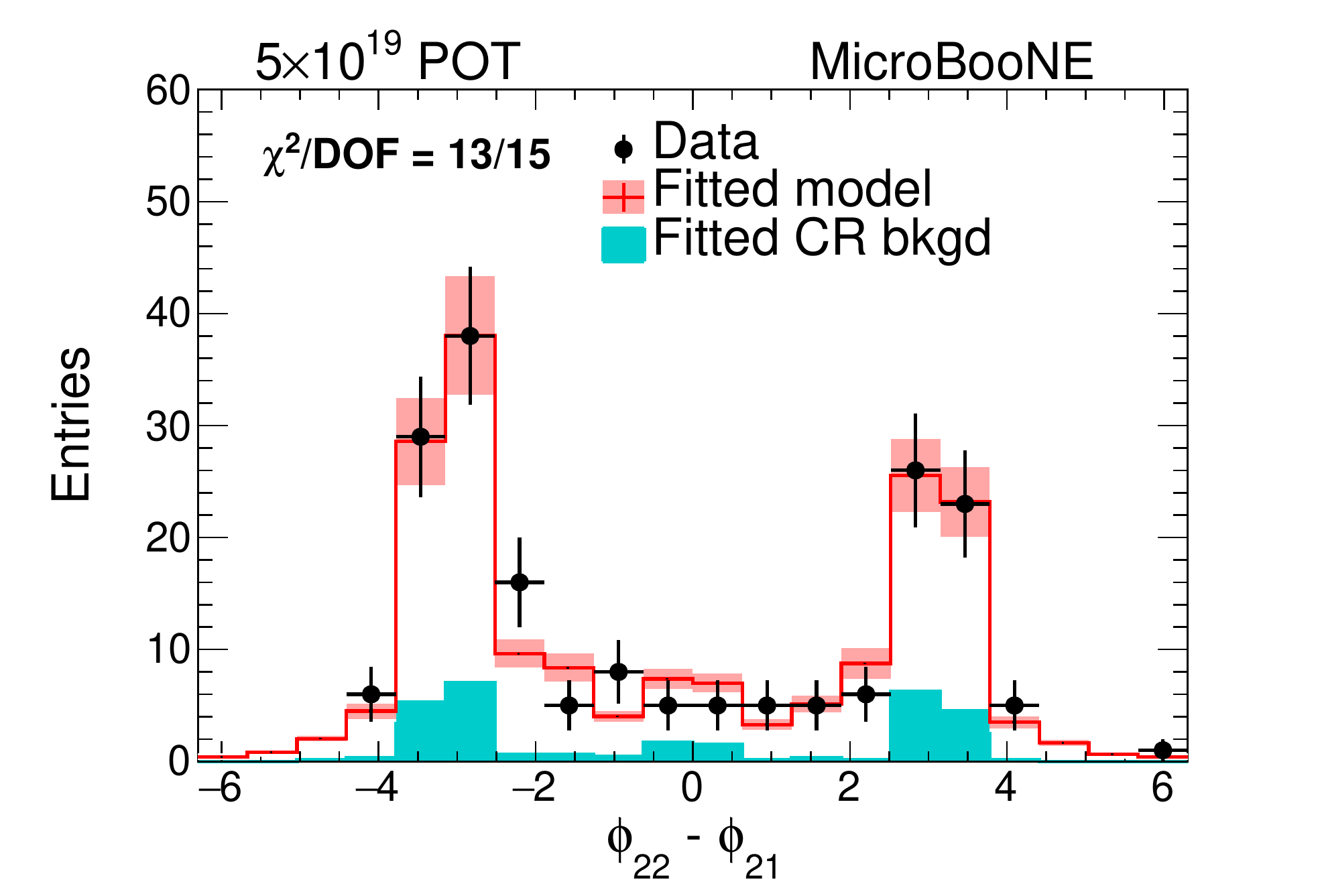}} \centering
\subfloat{\includegraphics[width=0.5%
\linewidth]{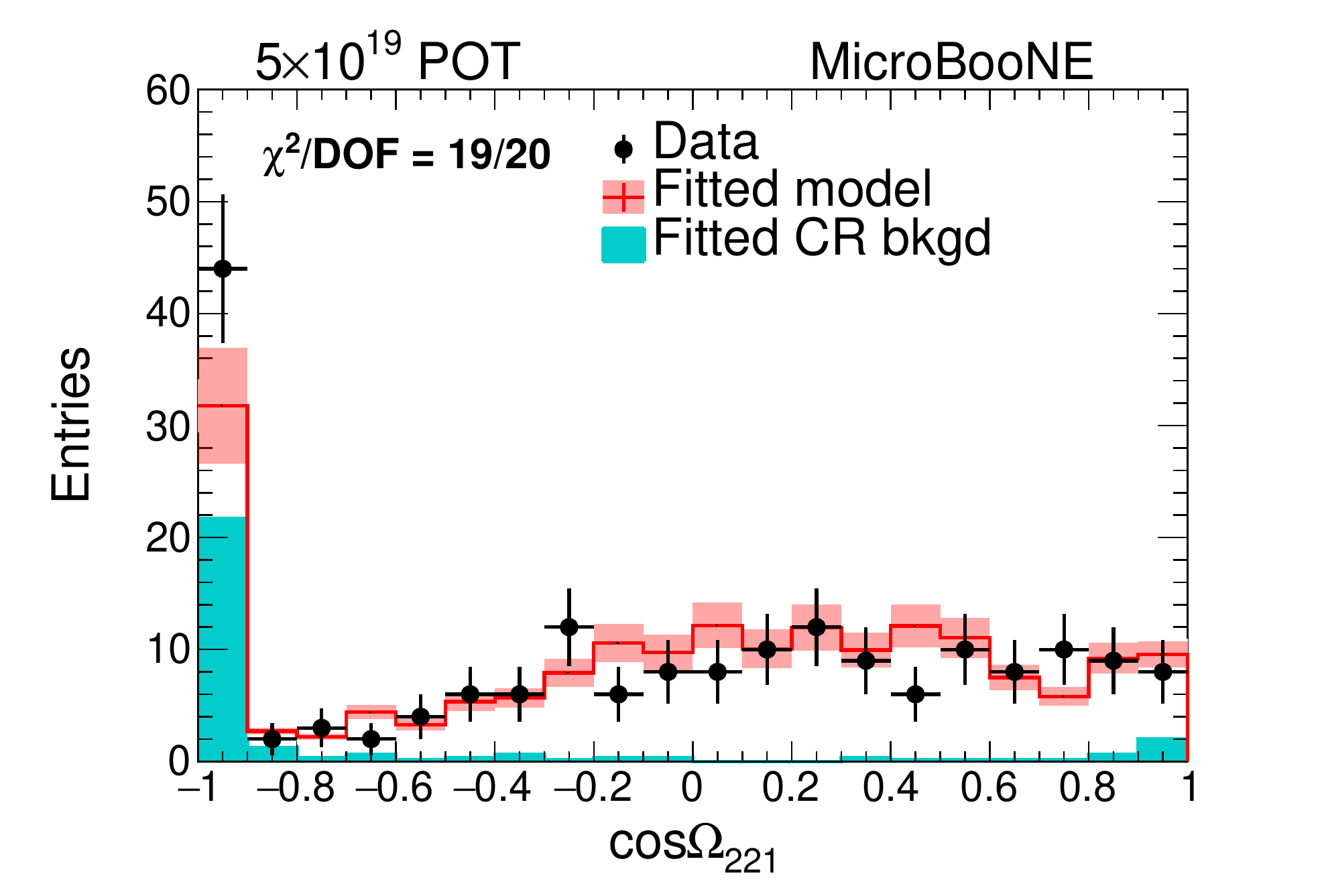}}
\caption{\textbf{Multiplicity = 2} $\protect\phi_{2} - \protect\phi_{1}$
distribution (left); $\cos\Omega_{21}$ distribution (right) from
neutrino-enriched sample for data and GENIE default MC.}
\label{Ddphi2}
\end{figure*}

The opening angle serves a useful role in identifying spurious two-track
events that result from the tracking algorithm \textquotedblleft
breaking\textquotedblright\ a single track into two tracks, most commonly in\ cosmic
ray events. \ Broken tracks produce values of $\cos \Omega _{221}$ very close
to $-1$. Figures~\textcolor{Blue}{\ref{DL2}} and \textcolor{Blue}{\ref{Dcostheta2}} show the distributions of ($L_{21}$ and $L_{22}$) and ($\cos \theta
_{21}$ and $\cos \theta _{22}$) from the
neutrino-enriched sample, compared to the GENIE\ default model. Figure~\textcolor{Blue}{\ref%
{DcosthetaOthers}} presents the distributions of $\cos \theta _{21}$ using
GENIE+MEC and GENIE+TEM models. This is the distribution where the agreement between data and GENIE default model, compared to the agreement between data and the other two models, is largest. Figures~\textcolor{Blue}{\ref{DsinMCS2}} and \textcolor{Blue}{\ref{Ddphi2}} show the distributions of ($\sin \Theta _{21}$ and $\sin \Theta _{22}$) and ($\phi _{22}-\phi _{21}$ and $\cos \Omega _{221}$) from the neutrino-enriched sample, compared to the GENIE\ default model. We have performed a test where we remove all events in the first bin of Fig.~\textcolor{Blue}{\ref{Ddphi2}}; we see no changes in the level of agreement between data or model in other kinematic distribution comparisons, and no statistically significant shifts in the observed multiplicity distributions.

\begin{figure*}[!hpt]
\begin{adjustwidth}{-2cm}{-2cm}
\centering
\subfloat{\includegraphics[width=.35\textwidth]{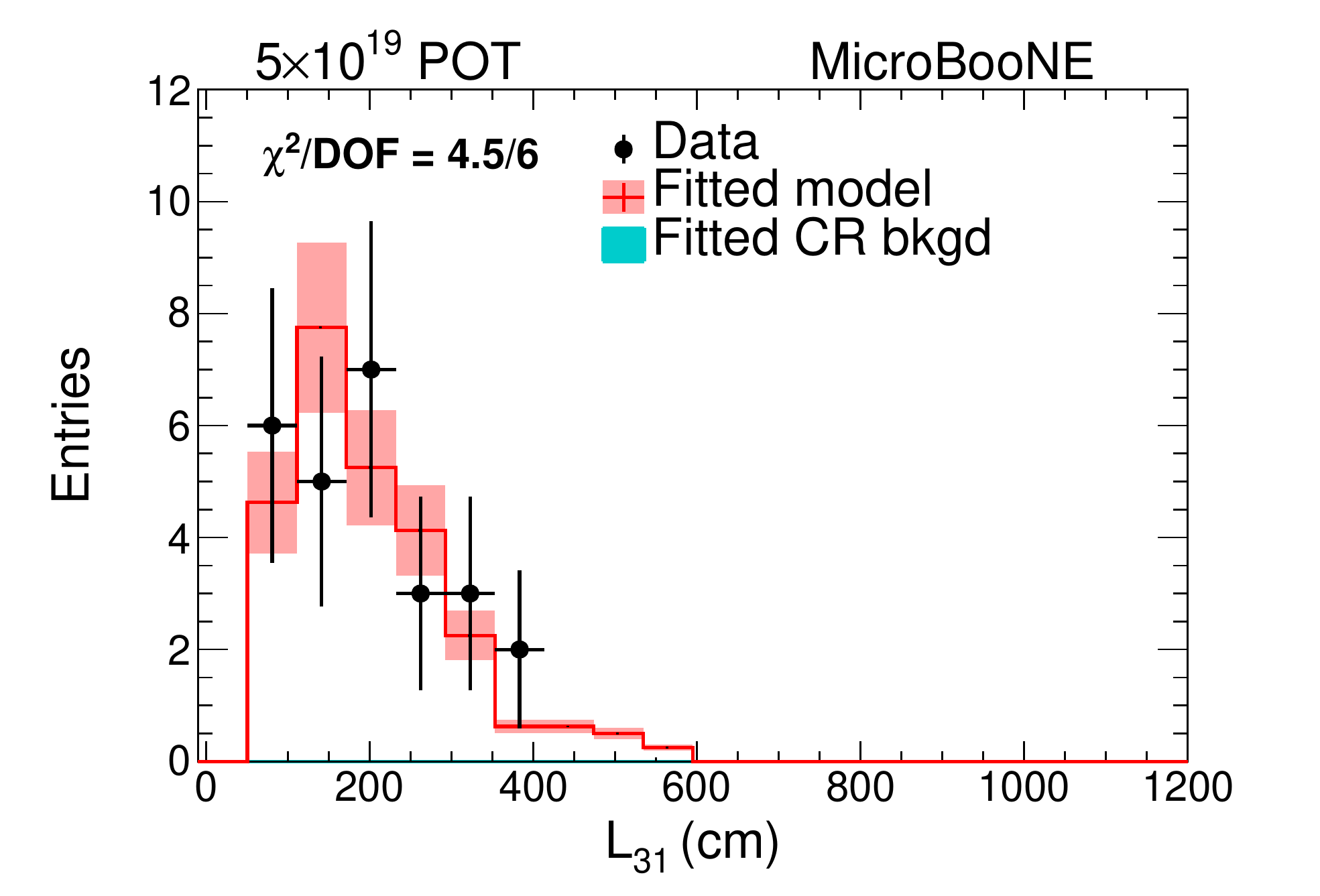}}
\subfloat{\includegraphics[width=.35\textwidth]{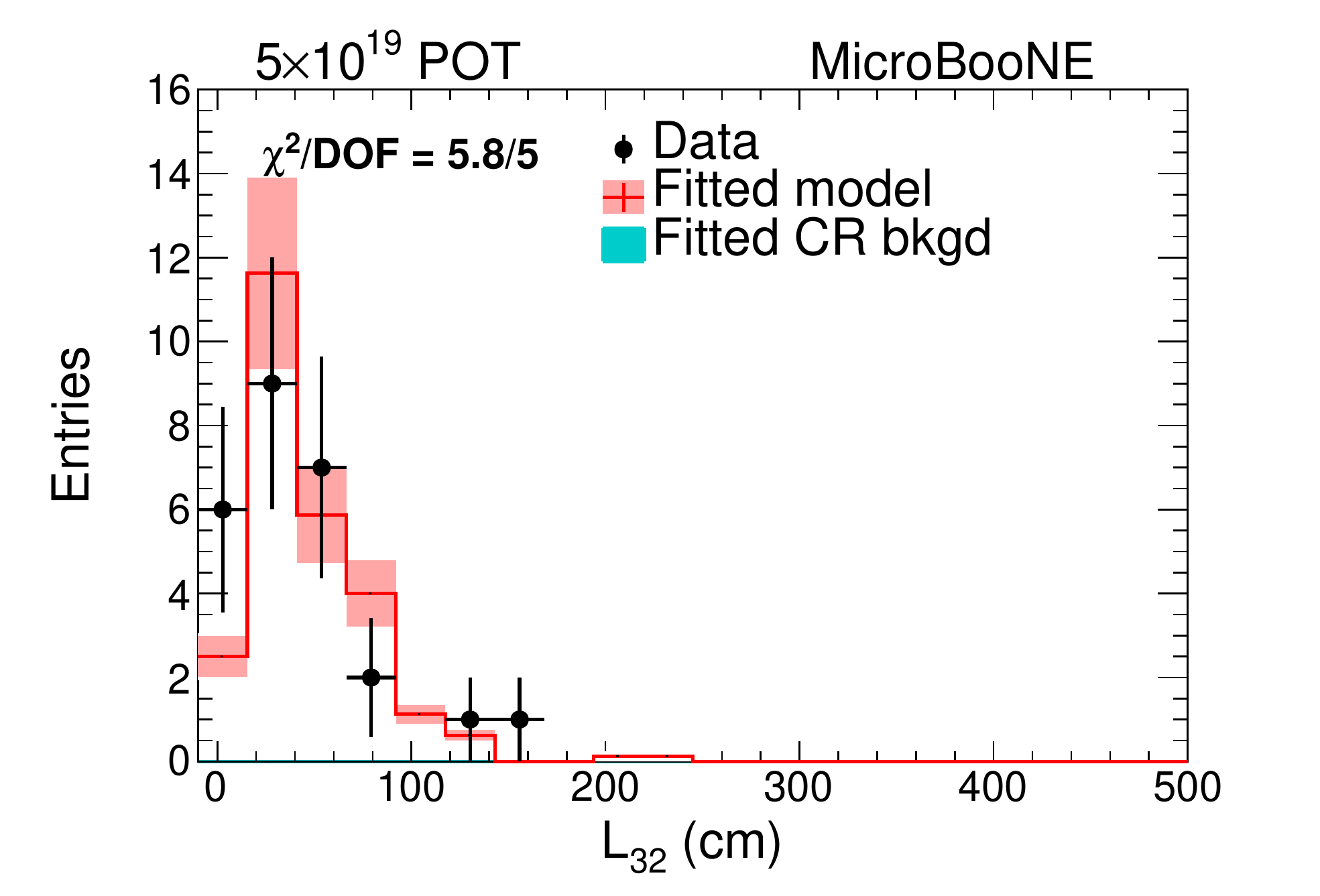}}
\subfloat{\includegraphics[width=.35\textwidth]{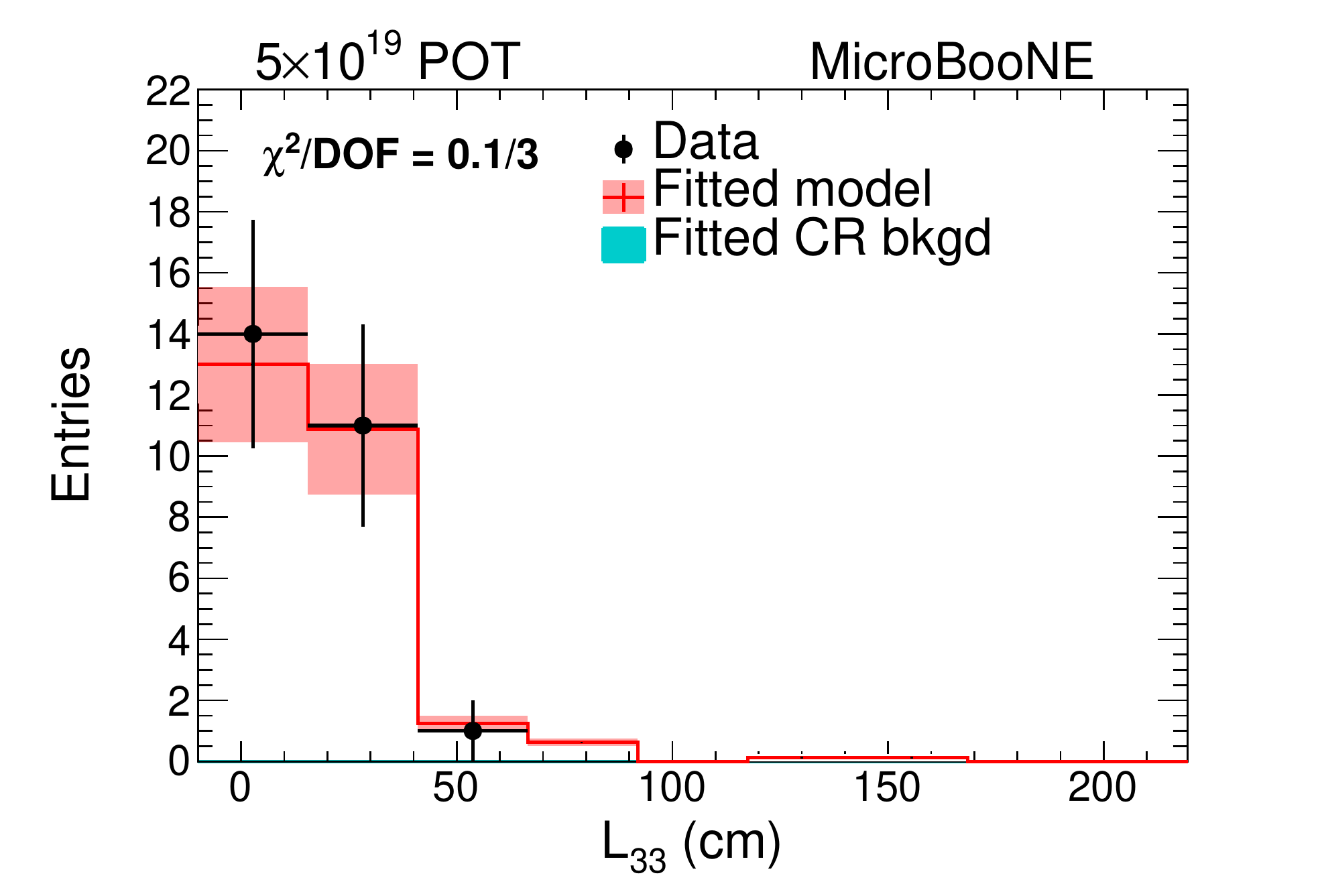}}
\caption{\textbf{Multiplicity = 3} Track length distribution for candidate muon (left); for second longest track (middle); for shortest track of the event (right) from neutrino-enriched sample for data and GENIE default MC.}
\label{DL3}
\end{adjustwidth}
\end{figure*}

\begin{figure*}[!hpt]
\begin{adjustwidth}{-2cm}{-2cm}
\centering
\subfloat{\includegraphics[width=.35\textwidth]{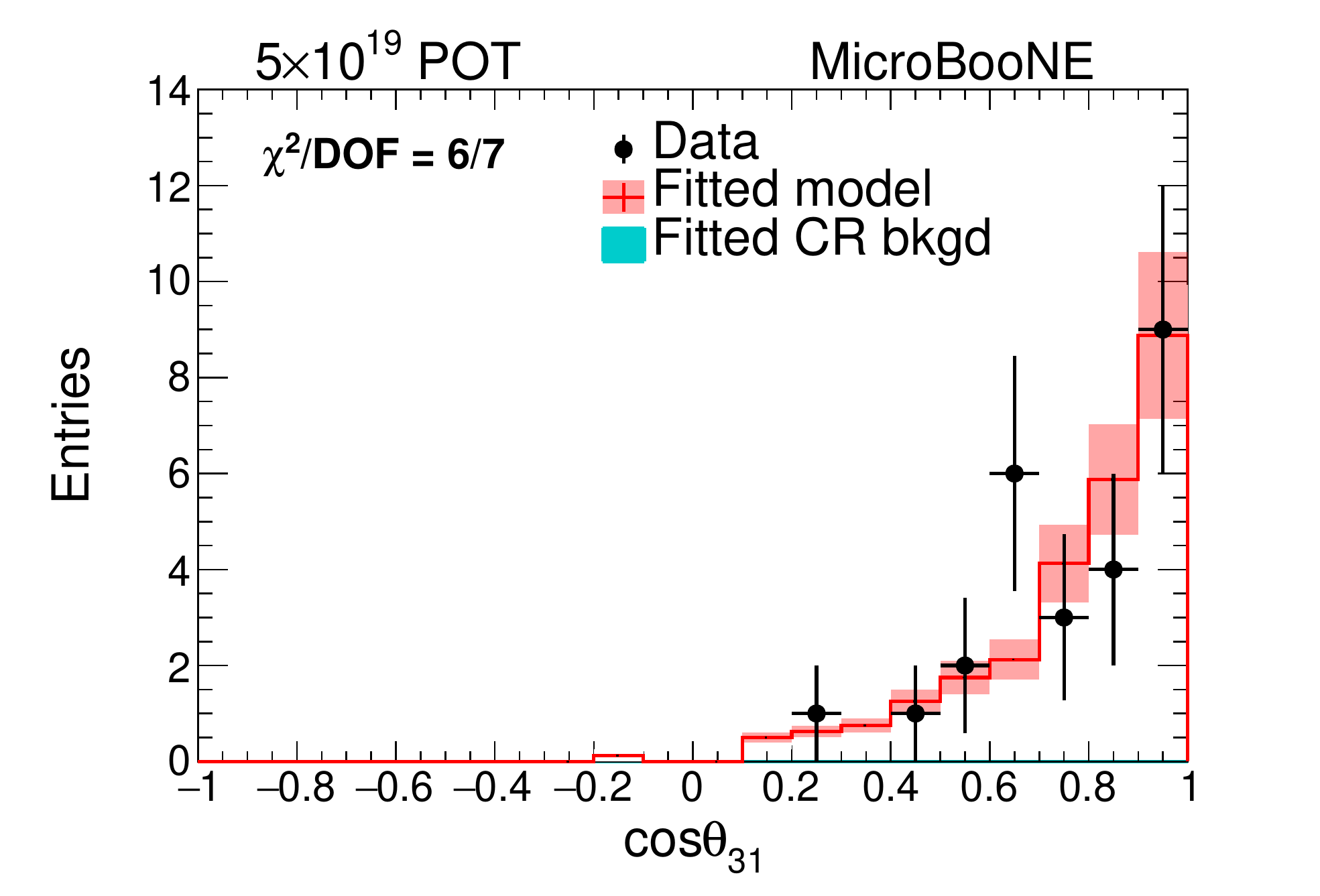}}
\subfloat{\includegraphics[width=.35\textwidth]{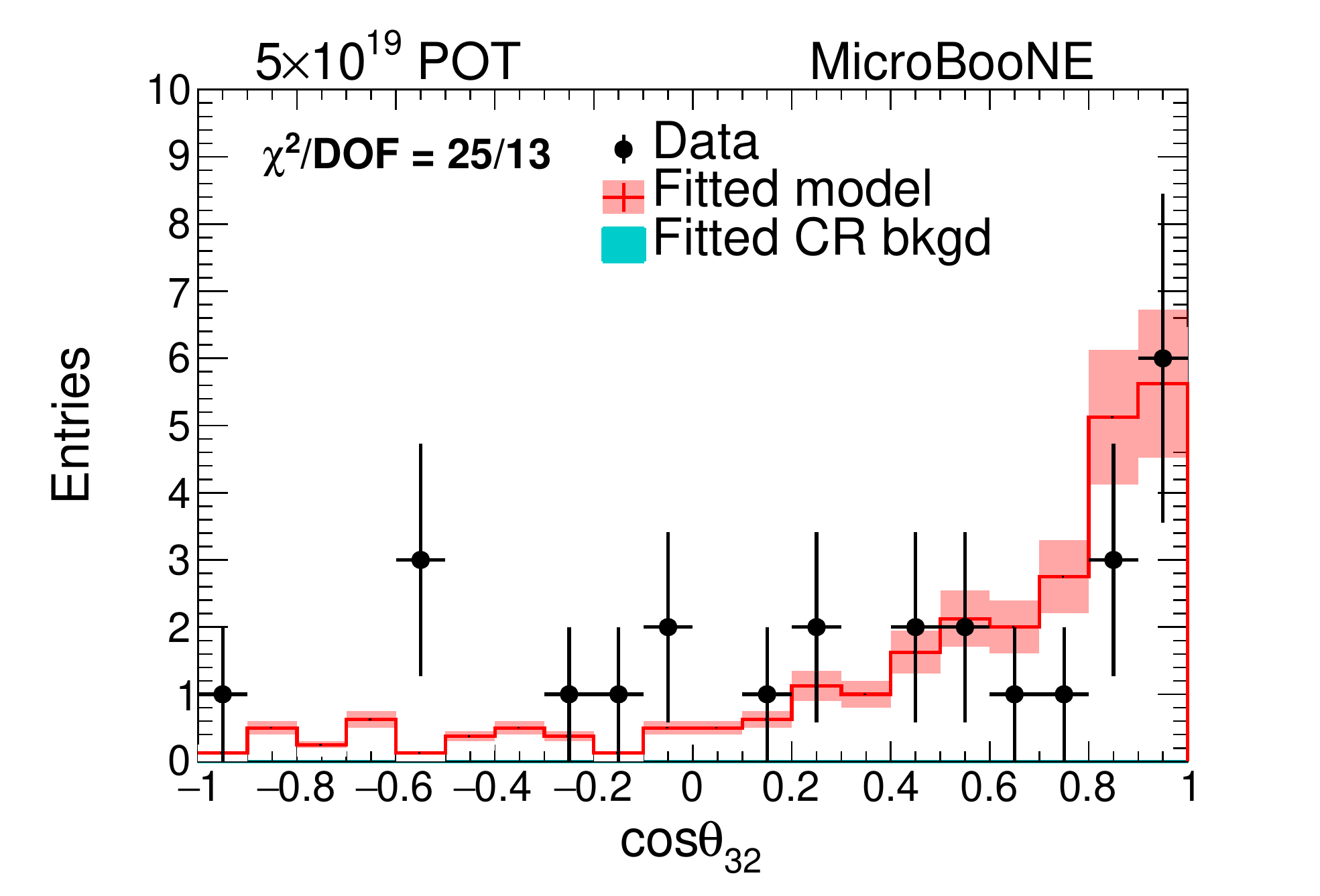}}
\subfloat{\includegraphics[width=.35\textwidth]{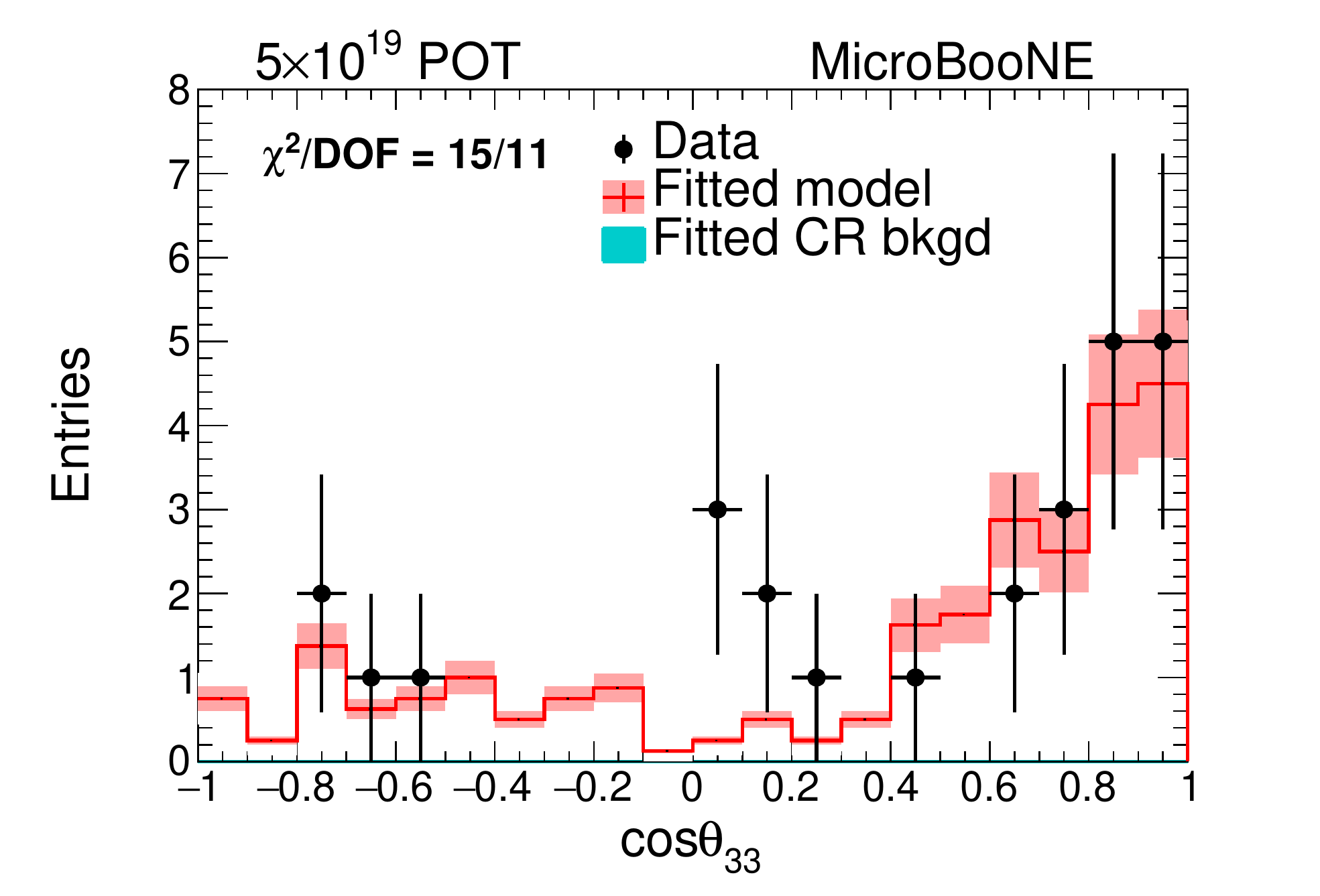}}
\caption{\textbf{Multiplicity = 3} $\cos\theta$ distribution for candidate muon (left); for second longest track (middle); for shortest track of the event (right) from neutrino-enriched sample for data and GENIE default MC.}
\label{Dcostheta3}
\end{adjustwidth}
\end{figure*}

\begin{figure*}[!hpt]
\begin{adjustwidth}{-2cm}{-2cm}
\centering
\subfloat{\includegraphics[width=.35\textwidth]{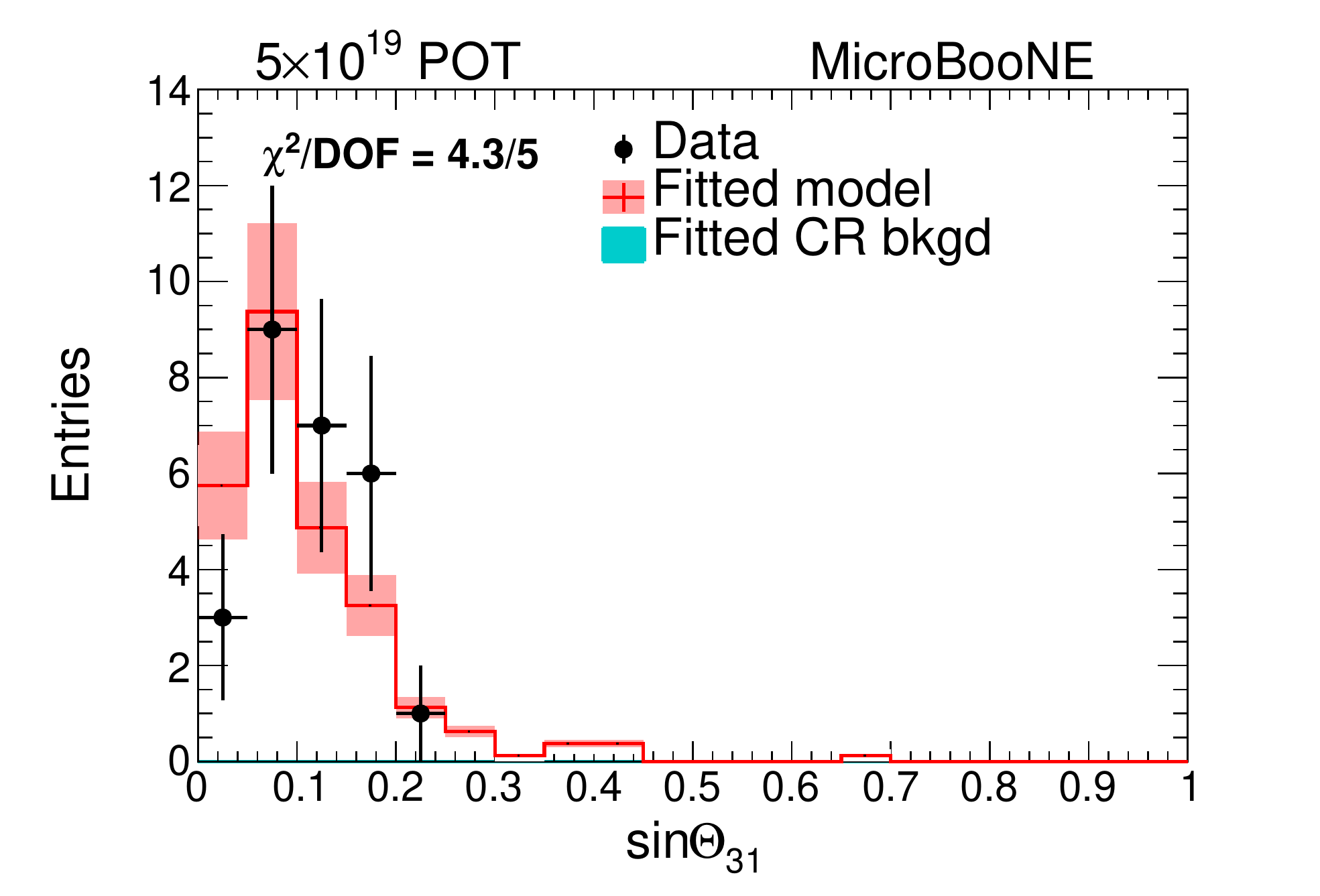}}
\subfloat{\includegraphics[width=.35\textwidth]{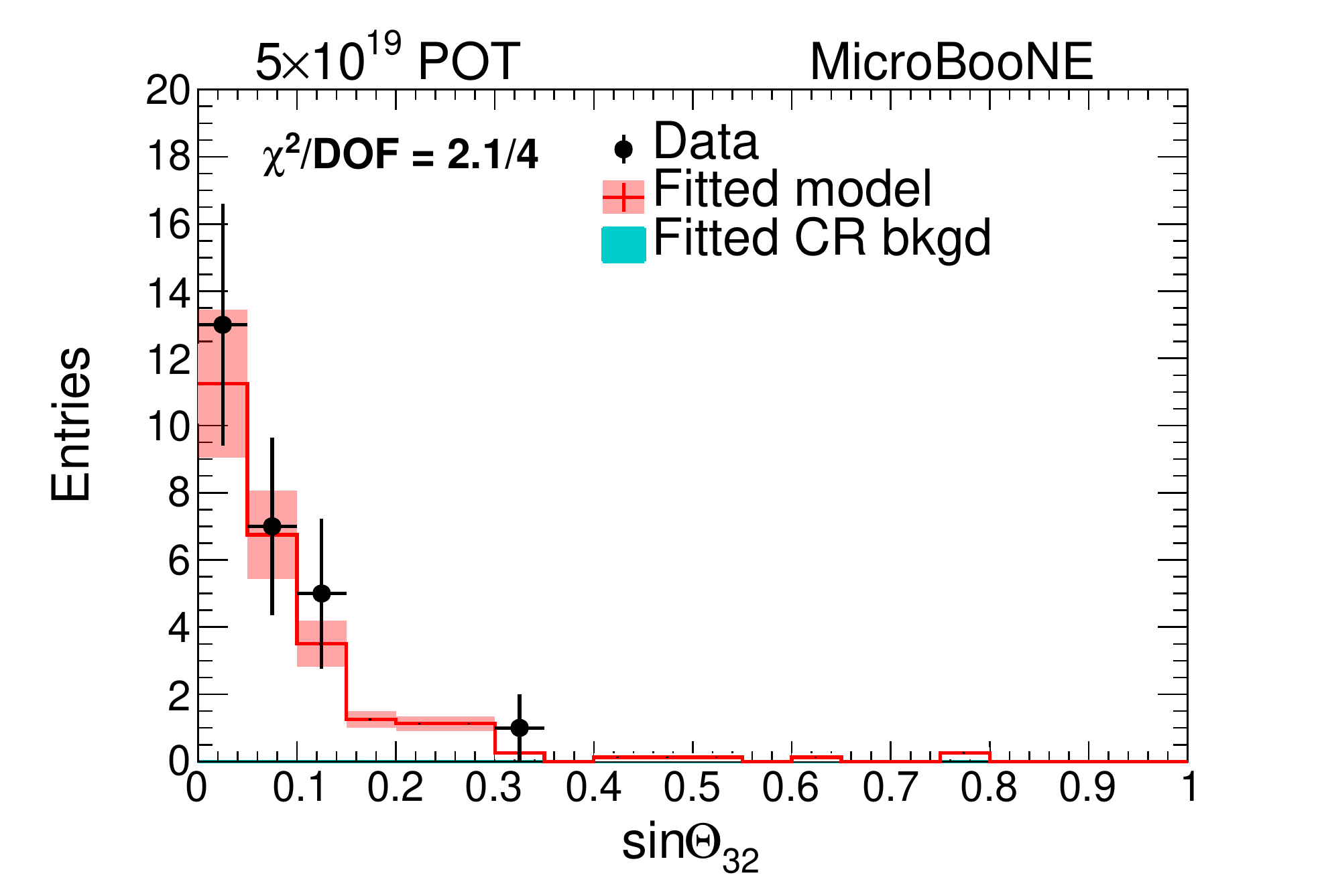}}
\subfloat{\includegraphics[width=.35\textwidth]{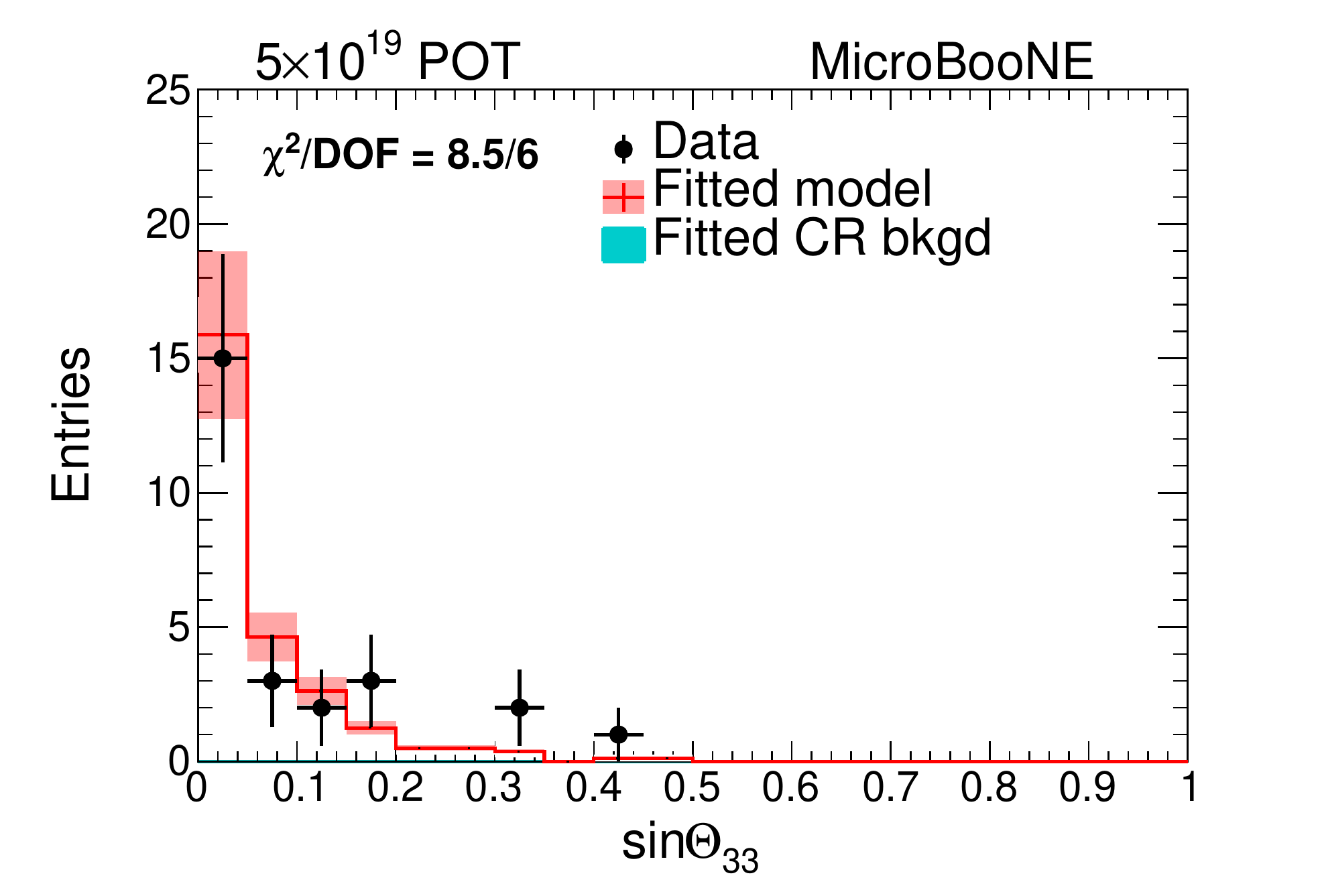}}
\caption{\textbf{Multiplicity = 3} $\sin\Theta$ distribution for candidate muon (left); for second longest track (middle); for shortest track of the event (right) from neutrino-enriched sample for data and GENIE default MC.}
\label{DsinMCS3}
\end{adjustwidth}
\end{figure*}

\begin{figure*}[!hpt]
\centering
\subfloat{\includegraphics[width=0.5%
\linewidth]{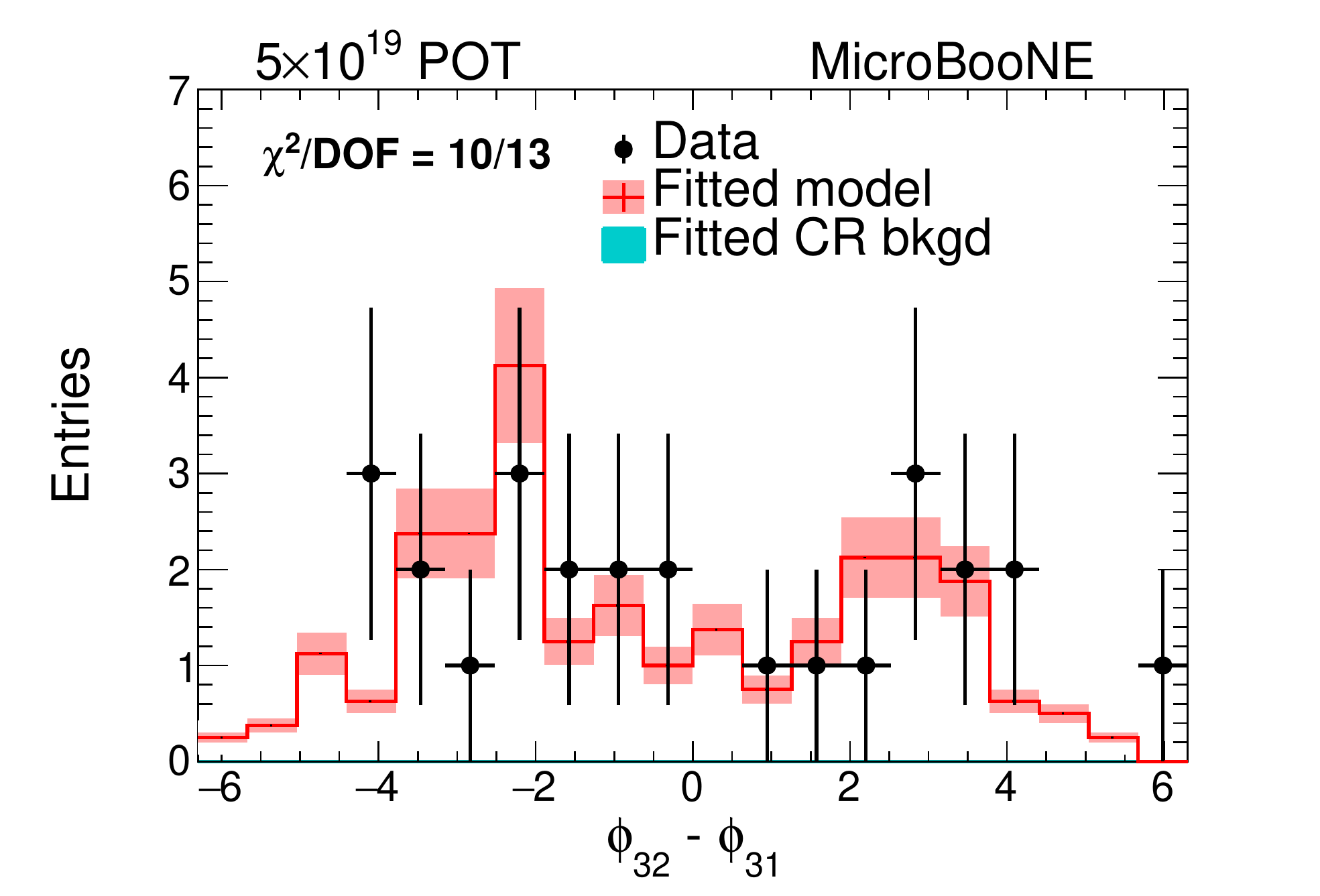}} \centering
\subfloat{\includegraphics[width=0.5%
\linewidth]{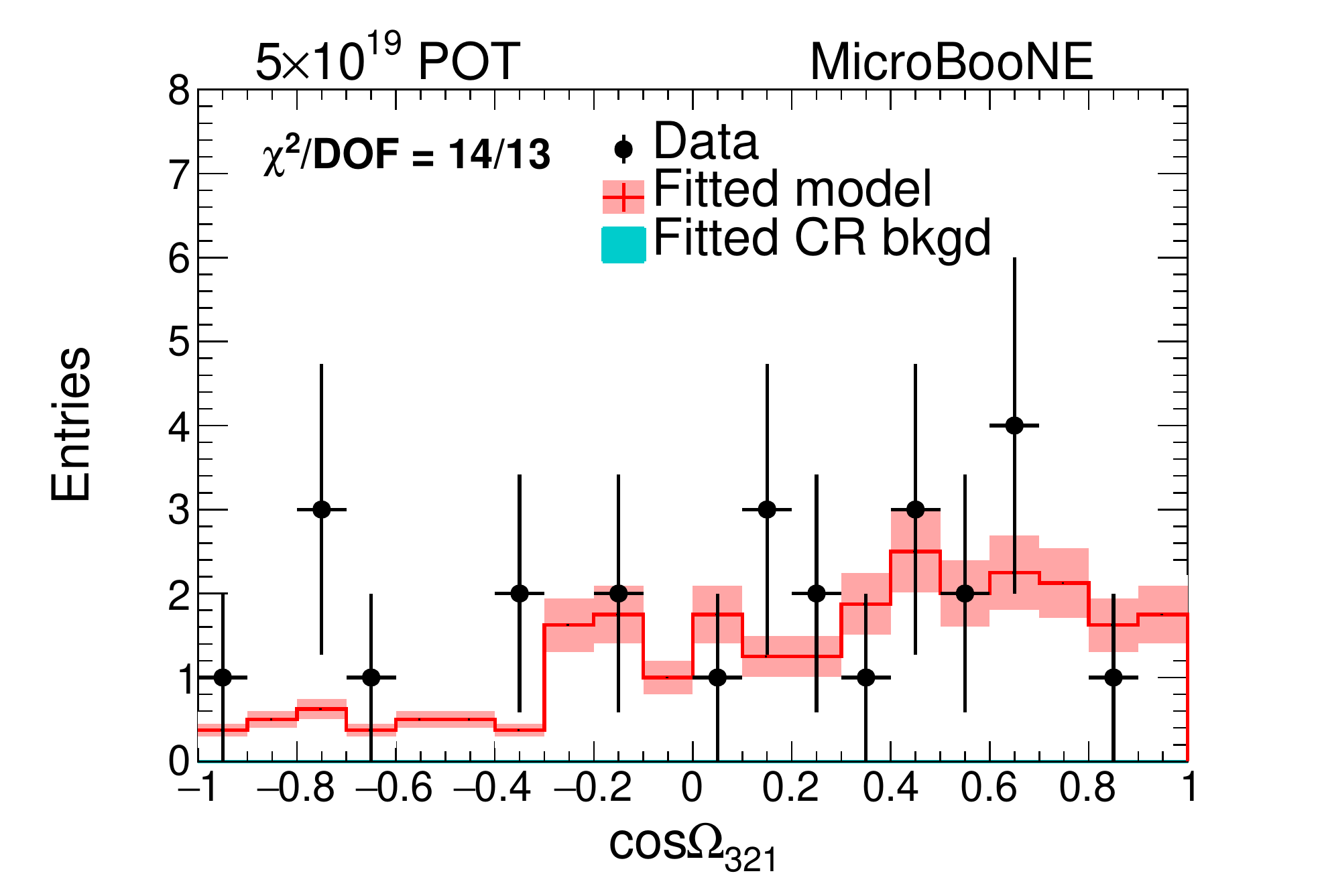}}
\caption{\textbf{Multiplicity = 3} $\protect\phi_{2} - \protect\phi_{1}$
distribution (left); $\cos\Omega_{21}$ distribution between first and second
track (right) from neutrino-enriched sample for data and GENIE default MC.}
\label{Ddphi31}
\end{figure*}

\begin{figure*}[!hpt]
\centering
\subfloat{\includegraphics[width=0.5%
\linewidth]{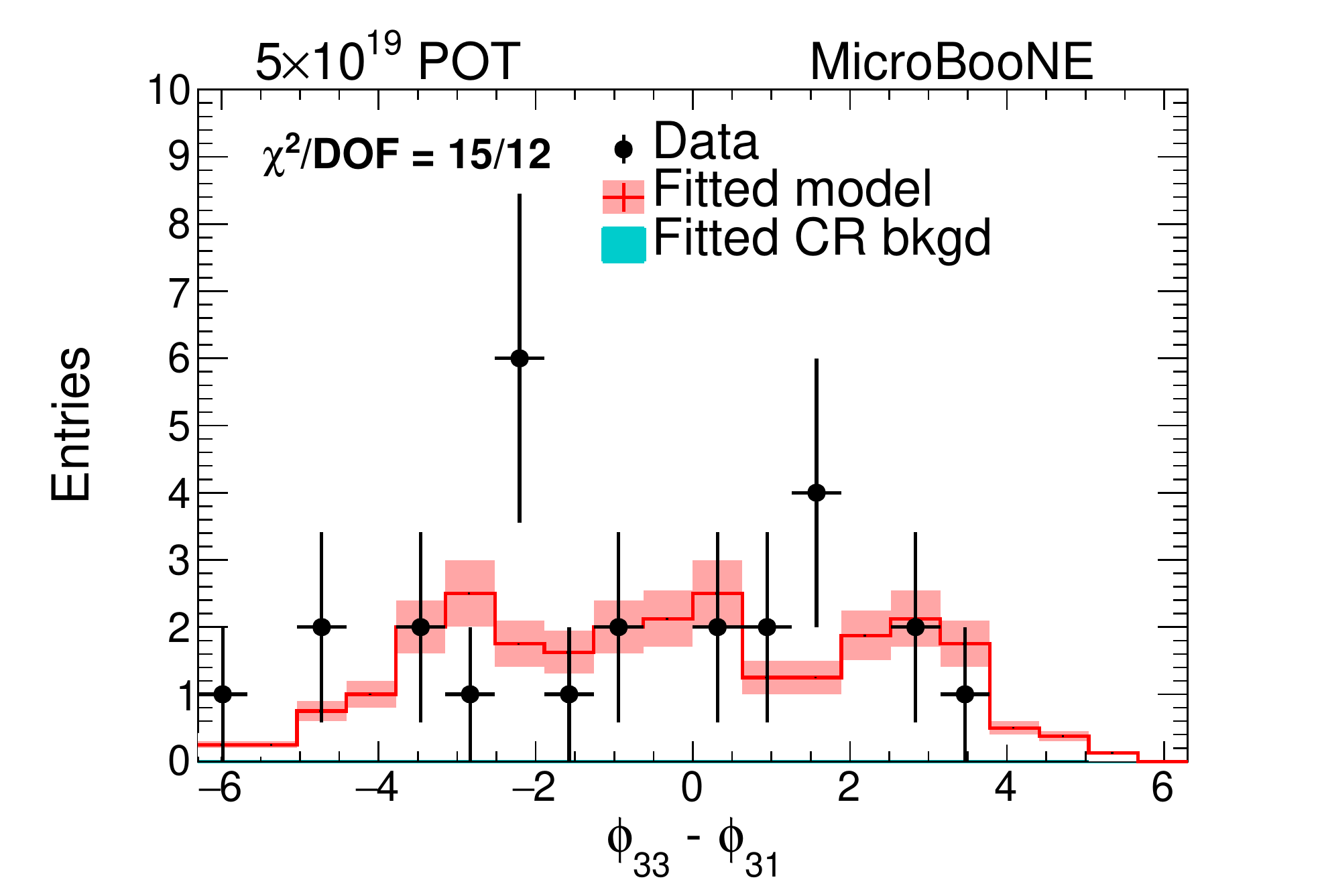}} \centering
\subfloat{\includegraphics[width=0.5%
\linewidth]{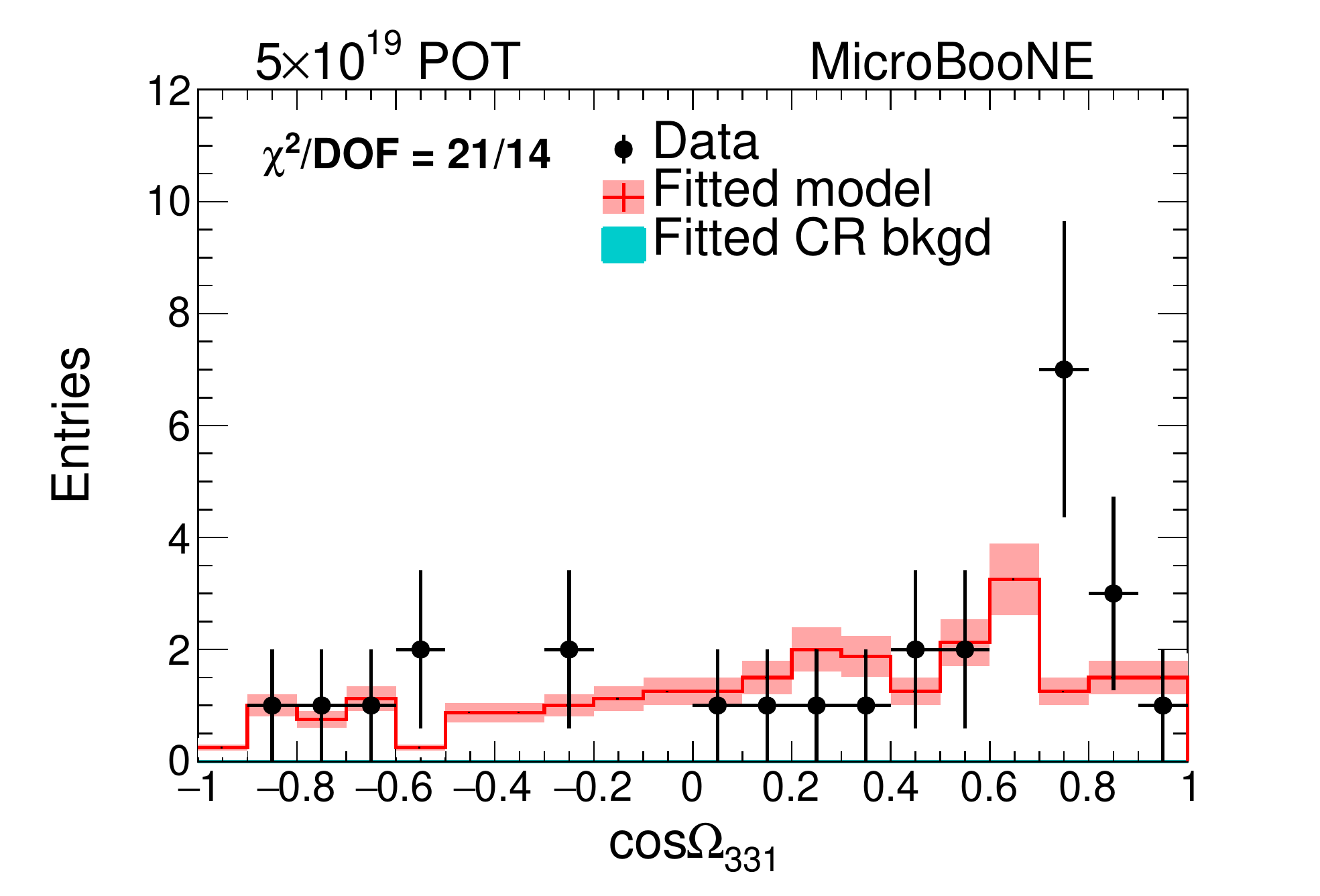}}
\caption{\textbf{Multiplicity = 3} $\protect\phi_{3} - \protect\phi_{1}$
distribution (left); $\cos\Omega_{31}$ distribution between first and third
track (right) from neutrino-enriched sample for data and GENIE default MC.}
\label{Ddphi32}
\end{figure*}

\begin{figure*}[!hpt]
\centering
\subfloat{\includegraphics[width=0.5%
\linewidth]{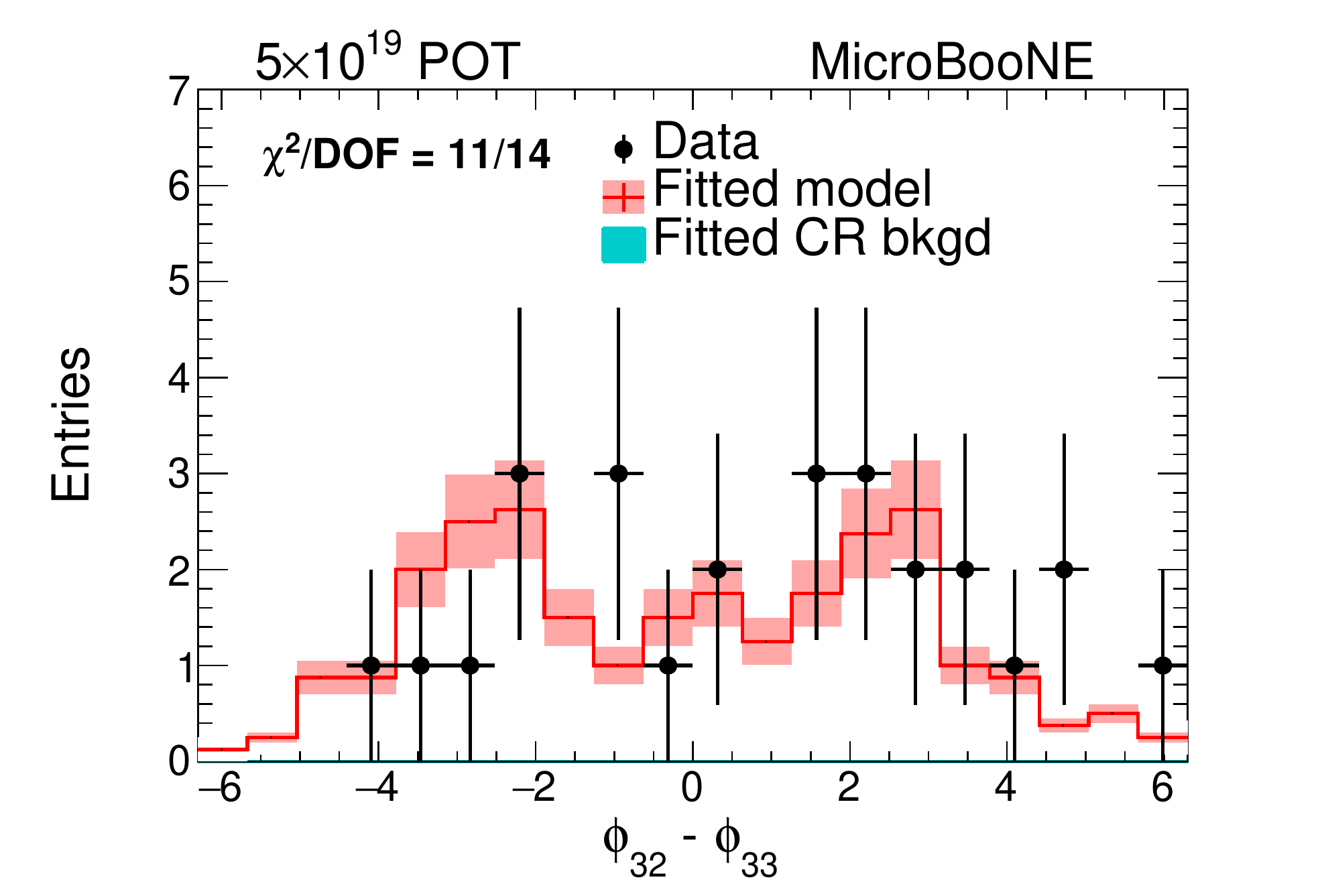}} \centering
\subfloat{\includegraphics[width=0.5%
\linewidth]{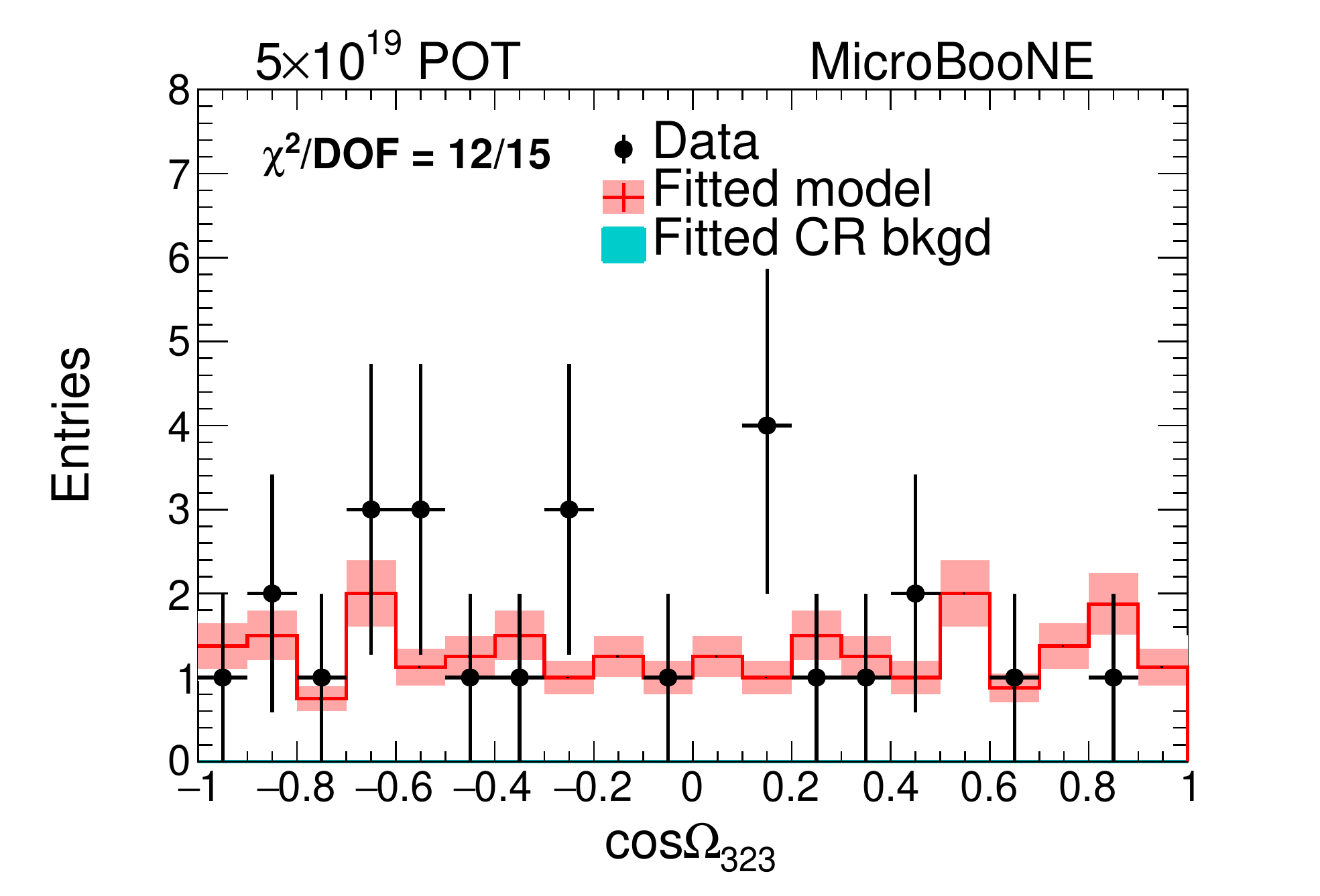}}
\caption{\textbf{Multiplicity = 3} $\protect\phi_{2}-\protect\phi_{3}$
distribution (left); $\cos\Omega_{23}$ distribution between second and third
track (right) from neutrino-enriched sample for data and GENIE default MC.}
\label{Ddphi33}
\end{figure*}

For three-track events, eleven dynamic variables exist. \ A straightforward
continuation of the previous choices leads to the choice of $L_{31}$, $\cos\theta_{31}$, $%
\sin\Theta_{31}$, $L_{32}$, $\cos\theta_{32}$, $\sin\Theta_{32}$, $\phi
_{32}-\phi_{31}$, $L_{33}$, $\cos\theta_{33}$, $\sin\Theta_{33}$, and $%
\phi_{33}-\phi_{31}$ as the eleven variables. \ Other azimuthal angle difference such as
\begin{equation}
\phi_{32}-\phi_{33}=\left( \phi_{32}-\phi_{31}\right) -\left( \phi
_{33}-\phi_{31}\right)
\end{equation}
are not independent. \ Figures~\textcolor{Blue}{\ref{DL3}}, \textcolor{Blue}{\ref{Dcostheta3}}, \textcolor{Blue}{\ref{DsinMCS3}}, %
\textcolor{Blue}{\ref{Ddphi31}}, \textcolor{Blue}{\ref{Ddphi32}}, and \textcolor{Blue}{\ref{Ddphi33}} show the distributions of ($%
L_{31}$, $L_{32}$, and $L_{33}$), ($\cos\theta_{31}$, $\cos\theta_{32}$, and 
$\cos\theta_{33}$), ($\sin\Theta_{31}$, $\sin\Theta_{32}$, and $\sin
\Theta_{33}$), ($\phi_{32}-\phi_{31}$ and $\cos\Omega_{321}$), ($%
\phi_{33}-\phi_{31}$ and $\cos\Omega_{331}$), ($\phi_{32}-\phi_{33}$ and $%
\cos \Omega_{323}$) from the neutrino-enriched sample, compared to
the GENIE\ default model.

\begin{figure*}[!hpt]
\centering
\subfloat{\includegraphics[width=0.5%
\linewidth]{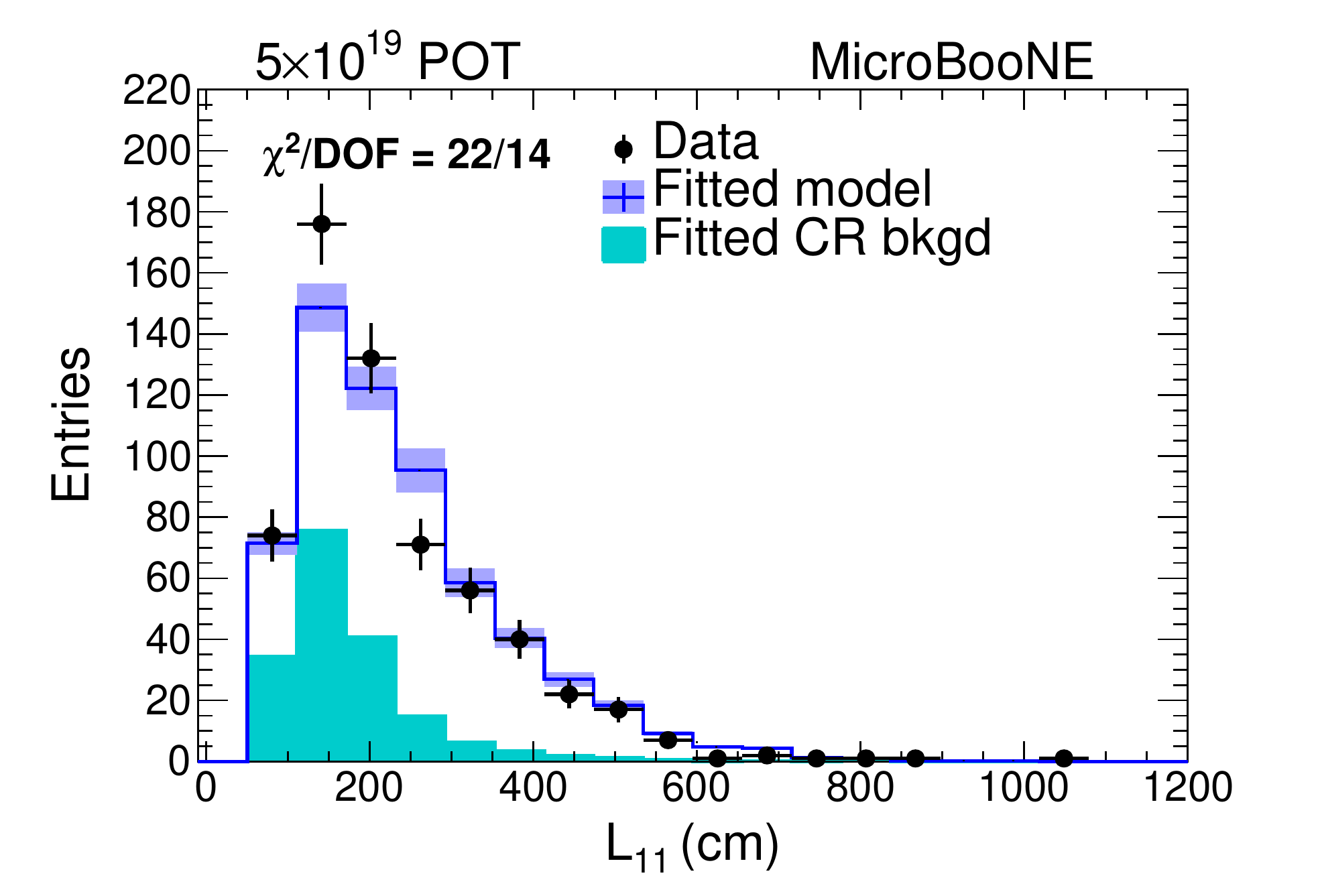}} \centering
\subfloat{\includegraphics[width=0.5%
\linewidth]{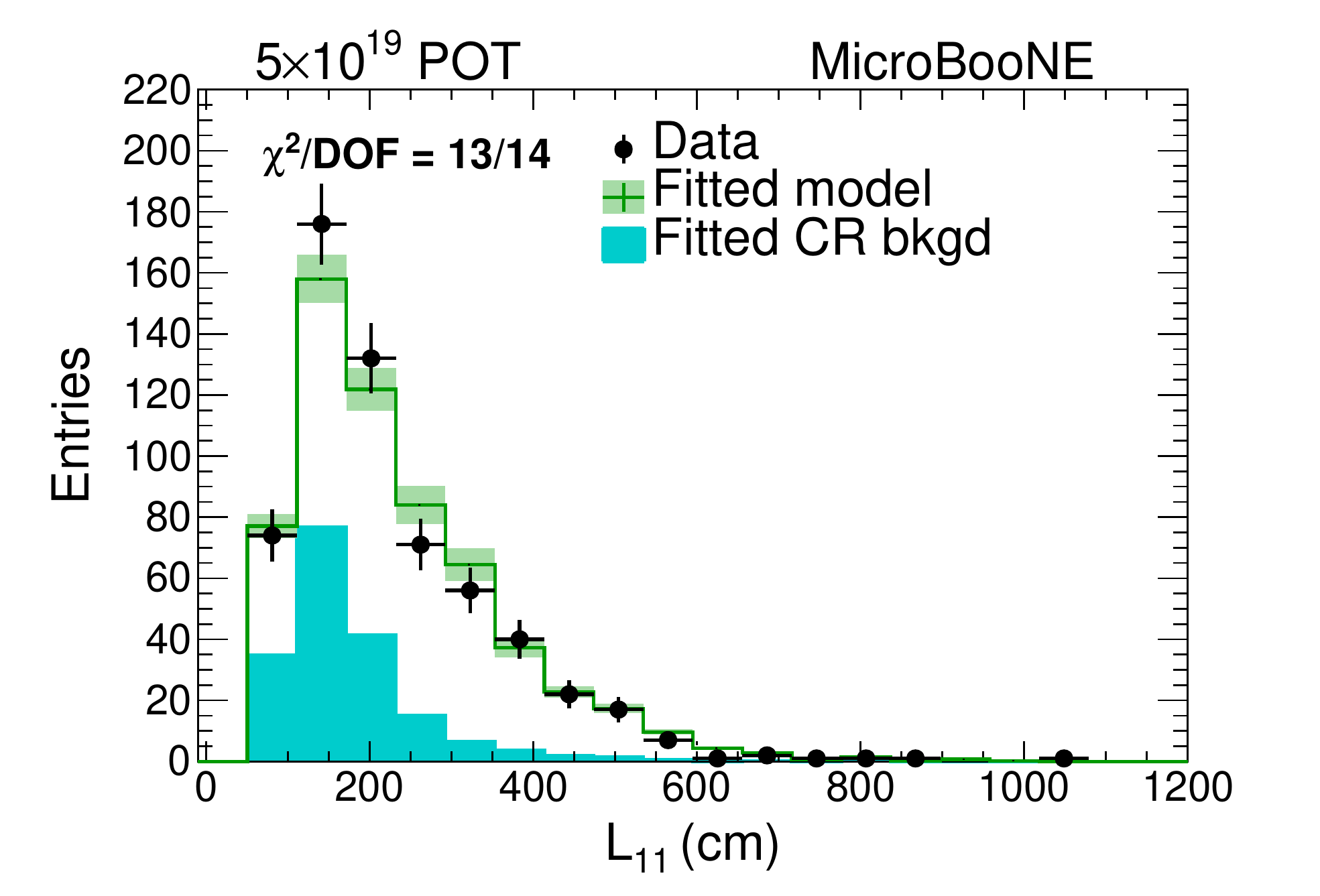}}
\caption{\textbf{Multiplicity = 1} candidate muon track length distribution
using GENIE+MEC model (left); using GENIE+TEM model (right) from
neutrino-enriched sample.}
\label{Dlenmult1Others}
\end{figure*}

\begin{figure*}[!hpt]
\centering
\subfloat{\includegraphics[width=0.5%
\linewidth]{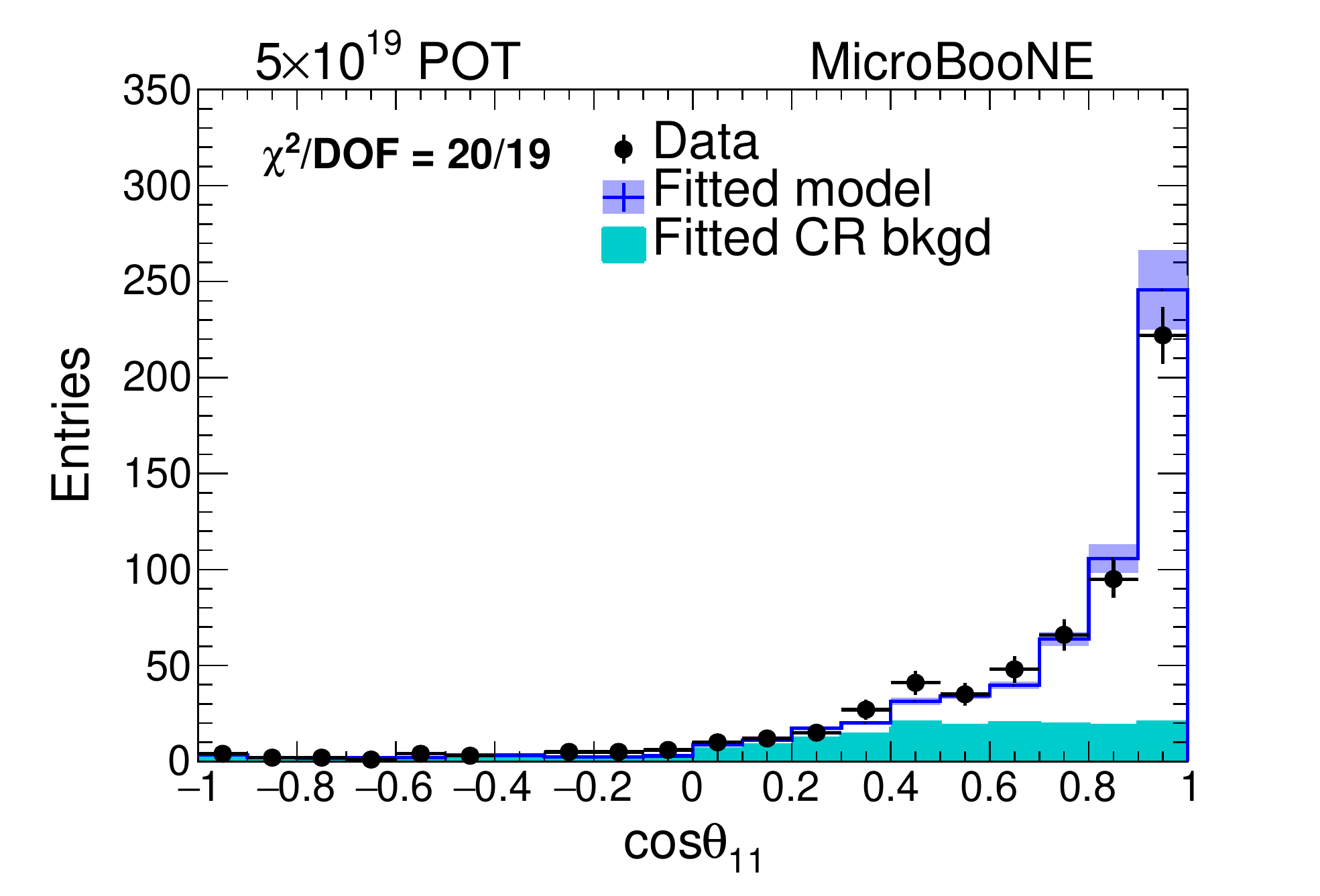}} \centering
\subfloat{\includegraphics[width=0.5%
\linewidth]{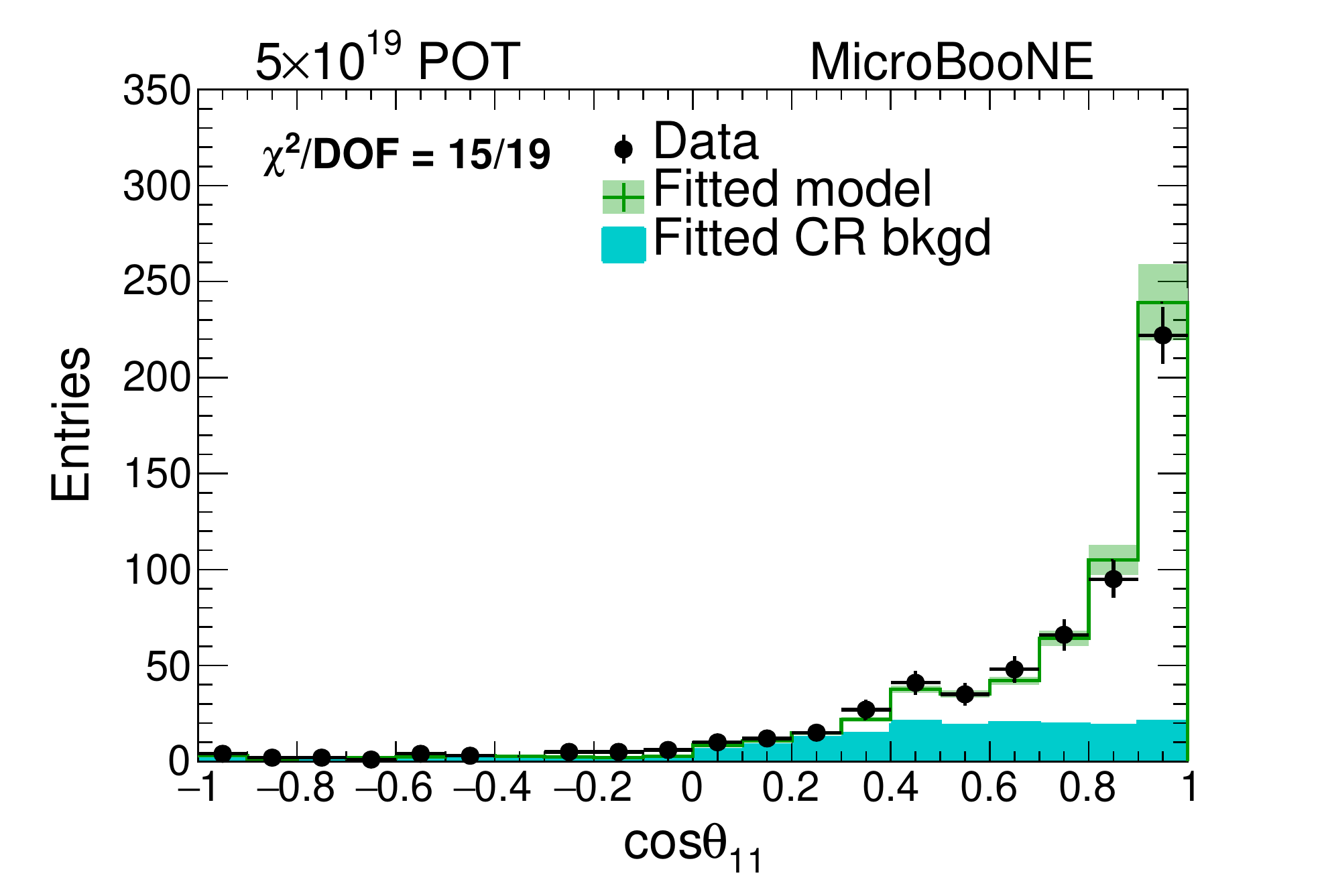}}
\caption{\textbf{Multiplicity = 1} candidate muon $\cos\protect\theta$
distribution using GENIE+MEC model (left); using GENIE+TEM model (right)
from neutrino-enriched sample.}
\label{Dthetamult1Others}
\end{figure*}

\begin{figure*}[!hpt]
\centering
\subfloat{\includegraphics[width=0.5%
\linewidth]{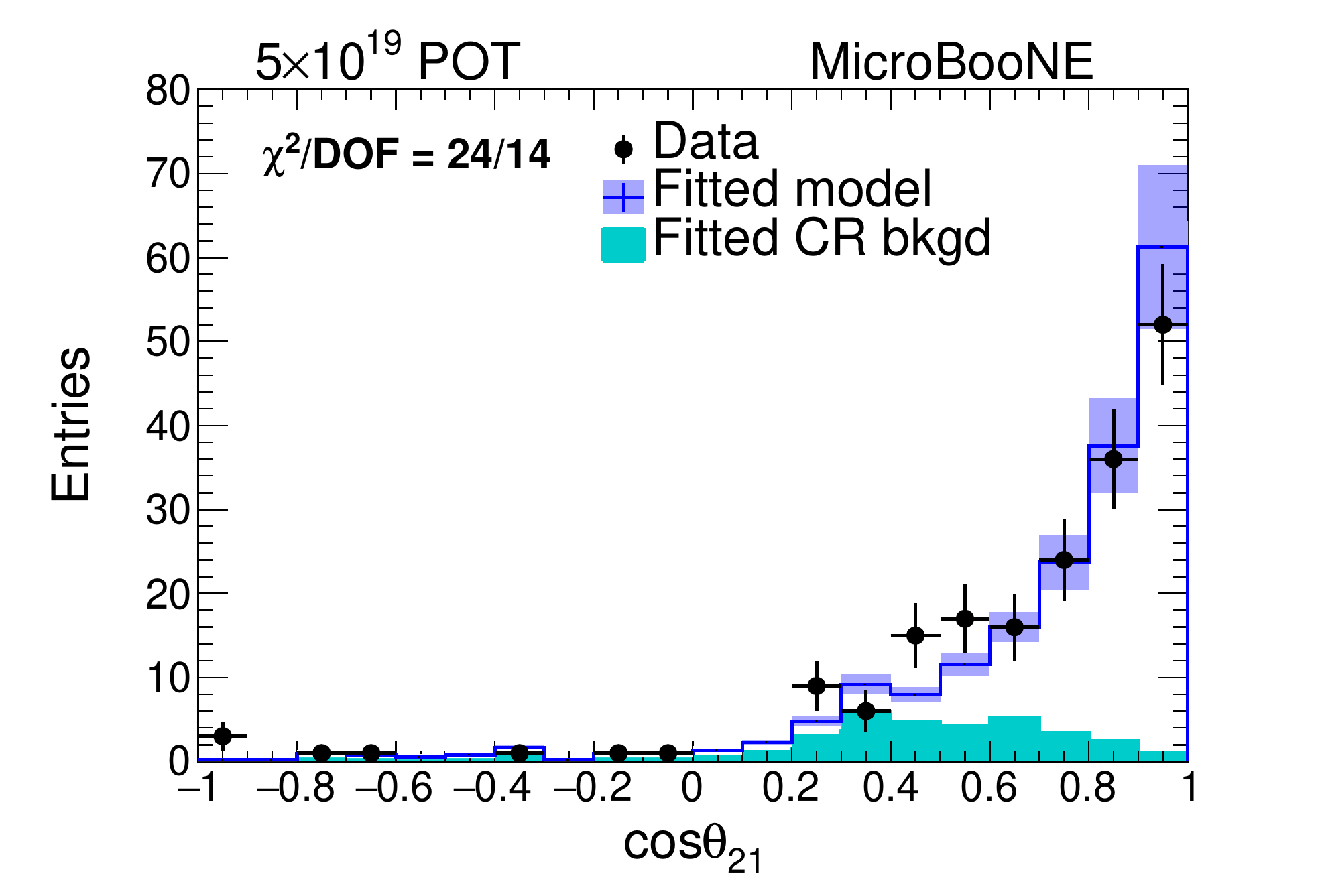}} \centering
\subfloat{\includegraphics[width=0.5%
\linewidth]{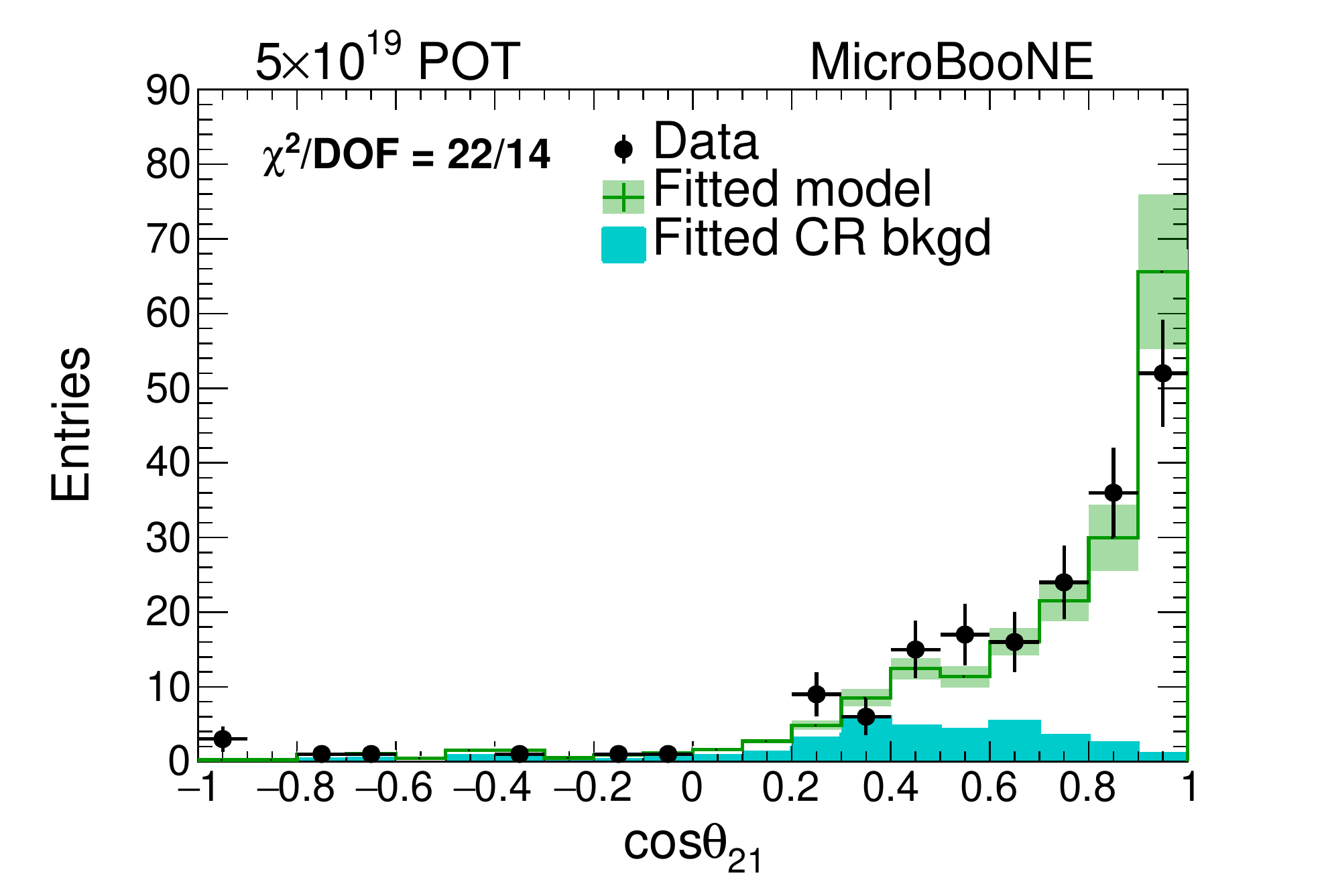}}
\caption{\textbf{Multiplicity = 2} $\cos\protect\theta$ distribution for
candidate muon using GENIE+MEC model (left); using GENIE+TEM model (right)
from neutrino-enriched sample.}
\label{DcosthetaOthers}
\end{figure*}

\subsection{$\protect\chi^{2}$ Tests for Kinematic Distributions}

We quantify agreement between model and observation through use of $\chi^{2}$
tests on the kinematic distributions described in Sec.~\textcolor{Blue}{\ref{Dynamic Variables}}. Ensemble tests have established the validity of the use of the $\chi^{2}$ criterion. We use only the \textquotedblleft neutrino-enriched\textquotedblright\ sample of events in which the candidate
muon passes both the $PH$\ and $MCS$ tests. \ Data are binned into
histograms, with a bin $k$ for a variable $x_{ij}$, $d_{ijk}$, and
compared to model predictions constructed by assuming that the number of
events in a bin $k$ of a variable $X_{ij}$, $m_{ijk}$, consists of
contributions from neutrino and CR background contributions. \ We shorten
the notation in Eq. \textcolor{Blue}{\ref{Model definition}} to%
\begin{equation}
m_{ijk}=M_{\nu,i}\hat{x}_{\nu,ijk}+M_{CR,i}\hat{x}_{CR,ijk},
\end{equation}
where $M_{\nu,i}$ and $M_{CR,i}$ are the number of neutrino and CR events,
respectively, predicted to be in the neutrino-enriched category for
multiplicity $i$ (as described in Sec.~\textcolor{Blue}{\ref{signal extraction}}); and $\hat{x}%
_{\nu,ijk}$ and $\hat{x}_{CR,ijk}$ the fraction of neutrino and CR events,
respectively, falling in the $k$ bin for variable $x_{ij}$ as predicted by
the GENIE model and the off-beam CR\ sample, respectively. \ The $\hat{x}%
_{\nu,ijk}$ and $\hat{x}_{CR,ijk}$ are shape distributions normalized to one:%
\begin{equation}
\sum_{k=1}^{bins}\hat{x}_{\nu,ijk}=\sum_{k=1}^{bins}\hat{x}_{CR,ijk}=1.
\end{equation}
We then construct a $\chi^{2}$ for $x_{ij}$ using a Poisson form appropriate
for the low statistics in many bins:%
\begin{equation}
\chi_{ij}^{2}=2\sum_{k=1}^{bins}\left( m_{ijk}-d_{ijk}-d_{ijk}\ln
m_{ijk}+d_{ijk}\ln d_{ijk}\right) . 
\end{equation}

\begin{table*}[!htb]
\caption{$\protect\chi^{2}$ test results for dynamically significant
variables for all three GENIE models. \ Only the uncorrelated statistical
uncertainties from data are used in forming the $\protect\chi^{2}$. \ Contributions
from systematic uncertainties are not included. The last five listed distributions are
not included in the total\textbf{\ }$\protect\chi^{2}/DOF$ since these
quantities can be expressed in terms of others.}
\label{tab:chi-square}
\begin{center}
\begin{ruledtabular}
\begin{tabular}
[c]{cccc}
& \multicolumn{3}{c}{\textbf{$\chi^{2}/DOF$}}\\
\textbf{Distributions} & \textbf{GENIE default} & \textbf{GENIE+MEC} &
\textbf{GENIE+TEM}\\\hline
$L_{11}$ & 19/14 & 22/14 & 13/14 \\
$L_{21}$ & 4.0/9 & 4.6/9 & 7.3/9 \\
$L_{22}$ & 10/7 & 8.4/7 & 16/7 \\
$L_{31}$ & 4.5/6 & 3.4/6 & 5.5/6 \\
$L_{32}$ & 5.8/5 & 3.9/6 & 6.5/6 \\
$L_{33}$ & 0.1/3 & 0.7/3 & 0.5/3 \\
$\cos \theta_{11}$ & 23/19 & 20/19 & 15/19 \\
$\cos \theta_{21}$ & 14/14 & 24/14 & 22/14 \\
$\cos \theta_{22}$ & 16/20 & 15/20 & 16/20 \\
$\cos \theta_{31}$ & 6.0/7 & 4.2/7 & 9.2/7 \\
$\cos \theta_{32}$ & 25/13 & 20/13 & 15/13 \\
$\cos \theta_{33}$ & 15/11 & 13/11 & 17/11 \\
$\sin \Theta_{11}$ & 24/20 & 21/20 & 25/20 \\
$\sin \Theta_{21}$ & 6.4/7 & 3.6/7 & 6.3/7 \\
$\sin \Theta_{22}$ & 2.4/7 & 3.4/7 & 2.4/6 \\
$\sin \Theta_{31}$  & 4.3/5 & 6.0/5 & 9.1/5 \\
$\sin \Theta_{32}$  & 2.1/4 & 2.5/4 & 1.6/4 \\
$\sin \Theta_{33}$  & 8.5/6 & 7.0/5 & 9.5/6 \\
$\phi_{22}-\phi_{21}$ & 13/15 & 12/15 & 14/15 \\
$\phi_{32}-\phi_{31}$ & 10/13 & 9.2/13 & 10/14 \\
$\phi_{33}-\phi_{31}$ & 15/12 & 13/12 & 8.7/11 \\
$\phi_{32}-\phi_{33}$ & 11/14 & 11/14 & 11/14 \\
$\cos \Omega_{221}$  & 19/20 & 13/20 & 13/20 \\
$\cos \Omega_{321}$ & 14/13 & 13/13 & 17/13 \\
$\cos \Omega_{331}$ & 21/14 & 16/14 & 12/14 \\
$\cos \Omega_{323}$ & 12/15 & 18/15 & 19/15 \\
\textbf{Total $\chi^{2}/DOF$} & \textbf{228.1/216} & \textbf{216.9/216} &\textbf{229.6/216}\\
\end{tabular}
\end{ruledtabular}
\end{center}
\end{table*}

Table~\textcolor{Blue}{\ref{tab:chi-square}} summarizes the results of these $\chi^{2}$
comparison tests for 21 independent kinematic variables to the three GENIE
models. \ Only bins with at least one data event and one model event were
used in the calculation of $\chi^{2}$. \ The number of degrees of freedom
associated with the $\chi^{2}$ test was set equal to the number of bins used
for that histogram minus one to account for the overall normalization adjustment. \ We note here that these tests for consistency are
defined at the level of statistical uncertainties only; systematic
uncertainties are not incorporated into the $\chi^{2}$ terms.

We summarize here salient features of Table~\textcolor{Blue}{\ref{tab:chi-square}} as follows:
\ All three models consistently describe the data, with summed $\chi^{2}$
per degree-of-freedom ($\chi^{2}/DOF$) of $228.1/216$, $216.9/216$, and $229.6/216$,
respectively, and corresponding  p-values of $%
P_{\chi^{2}}=27\%$, $47\%$, and $25\%$ for GENIE default, MEC, and TEM,
respectively. The total $\chi^{2}$ after including all dynamic and non-dynamic variable distributions is $714/652$. \ No tune of GENIE\ is superior to any other with any
meaningful statistical significance for the distributions we have
considered. \ The acceptable values of $\chi^{2}$ are consistent with the
hypothesis that the combination of a GENIE event generator, the MicroBooNE
BNB flux model, and the MicroBooNE GEANT-based detector simulation
satisfactorily describe the properties of neutrino events examined in this
analysis in a shape comparison. \ All elements of the MicroBooNE analysis chain thus appear to be
performing satisfactorily; and no evidence exists for missing systematic
effects that would produce data-model discrepancies outside the present
level of statistics.

Aggregating Table~\textcolor{Blue}{\ref{tab:chi-square}} different ways uncovers no
significant discrepancies. \ The $\chi^{2}$ tests on leading track $%
\cos\theta$ and $\sin\Theta$ yield satisfactory results for all
multiplicities. \ Combined $\chi^{2}/DOF$ for all distributions associated
with a particular multiplicity likewise exhibit adequate agreement. \ The
most poorly described single distribution is that for the length of the muon
candidate in multiplicity $1$ events. The $P_{\chi^{2}}$, while acceptable,
are $16\%$ and $8\%$ for the GENIE default and GENIE+MEC, respectively. \ The
GENIE+TEM model has $P_{\chi^{2}}=53\%$. \ 

The $\chi^{2}$ values for different distributions in a given multiplicity
are calculated using the same events, which gives rise to concerns about
correlations between different distributions. \ We have performed studies that verify that the $\chi^{2}$
values would be highly correlated if the model
and data disagreed by an overall normalization, but that otherwise the $%
\chi^{2}$ tests on different distributions exhibit independent behavior,
even when the same events are used. \ The $P_{\chi^{2}}$ values for
different distributions do not cluster near $0$ or $1$, which is consistent
with the view that the projections display approximately independent
statistical behavior.

In summary, all GENIE\ models successfully describe, through $\chi^{2}$
tests, the shapes of a complete set of dynamically significant kinematic
variables for observed charged particle multiplicity distributions 1, 2, and
3. \ The statistical power--the highest precision afforded by the available statistics with which the predictions can be tested--of these tests from the overall data statistics
available corresponds to approximately $4\%$, $7\%$, and $20\%$ for
multiplicity 1, 2, and 3, respectively.

\subsection{$\protect\chi^{2}$ Tests for Multiplicity Distribution}

\label{chisq-multiplicity}

While we find satisfactory agreement between GENIE\ models and kinematic
distribution shapes using $\chi^{2}$ tests that incorporate only statistical
uncertainties, the situation differs for the overall multiplicity
distribution. \ Here, we find statistical $\chi_{M}^{2}/$DOF$=30/4$, $%
22/4$, and $28/4$ for the default, MEC, and TEM GENIE\ models, respectively.
\ However, in the case of multiplicity, a significant systematic uncertainty
exists for tracking efficiency that must be taken into account before any
conclusion can be drawn.

We incorporate a tracking efficiency contribution to the $\chi^{2}$ test by
defining%
\begin{widetext}
\begin{align}
\chi_{M}^{2} &  =\sum_{i=1}^{2}\frac{\left(  D_{i}-K\hat{M}\left(
\delta\right)  _{i}\right)  ^{2}}{\sigma_{i}^{2}}+\sum_{i=3}^{5}2\left[
K\hat{M}\left(  \delta\right)  _{i}-D_{i}\ln\left(  K\hat{M}\left(
\delta\right)  _{i}\right)  -D_{i}+D_{i}\ln\left(  D_{i}\right)  \right]
\label{chisq-final}\\
&  +\sum_{i=1}^{5}2\left[  \hat{M}\left(  0\right)  _{i}-M_{i}\ln\left(
\hat{M}\left(  0\right)  _{i}\right)  -M_{i}+M_{i}\ln\left(  M_{i}\right)
\right]  +\left(  \frac{\delta}{0.15}\right)  ^{2}.\nonumber
\end{align}
\end{widetext}
Here $D_{i}$ is the number of neutrino events estimated by the signal
extraction procedure (Sec.~\textcolor{Blue}{\ref{signal extraction}}), and $\sigma_{i}$ is the
estimated uncertainty on $D_{i}$ using the signal extraction procedure. \ For
multiplicity $3$ and higher, the uncertainty on $D_{i}$ is purely statistical
as the CR\ background becomes negligible. \ The quantities $M_{i}$ are the
number of events in the MC sample with multiplicity $i$. \ Finite statistics
in the MC sample are incorporated by interpreting the $M_{i}$ as Poisson
fluctuations about their true values $\hat{M}\left( 0\right) _{i}$ in the
third term of Eq.~\textcolor{Blue}{\ref{chisq-final}}. \ This analysis does not absolutely
normalize MC\ to data, hence the relative normalization of data to MC is
allowed to float via the parameter $K$ in the first term of Eq.~\textcolor{Blue}{\ref%
{chisq-final}}. The normalization constant $K$, while not used directly in the model test,
is consistent with the predicted value from the default GENIE model.

As discussed in Sec.~\textcolor{Blue}{\ref{systematic errors}}, changing the per-track
efficiency by a constant fraction $\delta$ in the model would shift events
between multiplicities according to%
\begin{align}
\hat{M}\left( \delta\right) _{4} & =\left[ \hat{M}\left( 0\right) _{4}\right]
\left( 1-\delta\right) ^{3}, \\
\hat{M}\left( \delta\right) _{3} & =\left[ \hat{M}\left( 0\right) _{3}+3\hat{%
M}\left( 0\right) _{4}\delta\right] \left( 1-\delta\right) ^{2}, \\
\hat{M}\left( \delta\right) _{2} & =\left[ \hat{M}\left( 0\right) _{2}+2\hat{%
M}\left( 0\right) _{3}\delta+3\hat{M}\left( 0\right) _{4}\delta^{2}\right]
\left( 1-\delta\right) , \\
\hat{M}\left( \delta\right) _{1} & =\left[ \hat{M}\left( 0\right) _{1}+\hat{M%
}\left( 0\right) _{2}\delta+\hat{M}\left( 0\right) _{3}\delta^{2}+\hat{M}%
\left( 0\right) _{4}\delta^{3}\right] .
\end{align}
For the nominal model used in the MC\ simulation $\delta=0$. \ As discussed
in Sec.~\textcolor{Blue}{\ref{systematic errors}} we estimate the uncertainty on $\delta$ to
be $15\%$, and we introduce this into $\chi_{M}^{2}$ through the
\textquotedblleft pull term\textquotedblright\ $\left( \delta/0.15\right)
^{2}$.

We minimize $\chi_{M}^{2}$ with respect to the tracking efficiency pull
parameter $\delta$, the MC-to-data normalization $K$, and the five MC
statistical quantities $\hat{M}\left( 0\right) _{i}$, $i=1-5$. \ This
procedure yields%
\begin{align}
\chi_{M}^{2}/\text{DOF} & =6.4/3\text{ (default), }4.3/3\text{ (MEC), }5.8/3%
\text{ (TEM),} \\
\delta & =0.32\text{ (default)},0.27\text{ (MEC), }0.32\text{ (TEM).}
\end{align}
We find that a satisfactory $\chi^{2}$ value can be obtained for the
multiplicity distribution itself, albeit at the cost of a $\approx2\sigma$
pull in the parameter $\delta$. \ 

\section{\label{Discussion}Discussion}

\subsection{GENIE Predictions for Observed Multiplicity\ }

\label{expectations} \ 

At BNB energies, the nominal GENIE expectations for charged particle
multiplicities at the neutrino interaction point are 
$\approx \left(
80\%\right) $ $n=2$ (from quasi-elastic scattering, $\nu_{\mu}n\rightarrow
\mu^{-}p$, neutral pion resonant production $\nu_{\mu}n\rightarrow%
\mu^{-}R^{+}\rightarrow\mu^{-}p\pi^{0}$, and coherent pion production $%
\nu_{\mu }\mathrm{Ar}\rightarrow\mu^{-}\pi^{+}\mathrm{Ar}$); $\approx%
\left( 20\%\right) $ $n=3$ (resonant \ charged pion production $\nu_{\mu
}p\rightarrow\mu^{-}R^{++}\rightarrow\mu^{-}p\pi^{+}$); and $\approx%
\left( 1\%\right) $ $n\geq4$ (from multi-particle production processes
referred to as DIS). \ However, final state interactions (FSI)
of hadrons produced in neutrino scattering with the argon nucleus can
subtract or add charged particles that emerge from within the nucleus. These
multiplicities are further modified by the selection criteria.

The following list summarizes qualitative expectations for components of observed
multiplicities from particular processes. \ These components can include
contributions from the primary neutrino-nucleon scatter within the nucleus
and secondary interactions of primary hadrons with the remnant nucleus. \
Secondary charged particles are usually protons, which are expected to be
produced with kinetic energies that are usually too low for track
reconstruction in this analysis. \ However, more energetic forward-produced
protons from the upper \textquotedblleft tail\textquotedblright\ of this
secondary kinetic energy distribution may contribute. Note that the particle-type-dependent kinetic energy thresholds for charged particles entering our sample range from $31$ MeV for a $\pi ^\pm$ to $69$ MeV for a proton.

\begin{itemize}
\item Multiplicity $>3$, mainly predicted to be \textquotedblleft DIS
events\textquotedblright\ in which at least three short tracks are
reconstructed. \ \textquotedblleft DIS\textquotedblright\ is the usual term
for multi-particle final states not identified with any particular resonance
formation.

\item Multiplicity $=3$, mainly predicted to be $\mu^{-}p\pi^{+}$ events
from $\Delta$ resonance production in which all three tracks are
reconstructed. \ \textquotedblleft Feed down\textquotedblright\ from higher
multiplicity would be small due to the relatively small DIS cross section at MicroBooNE
energies.

\item Multiplicity $=2$, mainly predicted to be QE $\mu^{-}p$ events and
resonant $\mu^{-}p\pi^{0}$ events in which the proton is reconstructed, with
a sub-leading contribution from \textquotedblleft feed down" of resonant
charged pion production events where one track fails to be reconstructed.

\item Multiplicity $=1$, mainly predicted to be \textquotedblleft feed
down\textquotedblright\ from QE $\mu^{-}p$ and $\mu^{-}p\pi^{0}$ events in
which the proton is not reconstructed, with contributions from other higher
multiplicity topologies in which more than one track fails to be
reconstructed.
\end{itemize}

\begin{figure*}[!hpt]
\centering
\subfloat{\includegraphics[width=0.5\linewidth]{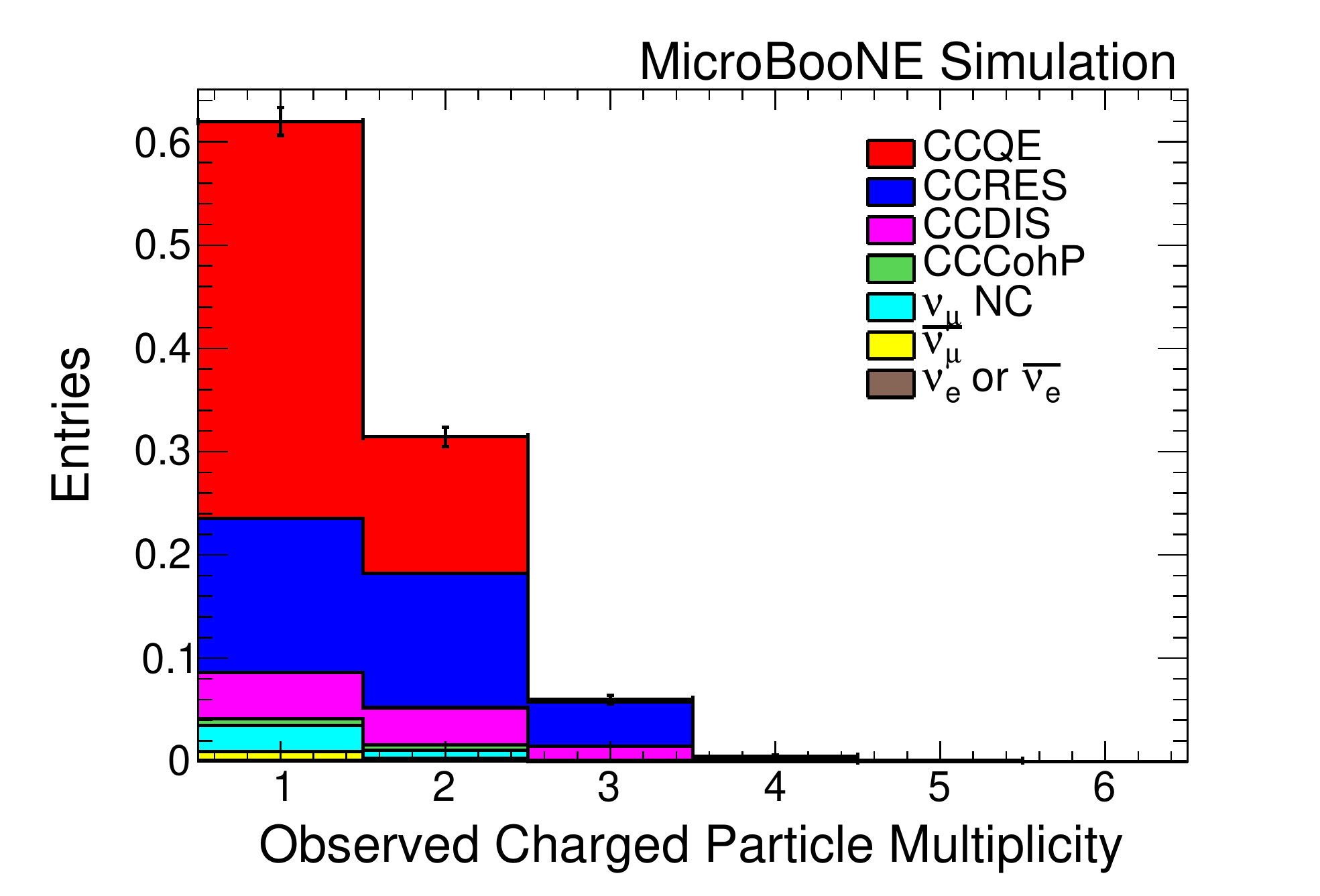}} \centering
\subfloat{\includegraphics[width=0.5\linewidth]{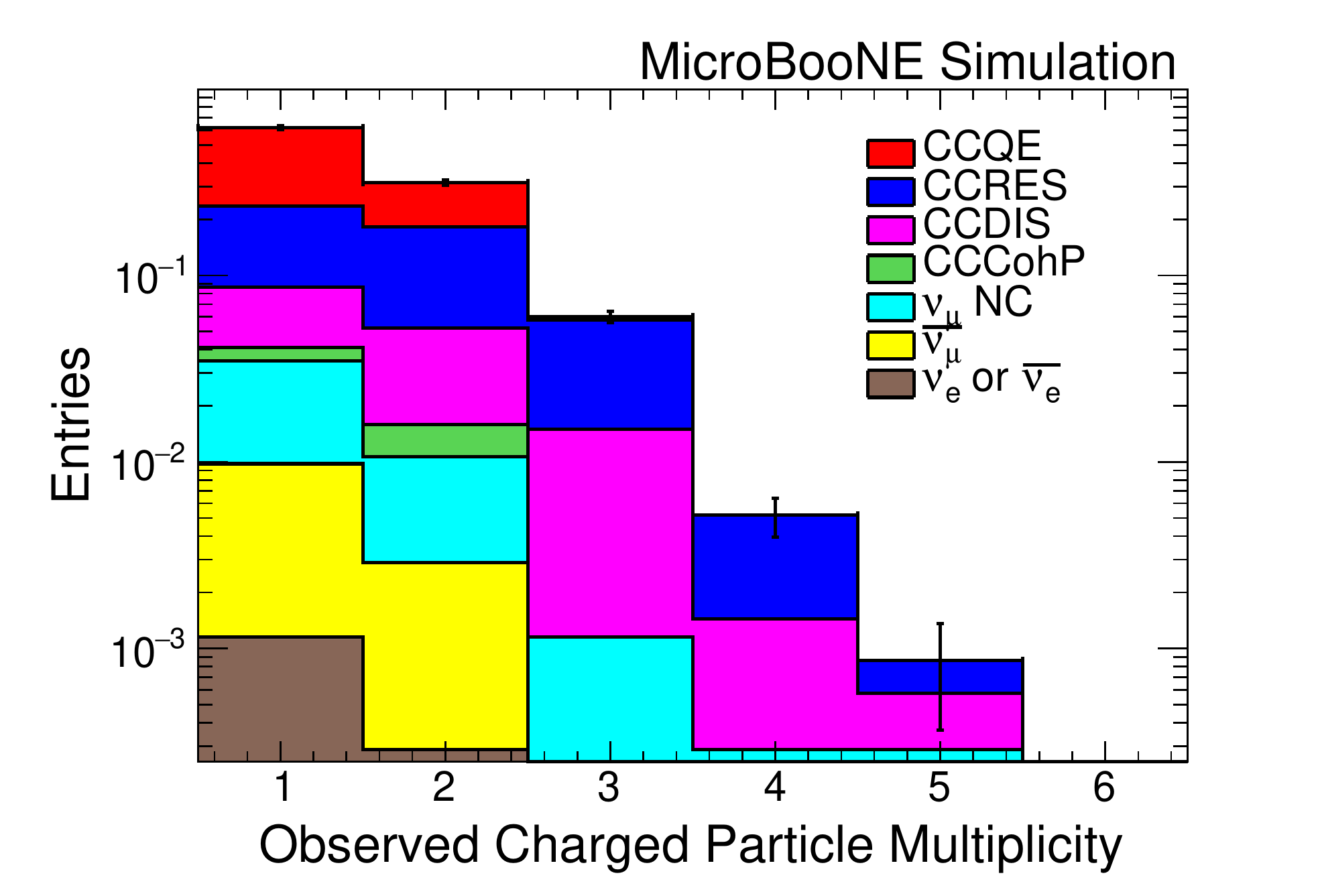}}
\caption{Observed (stacked) multiplicity distributions for different
neutrino interaction types from BNB-only default MC simulation in linear
scale (left); and in log y scale (right). Black error bars represent MC statistical uncertainties.}
\label{img:inttype_MC_MCtrue}
\end{figure*}

Figure~\textcolor{Blue}{\ref{img:inttype_MC_MCtrue}} illustrates these expectations from GENIE. \ We note that, as expected, the three-track topology
is dominated by resonant pion production in the default GENIE model, while
the two-track and one-track topologies are QE-dominated with non-negligible
resonance feed-down. The coherent pion production process ($\nu_\mu +$ Ar $\rightarrow$ $\mu^- \pi^+ +$ Ar) denoted by ``CCCohP" in this figure, as well as NC and $\nu_{e}$ and $\bar{\nu}_{e}$ scattering, only lead to small contributions. \ 

Our observation of discrepancy of data compared to simulation in three-track compared to two-track topologies, shown in Fig.~\textcolor{Blue}{\ref{img:final_mult_dist}}, is qualitatively similar to the low $\nu_\mu$ CC pion cross sections compared to GENIE\ reported by 
MINERvA~\textcolor{Blue}{\cite{McGivern:2016bwh}} using hydrocarbon targets at the somewhat
higher neutrino energy from the Fermilab Neutrinos at the Main Injector (NuMI) beam.
\ The T2K experiment~\textcolor{Blue}{\cite{Abe:2016aoo}} also reports a low pion production
cross section relative to GENIE expectations using water targets in a
neutrino beam with comparable energy to the BNB. \ However, MiniBooNE measured a charged pion production rate more in agreement with GENIE using mineral oil as a target in the same Fermilab BNB
as used by MicroBooNE~\textcolor{Blue}{\cite{Rodrigues}}.

MicroBooNE also observes more one-track events than GENIE\ predicts, as shown in Fig.~\textcolor{Blue}{\ref{img:final_mult_dist}}. \ This
corroborates ArgoNeuT's observation that approximately 35$\%$ of
neutrino interactions on argon targets with no pions detected in the final
state also contained no observable proton~\textcolor{Blue}{\cite{Partyka:2015rua}}.

While our observed multiplicity distribution disagrees with GENIE\
expectations and shows consistency with a number of other experiments, we
cannot, as noted in Sec.~\textcolor{Blue}{\ref{chisq-multiplicity}} definitively exclude an
alternate explanation of the discrepancy in terms of a tracking efficiency error at this time.

Our kinetic energy thresholds limit acceptance in such a way that protons
produced in FSI may not significantly contribute to the observed CPMD.
Furthermore, our analysis requires a forward-going long contained track as a
muon candidate, which restricts the final state phase space. Also, our analysis makes use of fully automated reconstruction. \ Therefore, results of this analysis should not be directly compared to the low energy proton multiplicity measurement reported by ArgoNeuT~\textcolor{Blue}{\cite{Acciarri:2014gev}}.

\subsection{GENIE Predictions for Kinematic Distributions}

Kinematic distributions for fixed multiplicity suffer much less from
tracking-related systematic uncertainties than the multiplicity
probabilities; hence GENIE expectations for the shapes of kinematic
distributions can be compared directly to data. \ The MEC and TEM tunes
of GENIE\ primarily change the normalization of QE-like event topologies
relative to resonance type topologies, and secondarily modify properties of
low energy final state protons that would usually not satisfy our acceptance
criteria. \ Shape comparisons would thus not be expected to differentiate MEC
and TEM from the GENIE\ default, and we have verified this expectation with our $%
\chi^{2}$ tests. \ Accordingly we confine the following discussion to the
default GENIE tune.

\begin{figure*}[!hpt]
\centering
\begin{tabular}{cc}
\subfloat{\includegraphics[width=.5%
\textwidth]{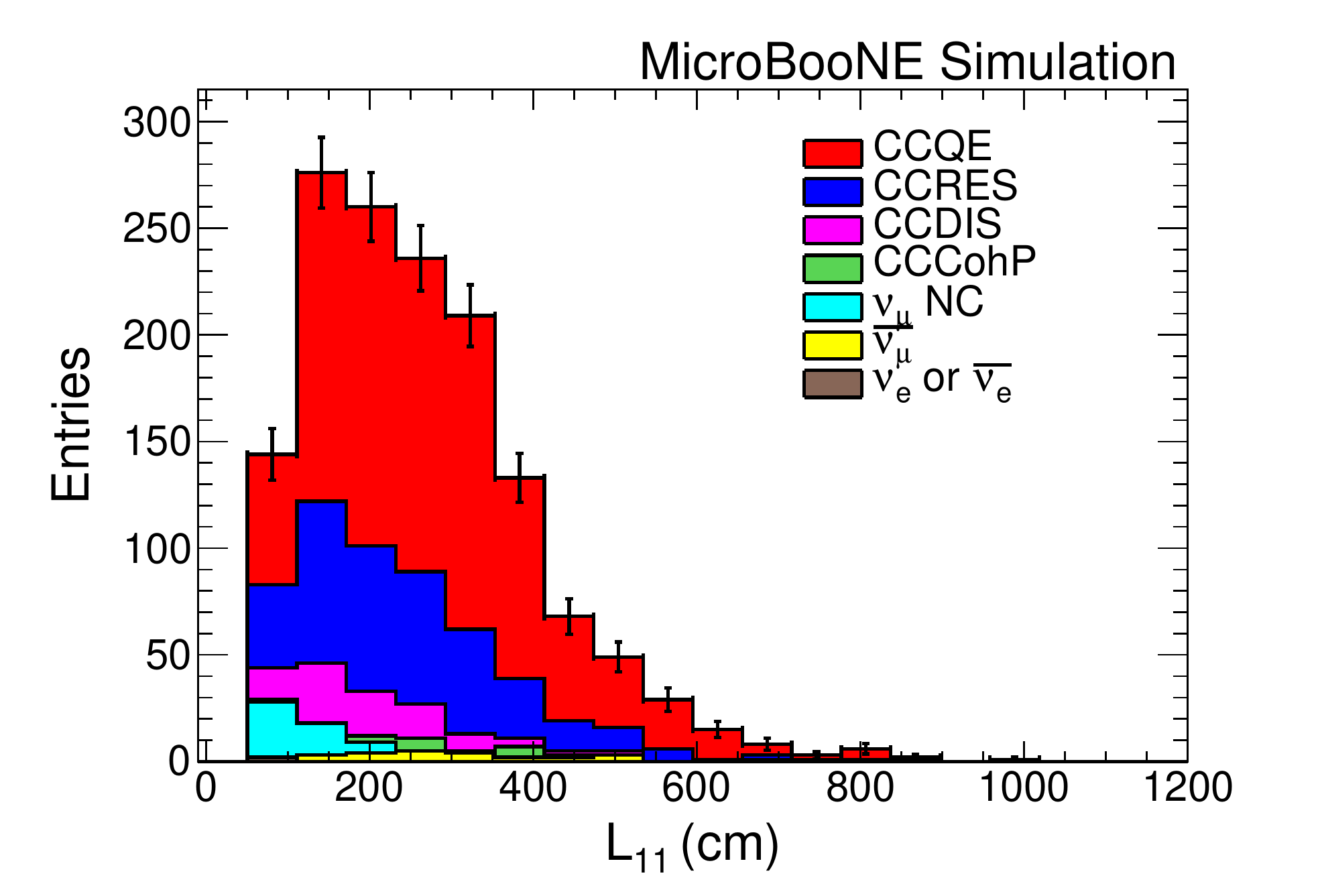}} & \subfloat{%
\includegraphics[width=.5\textwidth]{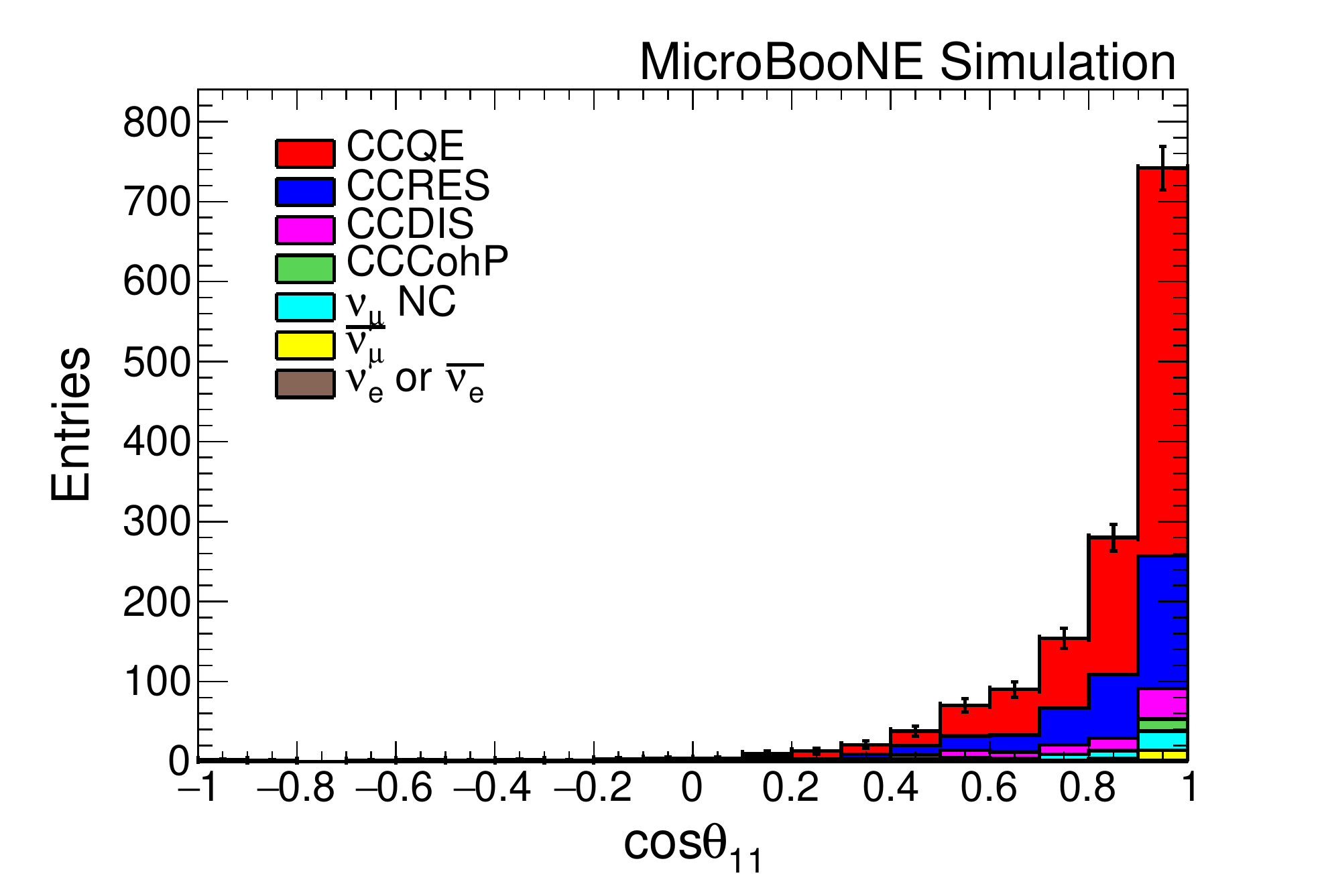}} \\ 
\multicolumn{2}{c}{\subfloat{\includegraphics[width=.5%
\textwidth]{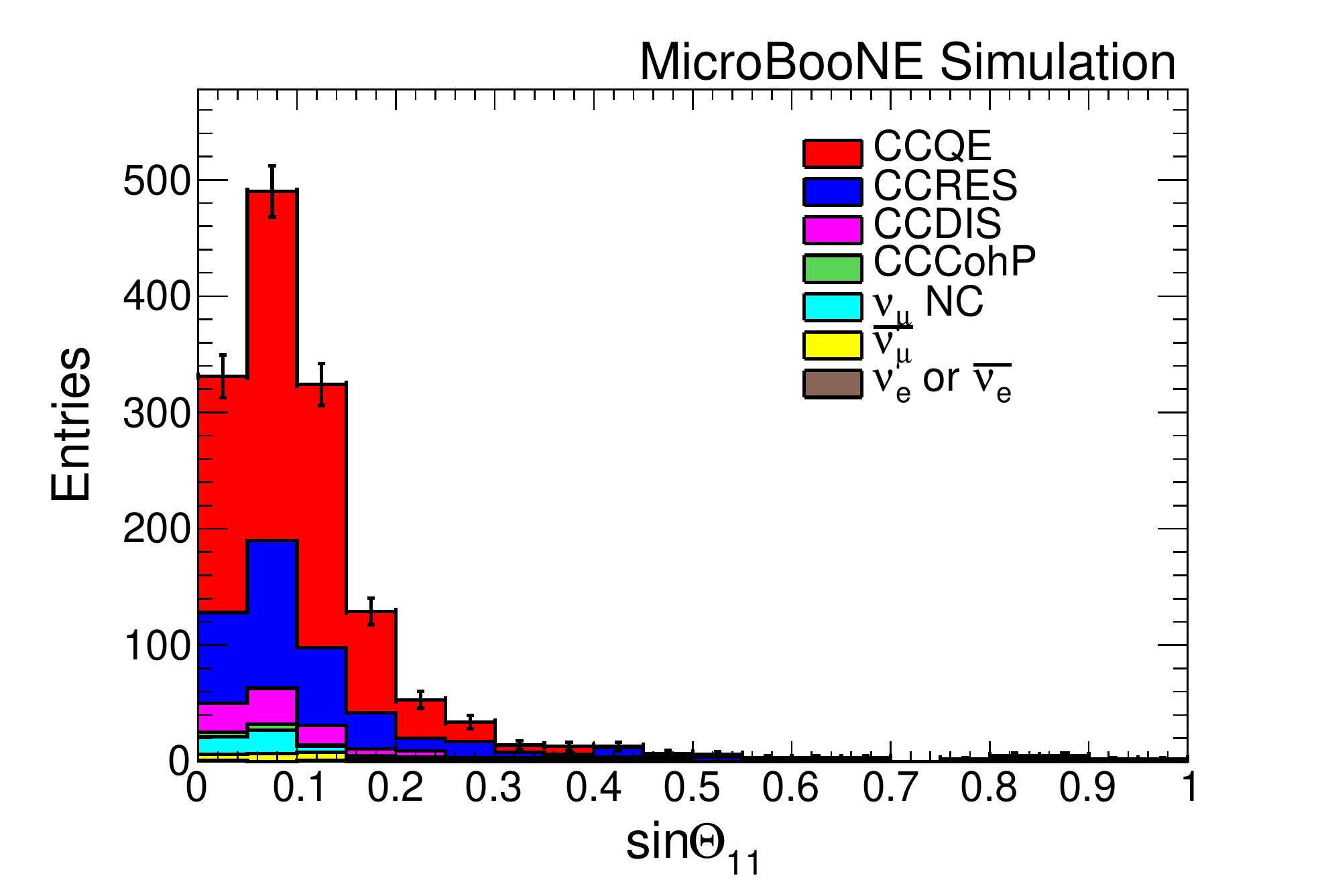}}}%
\end{tabular}
\caption{\textbf{Multiplicity = 1} GENIE default predictions for candidate muon
track length, $\cos\protect\theta$, and $\sin\Theta$ distributions. Black error bars represent MC statistical uncertainties.}
\label{Tmult1}
\end{figure*}

Figure~\textcolor{Blue}{\ref{Tmult1}} shows the predictions for reconstructed $L_{11}$, $\cos\theta_{11}$%
, and $\sin\Theta_{11}$ from the neutrino-enriched sample, using the GENIE\ default
model. The muon track candidate is only mildly affected by the details of
the recoiling hadronic system, and thus QE, RES, and DIS\ production
produce similar shape contributions to $L_{11}$, $\cos\theta_{11}$, and $%
\sin \Theta_{11}$.

\begin{figure*}[!hpt]
\centering
\subfloat{\includegraphics[width=0.5%
\linewidth]{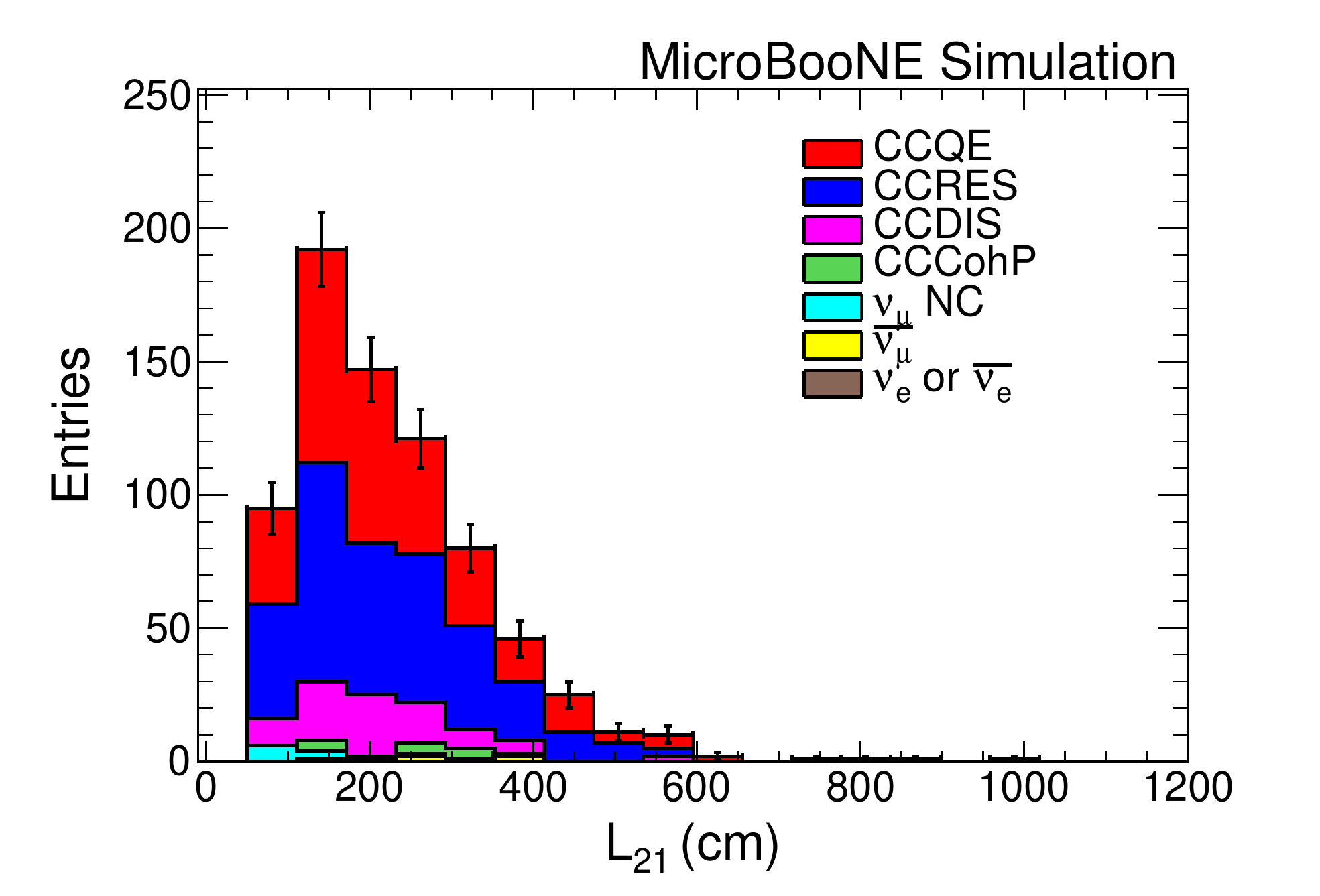}} \centering
\subfloat{\includegraphics[width=0.5%
\linewidth]{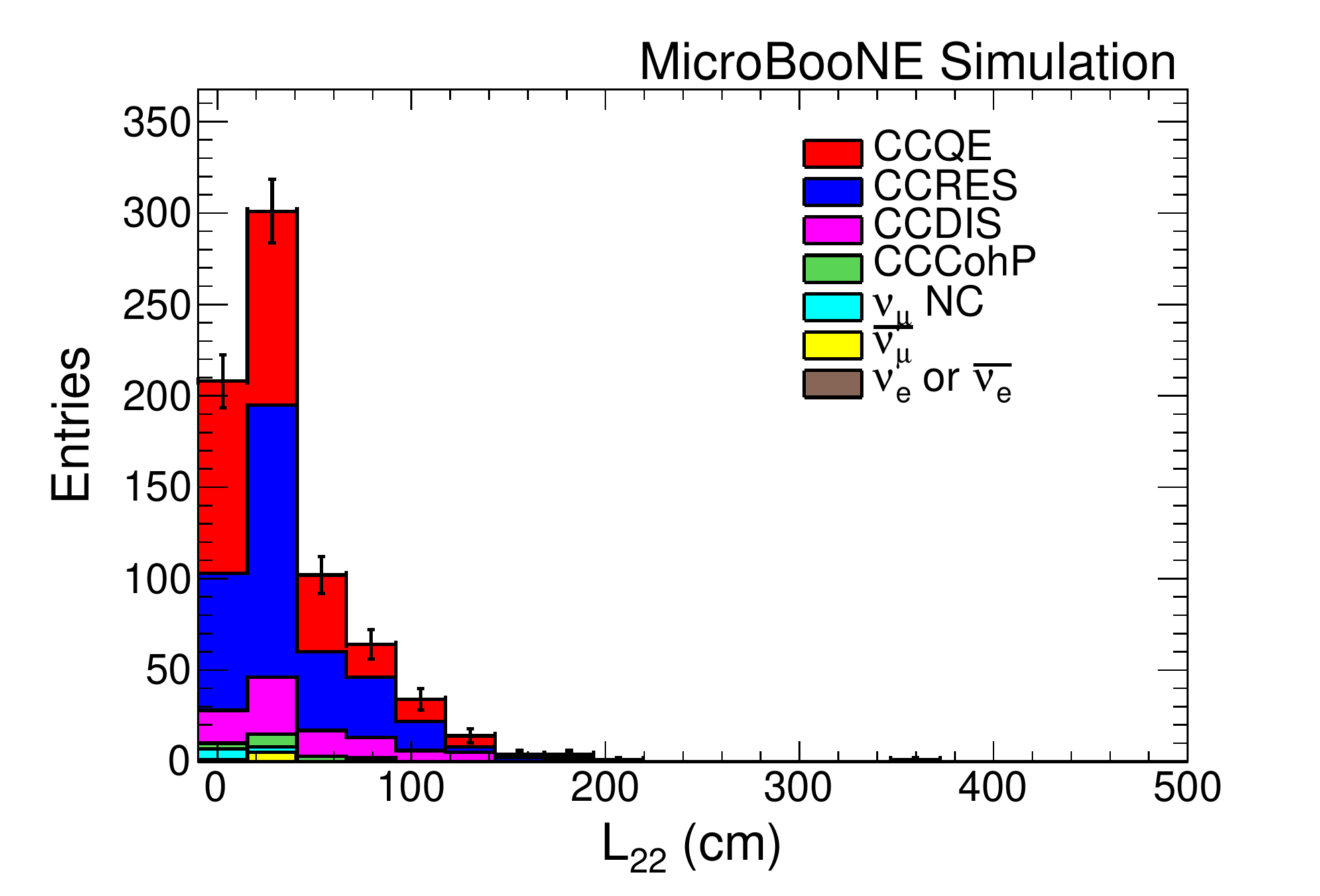}}
\caption{\textbf{Multiplicity = 2} Track length distribution for candidate
muon (left); for second track (right) from GENIE default MC. Black error bars represent MC statistical uncertainties.}
\label{TL2}
\end{figure*}

\begin{figure*}[!hpt]
\centering
\subfloat{\includegraphics[width=0.5%
\linewidth]{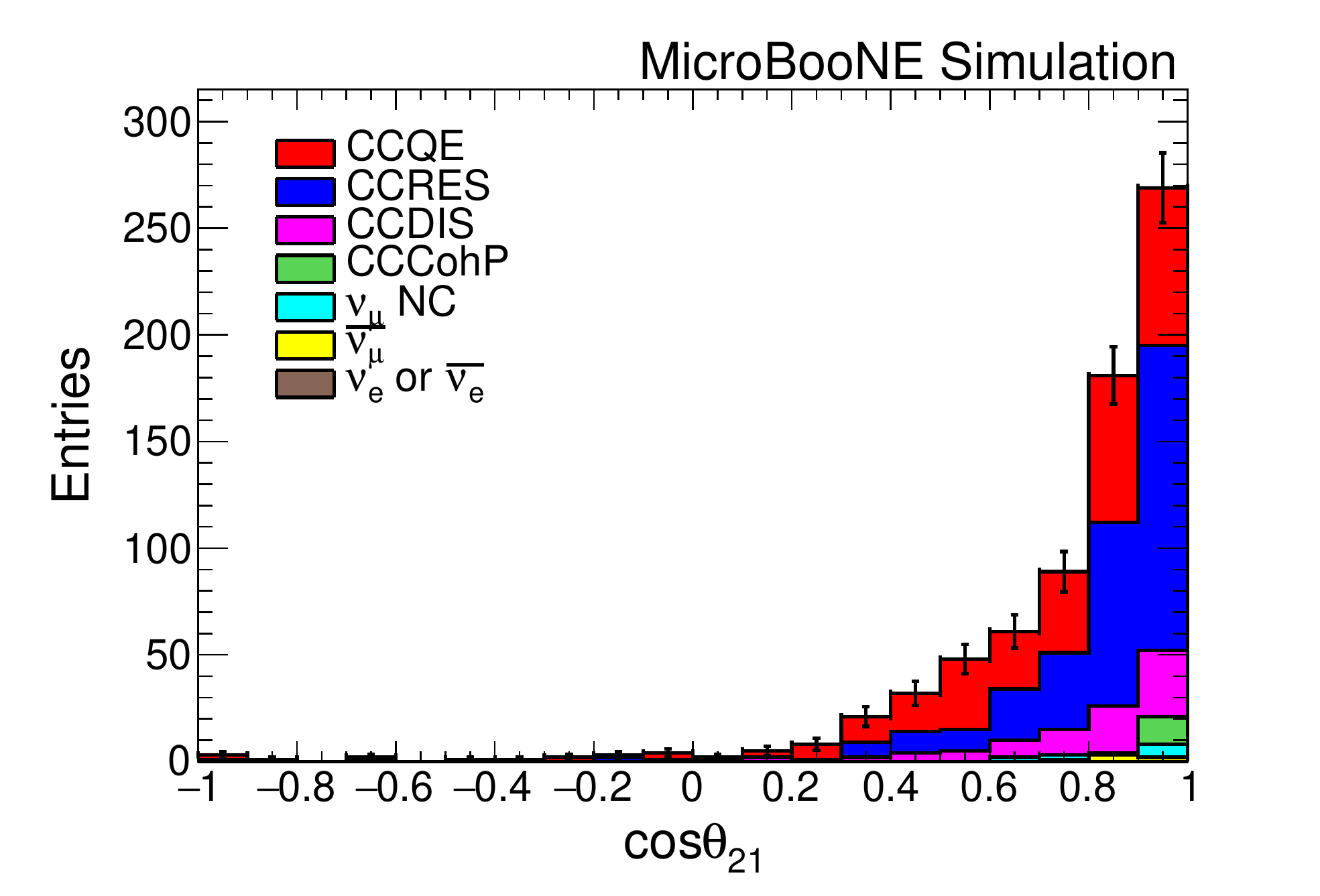}} \centering
\subfloat{\includegraphics[width=0.5%
\linewidth]{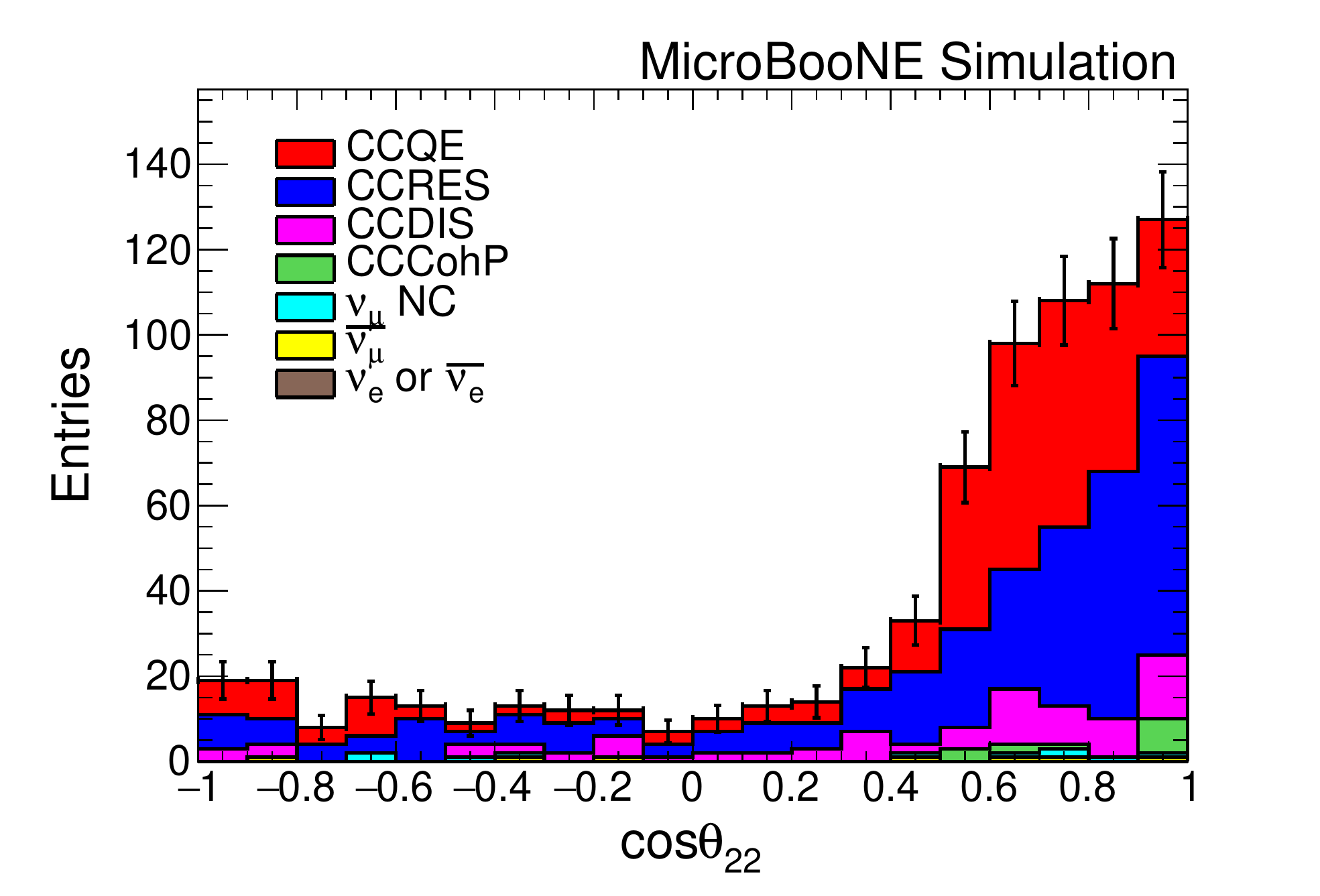}}
\caption{\textbf{Multiplicity = 2} Cosine of polar angle distribution for
candidate muon (left); for second track (right) from GENIE default MC. Black error bars represent MC statistical uncertainties.}
\label{Tcostheta2}
\end{figure*}

\begin{figure*}[!hpt]
\centering
\subfloat{\includegraphics[width=0.5%
\linewidth]{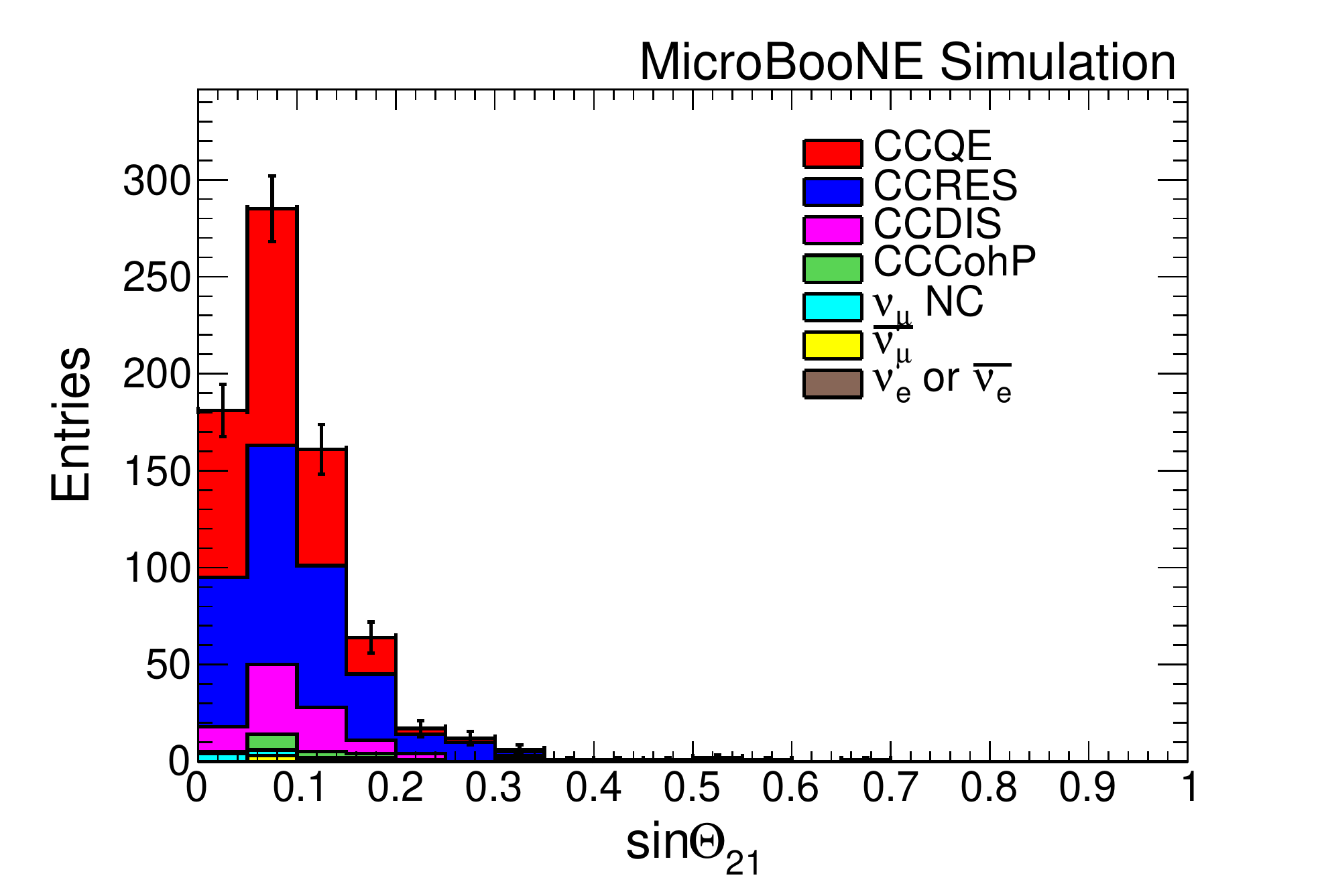}} \centering
\subfloat{\includegraphics[width=0.5%
\linewidth]{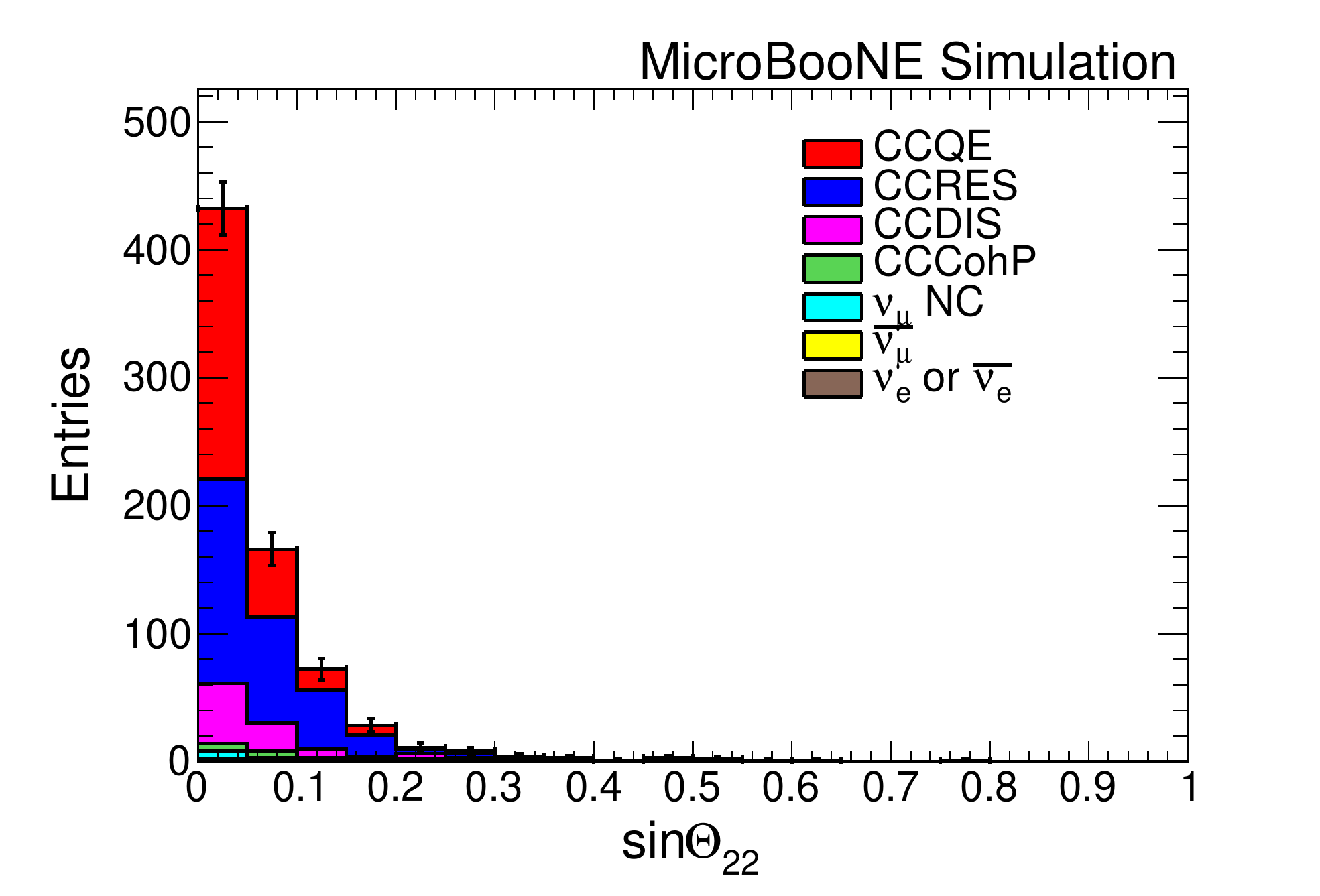}}
\caption{\textbf{Multiplicity = 2} Sin$\Theta$ for candidate muon (left);
for second track (right) from GENIE default MC. Black error bars represent MC statistical uncertainties.}
\label{TsinMCS2}
\end{figure*}

\begin{figure*}[!hpt]
\centering
\subfloat{\includegraphics[width=0.5%
\linewidth]{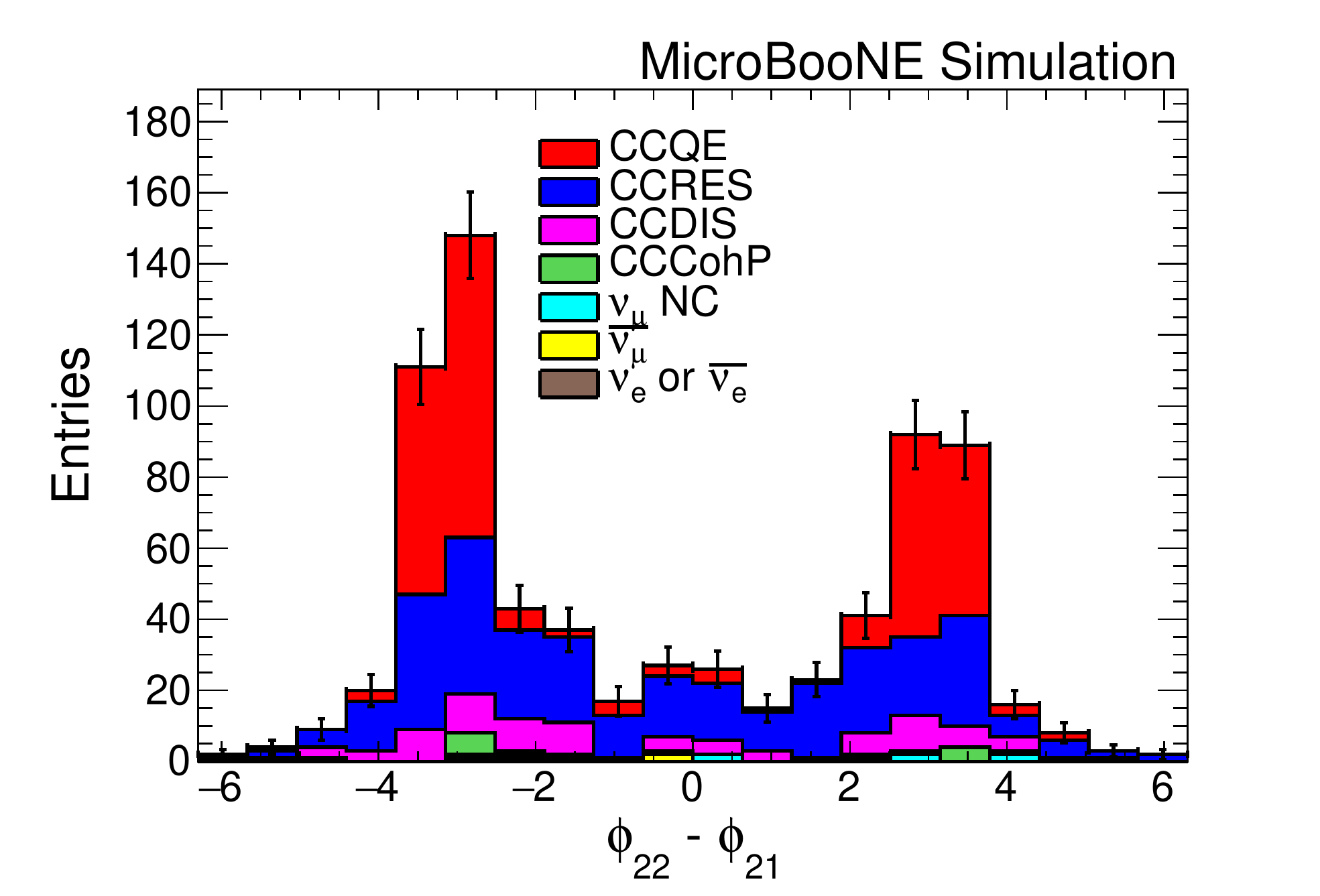}} \centering
\subfloat{\includegraphics[width=0.5%
\linewidth]{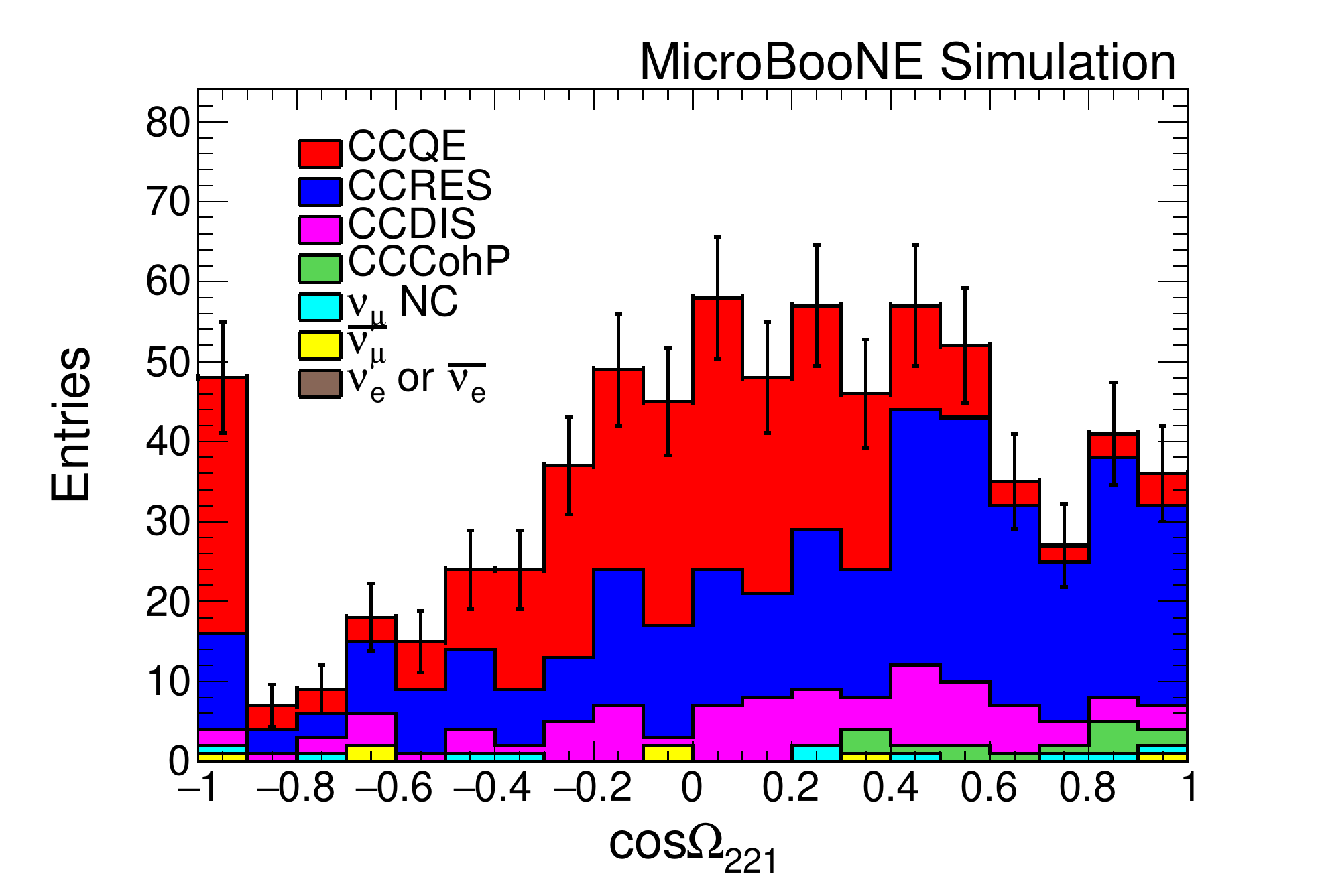}}
\caption{\textbf{Multiplicity = 2} $\protect\phi_{2} - \protect\phi_{1}$
distribution (left); Cosine of opening angle distribution (right) from GENIE default MC. Black error bars represent MC statistical uncertainties.}
\label{Tdphi2}
\end{figure*}

Figures~\textcolor{Blue}{\ref{TL2}}, \textcolor{Blue}{\ref{Tcostheta2}}, \textcolor{Blue}{\ref{TsinMCS2}}, and \textcolor{Blue}{\ref{Tdphi2}}
present the distributions of reconstructed ($L_{21}$ and $L_{22}$), ($\cos\theta_{21}$ and 
$\cos\theta_{22}$), ($\sin\Theta_{21}$ and $\sin\Theta_{22}$), and ($\phi
_{22}-\phi_{21}$ and $\cos\Omega_{221}$) respectively from the neutrino-enriched
sample, using the GENIE\ default model. \ There is again minimal difference
between QE, resonance, and DIS channels in track length, $\cos\theta$, or $%
\sin\Theta$ for the leading track. \ However, the QE channel produces
contributions to the distributions in $\cos\theta_{21}$ and $\cos\theta_{22}$
that are considerably less forward-peaked than the resonance channel
contributions. \ Distributions of these quantities in the data appear to be
consistent with this picture. We also note that the $\sin\Theta_{22}$
distribution receives a contribution from QE scattering peaked at small
values, consistent with expectations for a proton, and a broader
distribution more similar to that of the leading muon track candidate that
is consistent with the hypothesis that a charged pion can be reconstructed as
the second track in resonance contributions.

Striking differences between QE and RES contributions exist in the $%
\phi_{22}-\phi_{21}$ distribution between QE and resonance contributions in
the Fig.~\textcolor{Blue}{\ref{Tdphi2}}. \ The QE contributions demonstrate the clear $\phi
_{22}-\phi_{21}=\pm\pi$ peak expected for $2\rightarrow2$ scattering. \ The
gap between the $\pm\pi$ peaks is dominated by contributions from resonance
feed-down.

\begin{figure*}[!hpt]
\begin{adjustwidth}{-2cm}{-2cm}
\centering
\subfloat{\includegraphics[width=.35\textwidth]{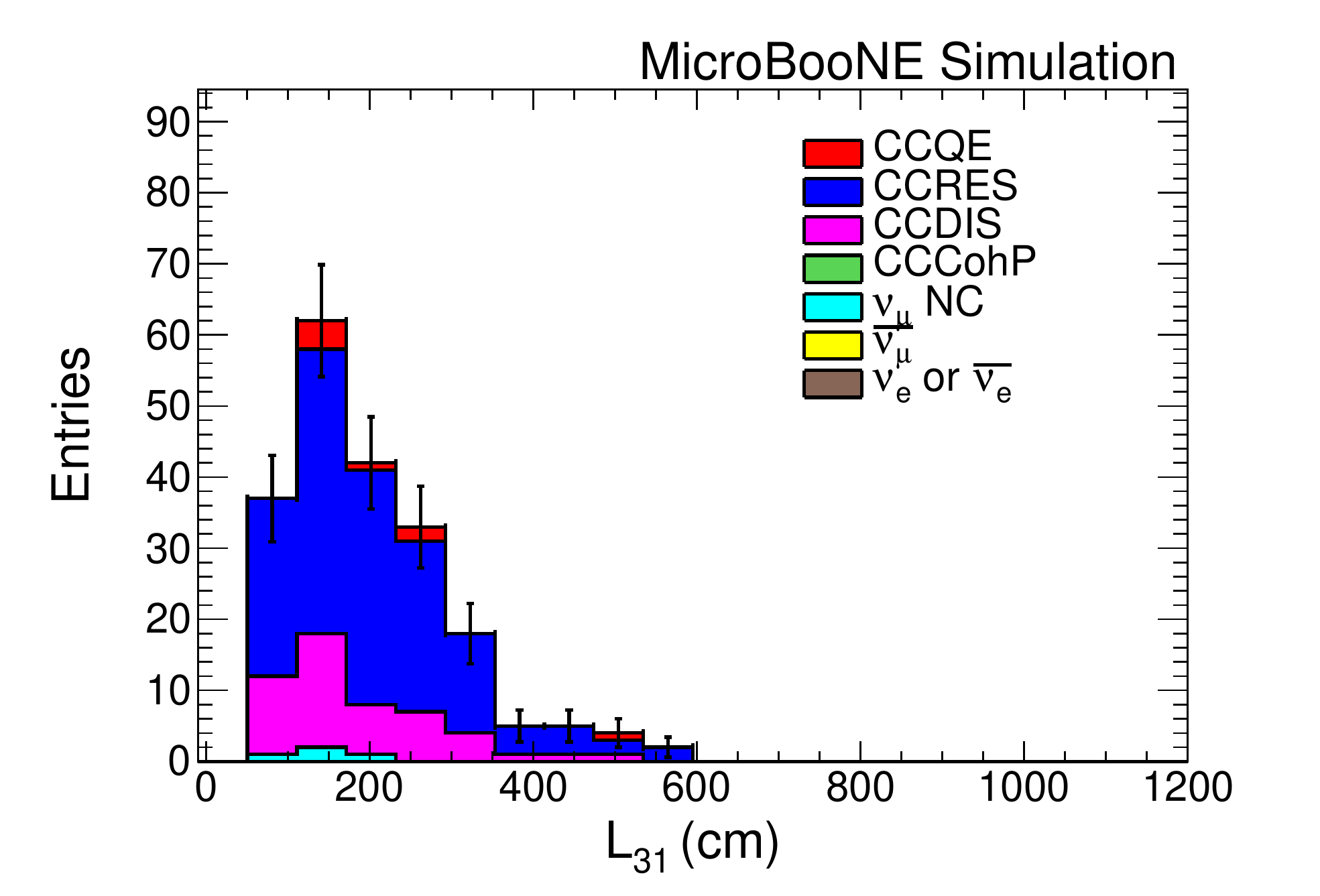}}
\subfloat{\includegraphics[width=.35\textwidth]{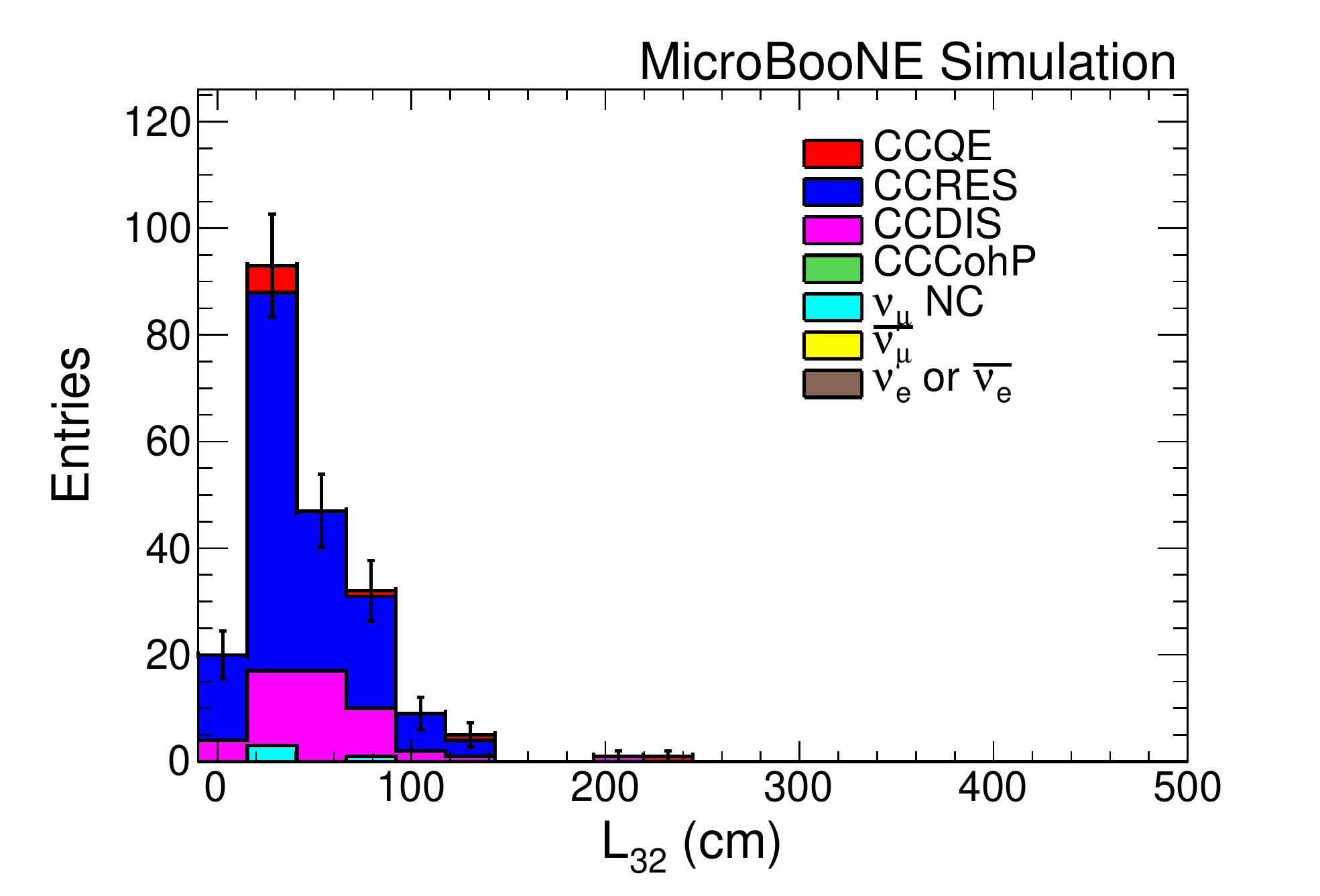}}
\subfloat{\includegraphics[width=.35\textwidth]{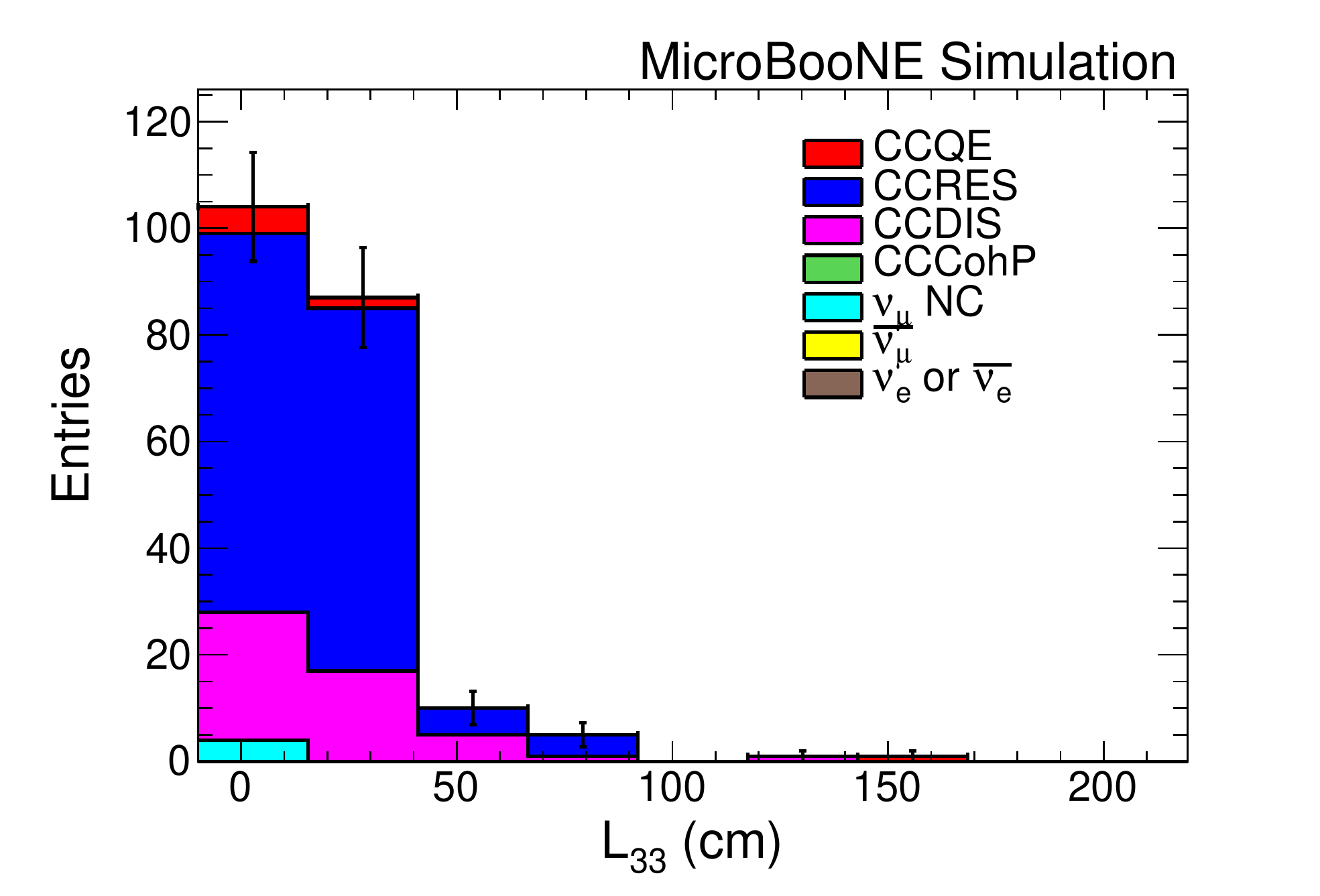}}
\caption{\textbf{Multiplicity = 3} Track length distribution for candidate muon (left); for second longest track (middle); for shortest track (right) from GENIE default MC. Black error bars represent MC statistical uncertainties.}
\label{TL3}
\end{adjustwidth}
\end{figure*}

\begin{figure*}[!hpt]
\begin{adjustwidth}{-2cm}{-2cm}
\centering
\subfloat{\includegraphics[width=.35\textwidth]{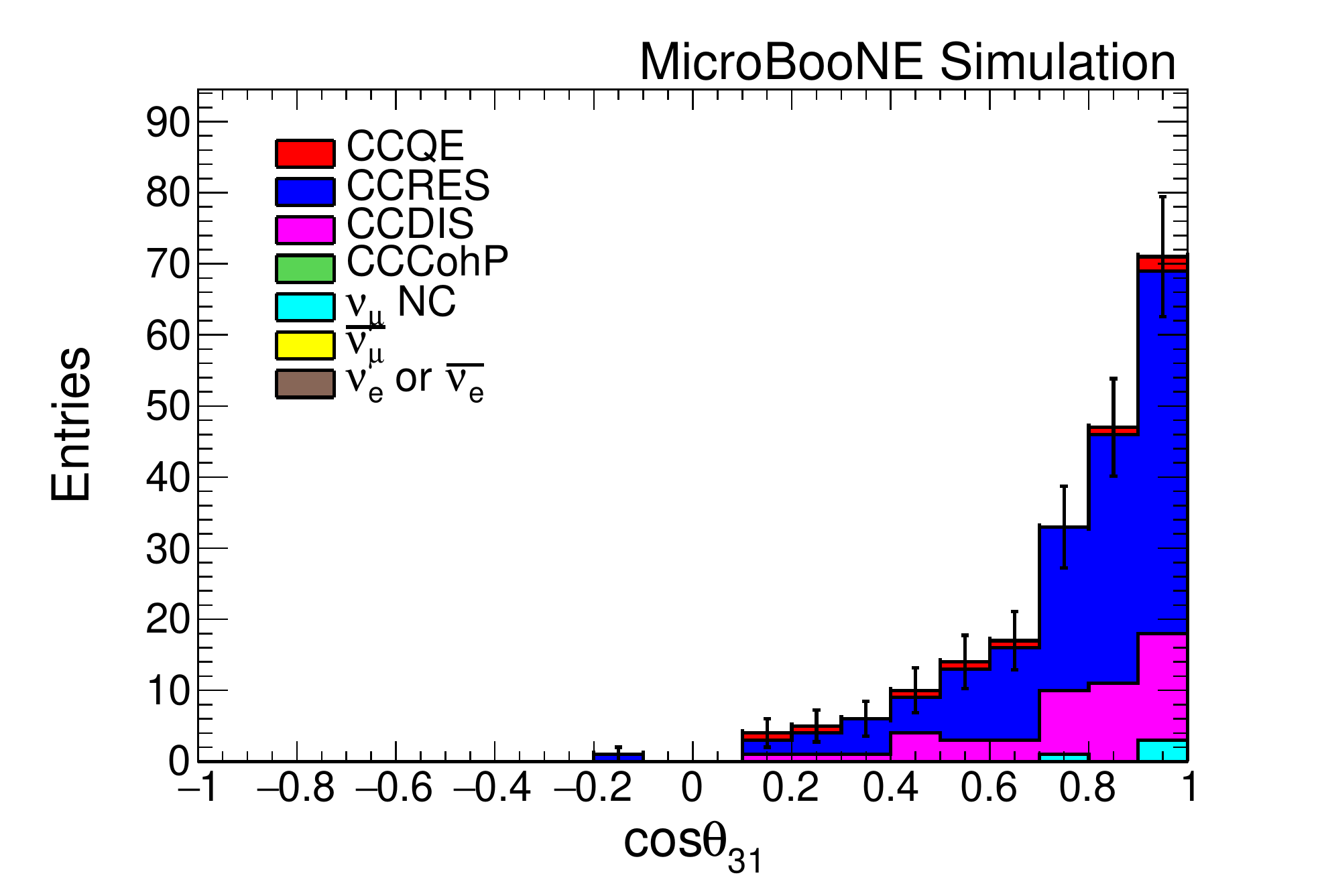}}
\subfloat{\includegraphics[width=.35\textwidth]{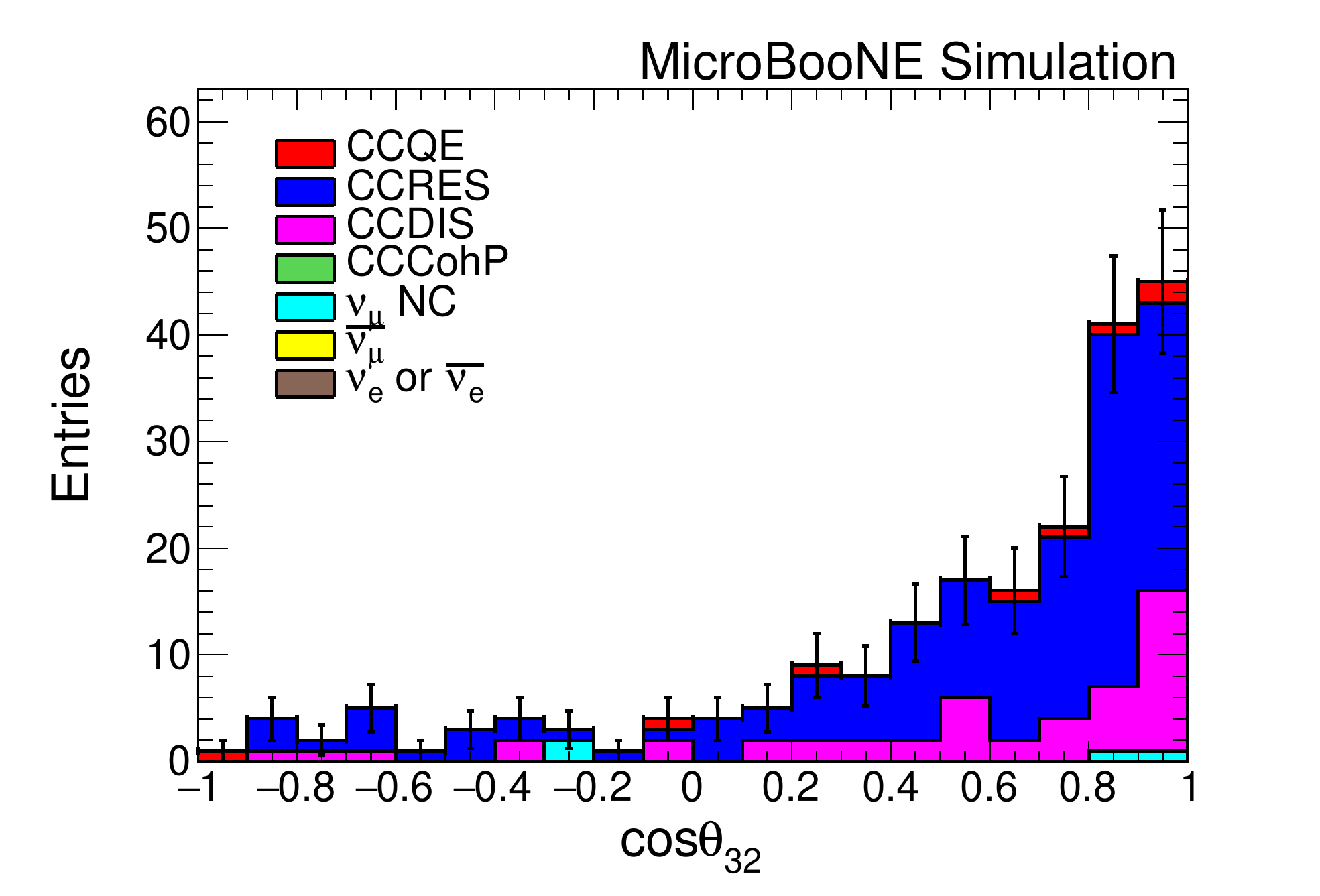}}
\subfloat{\includegraphics[width=.35\textwidth]{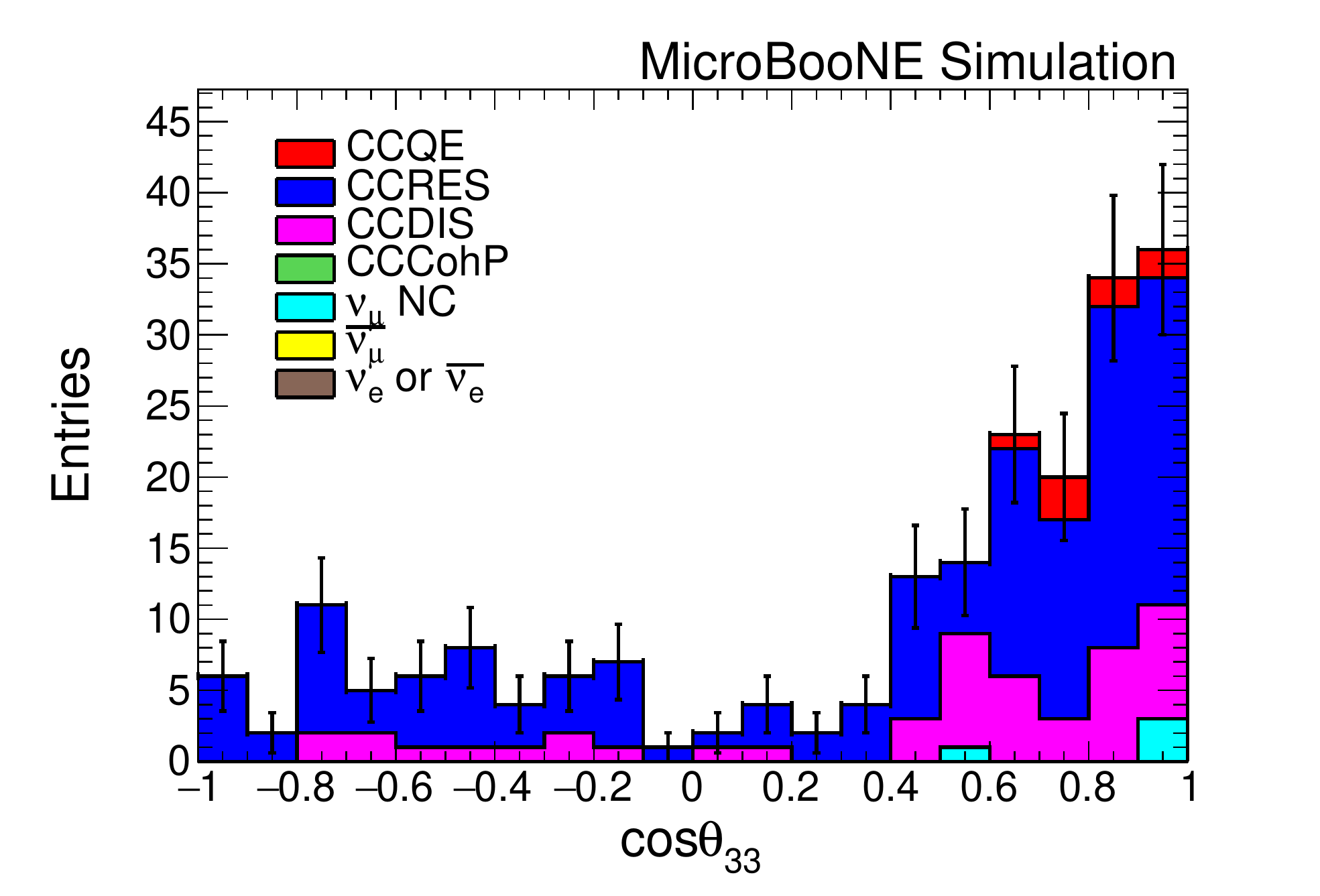}}
\caption{\textbf{Multiplicity = 3} Cosine of polar angle distribution for candidate muon (left); for second longest track (middle); for shortest track (right) from GENIE default MC. Black error bars represent MC statistical uncertainties.}
\label{Tcostheta3}
\end{adjustwidth}
\end{figure*}

\begin{figure*}[!hpt]
\begin{adjustwidth}{-2cm}{-2cm}
\centering
\subfloat{\includegraphics[width=.35\textwidth]{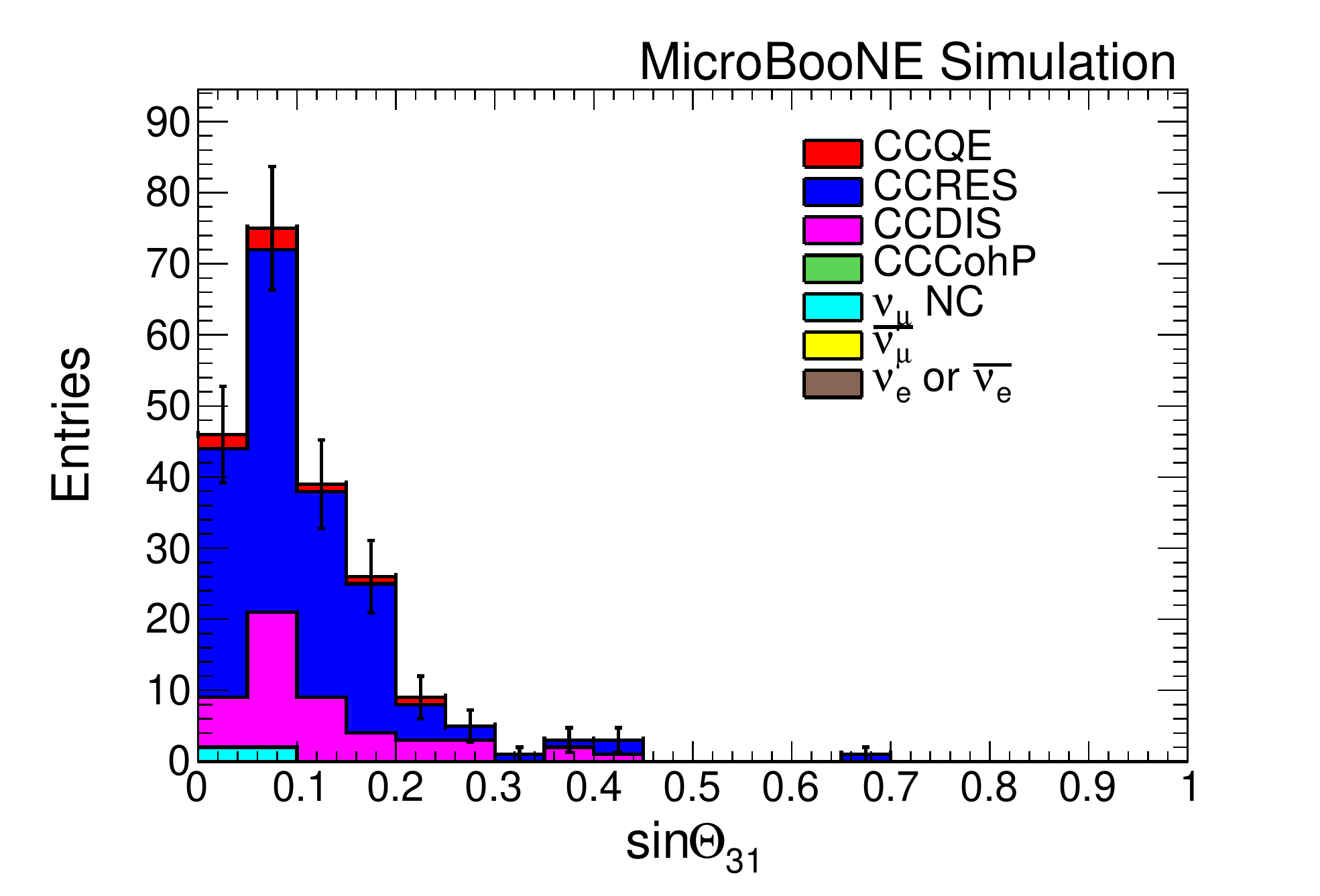}}
\subfloat{\includegraphics[width=.35\textwidth]{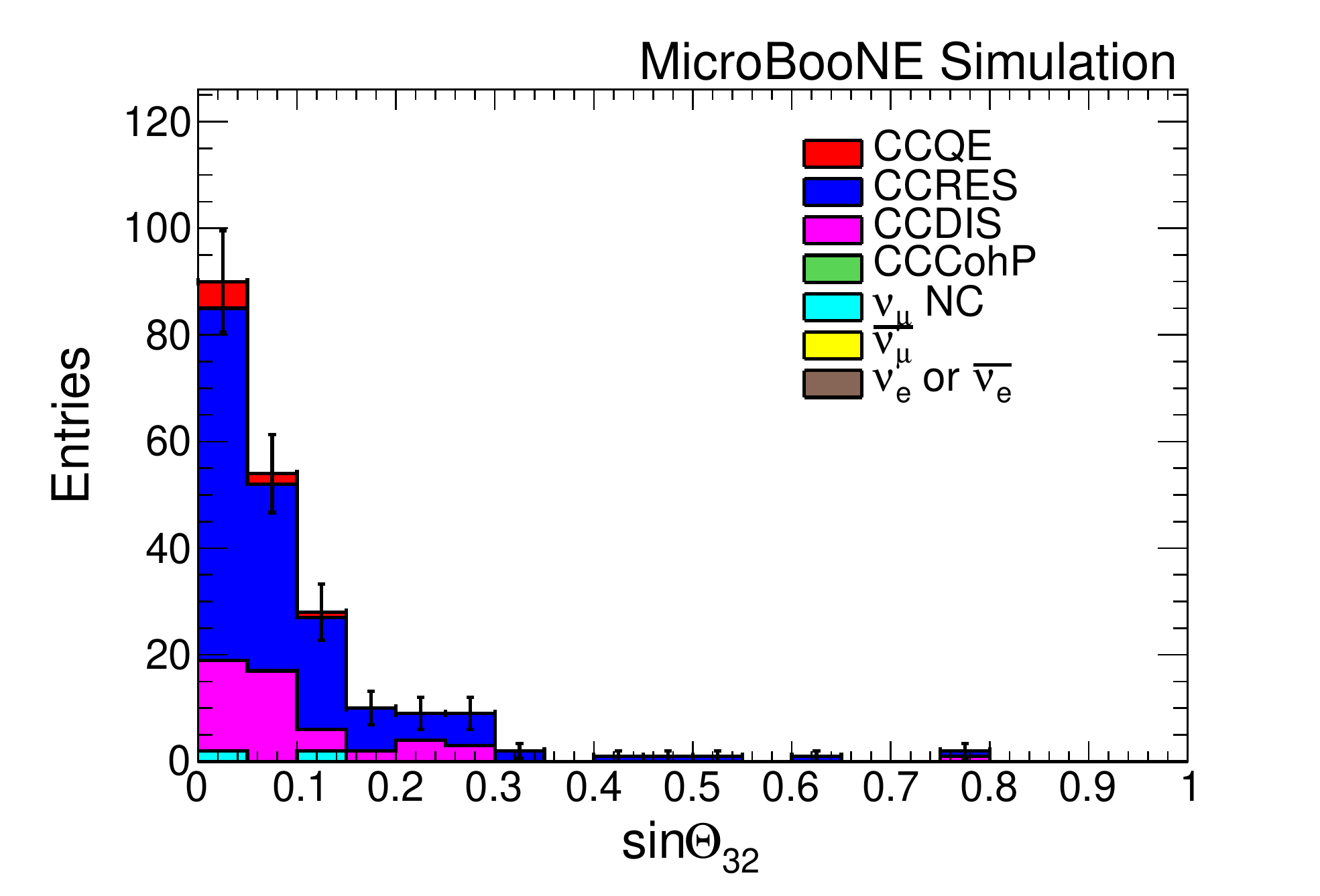}}
\subfloat{\includegraphics[width=.35\textwidth]{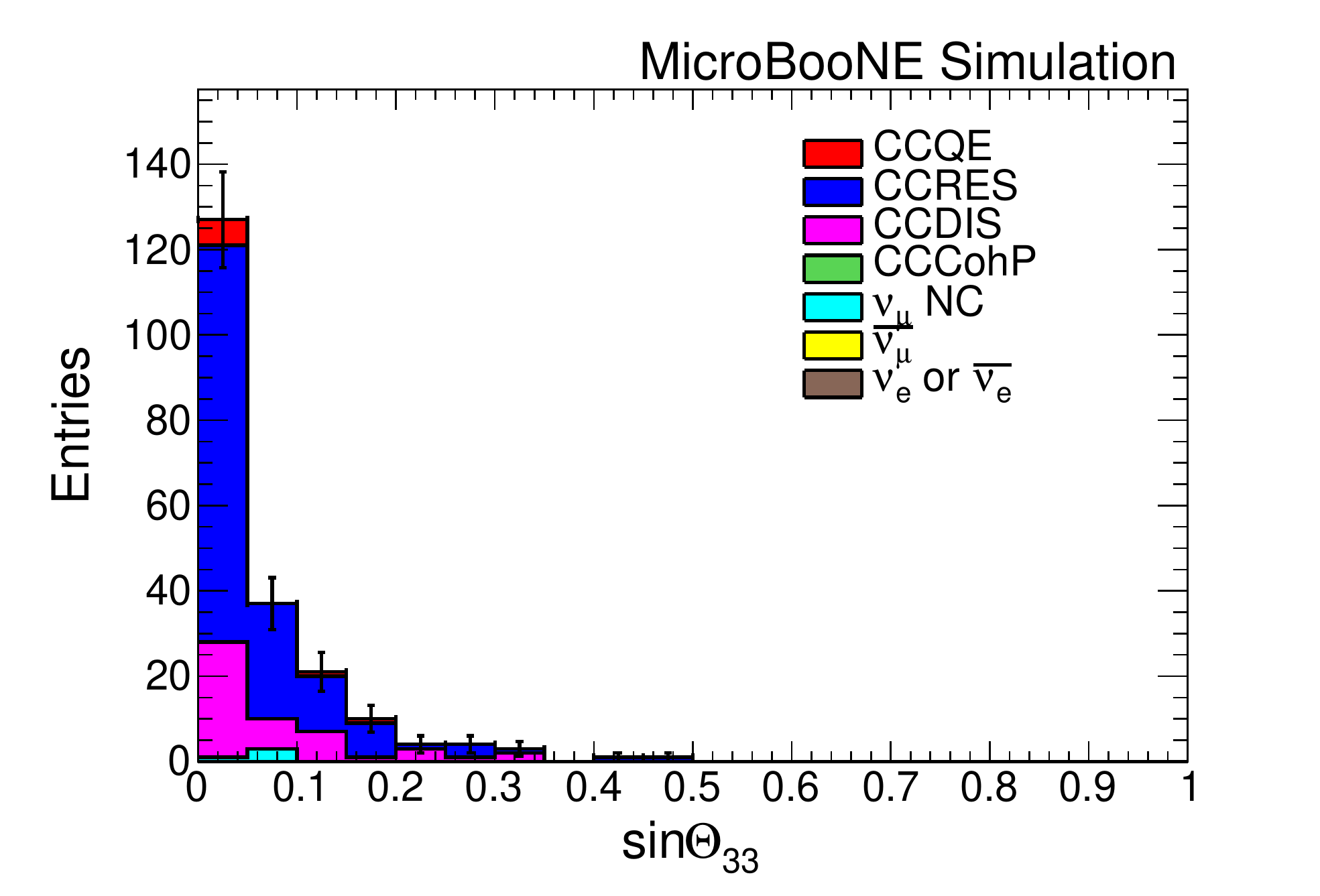}}
\caption{\textbf{Multiplicity = 3} Sin$\Theta$ distribution for candidate muon (left); for second longest track (middle); for shortest track (right) from GENIE default MC. Black error bars represent MC statistical uncertainties.}
\label{TsinMCS3}
\end{adjustwidth}
\end{figure*}

\begin{figure*}[!hpt]
\centering
\subfloat{\includegraphics[width=0.5%
\linewidth]{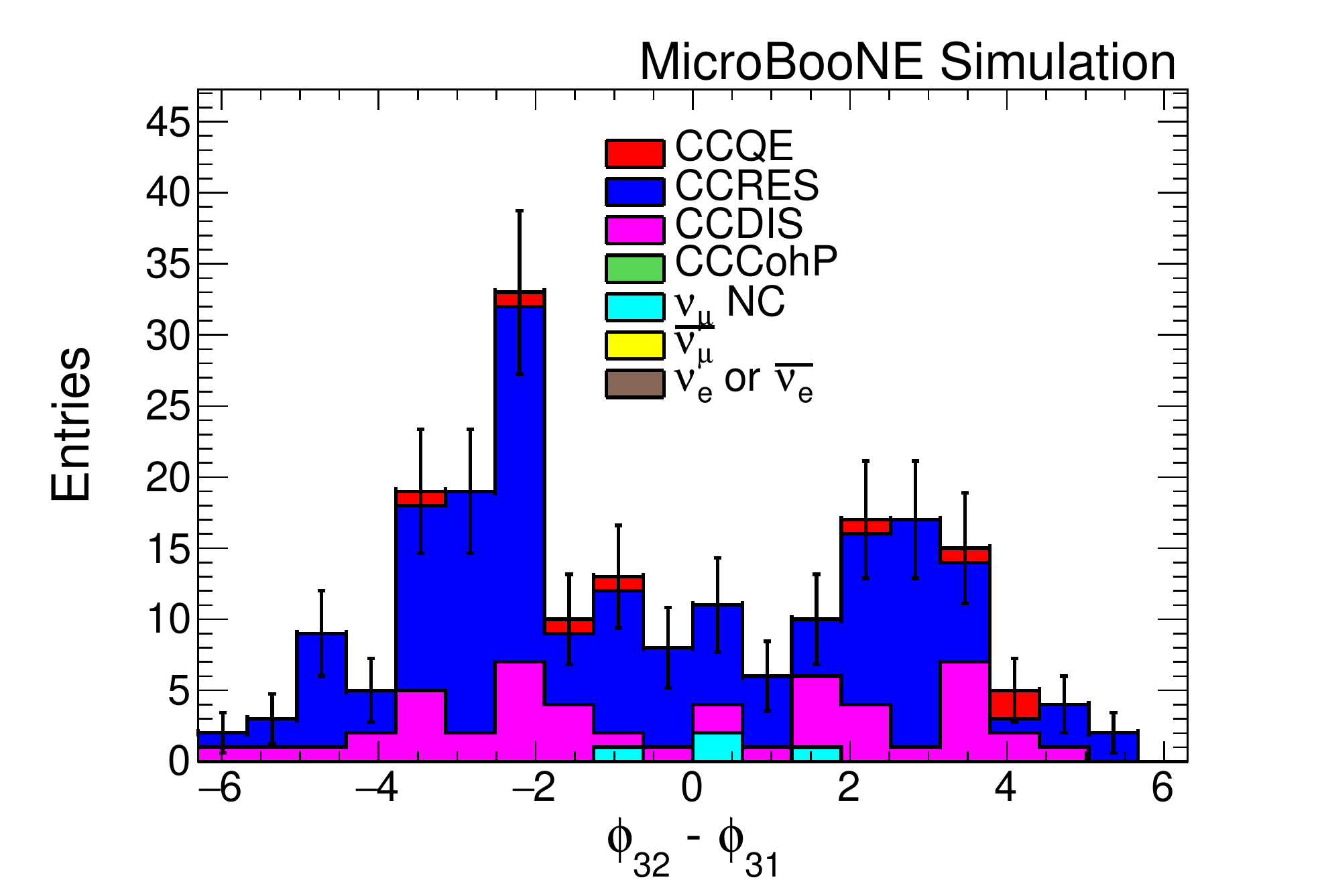}} \centering
\subfloat{\includegraphics[width=0.5%
\linewidth]{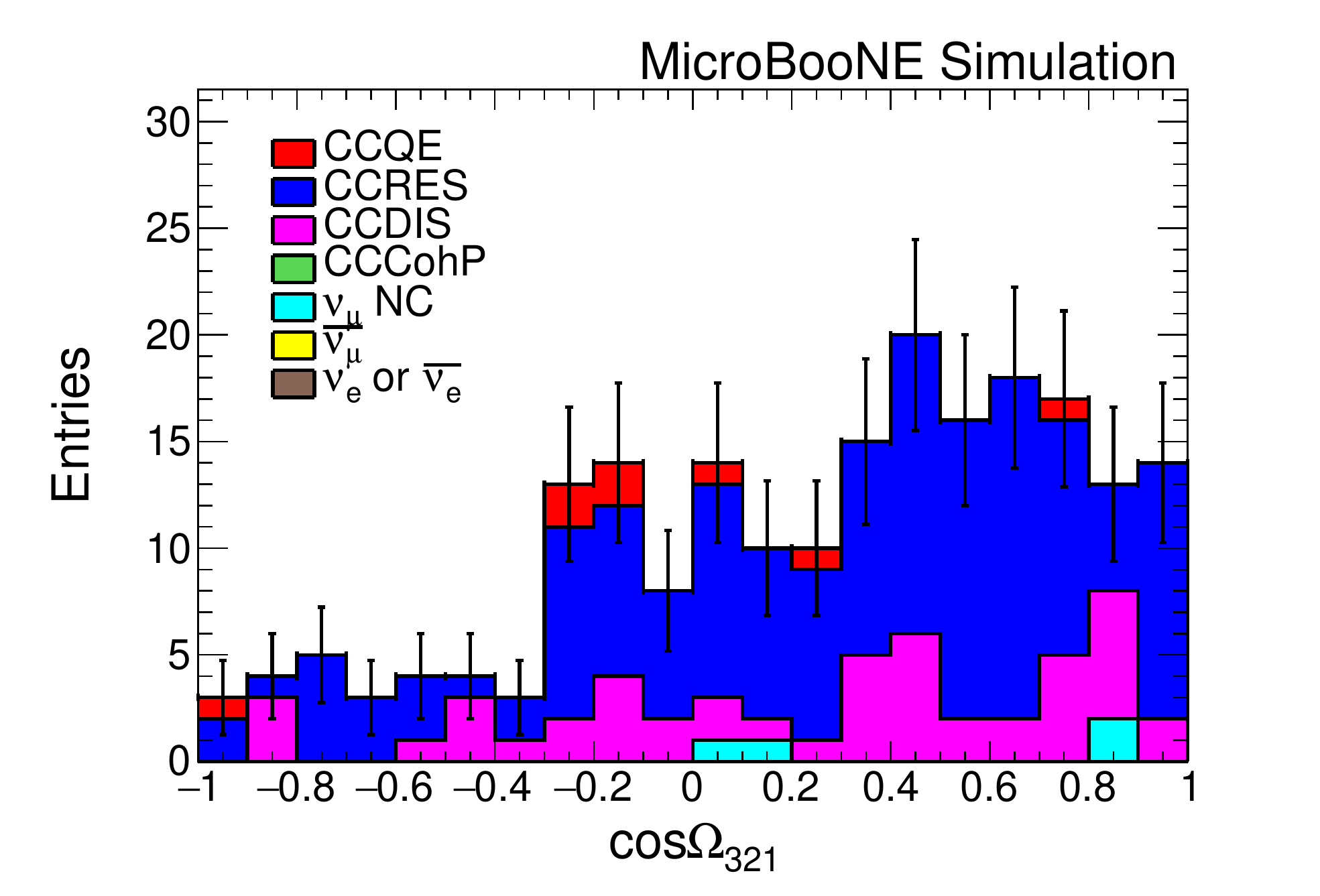}}
\caption{\textbf{Multiplicity = 3} $\protect\phi_{2} - \protect\phi_{1}$
distribution (left); Cosine of opening angle distribution between first and
second track (right) from GENIE default MC. Black error bars represent MC statistical uncertainties.}
\label{Tdphi31}
\end{figure*}

\begin{figure*}[!hpt]
\centering
\subfloat{\includegraphics[width=0.5%
\linewidth]{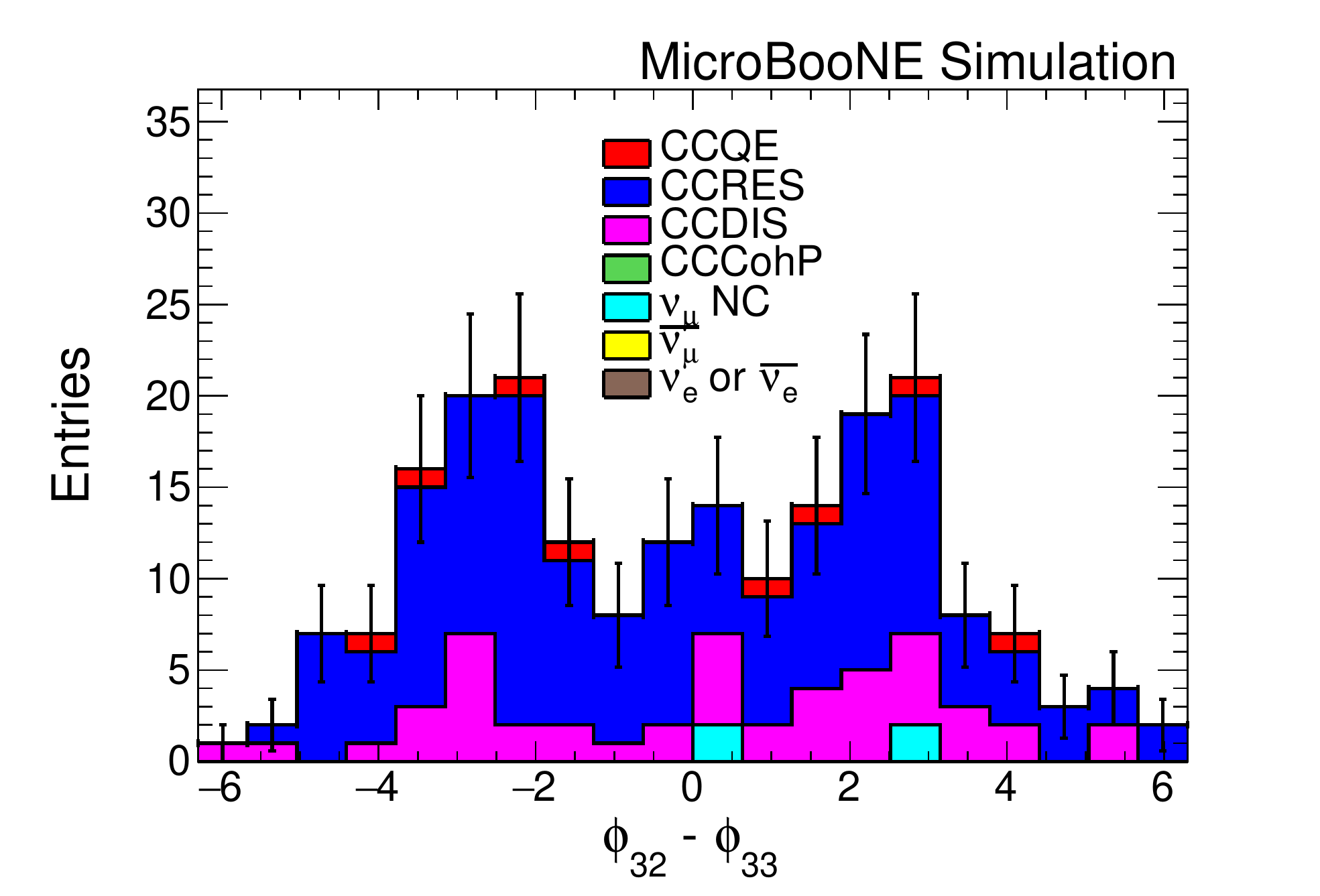}} \centering
\subfloat{\includegraphics[width=0.5%
\linewidth]{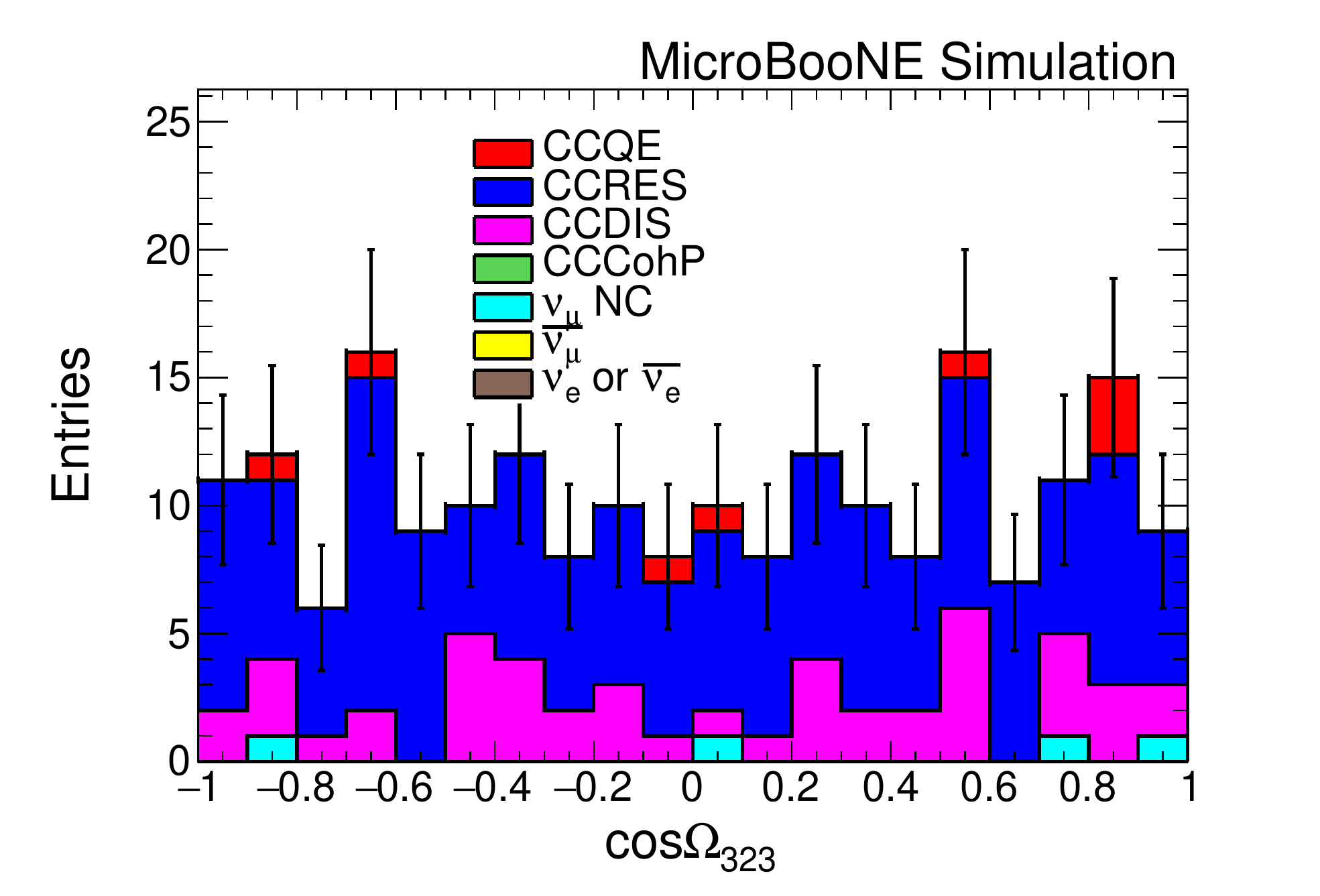}}
\caption{\textbf{Multiplicity = 3} $\protect\phi_{2} - \protect\phi_{3}$
distribution (left); Cosine of opening angle distribution between second and
third track (right) from GENIE default MC. Black error bars represent MC statistical uncertainties.}
\label{Tdphi33}
\end{figure*}

Figures~\textcolor{Blue}{\ref{TL3}}, \textcolor{Blue}{\ref{Tcostheta3}}, \textcolor{Blue}{\ref{TsinMCS3}}, \textcolor{Blue}{\ref{Tdphi31}}, \textcolor{Blue}{\ref%
{Tdphi32}}, and \textcolor{Blue}{\ref{Tdphi33}} present the reconstructed distributions of ($L_{31}$, $L_{32}$%
, and $L_{33}$), ($\cos\theta_{31}$, $\cos\theta_{32}$, and $\cos\theta_{33}$%
), ($\sin\Theta_{31}$, $\sin\Theta_{32}$, and $\sin \Theta_{33}$), ($%
\phi_{32}-\phi_{31}$ and $\cos\Omega_{321}$), ($\phi_{33}-\phi_{31}$ and $%
\cos\Omega_{331}$), ($\phi_{32}-\phi_{33}$ and $\cos \Omega_{323}$),
respectively from the neutrino-enriched sample, using the GENIE\ default model. \
The three-track sample in GENIE\ is dominated by resonance contributions,
and the data sample, although of limited statistics, has a CR\ background
consistent with zero. \ We can thus compare in detail GENIE\ predictions for
kinematic shape distributions to data. \ GENIE's predictions agree with
observations.

\begin{figure*}[!hpt]
\centering
\subfloat{\includegraphics[width=0.5%
\linewidth]{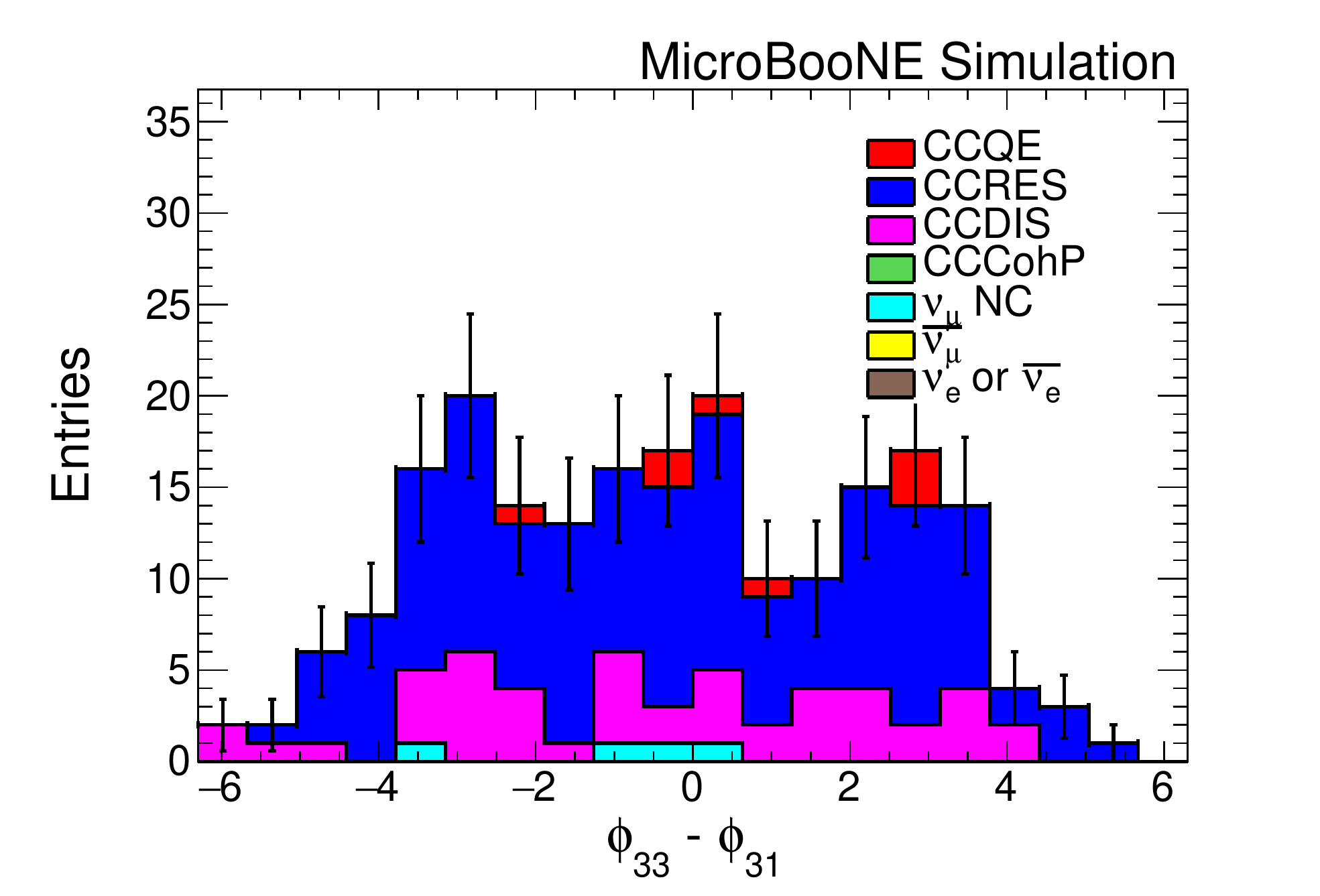}} \centering
\subfloat{\includegraphics[width=0.5%
\linewidth]{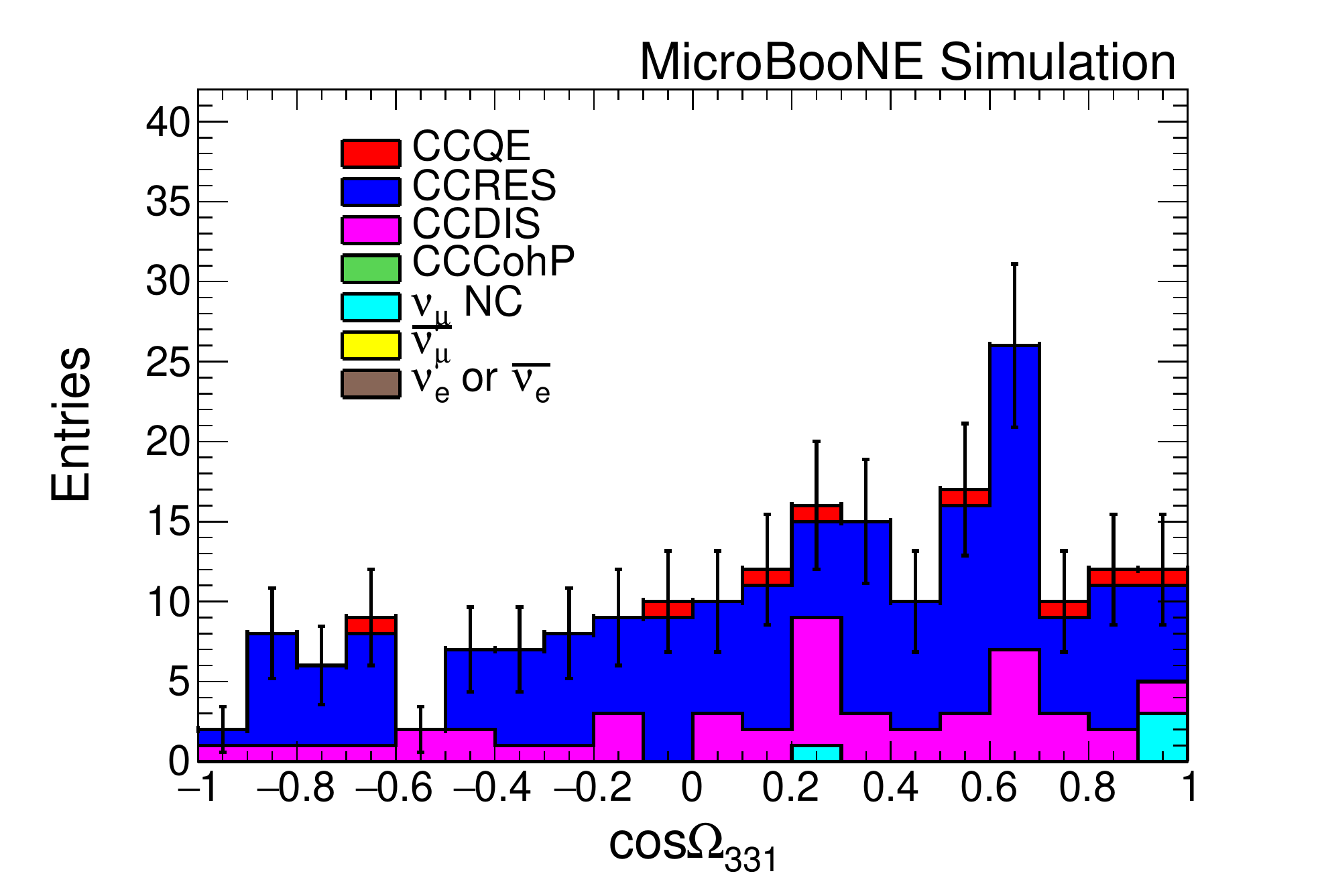}}
\caption{\textbf{Multiplicity = 3} $\protect\phi_{3} - \protect\phi_{1}$
distribution (left); Cosine of opening angle distribution between first and
third track (right) from GENIE default MC. Black error bars represent MC statistical uncertainties.}
\label{Tdphi32}
\end{figure*}

\section{\label{Conclusion}Summary}

We have completed an analysis that compares observed charged-particle multiplicities and observed kinematic distributions of charged
particles for fixed multiplicities in a restricted final state phase space
for neutrino scattering events in argon to
predictions from three GENIE\ tunes processed through the MicroBooNE
simulation and reconstruction chain. \ Our analysis takes into account
statistical uncertainties in a rigorous manner, and estimates the impact
of the largest expected systematic uncertainties. \ We observe that all
elements of the MicroBooNE\ measurement chain$-$detector performance, data
acquisition, event reconstruction, Monte Carlo event generator, detector
simulation, and flux modeling$-$perform well.

With particle-type-dependent kinetic energy thresholds of 31 MeV for $\pi ^\pm$ and $69$ MeV for protons, we find all three GENIE tunes consistently describe data in the shapes of
26 different kinematic distributions at fixed multiplicities. \ GENIE\
appears to over-predict the number of three-track events in data that would
be expected from resonant pion production, and to under-predict the number of
one-track events; however, we cannot rule out a higher than expected
tracking efficiency uncertainty as an alternative explanation for these
observations. \ Our study thus empirically supports the use of GENIE in
describing single-process (quasi-elastic, resonance) neutrino scattering on
argon, but not the predictions for the relative contributions of different
processes to the overall cross section. \ We find no significant differences
at this stage in the experiment between the default GENIE tune or
tunes that add MEC or TEM. \ Use of any of the three GENIE tunes for
future MicroBooNE analyses, or for physics studies of inclusive final states
performed for the SBN and DUNE experiments, receives empirical validation from this work.

As part of this analysis, we have developed a data-driven cosmic ray
background estimation method based on the energy loss profile and multiple
Coulomb scattering behavior of muons. \ Within the available Monte Carlo
statistics, we have shown that this method provides an unbiased estimate of
the number of neutrino events in a pre-filtered sample, and, given current
statistical precision, it is independent of the signal-to-background level, final state charged particle multiplicity, and other kinematic properties of
the final state particles. \ This method can be applied to a broad range
of charged current process measurements.

Significant improvements to MicroBooNE neutrino interaction property
measurements are anticipated in the future through incorporation of nearly
an order-of-magnitude more statistics, more fully developed reconstruction
tools$-$including momentum reconstruction, particle identification, and
lower kinetic energy thresholds for tracking$-$and the availability of a
recently installed external cosmic-ray tagger to the detector. \\

This material is based upon work supported by the following: the U.S. Department of Energy,
Office of Science, Offices of High Energy Physics and Nuclear Physics; the U.S. National Science
Foundation; the Swiss National Science Foundation; the Science and Technology Facilities Council
of the United Kingdom; and The Royal Society (United Kingdom). Additional support for the
laser calibration system and cosmic ray tagger was provided by the Albert Einstein Center for
Fundamental Physics. Fermilab is operated by Fermi Research Alliance, LLC under Contract No.
DE-AC02-07CH11359 with the United States Department of Energy.   

\end{document}